\pdfoutput=1
\documentclass[a4paper, 12pt, twoside, makeidx]{report}
\usepackage[english, brazil, french]{babel}
\usepackage[latin1]{inputenc}
\usepackage{graphicx}
\usepackage{epsfig}
\usepackage{amsmath}
\usepackage{amssymb}
\usepackage{amsthm}
\usepackage{dsfont}
\usepackage{booktabs}
\usepackage{stmaryrd}
\usepackage{longtable}
\usepackage{setspace}
\usepackage{slashed}
\usepackage{axodraw}
\usepackage{relsize}
\usepackage{pdfpages}
\usepackage{bm}  
\usepackage{epstopdf} 
\usepackage{mathtools}
\usepackage{multirow}
\usepackage{subfig}
\usepackage{cite}
\usepackage{makeidx}
\usepackage[overload]{textcase}

\makeindex
\usepackage{color}

\definecolor{greenyellow}   {cmyk}{0.15, 0   , 0.69, 0   }
\definecolor{yellow}        {cmyk}{0   , 0   , 1   , 0   }
\definecolor{goldenrod}     {cmyk}{0   , 0.10, 0.84, 0   }
\definecolor{dandelion}     {cmyk}{0   , 0.29, 0.84, 0   }
\definecolor{apricot}       {cmyk}{0   , 0.32, 0.52, 0   }
\definecolor{peach}         {cmyk}{0   , 0.50, 0.70, 0   }
\definecolor{melon}         {cmyk}{0   , 0.46, 0.50, 0   }
\definecolor{yelloworange}  {cmyk}{0   , 0.42, 1   , 0   }
\definecolor{orange}        {cmyk}{0   , 0.61, 0.87, 0   }
\definecolor{burntorange}   {cmyk}{0   , 0.51, 1   , 0   }
\definecolor{bittersweet}   {cmyk}{0   , 0.75, 1   , 0.24}
\definecolor{redorange}     {cmyk}{0   , 0.77, 0.87, 0   }
\definecolor{mahogany}      {cmyk}{0   , 0.85, 0.87, 0.35}
\definecolor{maroon}        {cmyk}{0   , 0.87, 0.68, 0.32}
\definecolor{brickred}      {cmyk}{0   , 0.89, 0.94, 0.28}
\definecolor{red}           {cmyk}{0   , 1   , 1   , 0   }
\definecolor{orangered}     {cmyk}{0   , 1   , 0.50, 0   }
\definecolor{rubinered}     {cmyk}{0   , 1   , 0.13, 0   }
\definecolor{wildstrawberry}{cmyk}{0   , 0.96, 0.39, 0   }
\definecolor{salmon}        {cmyk}{0   , 0.53, 0.38, 0   }
\definecolor{carnationpink} {cmyk}{0   , 0.63, 0   , 0   }
\definecolor{magenta}       {cmyk}{0   , 1   , 0   , 0   }
\definecolor{violetred}     {cmyk}{0   , 0.81, 0   , 0   }
\definecolor{rhodamine}     {cmyk}{0   , 0.82, 0   , 0   }
\definecolor{mulberry}      {cmyk}{0.34, 0.90, 0   , 0.02}
\definecolor{redviolet}     {cmyk}{0.07, 0.90, 0   , 0.34}
\definecolor{fuchsia}       {cmyk}{0.47, 0.91, 0   , 0.08}
\definecolor{lavender}      {cmyk}{0   , 0.48, 0   , 0   }
\definecolor{thistle}       {cmyk}{0.12, 0.59, 0   , 0   }
\definecolor{orchid}        {cmyk}{0.32, 0.64, 0   , 0   }
\definecolor{darkorchid}    {cmyk}{0.40, 0.80, 0.20, 0   }
\definecolor{purple}        {cmyk}{0.45, 0.86, 0   , 0   }
\definecolor{plum}          {cmyk}{0.50, 1   , 0   , 0   }
\definecolor{violet}        {cmyk}{0.79, 0.88, 0   , 0   }
\definecolor{royalpurple}   {cmyk}{0.75, 0.90, 0   , 0   }
\definecolor{blueviolet}    {cmyk}{0.86, 0.91, 0   , 0.04}
\definecolor{periwinkle}    {cmyk}{0.57, 0.55, 0   , 0   }
\definecolor{cadetblue}     {cmyk}{0.62, 0.57, 0.23, 0   }
\definecolor{cornflowerblue}{cmyk}{0.65, 0.13, 0   , 0   }
\definecolor{midnightblue}  {cmyk}{0.98, 0.13, 0   , 0.43}
\definecolor{navyblue}      {cmyk}{0.94, 0.54, 0   , 0   }
\definecolor{royalblue}     {cmyk}{1   , 0.50, 0   , 0   }
\definecolor{blue}          {cmyk}{1   , 1   , 0   , 0   }
\definecolor{cerulean}      {cmyk}{0.94, 0.11, 0   , 0   }
\definecolor{cyan}          {cmyk}{1   , 0   , 0   , 0   }
\definecolor{processblue}   {cmyk}{0.96, 0   , 0   , 0   }
\definecolor{skyblue}       {cmyk}{0.62, 0   , 0.12, 0   }
\definecolor{turquoise}     {cmyk}{0.85, 0   , 0.20, 0   }
\definecolor{tealblue}      {cmyk}{0.86, 0   , 0.34, 0.02}
\definecolor{aquamarine}    {cmyk}{0.82, 0   , 0.30, 0   }
\definecolor{bluegreen}     {cmyk}{0.85, 0   , 0.33, 0   }
\definecolor{emerald}       {cmyk}{1   , 0   , 0.50, 0   }
\definecolor{junglegreen}   {cmyk}{0.99, 0   , 0.52, 0   }
\definecolor{seagreen}      {cmyk}{0.69, 0   , 0.50, 0   }
\definecolor{green}         {cmyk}{1   , 0   , 1   , 0   }
\definecolor{forestgreen}   {cmyk}{0.91, 0   , 0.88, 0.12}
\definecolor{pinegreen}     {cmyk}{0.92, 0   , 0.59, 0.25}
\definecolor{limegreen}     {cmyk}{0.50, 0   , 1   , 0   }
\definecolor{yellowgreen}   {cmyk}{0.44, 0   , 0.74, 0   }
\definecolor{springgreen}   {cmyk}{0.26, 0   , 0.76, 0   }
\definecolor{olivegreen}    {cmyk}{0.64, 0   , 0.95, 0.40}
\definecolor{rawsienna}     {cmyk}{0   , 0.72, 1   , 0.45}
\definecolor{sepia}         {cmyk}{0   , 0.83, 1   , 0.70}
\definecolor{brown}         {cmyk}{0   , 0.81, 1   , 0.60}
\definecolor{tan}           {cmyk}{0.14, 0.42, 0.56, 0   }
\definecolor{gray}          {cmyk}{0   , 0   , 0   , 0.50}
\definecolor{black}         {cmyk}{0   , 0   , 0   , 1   }
\definecolor{white}         {cmyk}{0   , 0   , 0   , 0   }

\newcommand{\rag}{\right\rangle}
\newcommand{\lag}{\left\langle}
\newcommand{\nno}{\nonumber}

\newcommand{\qq[1]}{\langle \bar{#1} #1 \rangle}
\newcommand{\GG}{\langle g_s^2 G^2 \rangle}
\newcommand{\qGq[1]}{\langle \bar{#1} G #1 \rangle}

\newcommand{\qqqq[1]}{\langle \bar{#1} #1 \bar{#1} #1 \rangle}
\newcommand{\GGG}{\langle g_s^3 G^3 \rangle}
\newcommand{\DsxDsx}{D^\ast_s \bar{D}^\ast_s}
\newcommand{\DxDx}{D^\ast \bar{D}^\ast}
\newcommand{\DsxDso}{D^\ast_s \bar{D}^\ast_{s0}}
\newcommand{\DxDo}{D^\ast \bar{D}^\ast_{0}}
\newcommand{\DsoDso}{D^\ast_{s0} \bar{D}^\ast_{s0}}
\newcommand{\DoDo}{D^\ast_0 \bar{D}^\ast_{0}}
\newcommand{\BsxBso}{B^\ast_s \bar{B}^\ast_{s0}}
\newcommand{\BxBo}{B^\ast \bar{B}^\ast_{0}}
\newcommand{\BsoBso}{B^\ast_{s0} \bar{B}^\ast_{s0}}
\newcommand{\BoBo}{B^\ast_0 \bar{B}^\ast_{0}}
\newcommand{\DB}{D B}

\newcommand{\GGi}{\langle G^2 \rangle}
\newcommand{\GGGi}{\langle G^3 \rangle}

\makeatletter 
\def\cleardoublepage{\clearpage\if@twoside \ifodd\c@page\else%
    \hbox{}%
    \thispagestyle{empty}
    \newpage%
    \if@twocolumn\hbox{}\newpage\fi\fi\fi} 
\makeatother

\def\Img{\mbox{Im}~}
\def\Log{\mbox{Log}}
\def\Tr{\mbox{Tr}}
\def\Fbc{{\tilde{\cal F}}_{(\al,\be)}}
\def\Hbc{{\tilde{\cal H}}_{(\al)}}
\def\GeV{\nobreak\,\mbox{GeV}}
\def\MeV{\nobreak\,\mbox{MeV}}

\def\la{\lambda}

\def\ga{\gamma}

\def\de{\delta}
\def\al{\alpha}
\def\be{\beta}
\def\almax{\alpha_{max}}
\def\almin{\alpha_{min}}
\def\bemax{1 - \al}
\def\bemin{\beta_{min}}

\doublespace

\begin{document}

\thispagestyle{empty}

\begin{figure}[t]
  \vspace{-2.0cm}
  \centerline{
  \includegraphics[width=1.35\textwidth, height=1.25\textheight]{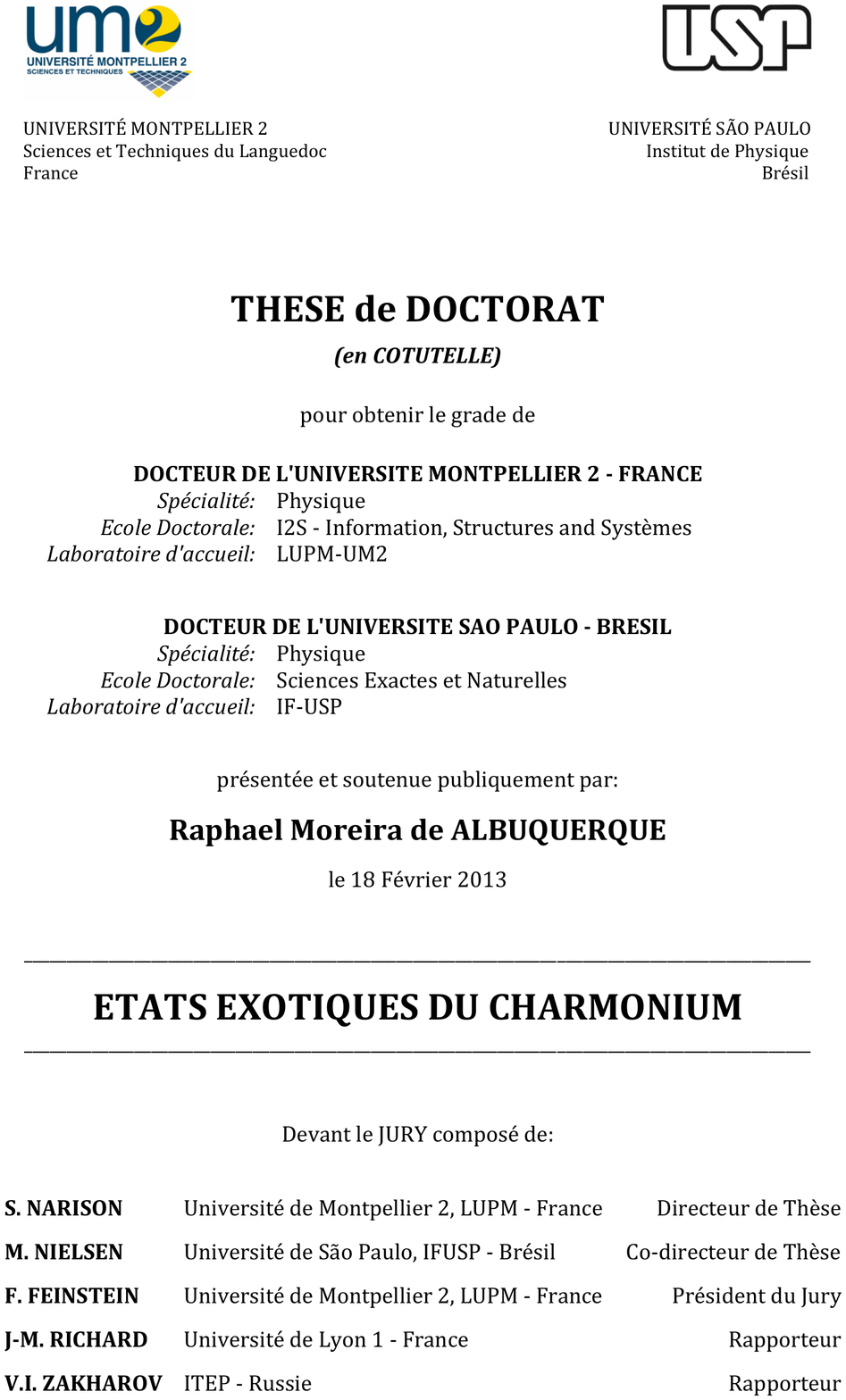}}
\end{figure}
\cleardoublepage

\thispagestyle{empty}

\begin{figure}[t]
  \vspace{-2.8cm}
  \centerline{
  \includegraphics[width=1.25\textwidth, height=1.35\textheight]{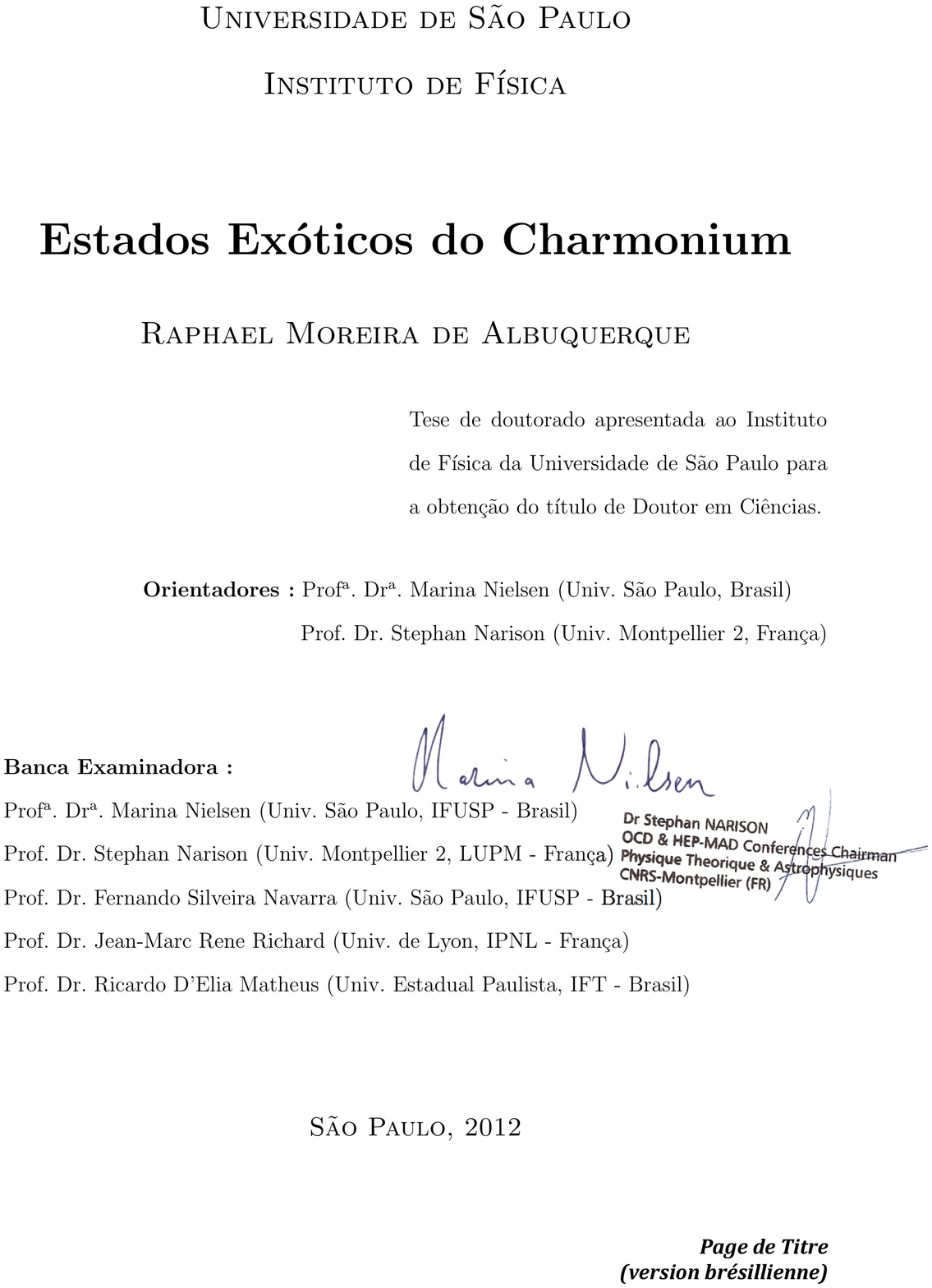}}
\end{figure}
\clearpage

\thispagestyle{empty}

\begin{figure}[t]
  \vspace{-2.8cm}
  \centerline{
  \includegraphics[width=1.3\textwidth, height=1.3\textheight]{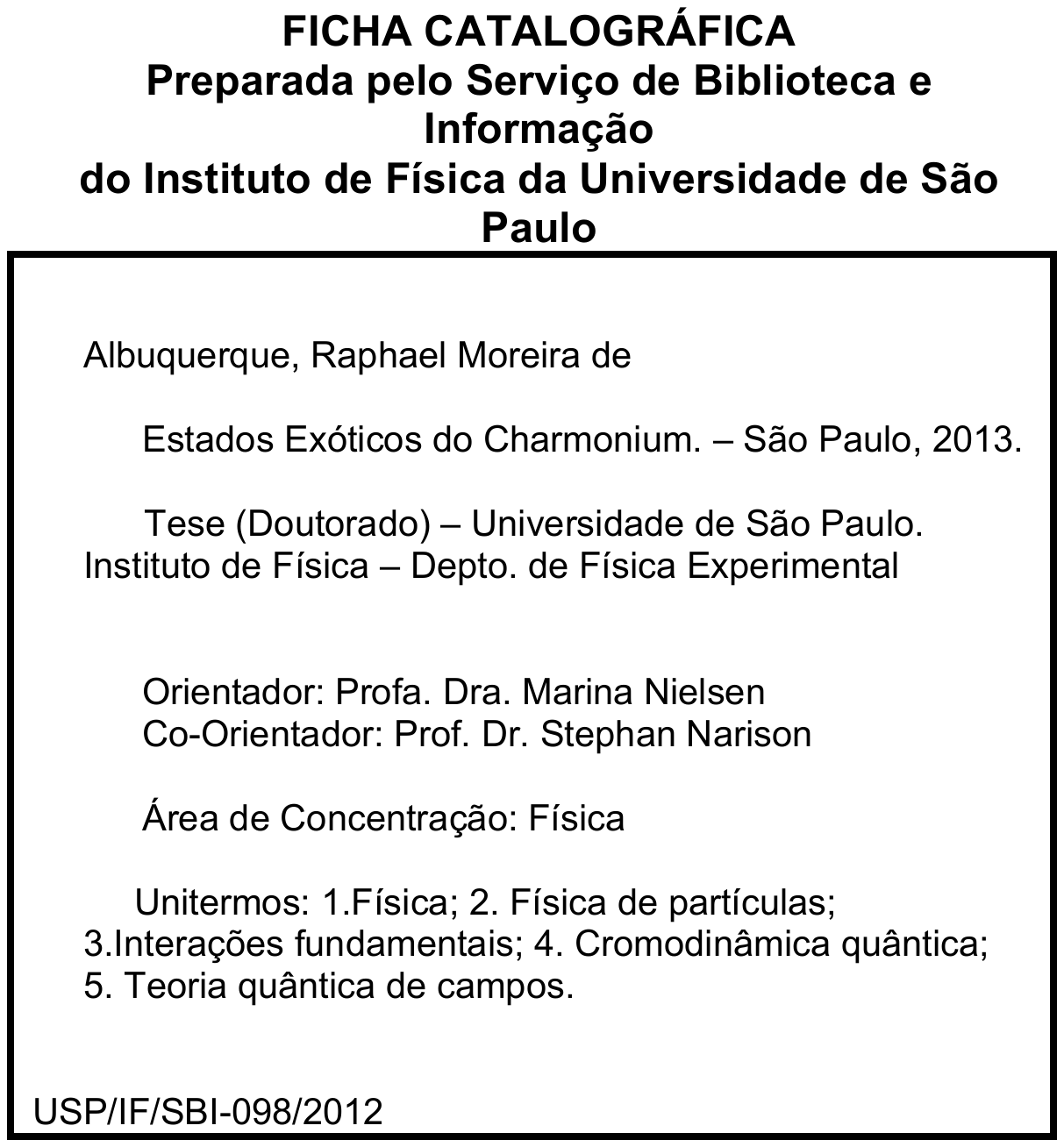}}
\end{figure}
\cleardoublepage

\thispagestyle{empty}
\vspace*{\fill}
\cleardoublepage

\thispagestyle{empty}
\vspace*{\fill}
\begin{center}
  {\bf Laboratoires d'accueil:}

  \vspace{1cm}
  \centerline{\epsfig{figure=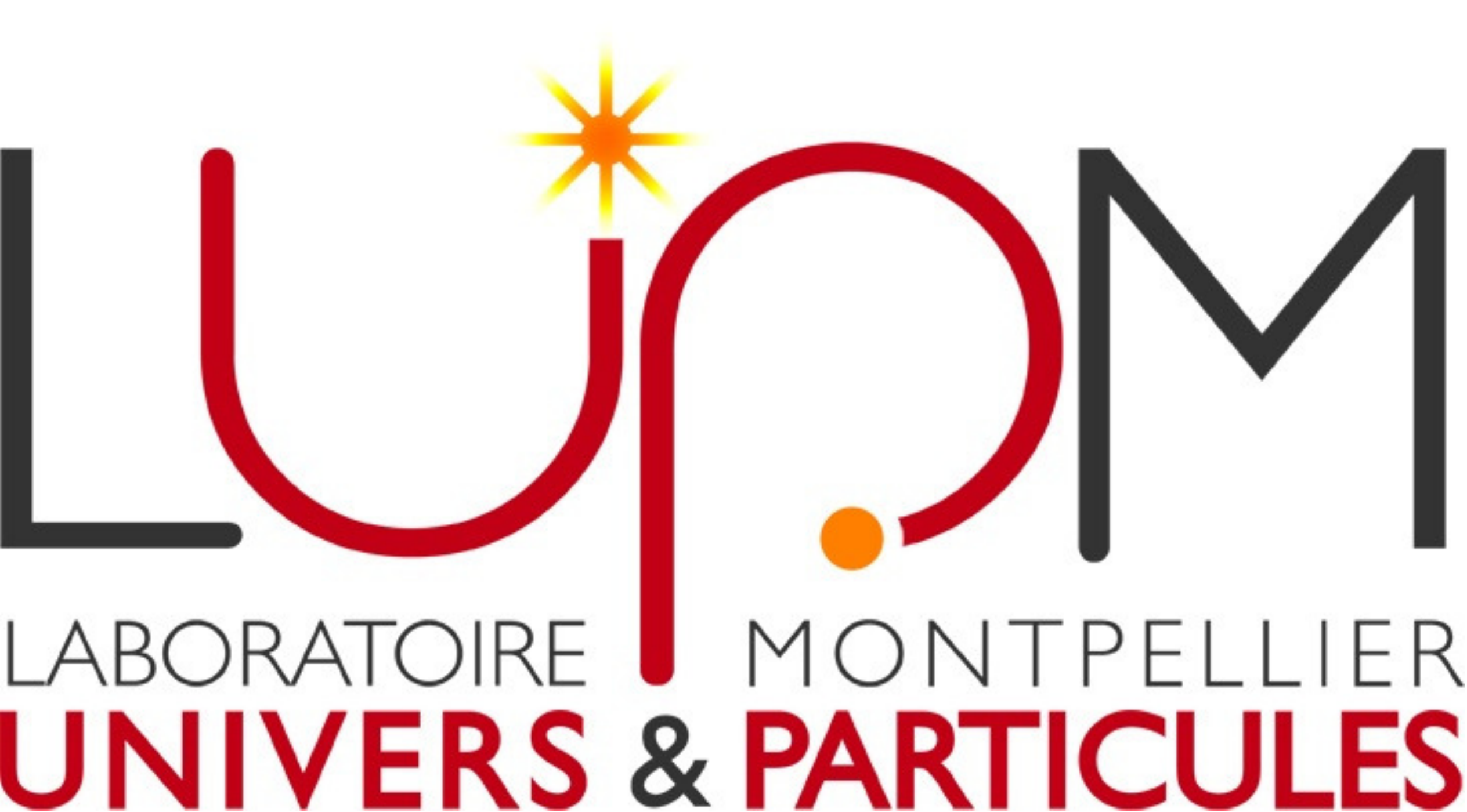,height=3.0cm}}
  \vspace{1cm}
  \centerline{\includegraphics[height=5.0cm]{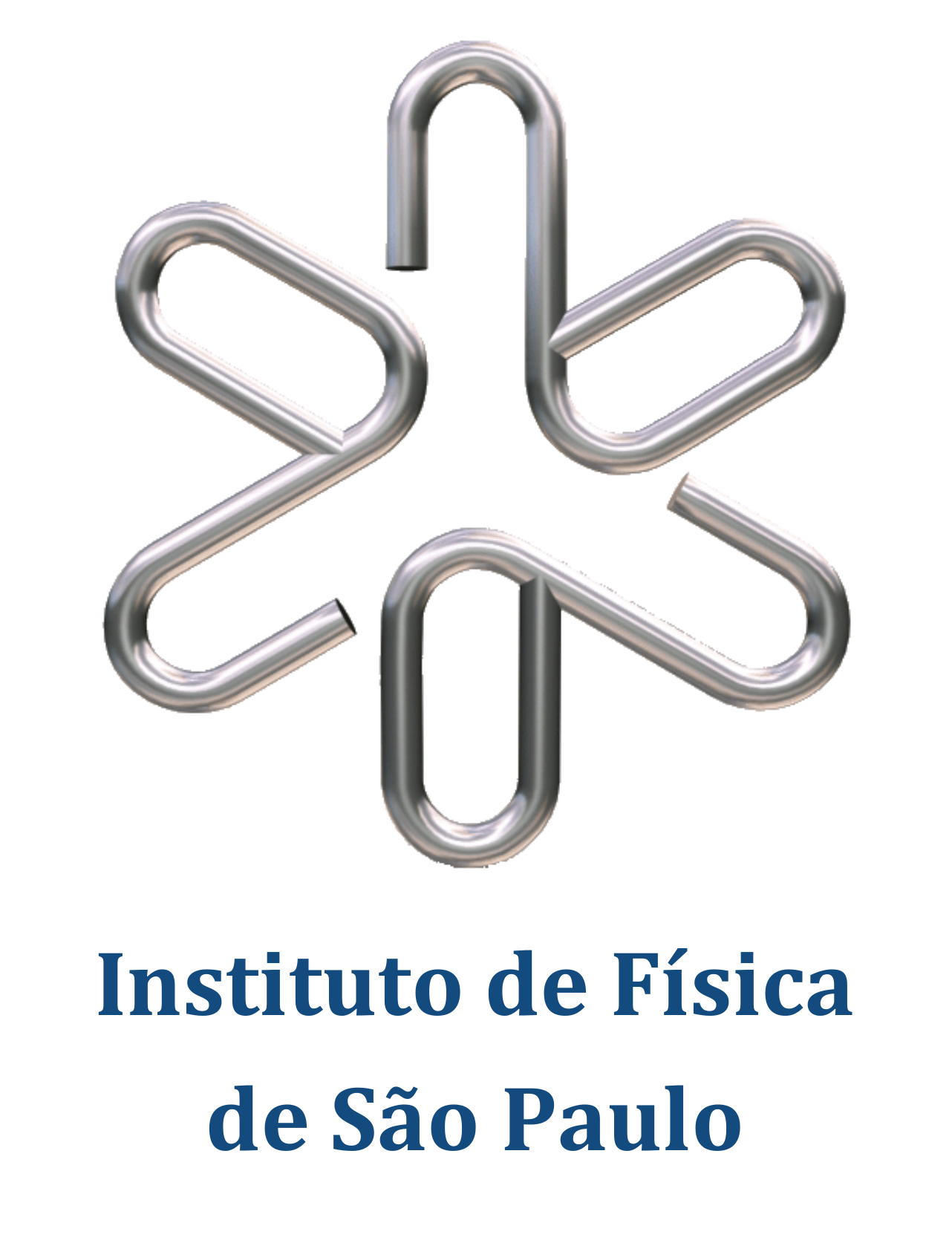}}
  \vspace{4cm}

  {\bf Le projet de recherche de cette thèse a été soutenu par:}

  \vspace{1cm}
  \centerline{\epsfig{figure=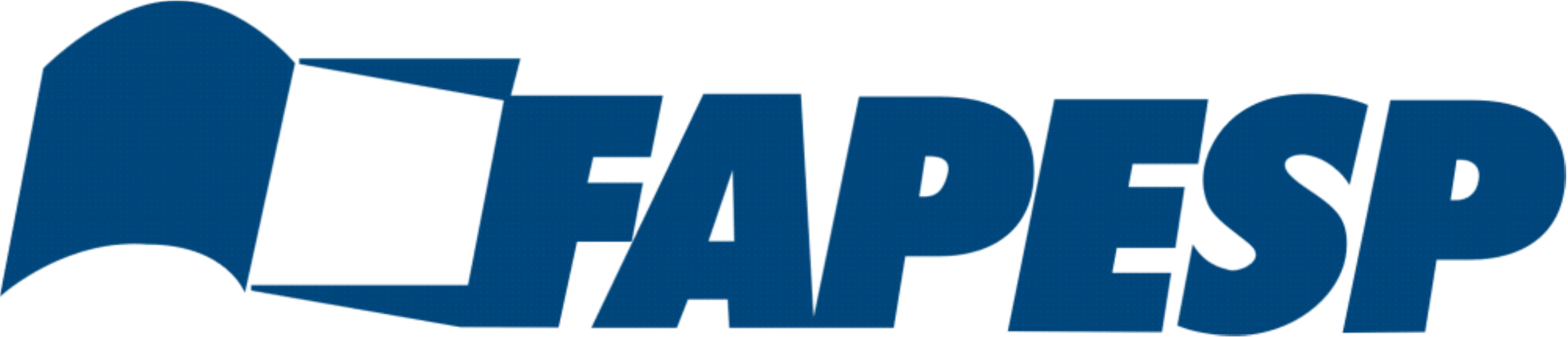,height=2.0cm}}

  \centerline{processus n: 2008/58289-8}
\end{center}

\vspace*{\fill}
\cleardoublepage

\thispagestyle{empty}
\vspace*{\fill}

\centerline{\bf Langue de Rédaction du Manuscrit }
\vspace{0.4cm}

Cette thèse a été réalisée sous la ``Convention de Cotutelle de Thèse'' 
approuvée par le Conseil Scientifique plénier du 7 juillet 2011. Cette 
Convention réglemente les relations entre l'Université de Montpellier 2 
Sciences et Techniques (Montpellier,  France) représentée par son Président  
Madame le Professeur Danièle HERIN, conformément à l'arrêté du 6 janvier 
2005, régissant les cotutelles internationales de thèses, et l'Université de São 
Paulo (São Paulo, Brésil) représentée par son Président Monsieur le Professeur
João Grandino Rodas, conformément à législation du pays concerné.

L'Article 2 du Titre II - Modalités Pédagogiques de cette Convention prévoit que:\\
{\it 
``... La thèse sera rédigée et soutenue en langue {\sc Anglais}. 
Lorsque les langues nationales des deux pays concernés sont différentes, le 
doctorant est tenu de compléter la thèse par un résumé écrit et oral dans l'autre 
langue. Le doctorant établira le résumé en langue: Anglais, Français et Portugais.''}

\vspace*{\fill}
\cleardoublepage

\thispagestyle{empty}
\vspace*{\fill}
\normalfont
\begin{flushright}
  To my beloved family,\\
  Ademir, Hilda, Fábio and Dany, \\
  and to my dearest love and best friend,\\
  Natalia.
\end{flushright}
\cleardoublepage

\rmfamily
\normalfont

\selectlanguage{french}
\begin{abstract}
Cette thèse a utilisé la méthode des règles de somme de QCD pour étudier la nature des 
résonances du charmonium suivantes: $Y(3930)$, $Y(4140)$, $X(4350)$, $Y(4260)$, 
$Y(4360)$ et $Y(4660)$. Il y a des fortes indications que ces états ont des structures 
hadroniques non conventionnelles (ou exotiques) lorsque leurs masses respectives et les 
modes de désintégration observés expérimentalement sont incompatibles avec ce qui 
est attendu pour l'état conventionnel du charmonium $(c \bar{c})$.
Le même phénomène se produit dans le secteur du bottomonium $(b \bar{b})$, où les 
nouveaux états $Y_b(10890)$ et $Y_b(11020)$ observées récemment pourraient indiquer 
l'existence de nouveaux états exotiques du bottomonium. De cette façon, on vérifie que 
l'état $Y(4140)$ peut être décrit soit par une structure moléculaire $\DsxDsx$ ($0^{++}$) 
ou par une mélange entre les états moléculaires $\DsxDsx$ ($0^{++}$) et $\DxDx$ ($0^{++}$). 
Les états $Y(3930)$ et $X(4350)$ ne peuvent pas être décrites par les courants moléculaires 
$\DxDx$ ($0^{++}$) et $\DsxDso$ ($1^{-+}$), respectivement. On vérifie également que 
la structure moléculaire $\psi^\prime \:f_0(980)$ ($1^{--}$) réproduit très bien la 
masse de l'état $Y(4660)$.

Une extension naturelle au secteur du bottomonium indique que l'état moléculaire 
$\Upsilon^\prime \:f_0(980)$ est un bon candidat pour l'état $Y_b(10890)$. 
On a également fait une estimation pour les états moléculaires possibles  
formées par des mésons $D$ et $B$, ce qui pourra être observé dans des expériences 
futures au LHC.

Une vaste étude, en utilisant le formalisme habituel des règles de somme et aussi 
le Double Rapport des Règles de Somme, est fait pour calculer les masses des baryons 
lourds en QCD. Les estimations pour les masses des baryons avec un ($Qqq$) et deux 
($QQq$) quarks lourds sont un excellent test pour la capacité de la méthode de 
règles de somme à prédire les masses des baryons qui n'ont pas encore été observés.
\let\thefootnote\relax\footnotetext{ \noindent {\bf Mots-clés:} QCD, Physique des Particules, 
Physique Hadronique, Phénoménologie, Baryons et Mesons, Règles de Somme de QCD.}

\end{abstract}
\cleardoublepage

\rmfamily
\normalfont

\selectlanguage{english}
\begin{abstract}
In this thesis, the QCD sum rules approach has been used to study the nature of the 
following charmonium resonances:  $Y(3930)$, $Y(4140)$, $X(4350)$, $Y(4260)$, 
$Y(4360)$ and $Y(4660)$. There is a strong evidence that these states have 
non-conventional (or exotic) hadronic structures since their respective masses 
and decay channels observed experimentally are inconsistent with expected 
for a conventional charmonium state $(c \bar{c})$.

The same phenomenon occurs on the bottomonium sector $(b\bar{b})$, where 
new states like $Y_b(10890)$ and $Y_b(11020)$ observed recently could indicate 
the existence of new bottomonium exotic states. In this way, one verifies that 
the $Y(4140)$ state could be described as a $\DsxDsx$ ($0^{++}$) molecular state
or even as a mixture of $\DsxDsx$ ($0^{++}$) and $\DxDx$ ($0^{++}$) molecular states. 
For the $Y(3930)$ and $X(4350)$ states, both cannot be described as a $\DxDx$ ($0^{++}$) 
and $\DsxDso$ ($1^{-+}$), respectively. From the sum rule point of view, the $Y(4660)$ state
could be described as a $\psi^\prime \:f_0(980)$ ($1^{--}$) molecular state.
The extension to the bottomonium sector is done in a straightforward way to demonstrate that 
the $\Upsilon^\prime \:f_0(980)$ molecular state is a good candidate for describing the 
structure of the $Y_b(10890)$ state. In the following, one estimates the mass of the exotic 
$B_c$-like molecular states using the sum rule approach, where these new exotic states 
would correspond to bound states of $D^{(*)}$ and $B^{(*)}$ mesons. All of these mass 
predictions could (or not) be checked in a near future experiments at LHC.

A large study using the Double Ratio of Sum Rules approach has been evaluated for 
studying the heavy baryon masses in QCD. The obtained results for the unobserved 
heavy baryons, with one ($Qqq$) and two ($QQq$) heavy quarks will be an excellent test for 
the capability of the sum rule approach in predicting their masses.
\let\thefootnote\relax

\end{abstract}
\cleardoublepage

\rmfamily
\normalfont

\selectlanguage{brazil}
\begin{abstract}
Nesta tese é utilizado o método das Regras de Soma da QCD para estudar a natureza 
dos seguintes estados ressonantes do charmonium: $Y(3930)$, $Y(4140)$, $X(4350)$, $
Y(4260)$, $Y(4360)$ e $Y(4660)$. Há fortes evidências de que estes estados possuam 
estruturas hadrônicas não convencionais (ou exóticas) uma vez que as suas respectivas 
massas e canais de decaimento observados experimentalmente são inconsistentes com o 
que é esperado para o estado ressonante convencional do charmonium $(c\bar{c})$.
O mesmo fenômeno ocorre no setor do bottomonium $(b\bar{b})$, onde os novos estados 
$Y_b(10890)$ e $Y_b(11020)$ observados recentemente poderiam indicar a existência de 
novos estados exóticos do bottomonium. Neste sentido, verifica-se que o estado $Y(4140)$ 
pode ser descrito ou por uma estrutura molecular $\DsxDsx$ ($0^{++}$) ou mesmo uma mistura entre 
os estados moleculares $\DsxDsx$ ($0^{++}$) e $\DxDx$ ($0^{++}$). Já os estados $Y(3930)$ 
e o $X(4350)$ não podem ser descritos por correntes moleculares $\DxDx$ ($0^{++}$) e 
$\DsxDso$ ($1^{-+}$), respectivamente. Verifica-se também que a estrutura molecular 
$\psi^\prime \:f_0(980)$ ($1^{--}$) descreve muito bem a massa do estado $Y(4660)$. 
Uma extensão ao setor do bottomonium indica que o estado molecular $\Upsilon^\prime \:f_0(980)$ 
é um bom candidato para descrever a estrutura do estado $Y_b(10890)$.
É feita também uma estimativa para os possíveis estados moleculares formados por mésons 
$D$ e $B$, que poderão ser observados em futuros experimentos realizados pelo LHC. 

Um amplo estudo, utilizando o formalismo das Regras de Soma e também da Dupla Razão 
das Regras de Soma, é feito para calcular as massas dos bárions pesados na QCD. As 
estimativas para as massas dos bárions com um ($Qqq$) e com dois ($QQq$) quarks pesados 
são um excelente teste para a capacidade do método das regras de soma em prever a 
massa dos bárions que ainda não foram observados.
\let\thefootnote\relax\footnotetext{ \noindent {\bf Palavras-chave:} QCD, Física de Partículas, 
Física Hadrônica, Fenomenologia, Bárions e Mésons, Regras de Soma da QCD.}

\end{abstract}
\cleardoublepage

\selectlanguage{english}
\pagestyle{headings}
\pagenumbering{roman}
\tableofcontents
\cleardoublepage

\pagenumbering{arabic}
\setcounter{page}{1}

\chapter{Introduction}
In particle physics, a hadron is a composite particle made up of quarks and gluons 
interacting via strong interactions. The hadrons are divided into two groups: mesons (made of a 
quark-antiquark pair, or $q\bar{q}$) and baryons (made of three quarks, or $qqq$). The 
strong interactions, responsible for keeping quarks together inside the hadrons, 
are described by a non-Abelian gauge theory called Quantum Chromodynamics 
(QCD) \cite{peskin}.

In QCD, the quarks and gluons are the fundamental particles which carry a 
non-vanishing color charge. These particles interact with each other through the
strong interactions that are mediated by gluons. Note the analogy with Quantum 
Electrodynamics (QED), where the particles with electromagnetic charge interact 
through photon exchange \cite{peskin}. 
However, QCD has some fundamental properties, such as asymptotic freedom and 
confinement, that make it a theory much more complex than QED.

In the early seventies, the physicists David Politzer, David J. Gross and Frank 
Wilczek \cite{gross} proved that in QCD, the effective strong coupling constant vanishes 
at short distances (high-energy regime) and increases at long distances (low-energy regime). 
The former case implies that QCD becomes a weakly coupled theory ($g_s \rightarrow 0$) 
and, in principle, can be evaluated using perturbative methods. For the latter case, however, 
QCD becomes a strongly coupled theory ($g_s \gg 1$) and the non-perturbative effects are 
expected to be the most valuable contribution to the formation of the physical observables. 
This running behavior of the strong coupling "constant" is associated to an important 
property in QCD known as asymptotic freedom.

Further analysis in QCD guarantees that the physical bound states must hold a color 
singlet combination in order to satisfy the Pauli exclusion principle \cite{peskin}. Since 
quark and gluons carry the color charge and rather cannot be described by color singlet 
states, then this could be an explanation for one has never seen them as free particles in 
the nature. This feature is commonly associated with the confinement mechanism, which 
ensures that quarks and gluons remain confined inside hadrons.
 
It is known that hadrons are color singlet states formed in the low-energy regime, and 
the strong interactions between them are due to the meson exchange (e.g. pions, kaons), 
instead of the gluon exchange. In this scenario, QCD is no longer able to adequately 
describe hadrons and their interactions in terms of degrees of freedom of quarks and gluons. 
Consequently, it is mandatory to construct physical models, which incorporate these 
fundamental properties of QCD and describe the hadrons in a consistent way with 
the experimental data.

One of these models, the so-called Constituent Quark Model (CQM), was proposed by the 
physicists Murray Gell-Mann \cite{gellmann} and George Zweig \cite{zweig}. The CQM is 
a successful model which is widely used to classify hadrons and calculate their masses 
and decay widths \cite{gisgur1}.
According to CQM, the mesons are particles composed of a quark-antiquark pair 
$(q_i \bar{q}_i)$, whereas the baryons are particles composed of three quarks 
$(\epsilon_ {ijk} \: q_i q_j q_k$). The latin indices represent the colors of the quark fields. 
In nature, mesons and baryons must be represented by color singlet states with a neutral 
color charge. 

It is worth mentioning that, in addition to these conventional structures, QCD 
would accept the existence of many other kinds of internal structures for hadrons, such as 
glueballs ($gg, ggg, \ldots$) \cite{meyer}, hybrid states ($q \bar{q} g, qqq g, \ldots$) \cite{hybrid}, 
hadronic molecules $(D \bar{D}, D^\ast \bar{D}, \ldots)$ \cite{molecules, torn, close}, tetraquark 
states ($qq \bar{q}\bar{q}$) \cite{jaffe, maiani}, among others. These kinds of structures are 
usually related to the exotic states. 
Indeed, although the current studies on particle physics reveal that the vast majority of 
observed hadrons are remarkably well described by CQM, recent researches on charmonium 
$(c \bar{c})$ spectroscopy, carried out by BaBar \cite{babar05} and Belle \cite{belle} 
collaborations, have indicated the possible existence of new mesons with a structure 
much more complex than the conventional $q\bar{q}$ system.

\section{Charmonium Exotic States}
At the beginning of the XXI century, two main research centers began performing 
experiments in order to test the CP symmetry violation in the Standard Model:
\begin{itemize}
\item BaBar collaboration at Stanford Linear Accelerator Collider, United States (SLAC),
\item Belle collaboration at High Energy Accelerator Research Organization, Japan (KEK).
\end{itemize}
Both collaborations analyzed $e^+ e^-$ collisions operating at a centre-of-mass energy 
($E_{CM}$) close to $10.6 \GeV$. In this $E_{CM}$, there is a large production of 
$B \bar{B}$ pairs, hence the name $B-$factories. The $B$ mesons which are produced 
decay predominantly into channels containing $c$ quarks and $\bar{c}$ antiquarks, thus 
often exhibiting a significant presence of charmed hadrons and charmonium ($c\bar{c}$) 
in the final states.

The charmonium production at the $B-$factories can also occur through a process called 
Initial State Radiation (ISR). By using ISR process, an energetic photon is emitted by either 
the initial electron or positron, it is possible to study not only the produced events at the collider 
nominal $E_{CM}$, but also at lower energies.
When the energy of irradiated photons $(\gamma_{ISR})$ is in the range between 
$4.0$ to $5.0 ~\GeV$, the $e^+ e^-$ annihilation occurs in an $E_{CM}$ which corresponds 
to the mass region allowed for the excited states of the charmonium $c \bar{c}$.

According to CQM, widely used in the study of hadron spectroscopy, it was expected that 
ISR processes could produce a large amount of phenomenological data on the $1^{--}$ 
family of excited states of the charmonium, including the $\psi(3770)$, $\psi(4040)$, 
$\psi(4160)$ and $\psi(4415)$. However, the first $1^{--}$ resonant states produced 
experimentally by $B-$factories were: $Y(4260)$, $Y(4360)$ and $Y(4660)$. In addition 
to having a mass incompatible with the spectrum predicted by CQM, these states also 
present a decay pattern totally unexpected for a conventional charmonium state.

Although in Ref.\cite{dzy}, the authors argue that the $Y(4360)$ and $Y(4660)$ are 
conventional $c\bar{c}$ states, the $3^3 D_1$ and $5^3 S_1$ states respectively,
their masses are inconsistent with predicted by CQM in \cite{gisgur1}. Besides, the absence 
of open charm production would be also inconsistent with a conventional $c\bar{c}$ structure.
A good review of these new states could be found in Refs.\cite{godfrey, nnl}.
Facing this impasse between CQM theoretical predictions and experimental data, it 
becomes necessary to build models where the structure for these new charmonium states goes 
beyond the conventional $c \bar{c}$ structure. Clearly, one needs further experimental data 
and theoretical development for a better understanding of the properties of these new states.

In the present work, one uses the QCD Sum Rules approach (QCDSR), proposed in 1979 
by M. Shifman, A.I. Vainshtein and V.I. Zakharov \cite{svz, rry, SNB}, to study whether these 
new charmonium states have an exotic structure, like molecular states.

\subsection{Y(4260), Y(4360) and Y(4660)}
In 2005, the experiments based on ISR processes, carried out by BaBar collaboration, indicated 
a resonant peak around $4300 \MeV$ produced in the decay channel \cite{babar3}:
\begin{equation*}
	e^+ \: e^- \longrightarrow \gamma_{\!_{ISR}} \:Y(4260) 
	\longrightarrow \gamma_{\!_{ISR}} \:J/\psi \:\pi^+ \:\pi^- ~~.
\end{equation*}
The full width measured for the intermediate $Y(4260)$ state was approximately of $90 \MeV$. 
The CLEO collaboration has also confirmed the observation of this state \cite{cleo1}. The proximity 
between the $Y(4260)$ mass and that obtained for $\psi(4153)$ state led some authors to 
identify them as the same particle. However, the hidden-charm decay channel and the total 
width observed for the $Y(4260)$ are not in agreement with expected by CQM predictions for 
the charmonium resonances, which should decay into channels containing predominantly $D$ 
mesons and would have a width larger than $90 \MeV$. Thus, Maiani {\it et al.} \cite{maiani1} have 
proposed that the $Y(4260)$ could be described as a tetraquark $[cs] [\bar{c} \bar{s}]$ state.

\begin{figure}[t]
    \subfloat[]{\includegraphics[width=7.2cm]{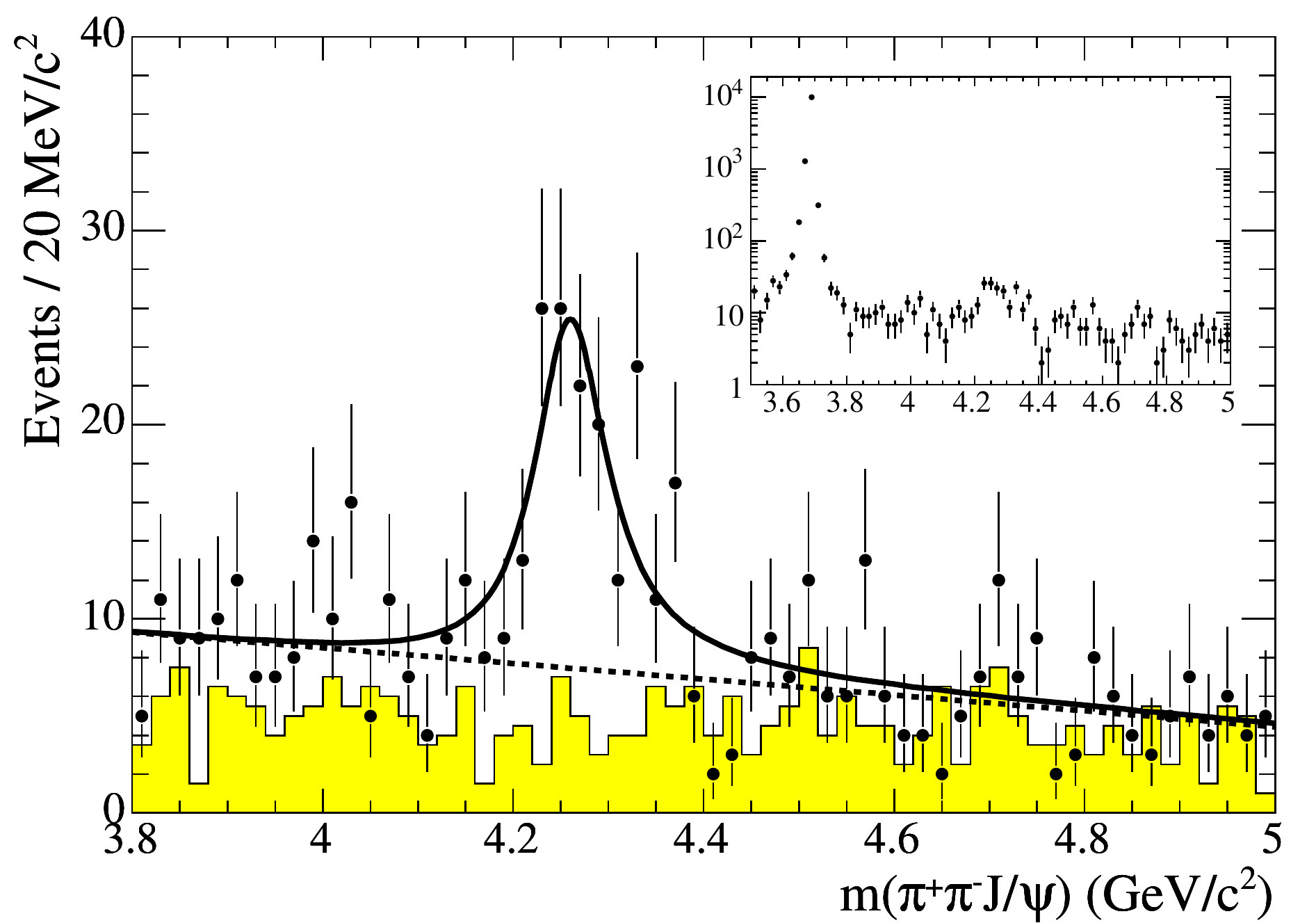}}  \vspace{-5.45cm}
    
    \hspace{8cm}
    \subfloat[]{\includegraphics[width=4.8cm, angle=-90]{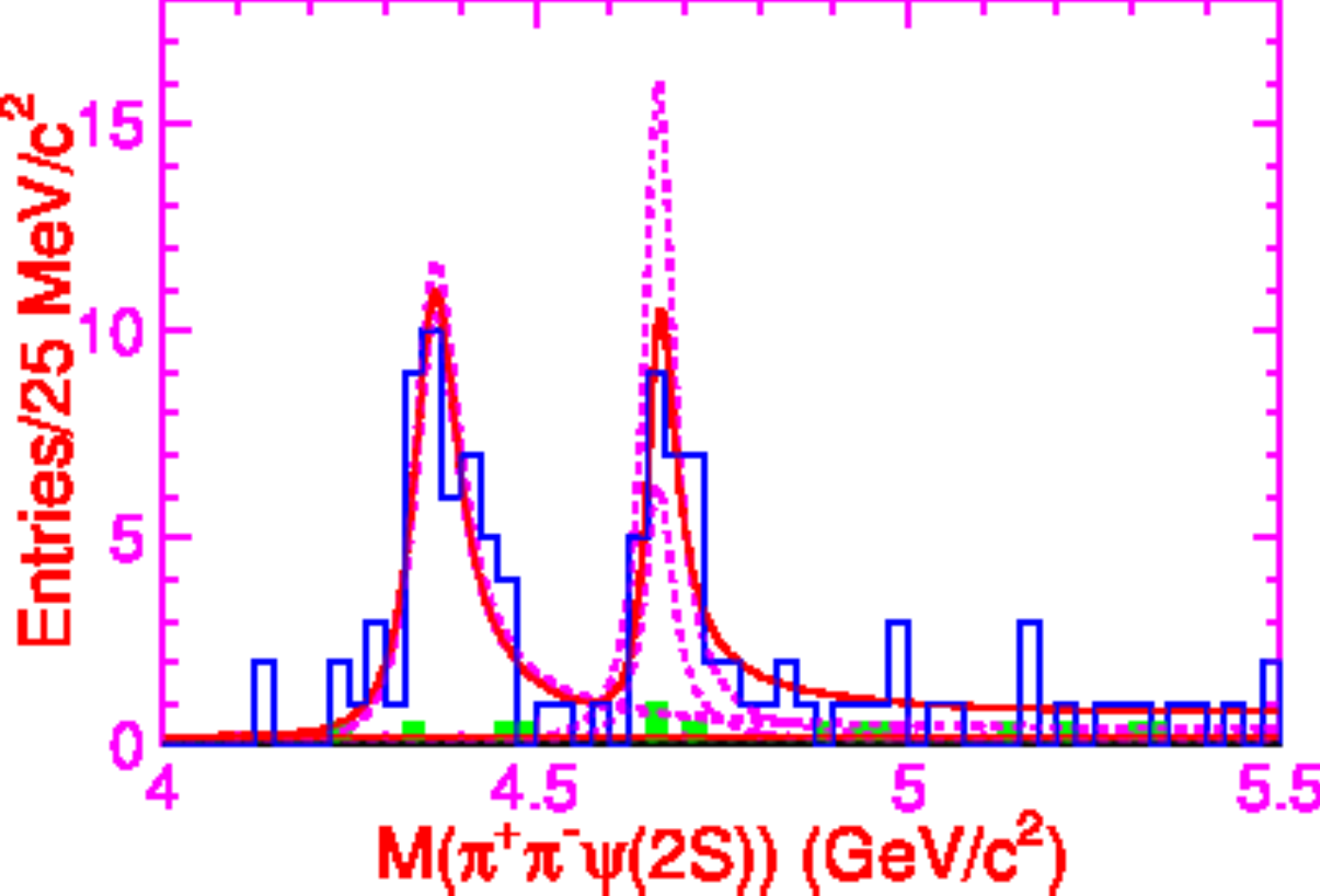}}
	\caption{\footnotesize{The invariant-mass distribution of \cite{babar05, belle}: 
	{\bf a)} $J/\psi \:\pi^+ \:\pi^-$ and 
	{\bf b)} $\psi(2S) \:\pi^+ \:\pi^-$.}}
    \label{bfact}
\end{figure}

Recently, Belle collaboration not only confirmed the existence of the $Y(4260)$ state 
\cite{belle1}, but also announced the discovery of two other vector mesons \cite{belle}, the 
$Y(4360)$ and $Y(4660)$, whose resonant peaks are around $4360 \MeV$ 
and $4660 \MeV$, respectively. Both states are produced through the following process:
\begin{equation*}
	e^+ \: e^- \longrightarrow \gamma_{\!_{ISR}} \:\psi' \:\pi^+ \:\pi^- .
\end{equation*}
with the same $J^{PC} = 1^{--}$ quantum numbers.
The measurements of the total widths for the $Y(4360)$ and $Y(4660)$ states are 
$74 \pm 18 \MeV$ and $48 \pm 15 \MeV$, respectively \cite{belle}. 
With these values, either $Y(4360)$ or $Y(4660)$ states are 
consistent with conventional charmonium states in this mass region. The invariant mass 
distributions obtained by Belle and BaBar collaborations are shown in Fig.(\ref{bfact}).

\begin{figure}[t]
\vspace{-0.6cm}
	\centerline{\epsfig{figure=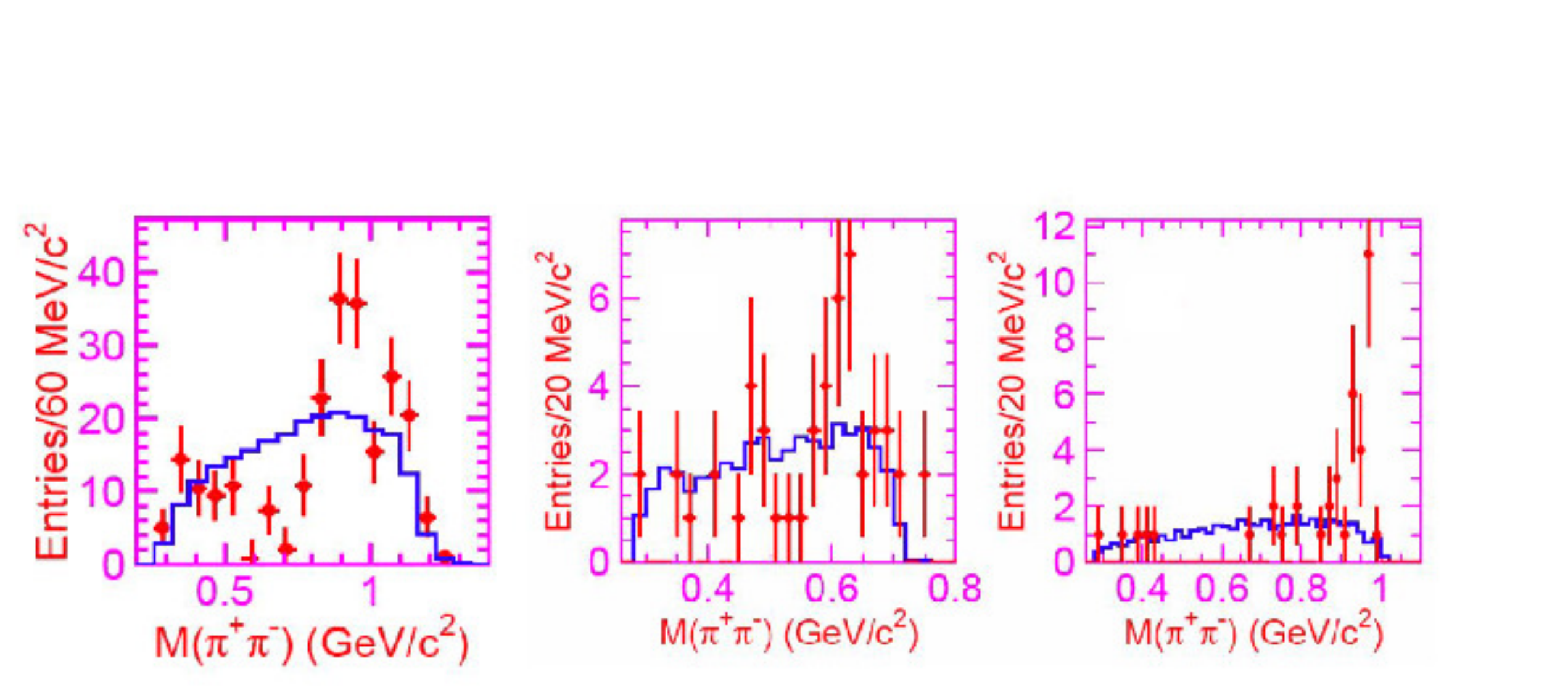,height=7cm}}
	\caption{\footnotesize{Dipion invariant mass spectrum of the decay channels \cite{babarconf}: 
	$Y(4260) \rightarrow J/\psi \:\pi^+ \:\pi^-$ (left), 
	$Y(4360) \rightarrow \psi(2S) \:\pi^+ \:\pi^-$ (middle) 
	and $Y(4660) \rightarrow \psi(2S) \:\pi^+ \:\pi^-$ (right).}}
	\label{dipion}
\end{figure}

Important information for understanding the structure of these states is which resonant 
intermediate state is responsible for producing the pair of pions in their respective decay 
channels. From the dipion invariant mass spectrum, which is shown in Fig.(\ref{dipion}), only 
the $Y(4660)$ state presents a well defined intermediate state consistent with $f_{0}(980)$ 
\cite{babarconf}. Due to this fact and the proximity of the mass of the system 
$\psi(2S)-f_{0}(980)$ with the $Y(4660)$ mass, in Ref.\cite{ghm}, the authors suggested that 
the $Y(4660)$ could be described by the $\psi(2S) \:f_0(980)$ molecular state.
There are also other interpretations for the $Y(4660)$ state like baryonium state \cite{qiao} or 
a conventional charmonium state $5^{3}S_1 ~c\bar{c}$ \cite{dzy}. For the other two mesons, 
$Y(4260)$ and $Y(4360)$, Fig.(\ref{dipion}) indicates that the $\pi\pi$ pairs in their respective 
decay channel are more consistent with the $\sigma(600)$ scalar meson \cite{babarconf, sigma} 
than the $f_0(980)$ meson.

In the present work \cite{rnr}, one uses the QCDSR approach to investigate if a correlation 
function based on $J/\psi \:f_0(980)$ or $J/\psi \:\sigma(600)$ molecular currents, with 
$J^{PC} = 1^{--}$, could describe the structure of at least one of these new charmonium states.

\subsection{$Y(4140)$}
The CDF collaboration found evidence for a new particle called $Y(4140)$ \cite{cdf09} 
produced by the particle accelerator Tevatron at FERMILAB. The signal of this particle, with 
significance of $3.8 \sigma$, was seen in the decay of $B$ mesons with a mass 
$M_{_{Y(4140)}} = (4143.0 \pm 2.9 \pm 1.2) \MeV$ and width 
$\Gamma = (11.7^{+8.3}_{-5.0} \pm 3.7) \MeV$. The final state observed contains muons 
and kaons pairs: 
\begin{eqnarray}
&&B^+ ~\rightarrow~ Y(4140) K^+ \nno\\
&& Y(4140) ~\rightarrow~ J/\psi \:\phi ~\rightarrow~ \mu^+ \mu^- K^+ K^-.  
\end{eqnarray}
\begin{figure}[t]
	\centerline{\epsfig{figure=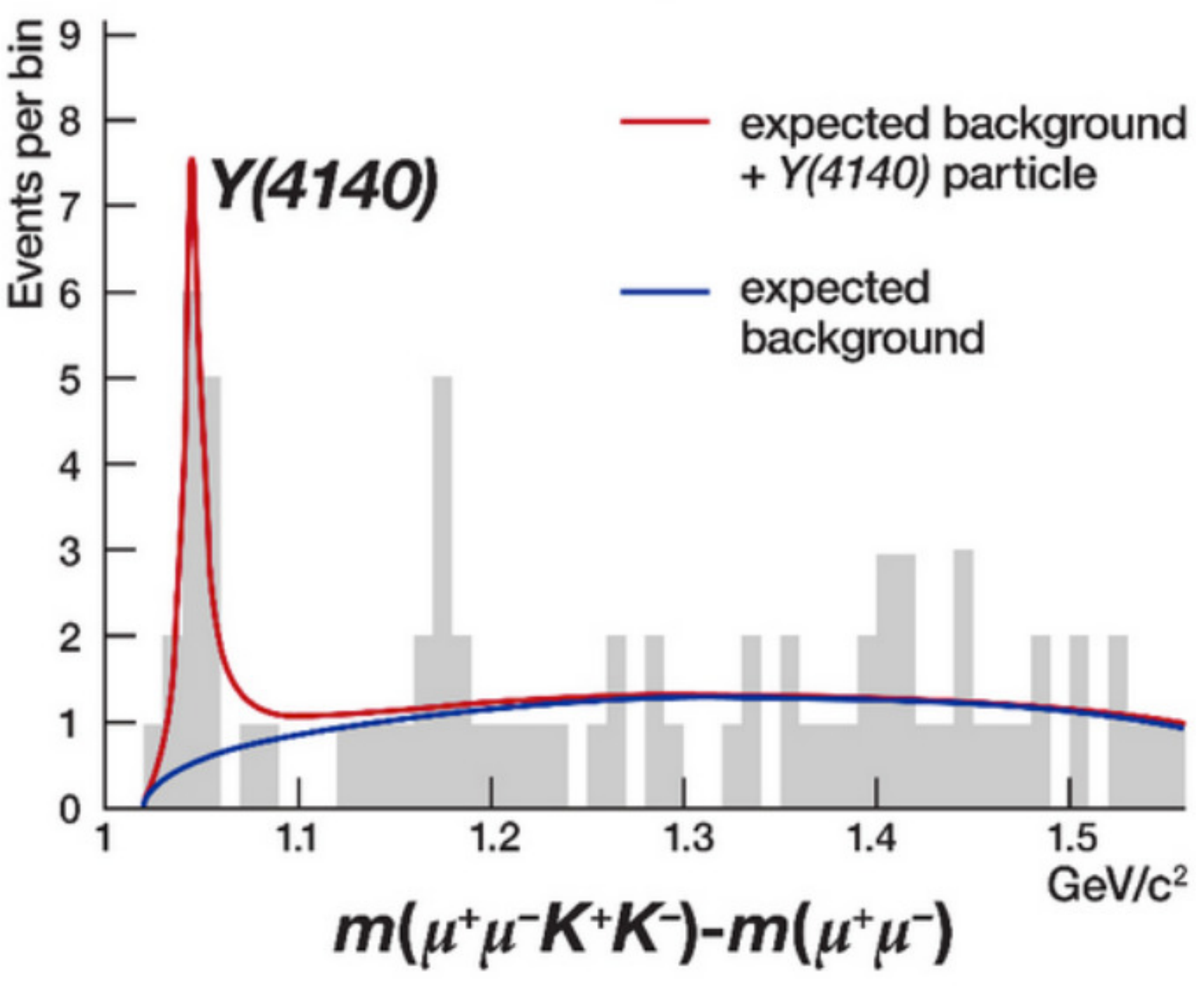,height=7cm}}
	\caption{\footnotesize{Evidence for observation of $Y(4140)$ state by Belle collaboration \cite{cdf09}. 
	One of the lines (blue) represents the expected background composed of mesons $K$ 
	$(M_{KK} \simeq 1.0 \GeV)$, while the other line with a peak (red) is the fit to events observed.}}
	\label{y4140evidence}
\end{figure}
Since the $Y(4140)$ state decays into two $I^G \left( J^{PC} \right) = 0^- \left( 1^{--} \right)$ vector mesons, 
it has positive $C$ and $G$ parities. There are already some theoretical interpretations for this structure. 
Its interpretation as a conventional $c\bar{c}$ state is complicated because, as pointed out by the 
CDF \cite{cdf09} and Belle \cite{belle3} collaborations, it lies well above the threshold for open 
charm decays and, therefore, a $c\bar{c}$ state with this mass would decay predominantly into an 
open charm pair with a large total width. 
Then, they concluded that the $Y(4140)$ is probably a $D^\ast_s \bar{D}^\ast_s$ molecular state 
with $J^{PC} = 0^{++}$ or $2^{++}$. 
In Ref.\cite{liu1}, the authors interpreted the $Y(4140)$ as the molecular partner of the charmonium-like 
state $Y(3930)$, which was observed by Belle and BaBar collaborations near the $J/\psi \:\omega$ 
threshold \cite{belle4, babar1}. In Ref.\cite{namit}, the authors have interpreted the $Y(4140)$ as an 
exotic hybrid charmonium with $J^{PC} = 1^{-+}$.

There are many other works about the possible theoretical interpretations for the $Y(4140)$ state and 
more discussions can be found in Refs.\cite{ramar2, zwang1, tanja1, liu2, liu3, tanja2}.

In the present work \cite{ramar2}, one uses the QCDSR approach to study if a two-point correlation 
function based on a $D^\ast_s \bar{D}^\ast_s$ molecular current, with $J^{PC} = 0^{++}$, can 
describe the new observed resonant structure $Y(4140)$.

\subsection{Y(3930)}
Another intriguing state is the so-called $Y(3930)$. The Belle collaboration \cite{belle4} has 
announced the observation of this state in the decay channel $B \rightarrow Y(3930) \:K$, with 
significance of $8.1 \sigma$. This signal has also been confirmed by BaBar collaboration 
\cite{babar1}. The experimental mass and total width are given by \cite{pdg}:  
$M_{_{Y(3930)}} = (3917.5 \pm 2.7) \MeV$ and $\Gamma = (27 \pm 10) \MeV$.

If it were a conventional charmonium state, the $Y(3930)$ state would decay predominantly 
into $D$ mesons since its mass is above the $DD$ threshold. However, this state decays 
dominantly through the following channel:
\begin{eqnarray}
  Y(3930) \rightarrow J/\psi \:\omega ~.
\end{eqnarray}

The $Y(3930)$ state has a mass of approximately $200 \MeV$ below the $Y(4140)$ mass. 
This difference could be related to the quarks constituting both particles. Note that the 
difference between the quark masses is in order of: $m_s - m_q \simeq 100 \MeV$. In this 
way, the $D^\ast \bar{D}^\ast$ molecular state, with $J^{PC} = 0^{++}$, is a good candidate 
to describe the $Y(3930)$ state.

In the present work \cite{ramar2}, one uses the QCDSR approach to study if a two-point 
correlation function based on a $D^\ast \bar{D}^\ast$ molecular current, with $J^{PC} = 0^{++}$, 
can describe the new observed resonant structure $Y(3930)$.

\subsection{X(4350)}
\begin{figure}[t]
	\centerline{\epsfig{figure=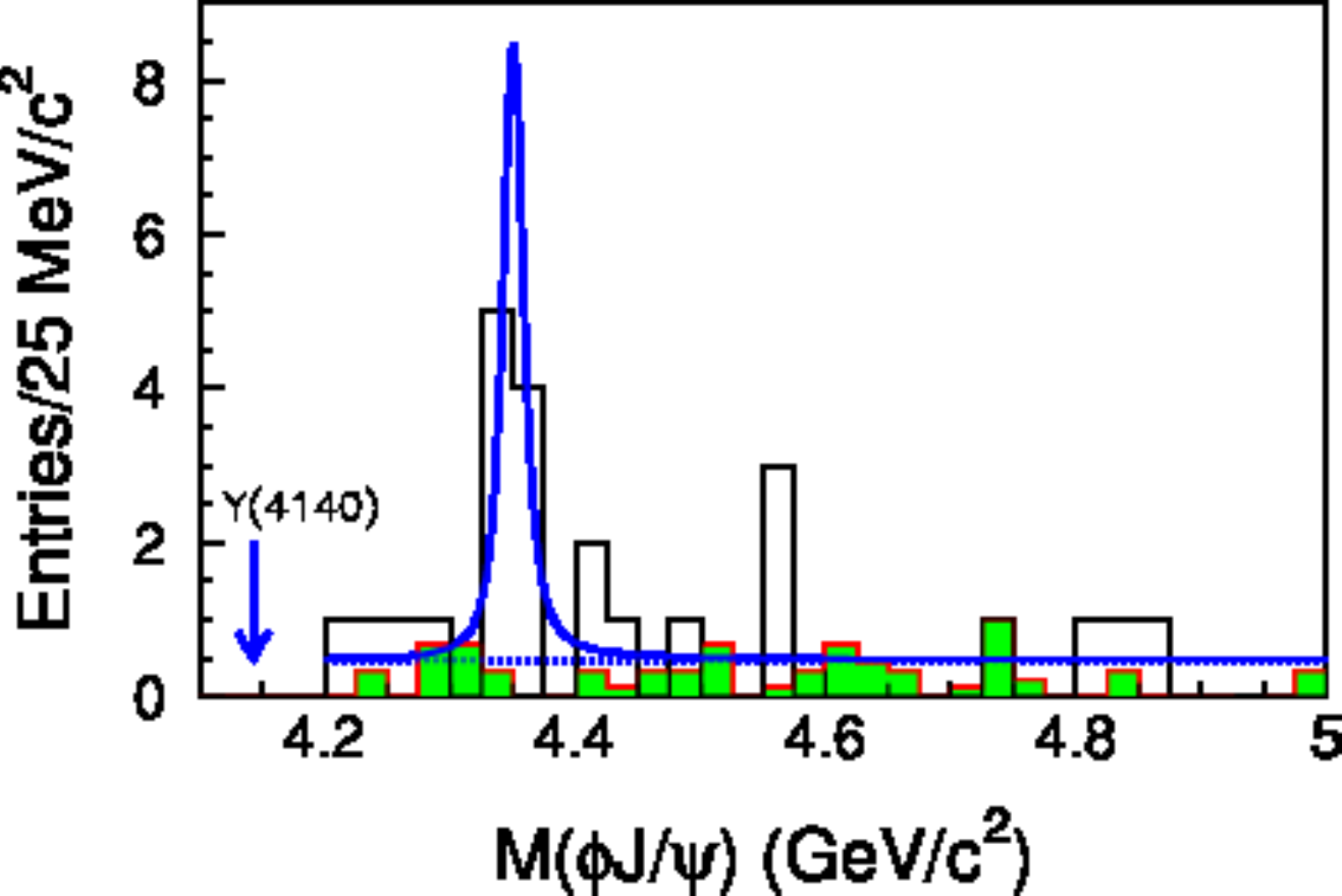,height=7cm}}
	\caption{\footnotesize{$J/\psi \:\phi$ invariant mass distribution \cite{belle3}.}}
	\label{IM_x4350}
\end{figure}

Searching for experimental evidence for the $Y(4140)$ state, the Belle collaboration \cite{belle3} 
found a new narrow structure in the $J/\psi \:\phi$ invariant mass spectrum at $4.35 \GeV$, see 
Fig.(\ref{IM_x4350}). The significance of the peak is $3.2 \sigma$ and, if interpreted as one resonance,
the mass and width of the state, called $X(4350)$ are: 
$M{_{X(4350)}} = (4350.6^{+4.6}_{-5.1}\pm0.7)~\MeV$ and 
$\Gamma = (13.3^{+7.9}_{-9.1} \pm4.1)~\MeV$.
The quantum numbers available for a state decaying into $J/\psi \:\phi$ are $J^{PC} = 0^{++}$, 
$1^{-+}$ and $2^{++}$. Among these quantum numbers, the $1^{-+}$ is not consistent with the 
CQM and it is considered as an exotic state \cite{nnl}. In Ref.\cite{belle3}, it was noted that 
the mass of the $X(4350)$ state is also consistent with the mass related to $cs\bar{c} \bar{s}$ 
tetraquark state, with $J^{PC}=2^{++}$, predicted by the authors in Ref.\cite{stancu}. Another 
work \cite{jrz2} suggests $D_s^{*+}\bar{D}_{s0}^{*-}$ molecular state as a good candidate for the
$X(4350)$. However, the state considered in Ref.\cite{jrz2} has $J^P = 1^-$ with no definite charge 
conjugation. A $D_s \bar{D}_{s0}$ molecular state with $J^{PC}=1^{--}$ was studied for the first 
time in Ref.\cite{ramar1}. The mass obtained was $(4.42 \pm 0.10) \GeV$, which is consistent 
with the $X(4350)$ mass, but it has not consistent quantum numbers. Analyzing the hadronic 
current used in Ref.\cite{ramar1} one can note the following molecular configuration: 
$D_s^{*+}\bar{D}_{s0}^{*-} + \bar{D}_s^{*-}D_{s0}^{*+}$, which results in a state with 
$J^{PC} = 1^{--}$. Thus, rearranging the terms of this hadronic current so that the charge 
conjugation becomes positive, one obtains a new molecular configuration
$D_s^{*+}\bar{D}_{s0}^{*-}-\bar{D}_s^{*-}D_{s0}^{*+}$, which allows to study the 
$D_s \bar{D}_{s0}$ molecular state, with $J^{PC}=1^{-+}$.

There are other interpretations for the $X(4350)$ state as indicated in Ref.\cite{xliu}, where the 
authors described it as an excited P-wave charmonium state, $\Xi_{c2}^{\prime\prime}$. In 
Ref.\cite{wang}, the authors interpreted it as a mixed charmonium-$D_s^*\bar{D}_s^*$ state.

In the present work \cite{rjm}, one uses the QCDSR approach to study if a two-point correlation function 
based on a $D^\ast_s \bar{D}^\ast_{s0}$ molecular current, with $J^{PC} = 1^{-+}$, can describe  
the resonant structure $X(4350)$ as suggested by Belle collaboration.

\subsection{$Y_b(10890)$}
In the bottom sector, the Belle collaboration announced the first observation of the decay channels 
\cite{chen}:
\begin{eqnarray}
  e^+ e^- &\rightarrow& \Upsilon(1S) \pi^+ \pi^-~~, ~~\Upsilon(2S) \pi^+ \pi^-
  ~~, ~~\Upsilon(3S) \pi^+ \pi^-~~, ~~\Upsilon(1S) K^+ K^- \nno ~.
\end{eqnarray}
which occurs at a centre-of-mass energy of about $10.87 \GeV$. It is likely that these decay 
channels contain as intermediate state the bottomonium resonant state $\Upsilon(5S)$. 
The total and partial decay widths observed are given by:
\begin{eqnarray}
  \Gamma &=& 55 \pm 28 \MeV \\
  \Gamma_{\Upsilon(1S) \pi^+ \pi^-} &=& 0.59 \pm 0.04(\mbox{stat}) \pm 0.09(\mbox{syst}) \MeV \\
  \Gamma_{\Upsilon(2S) \pi^+ \pi^-} &=& 0.85 \pm 0.07(\mbox{stat}) \pm 0.16(\mbox{syst}) \MeV
  \label{widthb}
\end{eqnarray}
When comparing these values with the respective widths of the decay channels for the first 
bottomonium resonant states, like $\Upsilon(2S)$, $\Upsilon(3S)$ and $\Upsilon(4S)$, it seems 
that they differ in at least two orders of magnitude from the values obtained in (\ref{widthb}). 
This result is not consistent with what expected for the conventional bottomonium resonant states. 
Therefore, the hypothesis that these decay channels are produced by the $\Upsilon(5S)$ state is 
not the most appropriate. Considering the experimental mass observed:
\begin{eqnarray}
  M_{Y_b} = (10888.4 \pm2.7\pm1.2) \MeV ~.
\end{eqnarray}
one usually calls this state as $Y_b(10890)$. 
Thereafter, these decay channels were also reproduced by experiments carried out by BaBar 
collaboration \cite{ybottom}, which has not only confirmed the existence of the state $Y_b(10890)$ 
as well as the existence of another new state in the bottomonium spectrum, the $Y_b(11020)$.

In the present work \cite{rnr}, one uses the QCDSR approach to study if the two-point correlation function 
based on a  $\Upsilon \:f_0(980)$ and $\Upsilon \:\sigma(600)$ molecular current, with 
$J^{PC} = 1^{--}$, can describe the new resonant structure $Y_b(10890)$.

\section{$B_c$ Meson Spectroscopy}
Another interesting sector investigated by CDF and D0 collaborations involves the $B_c$ meson 
spectroscopy: $B^+_c (c \bar{b})$ and $B^-_c (b\bar{c})$. Once the exotic structures are established 
in the charmonium spectroscopy, one expects that more exotic states could exist. One possibility 
would be the existence of the $B_c$-like molecules, whose formation is determined by the 
bound states of the $D^{(*)}$ and $B^{(*)}$ mesons.

In Ref.\cite{sunliu}, the one boson exchange (OBE) model was used to investigate hadronic molecules 
with both open charm and open bottom. These new structures were labelled as $B_c$-like molecules 
and were categorized into four groups ${\cal D}{\cal B}$, ${\cal D}^\ast{\cal B}^\ast$, ${\cal D}^\ast{\cal B}$
and ${\cal D}{\cal B}^\ast$, where these symbols represent the group of states: 
${\cal D}^{(*)} = \big[ D^{(*)}, \:D^{(*)+}, \:D^{(*)+}_s \big]$ for charmed mesons and 
${\cal B}^{(*)} = \big[ B^{(*)+}, \:B^{(*)0}, \:B^{(*)0}_s \big]$ for bottom mesons. 
These states were categorized using a hand-waving notation, with five-stars, four-stars, etc. A five-star 
state implies that a loosely molecular state probably exists. They found six five-star states, all of them 
isosinglets in the light sector, with no strange quarks.

In the present work \cite{DB}, one uses the QCDSR approach to check if some of the five-star states 
could be described by a two-point correlation function based on a ${\cal D}^{(*)}{\cal B}^{(*)}$ molecular 
current.

\section{Heavy Baryons in QCD}
The heavy baryons are composed by at least one heavy quark ($c$ or $b$) in their constitution. 
The first experimental evidence of these baryons were announced in the mid-1970s, with experiments 
carried out by Brookhaven National Laboratory (BNL) and European Organization for Nuclear Research 
(CERN, the old french acronym for Conseil Européen pour la Recherche Nucléaire) and others 
collaborations.

The heavy baryons spectroscopy is a research area of considerable interest for hadronic physics, 
since several of these states have been observed during the last years, see Tables 
(\ref{BarCharm}) and (\ref{BarBottom}). 
So it is extremely encouraging to use the available information - such as mass, quantum numbers, 
decay width, etc. - to build physical models to predict with an improved confidence level properties 
of the heavy baryons, which have not yet been observed experimentally
\cite{RICHARD,MARQUES,HUSSAIN,MARINA,zhu,SR,KORNER,EBERT,LATT,JENKINS}.

In the present work \cite{rnm, rnar}, one uses the Double Ratio of Sum Rules (DRSR) approach to 
estimate the following heavy baryon masses:
\begin{itemize}
\item with one heavy quark $(Qqq)$: 
	$\Lambda_Q$, $\Sigma_Q$, $\Xi_Q$ and $\Omega_Q$;
\item with two heavy quarks $(QQq)$: 
	$\Xi_{QQ}$ and $\Omega_{QQ}$.
\end{itemize}

\subsection{Baryons with One Heavy Quark $(Qqq)$}
The $\Lambda_b$ was the first bottom baryon observed and confirmed by several collaborations. 
The measurement of its mass was \cite{pdg}: $M_{\Lambda_b}=(5620.2\pm1.6)~\MeV$.
Only in 2007 - more than 15 years after the $\Lambda_b$ observation - the CDF collaboration 
\cite{TEV} has announced the observation of two other heavy baryons, the $\Sigma_b$ and 
$\Sigma_b^*$ baryons, in the decay channels $\Sigma_b^{(*)\pm}\to\Lambda^0\pi^\pm$, 
with masses given in Table \ref{tab2}.
Following these discoveries, D0 collaboration has announced the observation of the baryon $\Xi_b^-$ 
in the decay channel $\Xi_b^-\to J/\psi\Xi^-$, with a mass \cite{TEV}: 
$M_{\Xi_b}^{(D0)}=(5774\pm11\pm15)\mbox{ MeV}$. This observation was confirmed by CDF 
collaboration, but with a slightly higher mass \cite{TEV}: 
$M_{\Xi_b}^{(CDF)}=(5792.9\pm2.5\pm1.7) \mbox{ MeV}$.

\begin{table}[h]
\caption{\small Baryon masses for $\Sigma_b^{(*)}$ observed by CDF collaboration.}
\label{tab2}       
\begin{center}
\begin{tabular}{cl}
\hline\noalign{\smallskip}
Baryon  & Mass (MeV) \\ 
\noalign{\smallskip}\hline\noalign{\smallskip}
$\Sigma_b^+$   &   $5807.8^{+2.0}_{-2.2}\pm1.7$ \\ 
$\Sigma_b^-$   &   $5815.2\pm1.0\pm1.7$ \\ 
$\Sigma_b^{*+}$   &   $5829.0^{+1.6+1.7}_{-1.8-1.8}$ \\ 
$\Sigma_b^{*-}$   &   $5836.4\pm2.0^{+1.8}_{-1.7}$ \\ 
\noalign{\smallskip}\hline
\end{tabular}
\end{center}
\end{table}

The CDF result is in excellent agreement with the theoretical predictions made by Karliner {\it et al.} 
\cite{RICHARD}, $M_{\Xi_b}=(5795\pm5) \mbox{ MeV}$, and Jenkins \cite{jen}, $M_{\Xi_b}=(5805.7\pm8.1)\mbox{ MeV}$.

The $\Omega_b^-$ baryon was observed for the first time by $D0$ collaboration in the decay 
channel $\Omega_b^-\to J/\psi\Omega^-$, with a mass \cite{d02}: 
$M_{\Omega_b}^{(D0)}=(6165\pm10\pm13)\mbox{ MeV}$.
This value is much bigger than expected \cite{RICHARD} from different theoretical calculations.
However, a new observation for $\Omega_b^-$ baryon mass announced by CDF 
collaboration \cite{cdf3}: $ M_{\Omega_b}^{(CDF)} = (6054.4 \pm 6.8 \pm 0.9) \MeV$, 
is in a better agreement with the theoretical predictions presented in Table \ref{tab3}.

\begin{table}[h]
\caption{\small Theoretical predictions for the $\Omega_b$ baryon mass.}
\label{tab3}       
\begin{center}
\begin{tabular}{lcl}
\hline\noalign{\smallskip}
Theoretical Model & Ref.  & Mass (MeV) \\ 
\noalign{\smallskip}\hline\noalign{\smallskip}
QCDSR & \cite{MARINA} &   $5820\pm230$ \\ 
& \cite{zhu} &   $6036\pm81$ \\ 
Lattice & \cite{LATT}   &   $6006\pm10\pm19$ \\ 
$1/N_c$ & \cite{JENKINS}   &   $6039.1\pm8.3$ \\ 
Quark Models & \cite{kar2}  &   $6052.1\pm5.6$ \\ 
\noalign{\smallskip}\hline
\end{tabular}
\end{center}
\end{table}

Preliminary measurements of the $\Omega_b^-$ mass by LHCb collaboration \cite{lhc} are in 
a good agreement with CDF \cite{cdf3} measurements. The combination of the LHCb and 
CDF results indicates a large discrepancy with the $D0$ result.

In the present work \cite{rnm}, one uses the DRSR approach to study SU(3)-mass splittings for 
the spin $1/2^+$ and $3/2^+$ heavy baryons. The results for the baryon mass predictions can be 
compared, in a near future, to those values obtained experimentally and will be a good test 
of the sum rules approach.

\subsection{Baryons with Two Heavy Quarks $(QQq)$}
So far, the baryon spectroscopy has only one experimental evidence for a baryon with two heavy 
quarks, the $\Xi_{cc}$, made by SELEX collaboration \cite{selex}: 
$M_{\Xi_{cc}} = 3519 \pm 1 ~\MeV$. With the growing technological capability of the particle 
accelerators around the world, highlighting LHCb collaboration, there is an enormous expectation 
that new results for hadronic physics will be announced in the next few years.

In the present work \cite{rnar}, one uses the DRSR approach to predict the masses of these new 
spin $1/2^+$ and $3/2^+$ heavy baryons as well as their respective SU(3)-mass splittings.

\cleardoublepage

\chapter{QCD Sum Rules}\label{chap:QCDSR}
The QCD Sum Rules \cite{svz} consist of an analytical method in hadronic physics
that parameterizes the non-perturbative aspects of the QCD vacuum, in terms of the 
so-called quark and gluon condensates, in an attempt to explain various properties 
of the ground-state hadrons.

The main object in QCDSR is the two-point correlation function, which contains the 
hadronic current determined by the quantum numbers and the correct quark content 
in the structure of the hadron to be studied.
Comparing two possible descriptions of the correlation function one obtains the 
hadronic properties as follows: describing it using the theoretical development of QCD 
perturbative, in terms of quarks and gluons fields, and the inclusion of non-perturbative 
contributions parameterized by the condensates. This description is commonly known 
as the QCD side of the sum rules.
Furthermore, it is possible to describe the correlation function in terms of a spectral 
function separately containing the ground-state contribution of the hadron from its 
resonances. Considering experimental data, phenomenological parameters and 
general properties of the local field theory, this description characterizes the 
Phenomenological side of the sum rules.

In general, the great advantage of QCDSR is to analytically calculate the various hadronic 
parameters, such as the quark mass, decay constants, form factors, magnetic moments 
and others. Besides, one has by construction a physical model that simultaneously 
considers the non-perturbative effects of QCD vacuum at low energies and also all the 
theoretical information from the QCD perturbative at high energies. However, there are some 
limitations on the accuracy of the QCDSR due to the approximations made in the correlation 
function on both sides of the approach: QCD and Phenomenological side. Therefore, each 
sum rule must be carefully studied in order to verify its region of applicability.

Throughout the following sections, one discuss in more detail the most relevant properties 
of QCDSR to the present work, identifying the necessary steps for calculate the hadron masses.

\section{Correlation Function}

\begin{figure}[t] \vspace{-0.5cm}
  \begin{center}
  \includegraphics[width=14.0cm]{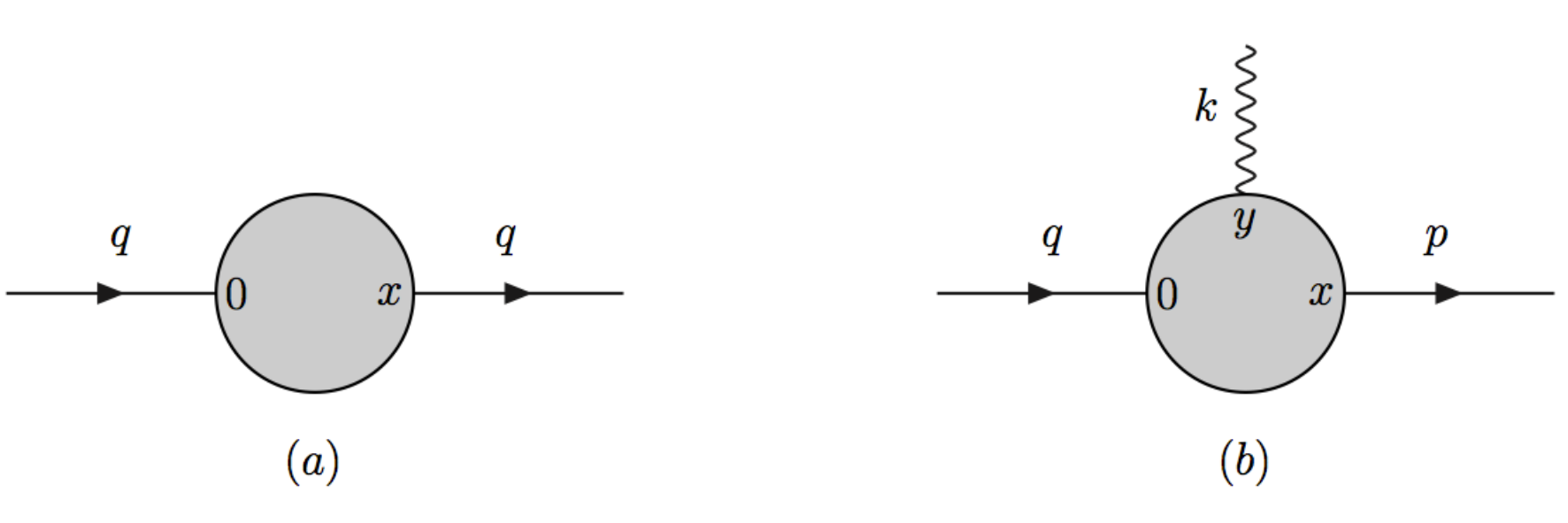}
  \caption{\footnotesize Representation of (a) the two-point and (b) the three-point correlation functions.}
  \label{diagfcorr}
  \end{center}
\end{figure}

In sum rules, the correlation function is defined as the functional expectation value of a 
time-ordered product of $n$-field operators at different positions. Particularly, the two-point 
correlation function, which contains the time-ordered product of two hadronic currents, is 
useful to estimate the hadron masses and decay constants. It is related to the 
amplitude of Fig.(\ref{diagfcorr}a) and can be evaluated through the expression:

\begin{equation}
  \Pi(q) = i \int d^4x ~e^{iq \cdot x}~ 
  \langle 0| \:T[\:j(x)\:j^\dagger(0)]\: | 0 \rangle \:
  \label{2point}
\end{equation}
where $q$ is the total four-momentum and $j(x)$ is the hadronic current determined by the 
quantum numbers and by the appropriate quark content in the structure of the hadron to be studied.

The calculation of the form factors and decay widths of a vertex interaction between three 
scalar particles $X$, $Y$ and $Z$ is done using the three-point correlation function shown 
in Fig.(\ref{diagfcorr}b) and whose expression is given by
\begin{equation}
  \Phi(q,k,p) = \int d^4x \int d^4y ~e^{ip \cdot x}~ e^{ik \cdot y}
  \langle 0| \:T[\:j_{_X}(x)\:j_{_Y}(y)\:j_{_Z}^\dagger(0)]\: | 0 \rangle \: ~~,
  \label{3point}
\end{equation}
where $p$ and $k$ are the loop four-momenta leaving the diagram, respectively, at the 
points $x$ and $y$. This expression is quite useful to study the decay channels  such as 
$Z \rightarrow X \:Y$.

Since Eqs.(\ref{2point}) and (\ref{3point}) could be used to study vector hadrons, 
one should separately analyze the Lorentz structures present in the correlation functions of 
both mesons and baryons.

For vector mesons described by a conserved current $j_\mu(x)$, the Lorentz structures of 
the correlation function to be considered are:
\begin{equation}
  \Pi_{\mu\nu}(q) = i \int d^4x ~e^{iq \cdot x}~ 
  \langle 0| \:T[\:j_\mu(x)\:j^\dagger_\nu(0)]\: | 0 \rangle \: = 
  - \left( g_{\mu\nu} - \frac{q_\mu q_\nu}{q^2} \right) \Pi_1(q^2) 
  \label{fc1}
\end{equation}
where $\Pi_1(q^2)$ is an invariant function and must contain the appropriate quantum numbers 
of the vector meson. The conservation of the current implies that the correlation function 
(\ref{fc1}) has only a transversal component in its structure. On the other hand, if the current is 
not conserved, one longitudinal component associated with a scalar state emerges, so that:
\begin{equation}
  \Pi_{\mu\nu}(q) = - 
  \left( g_{\mu\nu} - \frac{q_\mu q_\nu}{q^2} \right) \Pi_1(q^2) + \frac{q_\mu q_\nu}{q^2} \Pi_0 (q^2)
  \label{fc10}
\end{equation}
where both invariant functions, $\Pi_1(q^2)$ and $\Pi_0(q^2)$, are independent and contain 
respectively the quantum numbers of the vector and scalar state. Once the complete expression 
of $\Pi_{\mu \nu}(q)$ is calculated, it is possible to determine the two invariant functions using the 
following expressions: 
\begin{eqnarray}
 \label{proj1}
 \Pi_1(q^2) &=& - \frac{1}{3} \left( g^{\mu\nu} - \frac{q^\mu q^\nu}{q^2} \right) \Pi_{\mu\nu}(q)  \\
 \Pi_0(q^2) &=& \frac{ q^\mu q^\nu}{q^2} \:\Pi_{\mu\nu}(q) ~.
 \label{proj2}
\end{eqnarray}
Therefore, the two-point correlation function described by a non-conserved current allows to 
estimate the mass and decay constant of two states at the same time, one vector and other 
scalar state.

For the spin $1/2^+$ and $3/2^+$ baryons, the expressions are respectively given by
\begin{eqnarray}
  \label{fcbarion}
  \Pi^{_{1/2}}(q) &=& \slashed{q} F_1(q^2) + F_2(q^2) ~, \\
  \label{fcbarion*}
  \Pi^{_{3/2}}_{\mu\nu}(q) &=& g_{\mu\nu} \bigg[ \slashed{q} F_1(q^2) + F_2(q^2) \bigg] + \cdots ~,
\end{eqnarray}
where one must calculate both invariant functions $F_1(q^2)$ and $F_2(q^2)$.
In the case of spin $3/2^+$ baryons, many works in sum rules \cite{nlo} demonstrate that the 
structure $g_{\mu \nu}$ of the correlation function is enough for providing information of the 
ground-state baryons.

As previously discussed, in QCDSR the correlation function can be described in two ways. 
In the QCD side, one calculates the correlation function with the appropriate hadronic current 
in terms of the quark and gluon fields and the condensates.
In the Phenomenological side, one calculates it using experimental data, such as masses 
and decay constants, and parameterize the strength of the coupling between the current and 
all states of the hadron. Finally, the results for both descriptions can be compared in order to 
estimate the physical observables of the hadron.

\section{Local Operator Product Expansion}
In the QCD side, the calculation of the correlation function is based on the Local Operator Product 
Expansion (OPE) of the quark and gluon fields. This technique was introduced by K.G. Wilson in 
Ref.\cite{wilson}, where the OPE would describe analytically the complex structure of the QCD vacuum. 
Its application to the correlation function (\ref{2point}) is done in such a way that
\begin{equation}
  i \int d^4x ~e^{iq \cdot x}~ \langle 0| \:T[\:j(x)\:j^\dagger(0)]\: | 0 \rangle ~=~
  \sum_{d} C_{d}(q^2) \: \langle 0| {\hat{O}}_{d}(0) |0 \rangle
  \label{OPEexpansion}
\end{equation}
where the coefficients $C_d(q^2)$ are the so-called Wilson coefficients, which contain the 
effects of the QCD perturbative, at high energies. The vacuum expectation values (VEV) of the local 
operators ${\hat{O}}_d(0)$ parameterize the non-perturbative effects of the QCD vacuum
in terms of the condensates. The $d$-index represents the dimension of the local operator.
In QCDSR, the dimension-zero operator is given by ${\hat{O}}_0 = \hat{1}$ and the coefficient
$C_0 (q^2)$ represents the contributions from the QCD perturbative.
Considering only the lowest dimension local operators in the OPE, one obtains:
\vspace{-0.5cm}
\begin{eqnarray}
   \hat{O}_{3} &=& :\!\bar{q}(0) q(0)\!: ~~\equiv~~ \bar{q}q \nno \\
   \hat{O}_{4} &=& :\!g_s^2 \:G^{N}_{\alpha\beta}(0) G^{N}_{\alpha\beta}(0)\!: ~~\equiv~~ 
   	g_s^2 G^2 \nno \\
   \hat{O}_{5} &=& :\!\bar{q}(0) \: g_s \:\sigma^{\alpha\beta} G^{_N}_{\alpha\beta}(0)\: q(0)\!: 
   	~~\equiv~~ \bar{q}Gq \nno \\
   \hat{O}^q_{6} &=& :\!\bar{q}(0)q(0) \: \bar{q}(0) q(0)\!: ~~\equiv~~ \bar{q}q\bar{q}q \nno \\
   \hat{O}^G_{6} &=& :\!f_{_{NMK}} g_s^3 \:G^{N}_{\alpha\beta}(0) 
   	G^{M}_{\beta\gamma}(0) G^{K}_{\gamma\alpha}(0)\!: ~~\equiv~~ g_s^3 G^3
\end{eqnarray}
where the symbol $: :$ represents the normal ordering of the operators, $q(0)$ is the quark field, 
$G^{N}_{\alpha\beta}(0)$ is the gluon field tensor, $f_{_{NMK}}$ is the structure constant of 
the SU(3) group and 
$\sigma_{\alpha\beta}=\frac{i}{2} \left[ \gamma_{\alpha}, \gamma_{\beta} \right]$.
The VEV of these local operators \vspace{-0.5cm}
\begin{equation}
  \langle 0| \:\hat{O}_d\: |0 \rangle ~,
\end{equation}

\noindent gives rise to expressions for the quark condensate $\qq[q]$, gluon condensate 
$\GG$, mixed condensate $\qGq[q]$, four-quark condensate $\qqqq[q]$ and 
the triple gluon condensate $\GGG$. In general, considering the contributions 
up to dimension-six is enough for calculating the hadronic parameters.
Discussions on the condensates of higher dimension ($d > 6$), the difficulties to 
introduce them properly into the QCDSR and also their relevance to the OPE 
convergence can be found in Refs.\cite{rry, colan, niko}. \vspace{-0.2cm}

The Wilson's OPE is useful to transform non-local operators, $q_a(x) \: \bar{q}_b(0)$, 
into a sum of local operators $\hat{O}_d$, whose vacuum expectation values form the 
condensates. One obtains the main contributions when considering only the light quark field 
operators ($q=u,d,s$). In general, the contributions of the condensates formed by heavy 
quark field operators: $\langle \bar{Q}Q \rangle$, $\langle \bar{Q}GQ \rangle$, $\dots$ 
are negligible and do not contribute to the QCDSR. However, the relevance of these 
condensates should be studied carefully, especially in cases where the light quark 
condensates are missing in hadronic structure, which make them the principal source of 
non-perturbative effects of QCD vacuum.
By convention, hereafter the quark field operators $q_a(x)$ stand for light quarks and 
$Q_a(x)$ stand for heavy quarks. \vspace{-0.2cm}

\begin{figure}[t] \vspace{-1.0cm}
  \begin{center}
  \includegraphics[width=14.5cm]{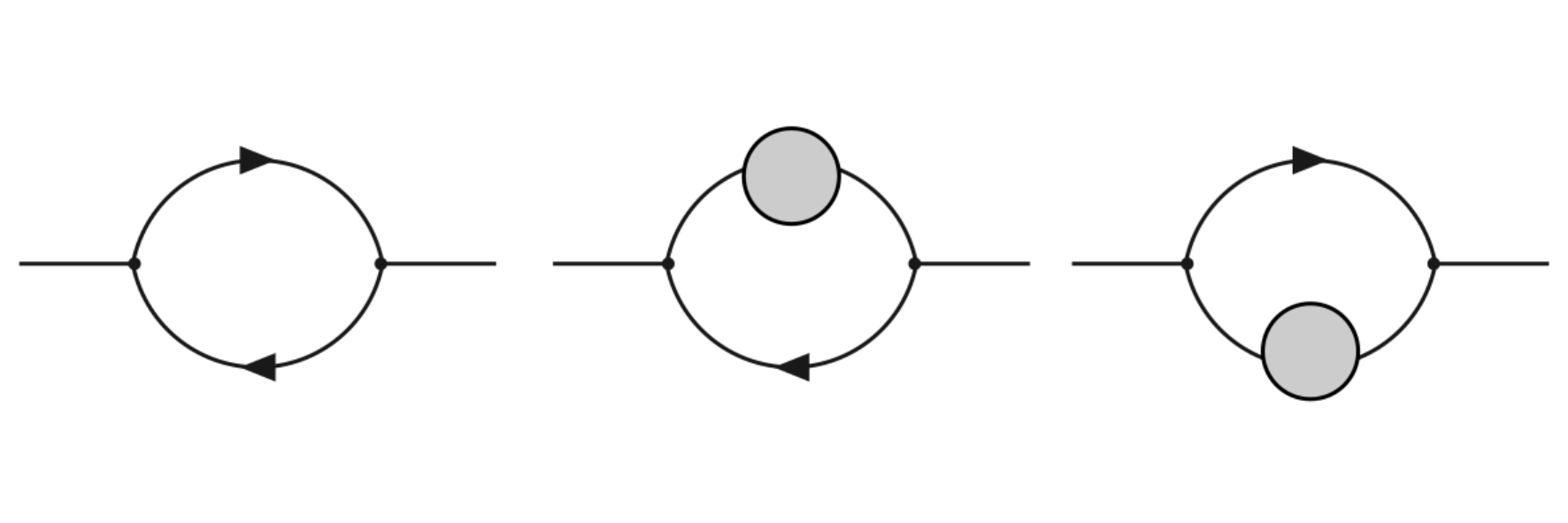}
  \vspace{-0.8cm}
  \caption{\footnotesize Diagrams related to the $\Pi^{(0)}(x)$ function of the scalar current,
  $j(x) = \bar{q}_a(x) q_a(x)$, where the first diagram is the perturbative contribution 
  and the other diagrams represent the non-perturbative effects from the QCD vacuum. 
  The gray blobs give rise to the condensates formed by the VEV of the quark fields: 
  $\langle 0| : q_{a}(x) \bar{q}_{b}(0) : |0 \rangle$.}
  \label{qq:DiagPERT}
  \end{center}  
\end{figure}

One uses the Eq.(\ref{OPEexpansion}) to calculate the contribution of leading order 
contributions to the OPE. However, it is possible to rewrite it in a more general form, 
where the contributions of the radiative corrections in the OPE can be introduced.
For this purpose, one should consider the following equation for the correlation function, 
in the QCD side \cite{pascual}:
\begin{equation}
  \Pi(q) = i \int d^4x ~e^{i q\cdot x} ~\langle 0| \:T [\:j(x)\:j^{\dagger}(0) ~
  e^{i \int \!d^4y \:{\cal L}_{_{int}}(y)} ]\: |0 \rangle
  \label{fcorr2}
\end{equation}
where the interaction Lagrangian between quarks and gluons ${\cal L}_{int}$ is given by
\begin{equation}
  {\cal L}_{_{int}}(y) = t^{_N}_{cd} \:g_s \:\bar{q}_c(y) \gamma^\alpha A^{N}_\alpha(y) q_d(y) ~~,
  \label{lint}
\end{equation}
the latin indices $(a,b,c,\ldots)$ represent the color charge of quark and gluon fields, 
the greek indices $(\alpha, \beta, \gamma, \ldots)$ represent the spinorial indices, 
$t^{_N}_{cd}$ are Gell-Mann matrices, $g_s^2 = 4\pi \al_s$ is the QCD strong coupling constant, 
$\gamma^\alpha$ are the Dirac matrices, $A^{N}_\alpha(y)$ the gluon fields and 
$N=(1,\ldots,8)$ are the SU(3) group generators. Expanding in series the exponential term 
contained in the VEV of Eq.(\ref{fcorr2}), one obtains
\vspace{-0.3cm}
\begin{eqnarray}
  \Pi(q) &=& \!\!\sum^{\infty}_{n=0} \frac{i^{n+1}}{n!} \!\!\int \!d^4x_0 \:d^4x_1 \ldots d^4x_n 
  ~e^{i q\cdot x_0} ~\langle 0| T[ j(x_0)\:j^{\dagger}(0) \:{\cal L}_{_{int}}(x_1) \ldots 
  {\cal L}_{_{int}}(x_n) ] |0 \rangle ~~~~~~~
  \label{seriesfc}
\end{eqnarray}
Explicitly, the first terms of this expansion are given by
\vspace{-0.3cm}
\begin{eqnarray}
  \Pi(q) &=& i \int \!d^4x ~e^{i q\cdot x} \Bigg\{
  \langle 0| T[ j(x)\:j^{\dagger}(0)] |0 \rangle + i \int \!d^4y~
  \langle 0| T[ j(x)\:j^{\dagger}(0) \:{\cal L}_{_{int}}(y)] |0 \rangle \nno\\
  && - \frac{1}{2} \iint \!d^4y \:d^4z ~\langle 0| T[ j(x)\:j^{\dagger}(0) 
  \:{\cal L}_{_{int}}(y) \:{\cal L}_{_{int}}(z)] |0 \rangle ~+~ \ldots \Bigg\} \nno\\
  &\equiv& i \int \!d^4x ~e^{i q\cdot x} \Bigg\{
  \Pi^{(0)}(x) + \Pi^{(1)}(x) + \Pi^{(2)}(x) + \ldots \Bigg\} ~~,
  \label{nlofc}
\end{eqnarray}
where the definition of $\Pi^{(n)}(x)$ functions is settled. As expected, Eq.({\ref{2point}}) is the 
leading order term of the expansion (\ref{nlofc}). The other terms include the non-perturbative 
contributions from the higher dimension condensates in the OPE and the radiative 
corrections up to $\mathcal{O}(g_s^2)$, as well.
To illustrate how the correlation function (\ref{nlofc}) is written in terms of the OPE, 
consider the scalar current: $j(x) = \bar{q}_a (x) q_a(x)$. Inserting into the VEV of the first 
term of the expansion (\ref{nlofc}), one obtains the time-ordered product of four quark fields.
Applying Wick's theorem, this may be expressed as:
\begin{eqnarray}
  \Pi^{(0)}(x) &=& \langle 0| T\left[ \bar{q}_{a, i}(x) q_{a, i}(x) 
  	\bar{q}_{b, j}(0) q_{b, j}(0) \right] |0 \rangle \nno\\
  &=& -~ \langle 0_p| T\left[ q_{a, i}(x) \bar{q}_{b, j}(0) \right] |0_p \rangle ~
  	\langle 0_p| T\left[ q_{b, j}(0) \bar{q}_{a, i}(x) \right] |0_p \rangle ~+ \nno\\
  && -~ \langle 0_p| T\left[ q_{b, j}(0) \bar{q}_{a, i}(x) \right] |0_p \rangle ~
  	\langle 0| : q_{a, i}(x) \bar{q}_{b, j}(0) : |0 \rangle ~+ \nno\\
  && -~ \langle 0_p| T\left[ q_{a, i}(x) \bar{q}_{b, j}(0)\right] |0_p \rangle ~
  	\langle 0| : q_{b, j}(0) \bar{q}_{a, i}(x) : |0 \rangle ~+ \nno \\
  && +~ \langle 0| : \bar{q}_{a, i}(x) q_{a, i}(x) \bar{q}_{b, j}(0) q_{b, j}(0) : |0 \rangle \nno \\
  &=& - \Tr \left[ S^0_{ab}(x) S^0_{ba}(-x) \right] - \Tr \left[
	{\cal S}^{q\bar{q}}_{ab}(x) S^0_{ba}(-x) \right] - \Tr \left[ 
  	S^0_{ab}(x){\cal S}^{q\bar{q}}_{ba}(-x) \right] + \ldots ~~~~~~~~
  \label{vac0}
\end{eqnarray}
where $\ldots$ represents terms which are proportional to the higher order condensates 
(e.g. the four-quark condensate $\qqqq[q]$ contribution). From now on, these kinds of 
contributions will be omitted. The notation $|0_p \rangle$ represents the perturbative 
vacuum and $|0 \rangle$ represents the physical QCD vacuum, $S^0_{ab}(x)$ is the 
perturbative propagator for free quarks defined as: 
\begin{equation}
  S^{0}_{ab}(x) ~=~ i\:\delta_{ab} \int\! \frac{d^4p}{(2\pi)^4} \left( 
  \frac{\slashed{p} + m_q}{p^2 - m^2_q + i\epsilon} \right) ~e^{-ip \cdot x}
\end{equation}
and ${\cal S}^{q\bar{q}}_{ab}(x)$ is the non-perturbative correction to the quark propagator and 
its expression is given by
\begin{eqnarray}
  {\cal S}^{q\bar{q}}_{ab}(x) &=& \langle 0| : q_{a}(x) \bar{q}_{b}(0) : |0 \rangle ~~.
\end{eqnarray}
For each trace in Eq.(\ref{vac0}), one has corresponding diagrams represented in 
Fig.(\ref{qq:DiagPERT}). If the QCD vacuum could be treated perturbatively, the VEVs of 
normal ordering of quark fields in Eq.(\ref{vac0}) would be zero and the time-ordered products 
of fields would be reduced to the usual perturbative propagators of the quantum field theory. 
However, the fact that the contributions of these VEVs are not zero introduces 
non-perturbative effects from the QCD vacuum to the sum rules.

Similarly, one does the same calculation for the other terms of the expansion (\ref{nlofc}). 
Keeping only the most relevant VEVs for the OPE and using the interaction Lagrangian 
(\ref{lint}), one obtains the following expression for the $\Pi^{(1)}(x)$:
\begin{eqnarray}
  \Pi^{(1)}(x) &=& i \:t^{_N}_{cd} ~ \int \!d^4y
  	\langle 0| T\left[ \bar{q}_a(x) q_a(x) \:\bar{q}_b(0) q_b(0) 
  	\:\bar{q}_c(y) \gamma^\alpha g_s A^{N}_{\alpha}(y) q_d(y) \right] |0 \rangle \nno\\
  &=& -i \:t^{_N}_{cd} ~ \int \!d^4y \bigg\{
  	\Tr \left[ S^0_{bc}(-y) \:\gamma^\alpha \:S^0_{da}(y-x)
	\:\langle 0| \!:\! q_a(x) g_s A^{N}_{\alpha}(y) \bar{q}_b(0) \!:\! |0 \rangle \right] ~+ \nno\\
  && +~ \:\Tr \left[ S^0_{ac}(x-y) \:\gamma^\alpha \:S^0_{db}(y)
  	\:\langle 0| \!:\! \bar{q}_a(x) g_s A^{N}_{\alpha}(y) q_b(0) \!:\! |0 \rangle \right] \bigg\}
  \label{pi1}
\end{eqnarray}
note that any other contraction of quark fields results in disconnected diagrams 
and, therefore, do not contribute to $\Pi^{(1)}(x)$. 

\begin{figure}[t] \vspace{-0.5cm}
  \begin{center}
  \includegraphics[width=13.5cm]{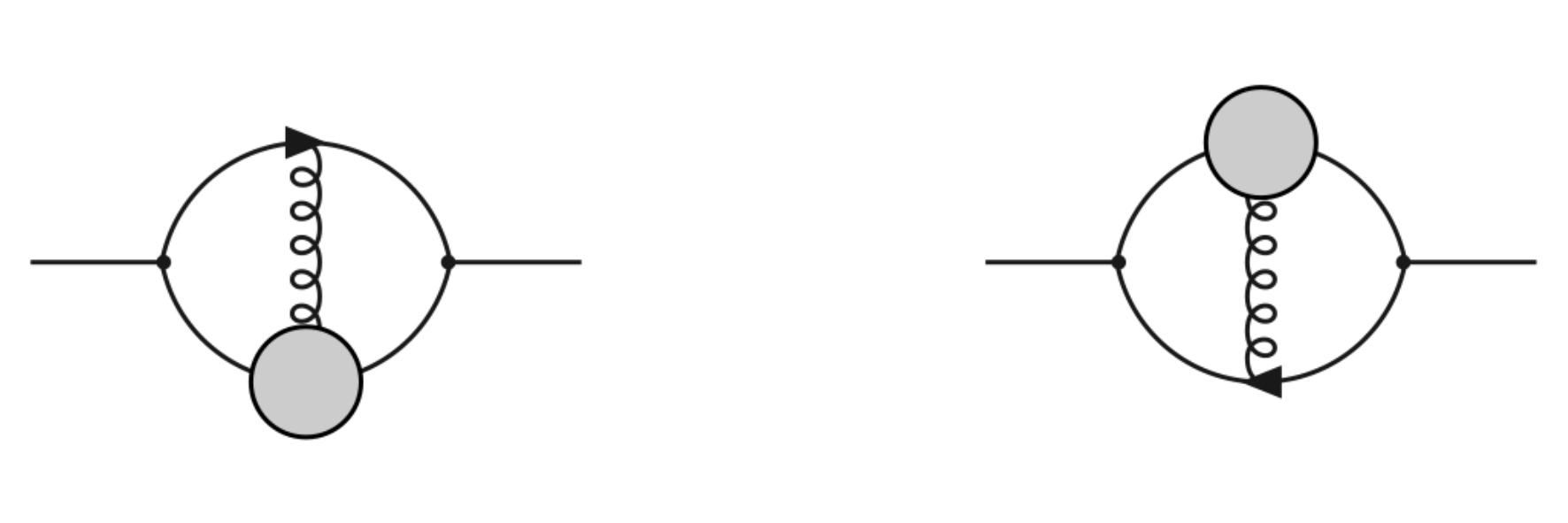}
  \vspace{-0.5cm}
  \caption{\footnotesize Diagrams related to the $\Pi^{(1)}(x)$ function of the scalar current 
  $j(x) = \bar{q}_a(x) q_a(x)$, where the gray blobs represent the non-perturbative 
  effects from the QCD vacuum, linked to the condensates formed by the VEV of the 
  quark and gluon fields: $\langle 0| : q_{a}(x) \bar{q}_{b}(0) : |0 \rangle$.}
  \label{qq:DiagQGQ}
  \end{center}  
\end{figure}

For the gluon field $A^N_\alpha(y)$, contained within the VEV of Eq.(\ref{pi1}), 
it is convenient to work with the fixed-point gauge or the Fock-Schwinger gauge \cite{fock}:
\begin{eqnarray}
  y^\alpha A^N_\alpha(y) = 0 ~.
  \label{fockschw}
\end{eqnarray}
so that, at the low-energy regime, it is possible to approximate the gluon field 
$A^N_\alpha(y)$ in terms of the totally antisymmetric gluon field strength tensor :
\vspace{-0.3cm}
\begin{equation}
  A^N_\alpha(y) \simeq -\frac{1}{2} \:G^N_{\alpha\beta}(0) \:y^\beta ~~.
  \label{AtoG}
\end{equation}
which helps in calculating the integrals of Eq.(\ref{pi1}). 
Finally, the $\Pi^{(1)}(x)$ is rewritten in the Fock-Schwinger gauge as:
\begin{eqnarray}
  \Pi^{(1)}(x) &=& \frac{i \:t^{_N}_{cd}}{2} ~ \int \!d^4y \bigg\{
  	\Tr \left[ S^0_{bc}(-y) \:\gamma^\alpha y^\beta \:S^0_{da}(y-x)
	\:\langle 0| \!:\! q_a(x) g_s G^{N}_{\alpha\beta}(0) \bar{q}_b(0) \!:\! |0 \rangle \right] ~+ \nno\\
  && +~ \:\Tr \left[ S^0_{ac}(x-y) \:\gamma^\alpha y^\beta \:S^0_{db}(y)
  	\:\langle 0| \!:\! \bar{q}_a(x) g_s G^{N}_{\alpha\beta}(0) q_b(0) \!:\! |0 \rangle \right] \bigg\} \nno\\
  &=& \Tr \left[ {\cal S}^{q\bar{q}G^N_{\al\be}}_{ab}(x) {\cal S}^{G_N^{\al\be}}_{ba}(-x) \right] +
   	 \Tr \left[ {\cal S}^{G_N^{\al\be}}_{ab}(x) {\cal S}^{q\bar{q}G^N_{\al\be}}_{ba}(-x) \right]
  \label{vac1}
\end{eqnarray}
where the following definitions are introduced:
\begin{eqnarray}
  \label{SG}
  {\cal S}^{G_N^{\al\be}}_{ab}(x) &=& \frac{i \:t^{_N}_{cd}}{2} ~ \int \!d^4y 
  \:S^0_{ac}(x-y) \:\gamma^\alpha y^\beta \:S^0_{db}(y) \\
  \label{SqqG}
  {\cal S}^{q\bar{q}G^N_{\al\be}}_{ab}(x) &=& 
  \langle 0| \!:\! \bar{q}_a(x) g_s G^{N}_{\alpha\beta}(0) q_b(0) \!:\! |0 \rangle
\end{eqnarray}
Observe that Eq.(\ref{SG}) is associated with the quark propagation through space and, 
in a certain point, it emits a non-perturbative gluon, $G_N^{\al\be}$. Since, this gluon field 
is defined in the Fock-Schwinger gauge, hence the label ``non-perturbative gluon''. In 
turn, Eq.(\ref{SqqG}) receives this emitted gluon forming the condensate from the VEV of 
the quark and gluon fields. The diagrams are represented in Fig.(\ref{qq:DiagQGQ}), 
where they are associated to each trace in Eq.(\ref{vac1}) 

Finally, one calculates the $\Pi^{(2)}(x)$ function of the expansion (\ref{nlofc}). Then:
\begin{eqnarray}
  \Pi^{(2)}(x) &=& -\frac{t^{N}_{cd} \:t^{M}_{ef}}{2} ~ \iint \!d^4y \:d^4z
      	\:\langle 0| T\big[ \bar{q}_a(x) q_a(x) \:\bar{q}_b(0) q_b(0) \:\bar{q}_c(y) \nno\\
  && \times~ \gamma^\alpha g_s A^{N}_{\alpha}(y) \:q_d(y) \bar{q}_e(z) 
  	\:\gamma^\beta g_s A^{M}_{\beta}(z) \:q_f(z) \big] |0 \rangle \nno\\ 
  &=& \frac{t^{N}_{cd} \:t^{M}_{ef}}{2} \iint \!d^4y \:d^4z ~
  g_s^2 \:\langle 0| T\left[ A^N_\alpha(y) \:A^M_\beta(z) \right] |0 \rangle  \nno\\
  && \times~ \bigg\{ \Tr \left[ S^0_{ba}(-x) \:S^0_{ae}(x-z) \:\gamma^\beta \:S^0_{fc}(z-y) 
  	\:\gamma^\alpha \:S^0_{db}(y) \right] \nno\\
  && +~ \:\Tr \left[ S^0_{ab}(x) \:S^0_{bc}(-y) \:\gamma^\alpha \:S^0_{de}(y-z) 
  	\:\gamma^\beta \:S^0_{fa}(z-x) \right] \nno\\
  && +~ \:\Tr \left[ S^0_{be}(-z) \:\gamma^\beta \:S^0_{fa}(z-x) \:S^0_{ac}(x-y) 
		\:\gamma^\alpha \:S^0_{db}(y) \right]  \bigg\}
  \label{vac2}
\end{eqnarray} \vfill

\begin{figure}[t]
  \begin{center}
  \includegraphics[width=15.5cm]{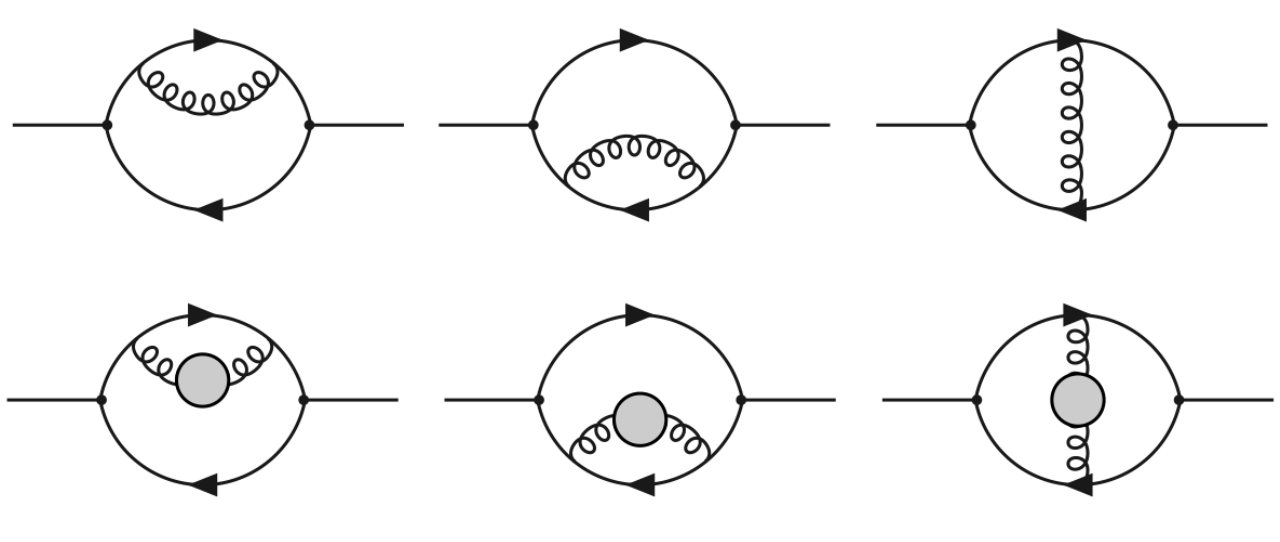}
  \caption{\footnotesize Diagrams related to the $\Pi^{(2)}(x)$ function of the scalar current 
  $j(x) = \bar{q}_a(x) q_a(x)$. The first three diagrams provide the NLO contributions, 
  given by the $\Pi^{(2)}_{NLO}$ function. The last ones, which contain the gray blobs, 
  provide the non-perturbative effects from the QCD vacuum, related to the condensates 
  formed by the VEV of the gluon fields: 
  $\langle 0| \!:\! g_s^2 G^N_{\alpha\rho}(0) \:G^M_{\beta\lambda}(0) \!:\! |0 \rangle$.}
  \label{qq:DiagGG}
  \end{center}  
\end{figure}

Using the definition of the time-ordered product of the gluon fields, one can deduce:
\vspace{-0.3cm}
\begin{eqnarray}
  g_s^2 \:\langle 0| T\left[ A^N_\alpha(y) \:A^M_\beta(z) \right] |0 \rangle
  &=& g_s^2 \:\langle 0_p| T\left[ A^N_\alpha(y) \:A^M_\beta(z) \right] |0_p \rangle +
  \langle 0| \!:\! g_s^2 A^N_\alpha(y) \:A^M_\beta(z) \!:\! |0 \rangle \nno \\
  &\equiv& \de_{NM} \:g_s^2 \: S^{GG}_{\alpha\beta}(y \!-\! z) + 
  \frac{y^\rho z^\lambda}{4} \:\langle 0| \!:\! g_s^2 G^N_{\alpha\rho}(0) 
  \:G^M_{\beta\lambda}(0) \!:\! |0 \rangle ~~~~~~~~
  \label{fullG}
\end{eqnarray}
Note that the expression $S^{GG}_{\alpha\beta}(y-z)$ is the perturbative gluon propagator, 
which is well defined in the coordinate space and the Feynman gauge, so that:
\begin{equation}
  S^{GG}_{\alpha\beta}(y-z) ~=~ - i \int\! \frac{d^4p}{(2\pi)^4} \left( 
  \frac{g_{\alpha\beta}}{p^2 + i\epsilon} \right) ~e^{-ip \cdot (y-z)} ~.
\end{equation}
Therefore, the VEV in Eq.(\ref{fullG}) contains the non-perturbative contributions of the 
gluon fields and, for this reason, it should be expressed in the Fock-Schwinger gauge 
(\ref{fockschw}). Inserting the definition (\ref{fullG}) into Eq.(\ref{vac2}), one obtains:
\begin{eqnarray}
  \Pi^{(2)}(x)
  &\!\!=\!\!& \Pi^{(2)}_{NLO}(x) + \Tr \left[ {\cal S}^{\GGi}_{ab}(x) S^0_{ba}(-x) \right] +
  \Tr \left[ S^0_{ab}(x) {\cal S}^{\GGi}_{ba}(-x) \right] \nno\\
  && +~ \frac{1}{2} \langle 0| \!:\! g_s^2 G^N_{\alpha\rho}(0) \:G^M_{\beta\lambda}(0) \!:\! |0 \rangle
  \:\Tr \left[ {\cal S}^{G_N^{\al\rho}}_{ab}(x) \:{\cal S}^{G_M^{\be\lambda}}_{ba}(-x) \right] ~~~~~~~~~
  \label{piGG}
\end{eqnarray}
where the definition of the non-perturbative correction to the gluon propagator was introduced 
and is given by
\begin{eqnarray}
  {\cal S}^{\GGi}_{ab}(x) &=& \frac{t^{N}_{cd} \:t^{M}_{ef}}{8} 
  \:\langle 0| \!:\! g_s^2 G^N_{\alpha\rho}(0) \:G^M_{\beta\lambda}(0) \!:\! |0 \rangle \nno\\
  && \times~ \iint \!d^4y \:d^4z ~y^\rho \: z^\lambda ~S^0_{ae}(x-z) \:\gamma^\beta \:S^0_{fc}(z-y) 
  	\:\gamma^\alpha \:S^0_{db}(y) ~~.
\end{eqnarray}
For simplicity, the $\Pi^{(2)}_{NLO}(x)$ function associated with the radiative corrections is 
not expressed explicitly in Eq.(\ref{vac2}) since, in the present work, the sum rules will be only 
evaluated with the leading order term in $\al_s$. The diagrams, related to each term of 
Eq.(\ref{vac2}), are shown in Fig.(\ref{qq:DiagGG}).

In quantum field theory, the calculation of a general fermion loop diagrams corresponds to 
evaluating traces of these fermion propagators. Then, notice that the $\Pi^{(n)}(x)$ functions, 
given by Eqs.(\ref{vac0}), (\ref{vac1}) and (\ref{vac2}), contain exactly the traces of the perturbative 
propagators as well as the non-perturbative propagators obtained from the VEVs. 
In general, it is possible to demonstrate that, for any hadronic current $j(x)$ used in QCDSR, the 
expression for $\Pi^{(n)}(x)$ function is always the same. Therefore, now it is convenient to 
introduce the definition of the full propagator of QCD, given by
\begin{eqnarray}
  {\cal S}^{QCD}(x) &=& S^0_{ab} + S^{GG}_{\al\be}(x) + {\cal S}^{q\bar{q}}_{ab}(x) + 
  {\cal S}^{G^{\al\be}_N}_{ab}(x) + {\cal S}^{q\bar{q}G^N_{\al\be}}_{ab}(x) + 
  {\cal S}^{\GGi}_{ab}(x) ~~,
  \label{FullProp}
\end{eqnarray}
where each term of this propagator gives the most relevant diagrams for a QCDSR calculation.

A detailed analysis of the calculation of the full propagator of QCD is presented in 
Appendix \ref{app:Propagators}. 
In summary, the expressions of non-perturbative light quark propagators, in the 
coordinate space, are given by
\begin{eqnarray}
  \label{qqexp}
  {\cal S}^{q\bar{q}}_{ab}(x)
  &=& - \frac{\de_{ab}}{12} ~\qq[q]  + \frac{i \:\de_{ab} \:\slashed{x}}{48} ~m_q \qq[q] 
  - \frac{\de_{ab} \:x^2}{192}\qGq[q] + \frac{i \:\de_{ab} \:x^2 \slashed{x}}{1152} ~m_q \qGq[q] + 
  \ldots ~~~~~~~~~ 
\end{eqnarray}

\begin{eqnarray}  
  \label{Gexp}
  {\cal S}^{G^{\al\be}_N}_{ab}(x) &=& - \frac{i \:t^N_{ab}}{32 \pi^2 \:x^2} ~ 
  ( \sigma_{\alpha\beta} \:\slashed{x} + \slashed{x} \:\sigma_{\alpha\beta} ) + \ldots \\
  \label{qGqexp}
  {\cal S}^{q\bar{q}G^N_{\al\be}}_{ab}(x) 
  &=& - \frac{t^{N}_{ab} \:\sigma_{\al\be}}{192} ~\qGq[q] -\frac{i \:t^N_{ab}}{768}
  ( \sigma_{\alpha\beta} \:\slashed{x} + \slashed{x} \:\sigma_{\alpha\beta} ) ~m_q \qGq[q]  
  + \ldots ~~~~ \\
  \label{GGexp}
  {\cal S}^{\GGi}_{ab}(x) &\simeq& 0 ~~.
\end{eqnarray}

In the case of non-perturbative heavy quark propagators, the heavy quark condensate 
contribution $\qq[Q]$ and the heavy mixed condensate $\qGq [Q]$ are given by 
\cite{BaganHQ}:
\begin{eqnarray}
  \label{QQ}
  \qq[Q] &\simeq& -\frac{1}{48 \pi^2 \:m_Q} \:\GG \\
  \label{QGQ}
  \qGq[Q] &\simeq& -\frac{5}{96 \pi^2 \:m_Q} \:\GGG ~.
\end{eqnarray}
As one can see, both condensate contributions are suppressed by the presence of the 
heavy quark mass, $m_Q$, in the denominator of Eqs.(\ref{QQ}) and (\ref{QGQ}), in such 
a way these contributions can be neglected during the calculation of the full propagator of 
QCD for heavy quarks. Therefore, the most relevant non-perturbative contribution involving 
heavy quarks comes from the gluon condensate (\ref{GGexp}). 
The expressions of the non-perturbative heavy quark propagators, in the momentum space, 
are given by
\begin{eqnarray}
  {\cal S}^{Q\bar{Q}}_{ab}(p) &\simeq& 0 \\
  {\cal S}^{G^{\al\be}_N}_{ab}(p) &=& - \frac{i \:t^N_{ab}}{4 (p^2 - m_Q^2)^2} ~ 
  \big[ \sigma_{\alpha\beta} (\slashed{p} + m_Q) + (\slashed{p} + m_Q) \sigma_{\alpha\beta} \big] \\
  {\cal S}^{Q\bar{Q}G^N_{\al\be}}_{ab}(p) &\simeq& 0 \\
  {\cal S}^{\GGi}_{ab}(x) &=& \frac{i \:\de_{ab} \:m_Q}{12(p^2 - m_Q^2)^3}
  \left[ 1 + \frac{m_Q(\slashed{p} + m_Q)}{p^2-m_Q^2} \right] ~\GG ~~.
\end{eqnarray}

\vfill

\section{QCD Side}
After choosing the hadronic current of interest, evaluating the OPE up to desired dimension and 
working with the most relevant condensates in the sum rules the next step is to express the 
correlation function (\ref{2point}) in terms of a dispersion relation:
\begin{equation}
  \Pi^{_{OPE}}(q) = \int\limits^{+\infty}_{t_q} \! ds \: \frac{\rho^{_{OPE}}(s)}{s - q^2} 
  \label{PiOPE}
\end{equation}
where $t_q$ is a kinematic limit, which usually corresponds to the square of 
the sum of the current quark masses of the hadron. The OPE spectral density, 
$\rho^{_{OPE}}(s)$, is defined in such a way the perturbative and condensate contributions are 
inserted through the following relation:
\vspace{-0.5cm}
\begin{equation}
  \rho^{_{OPE}}(s) ~=~ {\cal K}_0(s) ~+~ {\cal K}_{3}(s) \:\qq[q] ~+~ {\cal K}_{4}(s) \:\GG  ~+~ 
  {\cal K}_{5}(s) \:\qGq[q] ~+~ \ldots ~~~
  \label{rhoOPE}
\end{equation}
where: \vspace{-0.5cm}
\begin{eqnarray}
  {\cal K}_d(s) \equiv \frac{1}{\pi} \mbox{Im} [C_d(q^2)] ~.
\end{eqnarray}
There are some advantages of expressing the correlation function in terms of a 
dispersion relation. With that, one can do a better comparison between the QCD side 
and Phenomenological side of the sum rules and separate the ground-state contribution 
from the one related to the continuum.

\section{Phenomenological Side}
In the Phenomenological side, the correlation function (\ref{2point}) is evaluated in terms of 
hadronic parameters and the current is interpreted as the creation and annihilation operators
of a hadron. Using the definition of the time-ordered product, it is possible to rewrite the 
correlation function as:
\begin{equation}\label{fenpto}
  \Pi(q) = i \int d^4x ~e^{iq \cdot x}~ \left\{ 
  \langle 0 | \:\theta(x_0) \:j(x) j^\dagger(0)\: | 0 \rangle ~+~
  \langle 0 | \:\theta(-x_0) \:j^\dagger(0) j(x)\: | 0 \rangle \right\}
\end{equation}
Once more, for simplicity, one considers the scalar current $j(x) = \bar{q}_a(x)q_a(x)$.
Assuming that the hadrons $H(p) \equiv H_p$ created by the current $j(x)$ form a complete 
set of hadronic states, so it is possible to use an unitary projector through the completeness 
relation of these states: 
\begin{equation}
  \sum_{H_p} \frac{1}{(2\pi)^3} \int\! \frac{d^3 \vec{p}}{2p_0} ~ | \:H_p \: \rangle \langle \:H_p \: | = \hat{1} ,
\end{equation}
where the sum evolves the ground-state hadron and all its resonant states. Introducing 
this projector between the currents $j(x)$ and $j^\dagger(0)$ in both terms of Eq.(\ref{fenpto}), 
one obtains: \vspace{-0.5cm}
\begin{eqnarray}
  \Pi(q) &\!=\!& i \int\!\! d^4x  \:e^{iq \cdot x} ~\sum_{H_p} \int\! \frac{d^3 \vec{p}}{(2\pi)^3\: 2p_0} \bigg[ 
  \theta(x_0) \langle 0 | j(x) | H_p \rangle \langle H_p | j^\dagger(0) | 0 \rangle \nno \\
 && +~ \theta(-x_0) \langle 0 | j^\dagger(0) | H_p \rangle \langle H_p | j(x) | 0 \rangle \bigg]
 \label{fenA}
\end{eqnarray}
Expressing the current $j(x)$ in terms of a translation operator, $\hat{U} = {e^i p \cdot x}$, 
one gets:
\begin{equation}
  j(x) = e^{ip \cdot x} ~j(0)~ e^{-ip \cdot x}
\end{equation}
Inserting into Eq.(\ref{fenA}) the correlation function becomes:
\vspace{-0.5cm}
\begin{eqnarray}
  \Pi(q) &\!=\!& i \int\!\! d^4x  \:e^{iq \cdot x} ~\sum_{H_p}  
  \int\! \frac{d^3 \vec{p}}{(2\pi)^3 \:2p_0} \bigg[ 
  \langle 0 | j(0) | H_p \rangle \langle H_p | j^\dagger(0) | 0 \rangle 
  \:\theta(x_0) ~e^{-ip \cdot x} \nno \\
 && +~ \langle 0 | j^\dagger(0) | H_p \rangle \langle H_p | j(0) | 0 \rangle 
 \:\theta(-x_0)  ~e^{ip \cdot x} \bigg] \nno\\
 &=& i \!\int\!\! d^4x ~e^{i q \cdot x} ~\sum_{H_p} \!\!\int\! \frac{d^3 \vec{p}}{(2\pi)^3 \:2p_0} \bigg[
  \:\theta(x_0) ~e^{-i p \cdot x} + \theta(-x_0)  ~e^{i p \cdot x} \bigg]
  \left| \langle 0| j(0) |H_p \rangle \right|^2 ~~~.
 \label{fenB}  
\end{eqnarray}
Notice that it is possible to identify the Feynman propagator in Eq.(\ref{fenB}):
\begin{equation}
  \Delta_F(x) = \int\! \frac{d^3 \vec{p}}{(2\pi)^3 \:2p_0} \bigg[
  \:\theta(x_0) ~e^{-i p \cdot x} + \theta(-x_0)  ~e^{i p \cdot x} \bigg] ~=~
  i \int\! \frac{d^4p}{(2\pi)^4} \frac{e^{-i p \cdot x}}{p^2-E_H^2 + i \epsilon}
\end{equation}
where $E_H$ is the energy associated with the $| H_p \rangle$ state. Therefore, one gets
\begin{eqnarray}
 \Pi(q) &=& - \int\!\! d^4x ~e^{i q \cdot x} ~\sum_{H_p} \int\!\! \frac{d^4p}{(2\pi)^4} 
 \frac{e^{-i p \cdot x}}{p^2-E_H^2 + i \epsilon} \left| \langle 0| j(0) |H_p \rangle \right|^2 ~~~.
 \label{fen2}
\end{eqnarray}
Introducing the following identity into the Eq.(\ref{fen2}),
\begin{equation}
  \int\limits_0^{+\infty} \!ds ~\de(s - E^2_H) = 1 ~~,
\end{equation}
where the integral is well defined on the entire spectrum of the hadron $H$, one finally obtains 
the result:
\begin{eqnarray}
 \Pi(q) &=& - \int\limits^{+\infty}_0 \!\!ds \int\!\! d^4x ~e^{i q \cdot x} ~\sum_{H_p} \int\!\! \frac{d^4p}{(2\pi)^4} 
 \frac{e^{-i p \cdot x}}{p^2 - s + i \epsilon} \left| \langle 0| j(0) |H_p \rangle \right|^2 \de(s - E_H^2) \nno\\
 &=& - \int\limits^{+\infty}_0 \!\!ds ~\sum_{H_p} \int\!\! d^4p
 \frac{1}{p^2 - s + i \epsilon} \left| \langle 0| j(0) |H_p \rangle \right|^2 ~\de(s - E_H^2)  ~\de(q-p) \nno\\
 &=& \int\limits^{+\infty}_0 \!\!ds 
 \frac{1}{s - q^2 - i \epsilon} ~\sum_{H_q} \left| \langle 0| j(0) |H_q \rangle \right|^2 \de(s - E_H^2)\nno\\
 &\equiv& \int\limits^{+\infty}_0 \!\!ds 
 \frac{\rho(s)}{s - q^2 - i \epsilon}
 \label{Fen}
\end{eqnarray}
which contains the definition for the spectral density:
\begin{equation}
  \rho(s) = \sum_{H_q} \left| \langle 0| j(0) |H_q \rangle \right|^2 \de(s - E_H^2)  ~~.
  \label{rhofen1}
\end{equation}
One has to consider that, phenomenologically it is expected that the spectral density could be 
described by one ground-state, $H_0$, plus the continuum contribution formed by the 
resonant states, $H^\prime$. From the experimental data, one verifies that the resonant states 
only contribute to the spectral density after a certain point called as continuum threshold.
As one can see, this description could be obtained directly from the 
Eq.(\ref{rhofen1}) as follows:
\begin{eqnarray}
  \rho(s) &=& \left| \langle 0| j(0) |H_0 \rangle \right|^2 \:\de(s-M_H^2) ~+~ 
  \sum_{H^\prime} \left| \langle 0| j(0) |H^\prime \rangle \right|^2 \de(s - E_{H^\prime}^{2}) \nno\\ 
  &\equiv& \lambda^2 \:\de(s-M_H^2) ~+~ \theta(s-t_c) \:\rho^{cont}(s) ~~,
\end{eqnarray}
where $\lambda$ is the decay constant defined as the coupling between the current and the 
ground-state, $M_H$ is the ground-state mass and $t_c$ is the continuum threshold.
For values above this threshold, a very useful approximation of the continuum contribution to the 
spectral density is given by
\begin{equation}\label{so}
  \rho^{cont}(s) \simeq \rho^{_{OPE}}(s) ~~,
\end{equation}
where $\rho^{_{OPE}}(s)$ is exactly the spectral density given in the QCD side, see 
Eq.(\ref{rhoOPE}). As discussed in the previous section, the reason for using this 
approach relies on the quark-hadron duality principle. Thus,
\begin{equation}\label{rhofen}
  \rho(s) = \lambda^2 \:\de(s-M_H^2) ~+~ \theta(s-t_c) \:\rho^{_{OPE}}(s) ~.
\end{equation}
Finally, using this expression in Eq.(\ref{Fen}), one obtains the correlation function in the 
Phenomenological side:
\begin{equation}
  \Pi^{_{PHEN}}(q) = \frac{\lambda^2}{M_H^2 - q^2} ~+~ 
  \int\limits^{+\infty}_{t_c} \! ds \:\frac{\rho^{_{OPE}}(s)}{s - q^2} ~,
  \label{PiFen}
\end{equation}
which will be compared with the correlation function on the QCD side.

\section{Quark-Hadron Duality Principle}
In the QCDSR, the correlation function is commonly used to describe the properties of the 
ground-state hadrons. Since the hadrons are intrinsically connected to the low-energy physics, 
only the perturbative expansions in the QCD side - represented by the Wilson coefficients 
$C_d(q^2)$ - does not properly describe the correlation function $\Pi^{_{OPE}}(q)$. 
In this scenario, perhaps the approach in Eq.(\ref{so}) is not the most appropriate to be used in 
the Phenomenological side.
However, in QCDSR, one must consider the so-called Quark-Hadron Duality principle, 
which suggests the existence of an energy scale (usually in order of $1 \GeV$), where the 
hadronic spectrum is equivalently described by the correlation function in the QCD and 
Phenomenological sides, such that:
\begin{eqnarray}
  \Pi^{_{OPE}}(q) &\simeq& \Pi^{_{PHEN}}(q) ~.
  \label{duality}
\end{eqnarray}
Among many reasons, the most relevant ones that prevent Eq.(\ref{duality}) from be exact are 
the truncation of the OPE in QCD side and the crude approximation (\ref{rhofen}) to the 
spectral density in the Phenomenological side. Therefore, for extracting reliable results 
from the comparison between the two correlation functions, one should establish criteria 
which guarantee a good OPE convergence in the QCD side and simultaneously suppress 
the contributions of resonant states in the Phenomenological side.
A practical way of doing this is by using the so-called Borel transform, whose definition 
is given by \cite{svz, rry, colan}:
\begin{equation}	
  {\cal B}\! \left[ g(Q^2) \right] \equiv g(\tau) = 
  \lim\limits_{\tiny \begin{matrix} Q^2, n \rightarrow \infty \\ n/Q^2 = \tau \end{matrix}}
  \frac{(-1)^n (Q^2)^{n+1}}{n!} \left( \frac{\partial}{\partial Q^2} \right)^n \!g(Q^2)
\end{equation}
where $Q^2$ is the four-momentum of the particle in the Euclidean space $(Q^2 = -q^2)$. 
and $\tau$ is a free parameter of the sum rule. Consider, as an example, the Borel transform 
of the important functions below:

\begin{eqnarray}
	\label{borel1}
	{\cal B} \left[ (Q^2)^k \right] &=& 0 \\
	\label{borel2}
	{\cal B} \left[ \frac{(Q^2)^k}{(s + Q^2)^m} \right] &=& 
	\sum_{j=0}^{K} \frac{(-1)^{k+j} \:k!}{j!(m-j-1)!(k-j)!} \:s^{k-j} \:\tau^{m-j-1}  \:e^{-s \tau}
\end{eqnarray}
where $k,m$ are positive integer numbers, $s$ is an arbitrary variable (independent of 
$Q^2$) and the upper index $K$ in the series is given by
\begin{eqnarray}
  K &=& 
  \begin{cases}
    m-1, & \text{if } k \geq m\\
    k, & k < m
  \end{cases}
\end{eqnarray}
In the QCD side, the renormalization process introduces subtraction terms into (\ref{PiOPE}), 
which give rise to polynomials in $Q^2$. According to Eq.(\ref{borel1}), these terms are 
eliminated after Borel transform. Furthermore, the Wilson coefficients of the higher dimension 
condensates are proportional to terms like $1/(Q^2)^m$, where $m$ grows with the dimension 
of the operator. Then, by using Eq.(\ref{borel2}), one verifies that the contributions from higher 
dimension condensates are factorially suppressed. This is an indication that Borel transform 
helps to improve the OPE convergence.
In the Phenomenological side, the correlation function (\ref{PiFen}) is proportional to the term
$1/(s+Q^2)$. After using Borel transform on this term, one obtains:
\begin{equation}
	{\cal B} \left[ \frac{1}{s+Q^2} \right] = e^{-s \tau} 
\end{equation}
and the contributions from resonant states are exponentially suppressed.
Therefore, ensuring that the OPE contributions from the higher dimension condensates and 
the resonant states are removed, the comparison between both descriptions of the 
correlation function, in the QCD and Phenomenological sides, allows one to obtain reliable 
results for the properties of the ground-state hadrons.

\section{Evaluating the Mass in QCDSR}
Applying Borel transform in (\ref{PiOPE}) and (\ref{PiFen}), one obtains respectively:
\vspace{-0.4cm}
\begin{eqnarray}
  \label{borelope}
  \Pi^{_{OPE}}(\tau) &=& \int\limits^{+\infty}_{t_q} \! ds \: \rho^{_{OPE}}(s)~e^{-s \tau} \\
%
\label{borelfen}
  \Pi^{_{PHEN}}(\tau) &=& \la^2 ~e^{-M_{_H}^2 ~\tau} + 
  \int\limits^{+\infty}_{t_c} \! ds \:\rho^{_{OPE}}(s) ~e^{-s \tau} ~.
\end{eqnarray}
where $M_{_H}$ is the hadron mass. Thus, according to the quark-hadron duality
principle, the expressions for the correlation function (\ref{borelope}) and (\ref{borelfen}) 
now can be compared:
\vspace{-0.4cm}
\begin{equation}
  \la^2 ~e^{-M_{_H}^2 \:\tau} + \int\limits^{+\infty}_{t_c} \! ds \:\rho^{_{OPE}}(s) ~e^{-s\tau} =
  \int\limits^{+\infty}_{t_q} \! ds \: \rho^{_{OPE}}(s) ~e^{-s\tau}
\end{equation}	
considering that $\rho^{_{OPE}}(s)$ is continuous over the integration interval, then:
\vspace{-0.4cm}
\begin{eqnarray}\label{dual}
  \la^2 ~e^{-M_{_H}^2 \:\tau} + \int\limits^{+\infty}_{t_c} \! ds \:\rho^{_{OPE}}(s) ~e^{-s\tau} &=&
  \int\limits^{t_c}_{t_q} \! ds \: \rho^{_{OPE}}(s) ~e^{-s\tau} + \int\limits^{+\infty}_{t_c} \! ds 
  \: \rho^{_{OPE}}(s) ~e^{-s\tau} \nno \\
  \la^2 ~e^{-M_{_H}^2 \:\tau} &=& \int\limits^{t_c}_{t_q} \! ds \:\rho^{_{OPE}}(s) ~e^{-s\tau}	~.
\vspace{-0.4cm}
\end{eqnarray}
Taking the derivative, in both sides of the Eq.(\ref{dual}), with respect to $\tau$, one gets:
\vspace{-0.3cm}
\begin{equation}\label{deriv}
  \la^2 M_{_H}^2 ~e^{-M_{_H}^2 \:\tau} = \int\limits^{t_c}_{t_q} \! ds \:s\:\rho^{_{OPE}}(s) ~e^{-s\tau}	~.
\end{equation}
Dividing (\ref{deriv}) by (\ref{dual}), one finally obtains:
\vspace{-0.3cm}
\begin{equation}\label{massa}
  M_{_H}^2 = \frac{\int\limits^{t_c}_{t_q} \! ds 
  \:s\:\rho^{_{OPE}}(s) ~e^{-s\tau}}{\int\limits^{t_c}_{t_q} \! ds \:\rho^{_{OPE}}(s) ~e^{-s\tau}}
\end{equation}
which is the sum rule equation to determine the hadron mass.

\subsection{Borel Window}
\begin{figure}[t]
  \begin{center}
  \includegraphics[width=12.2cm]{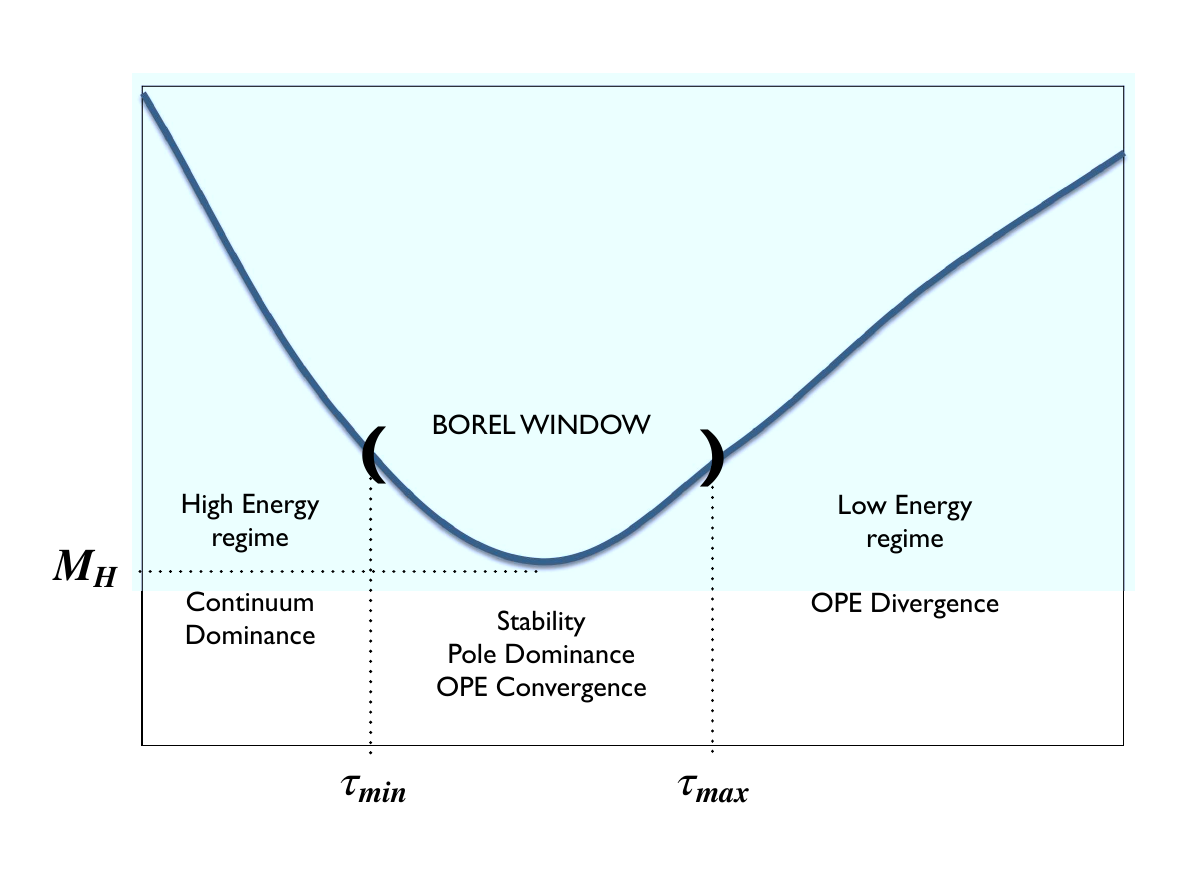}
  \caption{\footnotesize This figure shows the curve expected for the mass calculated using 
  the QCDSR. By definition, the Borel window is the region which contains $\tau$-stability, 
  pole dominance and a good OPE convergence. The point that defines the beginning of the
  pole dominance fixes the lower limit of Borel window, $\tau_{min}$. While the point, which 
  one can no longer guarantee a good OPE convergence, fixes the upper limit of Borel 
  window, $\tau_{max}$. One estimates the hadron mass, $M_H$, from a $\tau$-stability 
  region inside the Borel window.}
  \label{stability}
  \end{center}
\end{figure}

\subsubsection{Pole Dominance}
To extract the information on ground-state hadrons, it is crucial to work in a 
$\tau$-stability region, where the pole contribution is bigger than the continuum contribution.
With that, one tries to guarantee that the most part of the contribution in the mass equation 
(\ref{massa}) comes directly from the ground-state. As previously discussed, the approximation for 
the spectral density establishes that the continuum contributions vanishes below a 
certain value, the continuum threshold $t_c$. Therefore, defining the total contribution 
$\Pi_{cont}(\tau)$ as:
\begin{eqnarray}
	\Pi_{total}(\tau)	&=& 	\int\limits^\infty_{t_q} \! ds \:\rho^{_{OPE}}(s) ~e^{-s \tau} ~,
	\label{pitotal}
\end{eqnarray}
it is possible to separate the pole contributions, $\Pi_{pole}(\tau)$, from the continuum ones, 
$\Pi_{cont}(\tau)$. So that, the integral (\ref{pitotal}) can be rewritten as:
\begin{eqnarray}
	\Pi_{total}(\tau) &=&
	\int\limits^{t_c}_{t_q} \! ds \:\rho^{_{OPE}}(s) ~e^{-s \tau} + 
	\int\limits^{\infty}_{t_c} \! ds \:\rho^{_{OPE}}(s) ~e^{-s \tau} \nno\\
	&=& \Pi_{pole}(\tau) ~+~ \Pi_{cont}(\tau) ~~.
\end{eqnarray}

Note that the $\tau$ parameter is proportional to the inverse of the energy. Then, considering small 
values for $\tau$, it would correspond to the high-energy regime where the continuum contribution 
dominates. In order to avoid this region, one usually sets a lower bound in the $\tau$ space, 
$\tau_{min}$, beyond which one obtains the pole dominance.

\subsubsection{OPE Convergence}
Considering high values for $\tau$, which means working at the low-energy regime, the truncated 
OPE no longer provides a reasonable description for the ground-state hadron since the 
non-perturbative effects become extremely significant. In this region, others higher dimension 
condensates should be included during the sum rule calculation. One naively expects that fixing 
an upper bound in the $\tau$ space, $\tau_{max}$, it could be possible obtain a good OPE 
convergence.

Also note that, the Borel transform takes the terms proportional to the negative powers in 
$Q^2$ and transforms them into terms depending on increasing powers in $\tau$ parameter 
instead. This could be a strong indication that Borel transform improves the OPE convergence.
In general, one defines the $\tau_{max}$ value where the contribution of the higher dimension 
condensate in the OPE is smaller than $10\%$ to $25\%$ of the total contribution. 

Therefore, if the following condition is satisfied: $\tau_{min} < \tau_{max}$, it is possible to 
set a region where the QCDSR results are reliable. This region is the so-called Borel window. 
On the other hand, if the Borel window cannot be defined, then the pole dominance and OPE 
convergence are no longer guaranteed in a QCDSR calculation and the result obtained for 
the hadron mass is not reliable. 

In most cases, the OPE convergence is obtained only considering the OPE contributions 
up to dimension-five condensates. Then, one should analyze if the values $\tau_{min}$ and 
$\tau_{max}$ establish a good Borel window for a QCDSR calculation.

\subsubsection{$\tau$-Stability}
One expects that the hadron mass has a certain stability due to the free choice of $\tau$ 
parameter inside the Borel window. Then, Borel windows with a large $\tau$-instability 
could indicate that the obtained hadron mass is not reliable, and more improvements must 
be done for these QCDSR calculations. 
Sometimes, the inclusion of more condensates in the OPE could help to obtain improved  
$\tau$-behavior. Another good argument for $\tau$-stability comes from the sum rules applied 
to the harmonic oscillator \cite{pascual}, where the $\tau$-stability point on the curve of the 
ground-state mass, obtained with QCDSR, is the closest point to the exact value obtained 
from Quantum Mechanics. The qualitatively $\tau$-behavior of a QCDSR mass calculation 
is shown in Fig.(\ref{stability}).

\section{Finite Energy Sum Rules (FESR)}
There is also another compelling way to calculate hadron masses, known as Finite Energy 
Sum Rules.
The FESR are obtained by performing the expansion around $\tau = 0$ in Eq.(\ref{deriv}):
\begin{equation}
  \la^2 M_{_H}^2 ~\sum\limits^{+\infty}_{n=0} \frac{(-1)^n \tau^n}{n!} M_{_H}^{2n} = 
  ~\sum\limits^{+\infty}_{n=0}\frac{(-1)^n \tau^n}{n!}  \int\limits^{t_c}_{t_q} \! ds \:s^n\:\rho^{_{OPE}}(s)	~.
\end{equation}
Matching the polynomial coefficients in $\tau$, on both sides of the equation, one obtains $n$ 
equations:
\begin{equation}
  \la^2 M_{_H}^{2n+2} = \int\limits^{t_c}_{t_q} \! ds \:s^n\:\rho^{_{OPE}}(s)	~,
\end{equation}
with $n=(0,1,2,...)$. Finally, dividing two subsequent equations $n$ and $n+1$, one 
obtains the mass equation:
\begin{equation}
  M_{_H}^2 = \frac{\int\limits^{t_c}_{t_q} \! ds 
  \:s^{n+1} \:\rho^{_{OPE}}(s)}{\int\limits^{t_c}_{t_q} \! ds \:s^n\:\rho^{_{OPE}}(s)} ~.
  \label{mfesr}
\end{equation}

In general, the sum rules depend on the definition of the continuum threshold $t_c$ 
and the $\tau$ parameter. Note, however, that the FESR have an advantage over QCDSR 
since its result depends only on $t_c$ and no longer on $\tau$ parameter. Thus, one expects  
that the results obtained with the FESR could provide a more direct relationship between 
$M_{_H}$ and $t_c$, thereby decreasing the arbitrariness in determining these values. 
In principle, one also expects for a stability region for the $M_{_H} (t_c)$ function, which 
would consist in a good criterion for fixing the value of $t_c$ and $M_{_H}$, as well. 
However, this stability is not always reached in the FESR.

\section{Double Ratio of Sum Rules (DRSR)}
As already observed in this Chapter, the correlation function for baryons contains two 
invariant functions under Lorentz transformations, $F_1(q^2)$ and $F_2(q^2)$. 
In the QCD side, after applying Borel transform, the correlation function for 
baryons is given by
\begin{eqnarray}
  \Pi^{_{OPE}}(\tau) &=& {1\over\pi} \int\limits_{t_q}^{+\infty}ds~ e^{-s\tau}~ \bigg[ 
    \slashed{q} \:{\rm Im}F_{1}(s) ~+~ {\rm Im}F_{2}(s) \bigg],
  \label{bar1}
\end{eqnarray}
while in the Phenomenological side, one obtains:
\begin{eqnarray}
  \Pi^{_{PHEN}}(\tau) &=& (\slashed{q} + M_{_H}) \bigg[ \la^2 ~e^{-M_{_H}^2 ~\tau} + 
  \int\limits^{+\infty}_{t_c} \! ds \:\rho^{cont}(s)~e^{-s \tau} \bigg] ~~.
  \label{bar2}
\end{eqnarray}
Note that one can use the Dirac spinor sum relation for the spin $1/2^+$ baryons:
\begin{eqnarray}
  \sum_s u(q,s) \bar{u}(q,s) &=& \slashed{q} + M_{_H} 
\end{eqnarray}
and the Rarita-Schwinger sum relation for the spin $3/2^+$ baryons:
\begin{eqnarray}
  \sum_s u_\mu(q,s) \bar{u}_\nu(q,s) &=& (\slashed{q} + M_{_H}) \left(
  g_{\mu\nu} - \frac{1}{2}\gamma_\mu\gamma_\nu + 
  \frac{q_\mu \gamma_\nu- q_\nu \gamma_\mu}{3 M_{_H}} - \frac{2q_\mu q_\nu}{3 M_{_H}^2} \right)~.
\end{eqnarray}
Assuming now that, above the continuum threshold, the spectral densities in the 
Phenomenological side are given by the results obtained in the QCD side, then
\begin{eqnarray}
  (\slashed{q} + M_{_H}) \:\rho^{cont}(s) &=& 
  {1\over\pi} \bigg[ \:\slashed{q} \:{\rm Im} F_1(s) ~+~ {\rm Im} F_2(s) \bigg]  ~.
\end{eqnarray}
Using the quark-hadron duality principle and considering that the structures 
$\slashed{q}$ and $M_{_H}$ are independent of each other, it is possible to determine the 
following equations:
\begin{eqnarray}
\lambda^2~e^{-M_{_H}^2\tau} &=& {1\over\pi} \int\limits_{t_q}^{t_c}ds~e^{-s\tau}~{\rm Im}F_{1}(s)~, \\
\lambda^2 M_{_H} ~e^{-M_{_H}^2\tau} &=& {1\over\pi}\int\limits_{t_q}^{t_c}ds~ e^{-s\tau}~{\rm Im}F_{2}(s)~.
\label{dualbaryon}
\end{eqnarray}
From these relations, one determines three different ways to calculate the hadron mass:

\begin{eqnarray}
{\cal R}^q_i &=& {\int_{t_q}^{t_c}ds~s~e^{-s\tau}~{\rm Im}F_{i}(s)\over \int_{t_q}^{t_c}ds~
e^{-s\tau}~{\rm Im}F_{i}(s)}~,~~~~~(i=1,2)~,\\ &&\nno\\
{\cal R}^q_{21} &=& {\int_{t_q}^{t_c}ds~
e^{-s\tau}~{\rm Im}F_{2}(s)\over \int_{t_q}^{t_c}ds~
e^{-s\tau}~{\rm Im}F_{1}(s)}~,
\end{eqnarray}
where, at the $\tau$-stability point, one obtains:
\begin{equation}
  M_{_H} \simeq \sqrt{{\cal R}^q_i} \simeq {\cal R}^q_{21}~.  
\end{equation}
These equations are widely used in the calculations of baryon masses in sum rules. 
However, the results obtained with these equations lead to uncertainties in order of $15-20\%$ 
\cite{BAGAN, BAGAN2, BAGAN3}. The technique that could minimize these uncertainties 
is the Double Ratio of Sum Rules \cite{rnm, rnar, SNmad, SNDR, SNtcc}, which provides the 
baryon mass ratios through the equations:
\begin{equation}
  r^{sq}_i\equiv \sqrt{{\cal R}^s_i\over {\cal R}^q_i}~,~~~~~
  r^{sq}_{21} \equiv {{\cal R}^s_{21}\over {\cal R}^q_{21}} ~~,
  \label{DRSR}
\end{equation}
which contains the SU(3) symmetry breaking effects explicitly. As one can see through the 
following sections, these expressions are less sensitive to the choice of the heavy quark 
mass and to the value of the continuum threshold than the simple ratios 
${\cal R}_i$ and ${\cal R}_{21}$.

\cleardoublepage
\chapter{Charmonium and Bottomonium Exotic States}
The exotic structures, such as molecules, could be a possible explanation 
about the nature of the new states observed on charmonium spectroscopy, among them: 
$Y(3930)$, $Y(4140)$, $X(4350)$, $Y(4260)$, $Y(4360)$ and $Y(4660)$. There is a growing 
evidence that at least some of these states do not have conventional hadronic structures. It is 
remarkable that their masses and decay channels are not compatible with predictions from 
potential models for the conventional charmonium $(c \bar{c})$. 
Another interesting state is the $Y_b(10890)$, which also contains mass and decay channel 
incompatible with expected for a conventional bottomonium $(b \bar{b})$ state and could 
be an indication of new exotic states in this sector.
In the present work, one uses the QCD sum rule approach to test if these new observed 
states can be interpreted as molecular states.

\section{QCD Parameters}
For each exotic state studied in this section, one uses the numerical values for the 
QCD parameters listed in Table (\ref{TabParam}).

\begin{table}[hbt]
  \setlength{\tabcolsep}{1.5pc}
  \caption{\small QCD input parameters. For the heavy quark masses, one uses the range 
  spanned by the running $\overline{MS}$-scheme mass and the on-shell mass from QCDSR. 
  For a consistent comparison with other sum rule results, one considers the same values 
  used in the literature for the condensates and their respective ratios.}
  \begin{center}\vspace*{-1cm}
  \begin{tabular}{lll}
  &\\
  \hline \hline
  Parameters & Values & Refs.\\
  \hline \hline
  $m_b$ & $(4.17 - 4.70)$ \GeV & \cite{rnm, rnar, rnr, nnl, SNB, pdg, SNmad, SNmass, SNgg, SNHmass}\\
  $m_c$ & $(1.23 - 1.37)$ \GeV & \cite{rnm, rnar, rnr, nnl, SNB, pdg, SNmad, SNmass, SNgg, SNHmass}\\
  $\hat{m}_s$ & $(0.114 \pm 0.021)$ \GeV & \cite{rnm, rnar, rnr, nnl, SNB, pdg, SNmad, SNmass}\\ 
  $\qq[q]$ & $-(0.23\pm 0.03)^3 \GeV^3$& \cite{rnr, SNB, ricnar, nnl, morita, ricnav, NNN}\\
  $\GG$ & $(0.88 \pm 0.25)$ \GeV$^4$ & \cite{rnm, rnar, rnr, fesr, BB, SNTAU, SNB, SNgg, ricnar, nnl, morita, 
  ricnav, NNN, SNmass, LNT, SNI, YNDU, SNHmass, JAMI2, SNmad}\\
  $\GGG$ & $(0.58 \pm 0.18)$ \GeV$^6$ & \cite{SNB, SNmass, SNHmass, SNgg, SNmad}\\
  $\kappa \equiv \qq[s]/\qq[q]$ & $(0.74\pm 0.06)$ & \cite{rnr, rnm, rnar, SNmad}\\
  $m_0^2 \equiv \qGq[s]/\qq[s]$ & $(0.8 \pm 0.2)$ \GeV$^2$ & \cite{SNB, rnr, ricnar, nnl, morita, 
  ricnav, NNN, SNmass, JAMI2, HEID, rnm, rnar}\\ 
  $\rho \equiv \qqqq[s]/\qq[s]^2$ & $(0.5 - 2.0)$ & \cite{SNB, rnr, nnl, rnm, rnar, SNmad, 
  SNTAU, JAMI2, LNT}\\
  $\Lambda(n_f = 4)$ & $(324 \pm 15)$ \MeV & \cite{SNTAU, pdg, SNB, SNmad, rnm, rnar} \\
  $\Lambda(n_f = 5)$ & $(194 \pm 10)$ \MeV & \cite{SNTAU, pdg, SNB, SNmad, rnm, rnar} \\
  \hline \hline
  \end{tabular}
  \end{center}
  \label{TabParam}
\end{table}

\section{Molecular States}
\subsection{$D^\ast_s \bar{D}^\ast_s ~(0^{++})$}
The most recent acquisition for this list of peculiar states is the narrow structure $Y(4140)$ 
observed by the CDF Collaboration \cite{cdf09} in the decay  
$B^+ \rightarrow Y(4140) K^+ \rightarrow J/\psi \:\phi \:K^+$.
The mass and width of this structure are $M = (4143.0 \pm 2.9 \pm 1.2) \MeV$ and 
$\Gamma = (11.7^{+8.3}_{-5.0} \pm 3.7) \MeV$.

As discussed previously, the $Y(4140)$ state cannot be interpreted as a conventional 
$c \bar{c}$ state. Then, one uses the QCDSR to study if a two-point correlation function 
based on a $D^\ast_s \bar{D}^\ast_s$ molecular current, with $J^{PC} = 0^{++}$, could 
describe this new observed resonance structure.
The starting point for constructing a QCD sum rule to evaluate the mass of the ground-state 
hadron is to establish the current, $j(x)$, that contains all the information about the 
hadron of interest, like quantum numbers and quark contents. 
Then, a possible current that couples with a $D^\ast_s \bar{D}^\ast_s$ molecular state, 
with $J^{PC} = 0^{++}$, is given by 
\begin{equation}
  j_{_{D^\ast_s \bar{D}^\ast_s}}=(\bar{s}_a\gamma_\mu c_a)(\bar{c}_b\gamma^\mu s_b)
  \label{Jy4140}
\end{equation}
where $a$ and $b$ are color indices. Parameterizing the coupling of the scalar state, 
$Y \equiv D^\ast_s \bar{D}^\ast_s$, to the current (\ref{Jy4140}) in terms of the parameter 
$\lambda$:
\begin{equation}
  \langle 0| j_{_{D^\ast_s \bar{D}^\ast_s}} | Y \rangle = \lambda_{_{D^\ast_s \bar{D}^\ast_{s}}} ~,
\end{equation}
the correlation function in the Phenomenological side can be written as 
\begin{equation}
  \Pi^{_{PHEN}}(q^2) = {\lambda_{_{D^\ast_s \bar{D}^\ast_s}}^2\over
  M_Y^2-q^2}+\int\limits_{t_q}^\infty ds\, {\rho^{_{OPE}}_{_{D^\ast_s \bar{D}^\ast_s}}(s)\over s-q^2} ~.
  \label{FENy4140} 
\end{equation}
It is important to notice that there is no one-to-one correspondence between the current and the 
state since the current, in Eq.(\ref{Jy4140}), can be rewritten as a sum over tetraquark-like 
currents, by Fierz transformations. However, the parameter 
$\lambda_{_{D^\ast_s \bar{D}^\ast_{s}}}$ gives a measure of the strength of the coupling 
between the current and the state. 

\begin{figure}[t] \vspace{-0.4cm}
  \begin{center}
  \includegraphics[width=5.0cm]{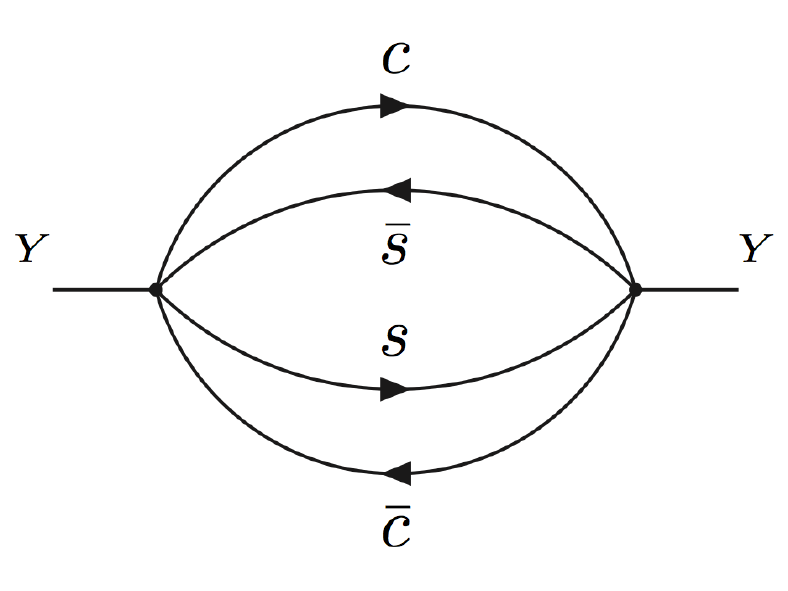}
  \caption{\footnotesize Representation of the perturbative contribution for 
  $D^\ast_s \bar{D}^\ast_s ~(0^{++})$ molecular state.}
  \label{Fig4140}
  \end{center}
  \vspace{-0.4cm}
\end{figure}

In the QCD side, inserting the current (\ref{Jy4140}) into the two-point correlation function, 
one obtains
\begin{eqnarray}
  \Pi^{_{OPE}}(q^2) &=& \frac{i}{(2\pi)^8} \int\! d^4x \:d^4p_1 \:d^4p_2 ~e^{ix \cdot (q-p_1-p_2)} ~ \nno\\
  && ~\times
  \Tr \left[ {\cal S}^c_{ab}(p_1) \ga_\nu {\cal S}^s_{ba}(-x) \ga_\mu \right] ~
  \Tr \left[ {\cal S}^s_{cd}(x) \ga^\nu {\cal S}^c_{dc}(-p_2) \ga^\mu \right] ~
\end{eqnarray}
where ${\cal S}^c(p)$ and ${\cal S}^s(x)$ are the propagators for the $c$- and $s$-quarks, respectively. 
One uses the light quark propagators in the coordinate-space while the heavy quark 
propagators are evaluated in the momentum-space, via Fourier transform:
\begin{equation}\label{hprop}
 {\cal S}^{c}_{ab}(x) ~=~ \int \frac{d^4{p_1}}{(2\pi)^4} ~e^{-i p_1 \cdot x}~ {\cal S}^{c}_{ab}(p_1) ~.
\end{equation}	

For simplicity, the molecule diagrams are constructed so that the two top lines represent the quark 
and antiquark propagators of a $D^\ast_s$ meson and the two lower lines represent the quark and 
antiquark of a $\bar{D}^\ast_s$ meson. An example of these diagrams can be seen in 
Fig.(\ref{Fig4140}).
To calculate all OPE contributions to the correlation function, one uses the expressions for the 
light/heavy propagators contained in Tables (\ref{tabSVZq}) and (\ref{tabSVZQ}).

For the scalar $\DsxDsx ~(0^{++})$ molecular state, the QCD sum rule calculation is done 
considering the OPE terms up to dimension-eight condensates, working at leading order in 
$\alpha_s$ in the operators and keeping terms which are linear in the strange quark mass. 
Finally, the contribution for each dimension in the OPE is calculated so that the expression 
for the spectral density $\rho^{_{OPE}}_{_{\DsxDsx}}(s)$ is given by 
\begin{equation}	
  \rho^{_{OPE}}_{_{\DsxDsx}}(s) = 
  \rho^{pert}_{_{\DsxDsx}}(s) + \rho^{\qq[s]}_{_{\DsxDsx}}(s) + \rho^{\GGi}_{_{\DsxDsx}}(s) + 
  \rho^{\qGq[s]}_{_{\DsxDsx}}(s) + \rho^{{\qq[s]}^2}_{_{\DsxDsx}}(s) + 
  \rho^{\GGG}_{_{\DsxDsx}}(s) + \rho^{\qq[s] \qGq[s]}_{_{\DsxDsx}}(s) ~~,
\end{equation}
where each term of the spectral density is explicitly shown below:
\begin{eqnarray}
  \rho^{pert}_{_{\DsxDsx}}(s)&=&{3\over 2^{9} \pi^6}\int\limits_{\almin}^{\almax}{d\al\over\alpha^3}
  \int\limits_{\bemin}^{1-\al}{d\be\over\be^3}(1-\al-\be) {\cal F}^{\:3}_{(\al,\be)}
  \left[{\cal F}_{(\al,\be)} - 4m_s m_c \beta \right], \\
  \rho^{\qq[s]}_{_{\DsxDsx}}(s) &=& {3\qq[s] \over 2^{5}\pi^4} 
  \int\limits_{\almin}^{\almax}{d\al\over\al}\Bigg\{
  {m_s {\cal H}^{\:2}_{(\al)} \over 1-\al} - m_c\int\limits_{\bemin}^{1-\al} \frac{d\be}{\al\be} {\cal F}_{(\al,\be)}
  \Big[ {\cal F}_{(\al,\be)} - 4m_s m_c \al \Big]\Bigg\}, \\
  \rho^{\GGi}_{_{\DsxDsx}}(s)&=&{m_c^2\GG \over2^{8}\pi^6}\int\limits_{\almin}^{
  \almax}{d\al\over\alpha^3}\int\limits_{\bemin}^{1-\al}{d\be}(1-\al-\be) {\cal F}_{(\al,\be)}, 
\end{eqnarray}

\begin{eqnarray}
  \rho^{\qGq[s]}_{_{\DsxDsx}}(s) &=& \frac{\qGq[s]}{2^{6} \pi^4} \Bigg\{
  m_s (8m_c^2 - s) \:v - 3m_c \int\limits^{\almax}_{\almin} \!\!\frac{d\al}{\al} \:{\cal H}_{(\al)} \Bigg\}, \\
  \rho^{\qq[s]^2}_{_{\DsxDsx}}(s)&=&{m_c \:\rho \qq[s]^2\over 8\pi^2}\bigg[ \left(
  2m_c-m_s\right) v - m_s m_c^2 \int\limits_0^1{d\al\over\al}~\delta\left(
  s-{m_c^2\over \al(1-\al)}\right)\bigg] ,\\ 
  \rho^{\GGGi}_{_{\DsxDsx}}(s)&=& \frac{\GGG}{2^{11}\pi^6}\int\limits_{\almin}^{
  \almax}\!\!{d\al\over\alpha^3}\int\limits_{\bemin}^{1-\al}\!\!\frac{d\be}{\be^3}
  (1 \!-\! \al \!-\! \be) \bigg[ 2m_c^2(\al^4 \!+\! \be^4) + (\al^3 \!+\! \be^3) {\cal F}_{(\al,\be)} \bigg], ~~~~ \\
  \rho^{\qq[s] \qGq[s]}_{_{\DsxDsx}}(s)&=&-{m_c \qq[s] \qGq[s] \over 96\pi^2}\int\limits_0^1
  \frac{d\al}{\al^2(1-\al)}
  \bigg\{ 12m_c \al \Big( \al(1-\al) + m_c^2 \:\tau \Big) ~+ \nno \\
  && -~ m_s \bigg[ 2\al(8 \!-\! 5\al) \Big( \al(1\!-\! \al)+ m_c^2 \:\tau \Big) + 
  5m_c^4 \:\tau^2 \bigg] \bigg\} ~\delta \!\left( \!s-{m_c^2 \over \al(1\!-\!\al)}\!\right) ~~~~~~
\label{eq:DsxDsx}
\end{eqnarray}
For simplicity, some variables are defined as follows:
\begin{eqnarray}
\addtolength{\fboxsep}{10pt} 
\boxed{ \begin{gathered} 
  \begin{array}{rcl} 
    {\cal F}_{(\al,\be)} &=& m_c^2(\al+\be) -\al \be s \\
    {\cal H}_{(\al)} &=& m_c^2 - \al(1-\al)s \\
    v &=& \sqrt{1-4m_c^2/s} \\
    {\cal L}_v &=& \displaystyle \Log \left( \frac{1+v}{1-v} \right)
  \end{array}
\end{gathered}} 
\label{def:variables}
\end{eqnarray}
and the integration limits are given by 
\begin{eqnarray}
\addtolength{\fboxsep}{10pt} 
\boxed{ \begin{gathered} 
  \begin{array}{rcl} 
    \almax &=& (1+v)/2 \\ 
    \almin &=& (1-v)/2 \\
    \be_{max} &=& \bemax \\ 
    \bemin &=& {\al m_c^2/( s\al-m_c^2)} ~~.  
  \end{array}
\end{gathered}} 
\label{def:limits}
\end{eqnarray}

\subsubsection{Numerical Results}
To evaluate the mass of the $\DsxDsx ~(0^{++})$ molecular state, one uses the Eq.(\ref{massa}) 
with the parameters given in Table (\ref{TabParam}).  

\begin{figure}[t]
    \hspace{-1.1cm}
    \subfloat[]{\includegraphics[width=7.5cm]{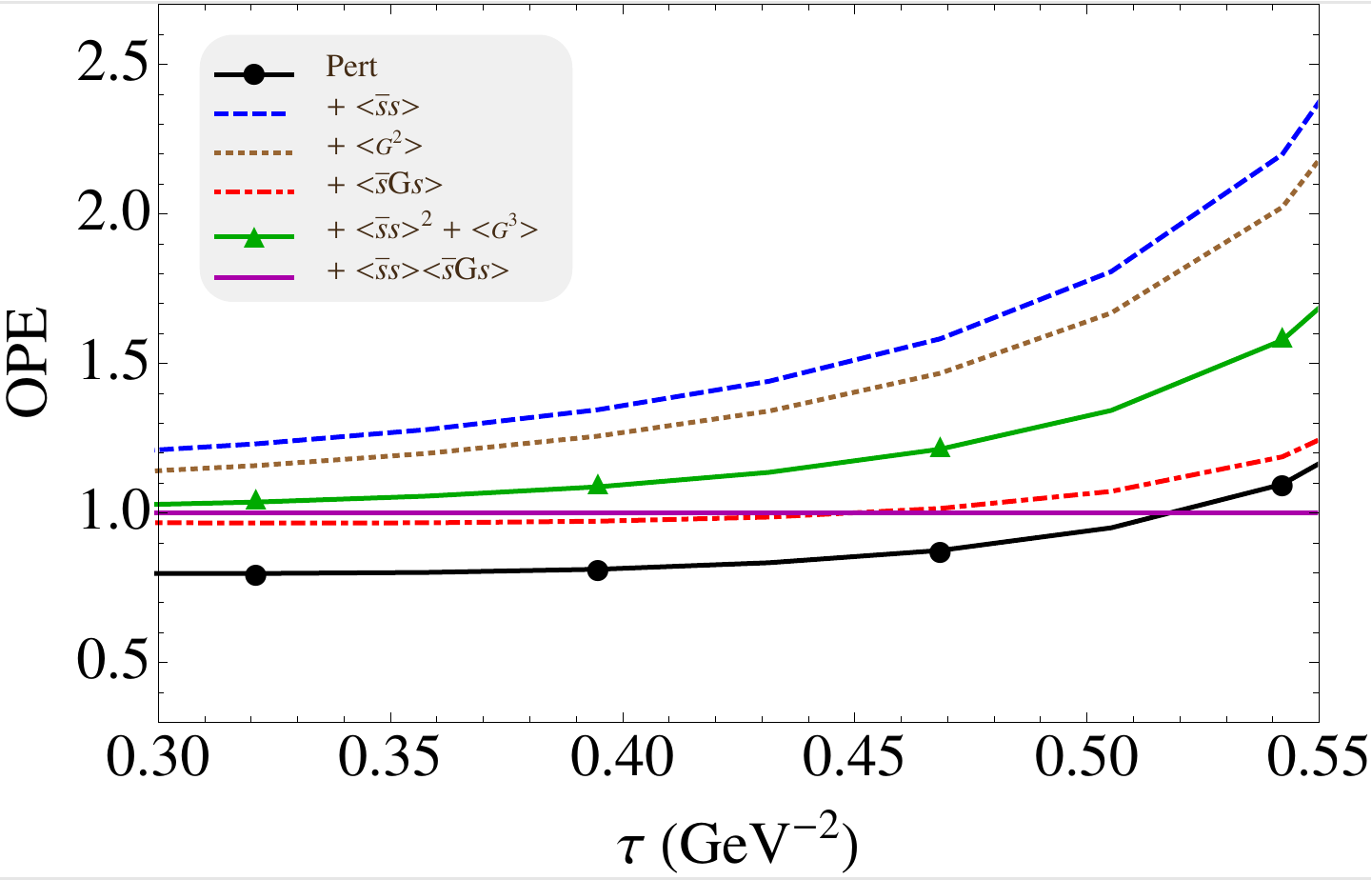}}\\ \vspace{-1.4cm}

    \hspace{-1.1cm}
    \subfloat[]{\includegraphics[width=7.5cm]{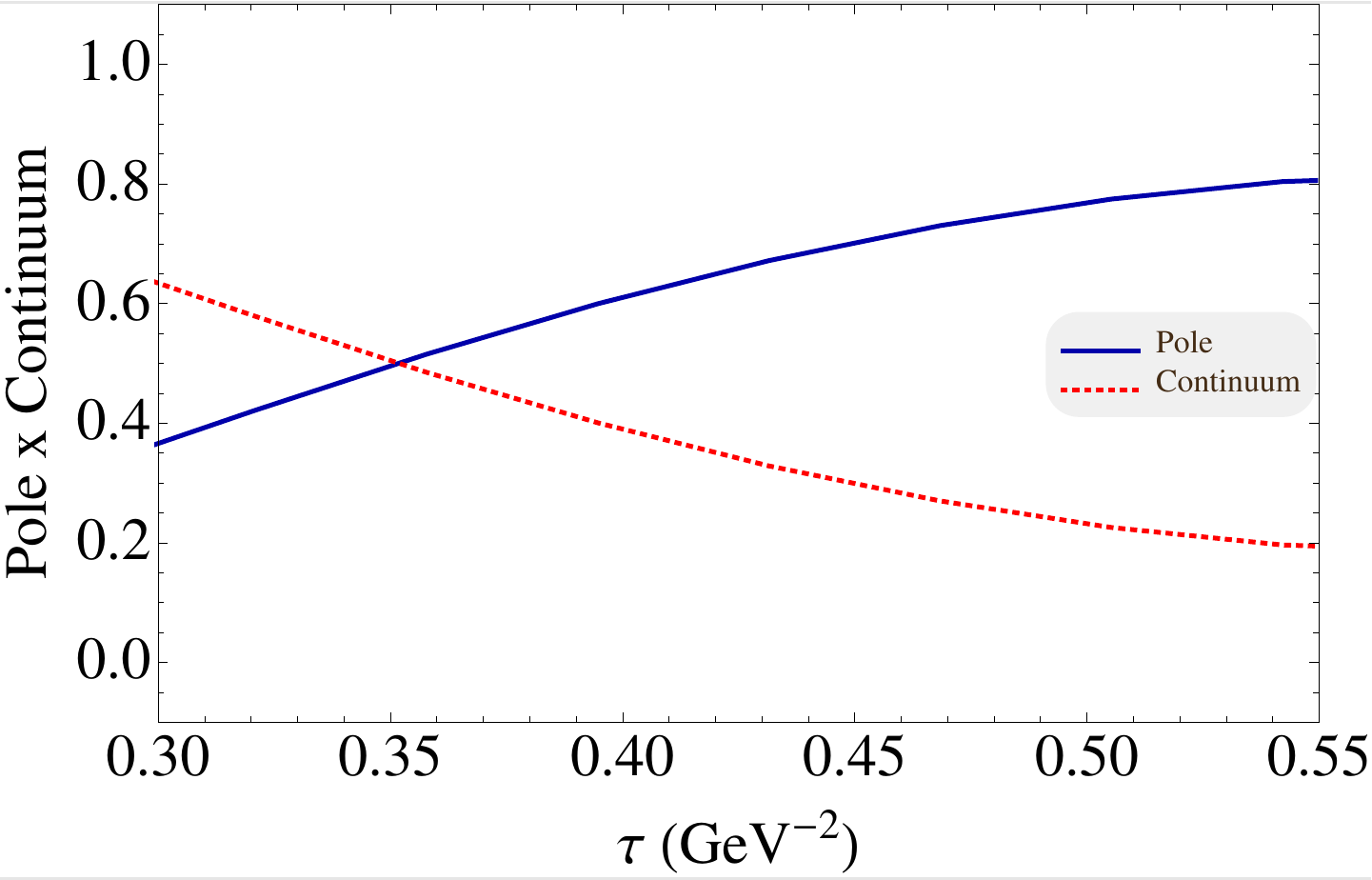}} \hspace{0.5cm}
    \subfloat[]{\includegraphics[width=10cm]{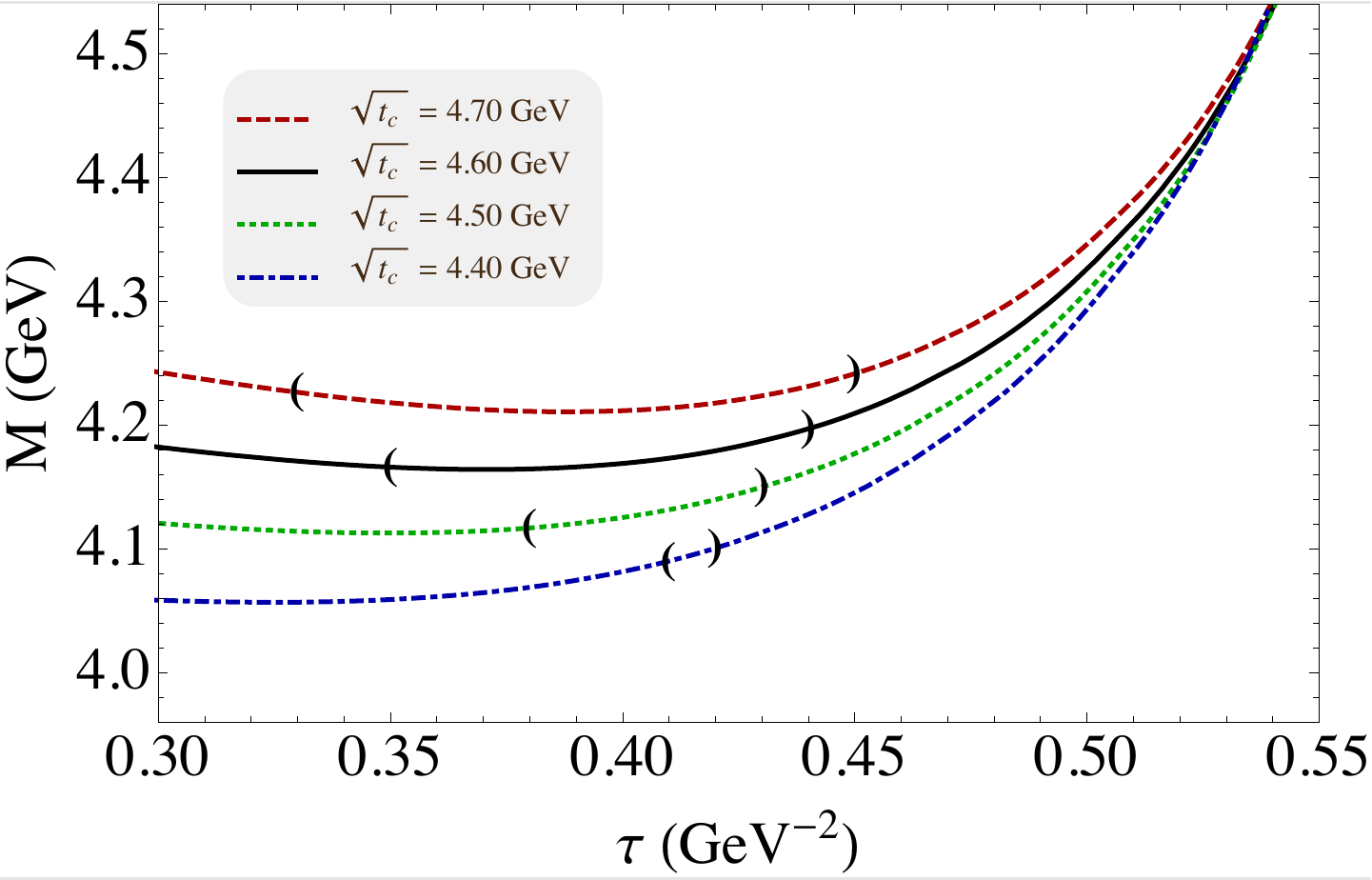}}    
\caption{\footnotesize $D^\ast_s \bar{D}^\ast_s ~(0^{++})$ molecule, considering the OPE contribution 
up to dimension-eight condensates and $m_c=1.23 \GeV$.
{\bf (a)} OPE Convergence in the region $0.30 \leq \tau \leq 0.55~\GeV^{-2}$ for
$\sqrt{t_c} = 4.60 \GeV$. The lines show the relative contributions starting with the perturbative 
contribution and each other line represents the relative contribution after adding of one extra 
condensate in the expansion: $+ \qq[s]$, $+ \GG$, $+ \qGq[s]$, $+ \qq[s]^2 + \GGGi$ and 
$+ \qq[s] \qGq[s]$.
{\bf (b)} Pole vs$.$ Continuum contribution, for $\sqrt{t_c} = 4.60 \GeV$.
{\bf (c)} The mass as a function of the sum rule parameter $\tau$, for different values of $\sqrt{t_c}$. 
The parentheses indicate the upper and lower limits of a valid Borel window.}
\label{fig:DsxDsx}
\end{figure}

In Fig.(\ref{fig:DsxDsx}a), one can see the relative contribution of all the terms in the QCD side 
of the sum rule, for $\sqrt{t_c} = 4.60 \GeV$. From this figure, it is possible to check that, for 
$\tau \leq 0.44 \GeV^{-2}$, the relative contribution of the dimension-eight condensate is less than 
$15\%$ of the total contribution. Therefore, one fixes the maximum value for $\tau$ in the Borel
window as $\tau_{max} = 0.44 \GeV^{-2}$.
On the other hand, one determines the minimum value for $\tau$ by imposing that the pole 
contribution must be bigger than the continuum contribution. This condition is satisfied when 
$\tau \geq 0.35 \GeV^{-2}$, as could be seen in Fig.(\ref{fig:DsxDsx}b).
Then, one fixes the minimum value in the Borel window as $\tau_{min} = 0.35 \GeV^{-2}$.

In Fig.(\ref{fig:DsxDsx}c), one can see the mass as a function of $\tau$, for different values of 
$\sqrt{t_c}$. For each case, the valid Borel window is indicated through the parentheses. 
The allowed values for the continuum threshold are defined in the region  
$4.40 \leq \sqrt{t_c} \leq 4.70 \GeV$, where the optimal choice for $\sqrt{t_c}$ is determined 
by the one that provides improved mass-stability as a function of $\tau$. 
Then, from Fig.(\ref{fig:DsxDsx}c), the optimized value for continuum threshold is given by 
\begin{equation}
  \sqrt{t_c} = 4.55 \pm 0.15 \GeV ~~.
  \label{tcDsxDsx}
\end{equation}
Notice that the sum rule for the $D^\ast_s \bar{D}^\ast_s$ molecular state does not allow values
for the continuum threshold less than $\sqrt{t_c} < 4.40 \GeV$. Below this value, there is no longer a 
valid Borel window since it is not possible to guarantee, at the same time, good OPE convergence 
and pole dominance over the continuum contributions.

Calculating the mass, for each value of $\sqrt{t_c}$, and taking into account the uncertainties from 
other QCD parameters (see Table \ref{TabParam}), one finally arrives at
\begin{equation}
  M_{_{D^\ast_s D^\ast_s}} = (4.19 \pm 0.13) \GeV ~~,
\end{equation}
which is in an excellent agreement with the mass of the narrow structure $Y(4140)$ observed by 
CDF collaboration. Notice that the central value for the mass is below of the meson-meson threshold,  
$E_{th}\left[D^\ast_s D^\ast_s\right] \simeq 4.22 \GeV$, where the notation 
$E_{th}\left[M_1 \:M_2\right]$ stands for the corresponding energy to the sum of the masses of 
$M_1$ and $M_2$ mesons. 
However, considering the uncertainties, this molecular state could not correspond to a bound state. 
One can also deduce, from Eq.(\ref{massa}), the value of the parameter 
\begin{equation}
  \lambda_{_{D_s^\ast D_s^\ast}} = (0.044 \pm 0.011) \GeV^5 ~~,
  \label{coupDsxDsx}
\end{equation}
which gives the strength of the coupling between the current (\ref{Jy4140}) and the 
$D_s^\ast \bar{D}_s^\ast ~(0^{++})$ scalar molecular state.

\subsection{$D^\ast \bar{D}^\ast ~(0^{++})$}
From the above study, it is easy to get results for the $D^\ast \bar{D}^\ast$ molecular 
state, with $J^{{PC}} = 0^{++}$. For this, it is enough to take the limit $m_s \rightarrow 0$ and
make the following changes in the quark condensate $\qq[s] \rightarrow \qq[q]$ and the 
mixed condensate $\qGq[s] \rightarrow \qGq[q]$ in the spectral density equations for the 
$D^\ast_s \bar{D}^\ast_s$ molecular state. These changes will be useful to test if the new 
resonance structure $Y(3930)$ could be described by the $D^\ast \bar{D}^\ast$ molecular 
state, with $J^{{PC}} = 0^{++}$. In such a case, the hadronic current is obtained directly from 
Eq.(\ref{Jy4140}) and is given by 
\begin{eqnarray}
    j_{_{D^\ast \bar{D}^\ast}}=(\bar{q}_a\gamma_\mu c_a)(\bar{c}_b\gamma^\mu q_b) ~.
  \label{Jy3930}
\end{eqnarray}

\subsubsection{Numerical Results}
\begin{figure}[t]
    \hspace{-1.1cm}
    \subfloat[]{\includegraphics[width=7.5cm]{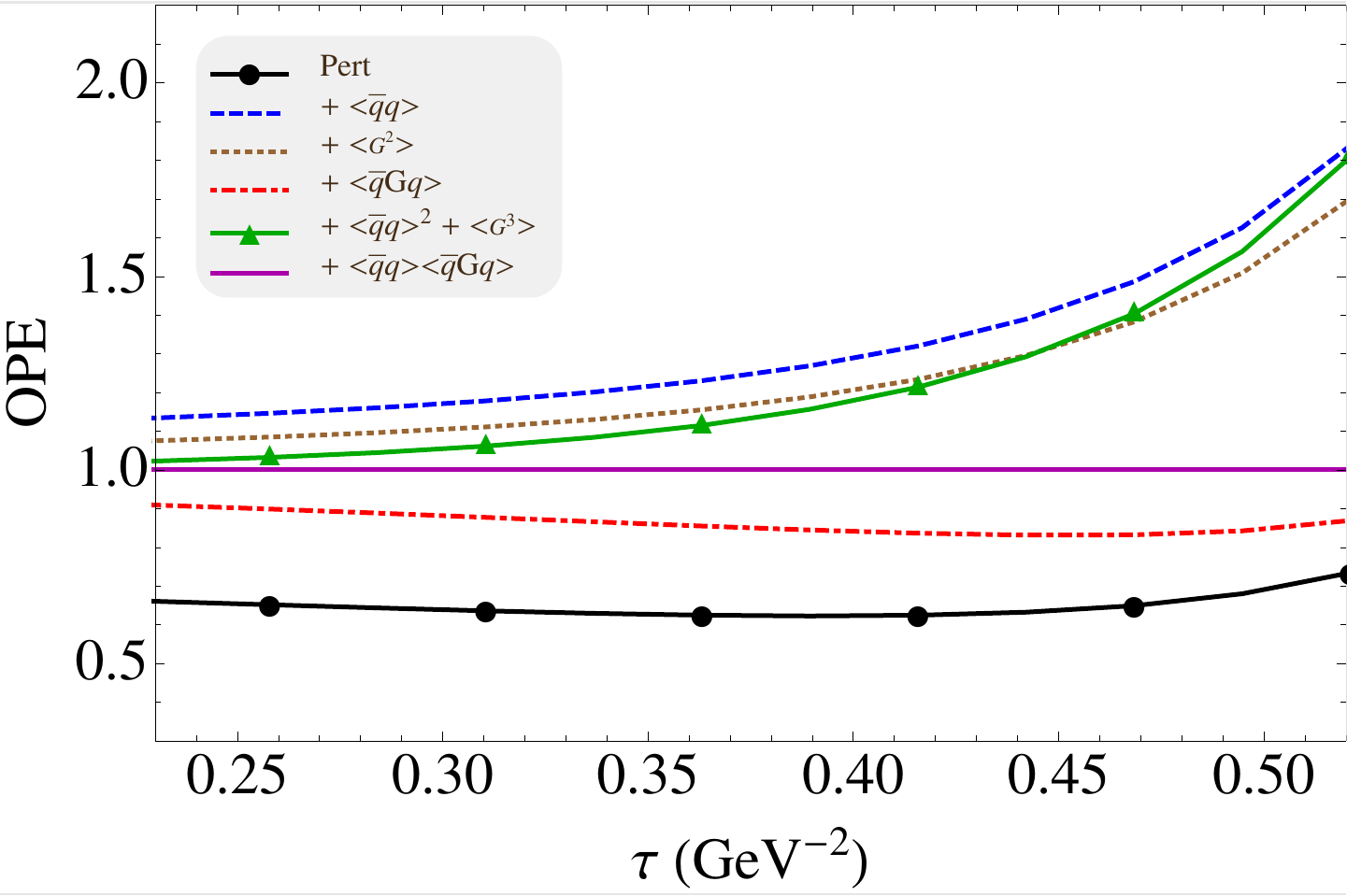}}\\ \vspace{-1.4cm}

    \hspace{-1.1cm}
    \subfloat[]{\includegraphics[width=7.5cm]{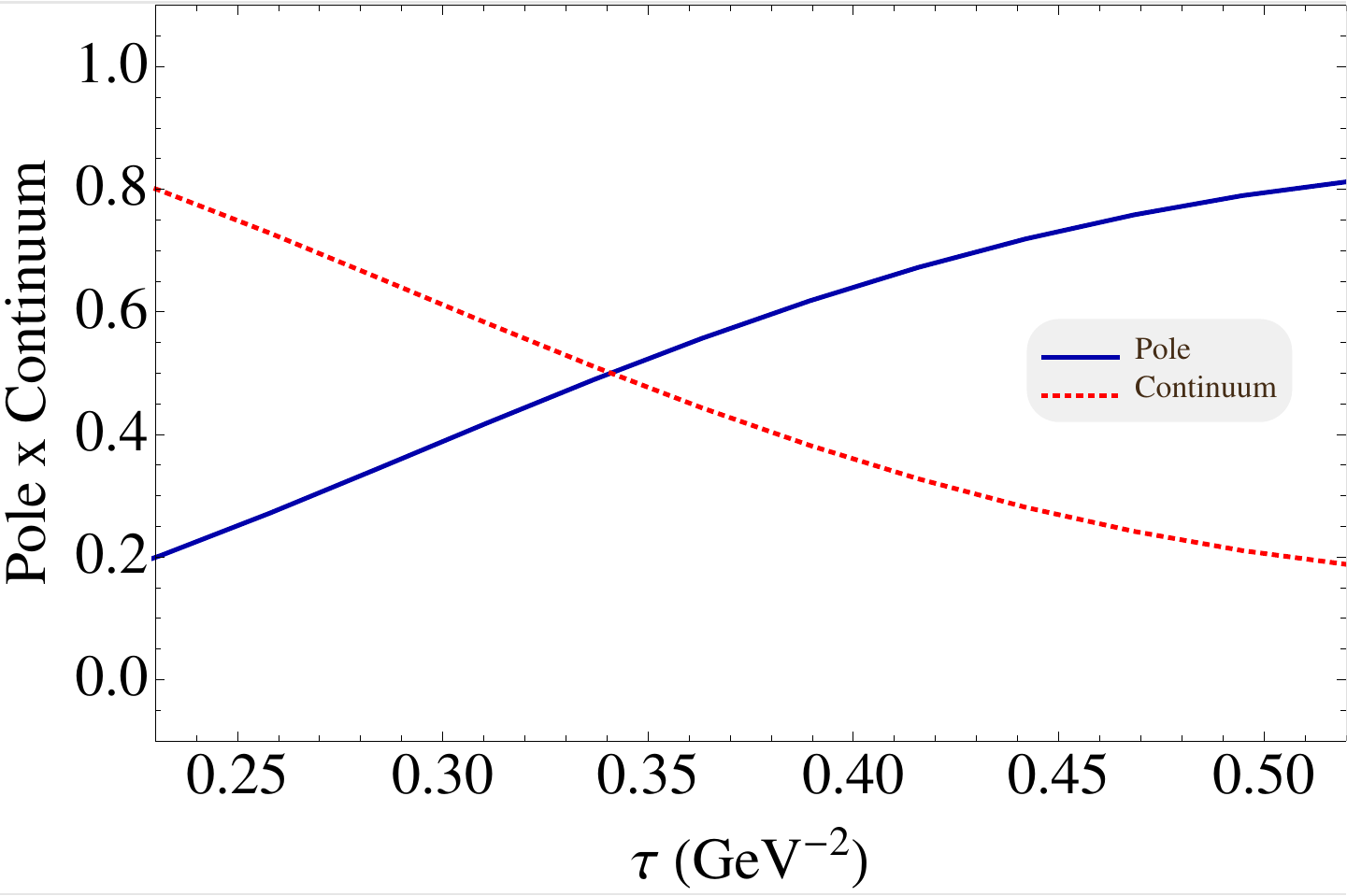}} \hspace{0.5cm}
    \subfloat[]{\includegraphics[width=10cm]{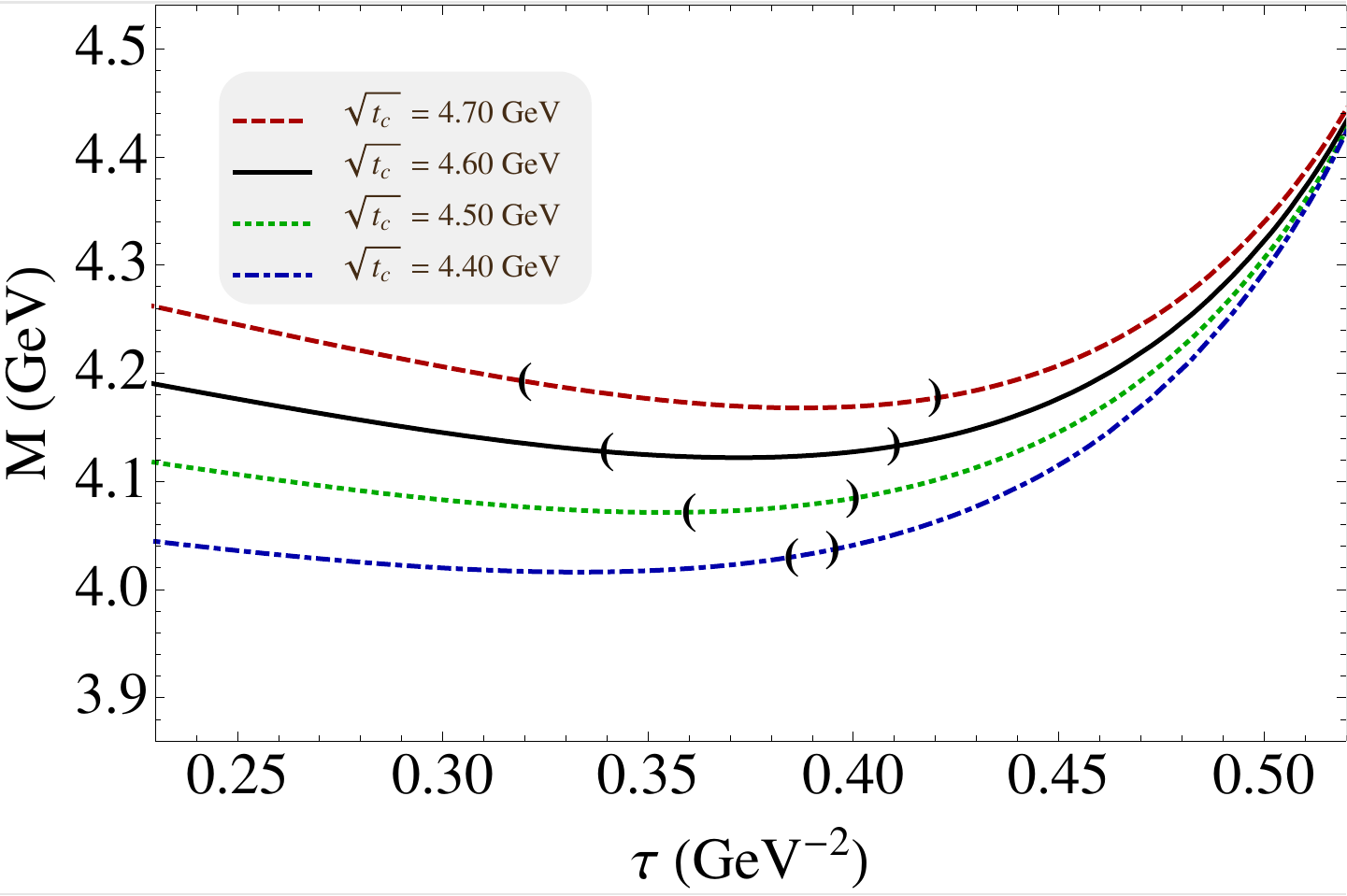}}    
\caption{\footnotesize $D^\ast \bar{D}^\ast ~(0^{++})$ molecule, considering the OPE contribution 
up to dimension-eight condensates and $m_c=1.23\GeV$.
{\bf (a)} OPE Convergence in the region $0.23 \leq \tau \leq 0.52~\GeV^{-2}$ for
$\sqrt{t_c} = 4.60 \GeV$. The lines show the relative contributions starting with the perturbative 
contribution and each other line represents the relative contribution after adding of one extra 
condensate in the expansion: $+ \qq[q]$, $+ \GG$, $+ \qGq[q]$, $+ \qq[q]^2 + \GGGi$ and 
$+ \qq[q] \qGq[q]$.
{\bf (b)} Pole vs$.$ Continuum contribution, for $\sqrt{t_c} = 4.60 \GeV$.
{\bf (c)} The mass as a function of the sum rule parameter $\tau$, for different values of $\sqrt{t_c}$. 
The parentheses indicate the upper and lower limits of a valid Borel window..}
\label{fig:DxDx}
\end{figure}

According to the relation
\begin{equation}
\kappa \equiv \qq[s]/\qq[q] = 0.74 \pm 0.06  ~,
\end{equation}
the exchange of the strange quark condensate $\qq[s]$ by the $\qq[q]$, in the 
expressions of the spectral density for the $\DsxDsx ~(0^{++})$ molecular state, leads to an 
increase of around $25\%$ in the quark and mixed condensates contributions to the OPE 
of the $D^\ast \bar{D}^\ast ~(0^{++})$ sum rule. As a consequence, one gets a worse 
OPE convergence and the upper limit of the Borel window, $\tau_{max}$, has to be defined 
when the relative contribution from the dimension-eight condensate is less than $20\%$ of 
the total contribution.
Following this criterion, from Fig.(\ref{fig:DxDx}a), one verifies that a good OPE convergence, 
for $\sqrt{t_c} = 4.60 \GeV$, is obtained assuming lower values for $\tau$ parameter than 
$\tau_{max} = 0.41 \GeV^{-2}$.

From Fig.(\ref{fig:DxDx}b), the pole contribution is bigger than the continuum contribution 
when $\tau \geq 0.34 \GeV^{-2}$, for $\sqrt{t_c} = 4.60 \GeV$. This minimum value for $\tau$ 
defines the lower limit of the Borel window as $\tau_{min} = 0.34 \GeV^{-2}$.

Considering the following values for the continuum threshold $4.40 \leq \sqrt{t_c} \leq 4.70 \GeV$ 
and taking into account the uncertainties, as indicated in Table (\ref{TabParam}), the obtained mass 
is given by 
\begin{equation}
    M_{_{D^\ast D^\ast}} = (4.15 \pm 0.14) \GeV ~~.
\end{equation}
Notice that the central value for the mass is approximately $130 \MeV$ above the meson-meson 
threshold $E_{th}\left[D^\ast D^\ast\right] \simeq 4.02 \GeV$. This could be an indication that there is 
a repulsive interaction between the two $D^\ast$ mesons, probably caused by strong interactions
effects. However, considering the uncertainties, it is still possible that the $D^\ast \bar{D}^\ast$ 
molecular state corresponds to a bound state. In any case, the obtained mass is not 
compatible with the one observed for the $Y(3930)$ charmonium-like state: 
$M_{_{Y(3930)}} = 3914.6 \:^{+3.8}_{-3.4} \pm 2.0 ~\MeV$.

\begin{figure}[t]
\centering
\includegraphics[width=10cm]{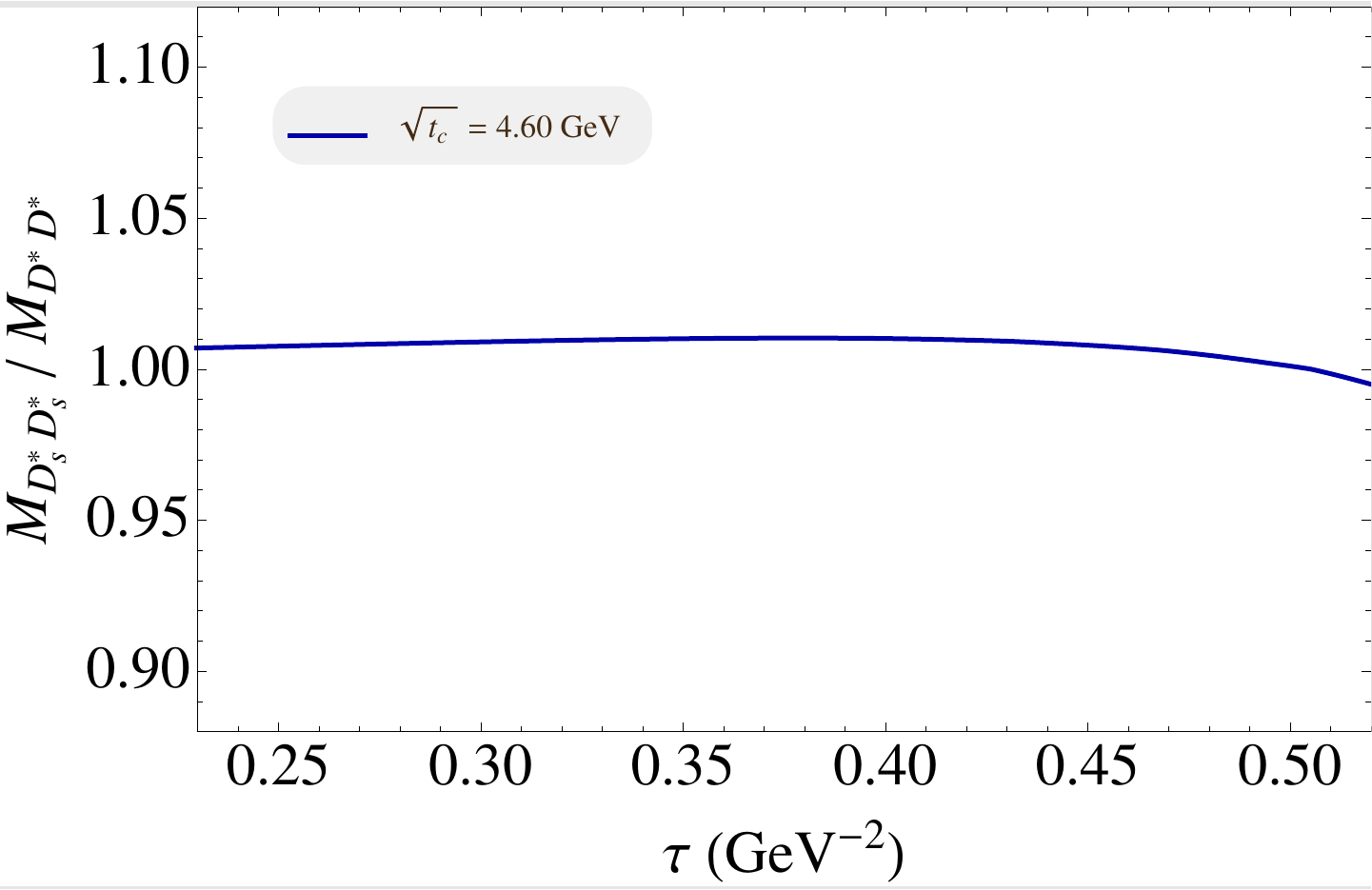}
\caption{\footnotesize The mass ratio for the $D^\ast_s \bar{D}^\ast_s ~(0^{++})$ and 
$D^\ast \bar{D}^\ast ~(0^{++})$ molecular states, for $\sqrt{t_c}=4.60~\GeV$.}
\label{DsxDsx_DxDx}
\end{figure}

In Fig.(\ref{DsxDsx_DxDx}), one can see the mass ratio for the $D^\ast_s \bar{D}^\ast_s$ and 
$D^\ast \bar{D}^\ast$ states, as a function of $\tau$, for $\sqrt{t_c} = 4.60 \GeV$. From this figure, 
it is remarkable that the mass of the $D^\ast \bar{D}^\ast$ molecular state is smaller $0.5\%$ than 
one obtained for the $D^\ast_s \bar{D}^\ast_s$ state.
This result for the mass difference is totally unexpected since, in general, each strange quark 
adds approximately $100 \MeV$ to the mass of the particle. Therefore, one would naively expect
that the mass obtained for the $D^\ast_s \bar{D}^\ast_s$ state should be around 
$200 \MeV$ heavier than the obtained mass for the $D^\ast \bar{D}^\ast$ state.

It is still possible to extract another relevant information from the sum rule of the $\DsxDsx ~(0^{++})$ 
state. By exchanging the strange quark ($s$) for an isospin quark ($q$), in the internal structure 
of the molecule, could give rise to an effect widely studied in potential models for the nucleon: 
the pion-exchange interactions \cite{tokyo, friar, jmarc}. In this case, such interaction could 
explain the high value for the mass, even removing the quark $s$ from the current (\ref{Jy4140}). 
Thus, the molecular description $D^\ast \bar{D}^\ast ~(0^{++})$ for the $Y(3930)$ state would 
remain valid.
The value of the parameter $\lambda_{_{D^\ast D^\ast}}$ is estimated as
\begin{equation} \vspace{-0.2cm}
  \lambda_{_{D^\ast D^\ast}} = (0.048 \pm 0.015) \GeV^5 ~~.
  \label{coupDxDx}
\end{equation}
Therefore, comparing the results in Eqs.(\ref{coupDsxDsx}) and (\ref{coupDxDx}), one concludes 
that the currents couple with similar strength to the corresponding states, and that both, 
$\DsxDsx$ and $\DxDx$ scalar molecular currents describe scalar mesons with masses 
compatible with the recently observed $Y(4140)$ narrow structure. 
The fact that the $Y(4140)$ was observed in the decay $Y(4140) \rightarrow J/\psi \phi$ could indicate 
that the $\DsxDsx$ assignment is more compatible with its quark content. However, the $\DxDx$ 
assignment cannot be excluded since they have the same quantum numbers of the $Y(4140)$ 
state. Another interesting interpretation is that the $Y(4140)$ could be a mixture of these 
two molecular states.

The QCDSR results for the $D^\ast_s \bar{D}^\ast_s$ and $D^\ast \bar{D}^\ast$ molecular 
states are published in Ref.\cite{ramar2}.

\subsection{$D^\ast_s \bar{D}^\ast_{s0} ~(1^{-+})$}
As discussed in the Introduction, the Belle Collaboration \cite{belle3} observed another possible 
candidate for an exotic state, the $X(4350)$ state, in the decay channel 
$\gamma\gamma \to X(4350) \to \phi J/\psi$. The possible quantum numbers for a state 
decaying into $J/\psi\phi$ are $J^{PC}=0^{++},~1^{-+}$ and $2^{++}$. In the present 
work, the QCDSR approach is used to study if a correlation function based on a 
$D^\ast_s \bar{D}^\ast_{s0}$ current, with $J^{{PC}} = 1^{-+}$, could describe this new 
observed resonance structure. A possible current is given by 
\begin{equation}
  j^\mu_{_{D^\ast_s \bar{D}^\ast_{s0}}}={1\over\sqrt{2}}\Big[
  (\bar{s}_a\gamma^\mu c_a)(\bar{c}_bs_b)-
  (\bar{c}_a\gamma^\mu s_a)(\bar{s}_bc_b)\Big] ~~.
  \label{Jx4350}
\end{equation}
In the Phenomenological side, one parameterizes the coupling of the exotic 
state, $X \equiv D^\ast_s \bar{D}^\ast_{s0}$, to the current in Eq.(\ref{Jx4350}) in 
terms of the decay constant parameter
$\lambda_{_{D^\ast_s \bar{D}^\ast_{s0}}}$. Then, 
\begin{equation}
  \langle 0| j^\mu_{_{D^\ast_s \bar{D}^\ast_{s0}}} | X \rangle = 
  \lambda_{_{D^\ast_s \bar{D}^\ast_{s0}}} \epsilon^\mu ~.
\end{equation}

In the QCD side, one inserts the current (\ref{Jx4350}) into the correlation function 
in order to obtain 
\begin{eqnarray}
  \Pi^{_{OPE}}_{\mu\nu}(q) &=& -\frac{i}{2^9\:\pi^8} 
  \int\! d^{4}x \:d^{4}p_1 \:d^{4}p_2 ~e^{ix \cdot (q-p_1-p_2)} \nno \\ && \times \left\{ \:
  \begin{array}{clc}
    \ & \Tr[ \ga_\mu \:{\cal S}^c_{ab}(p_1) \:{\cal S}^s_{ba}(-x) ]\: \cdot
    \Tr[ {\cal S}^s_{cd}(x)\: \ga_\nu \:{\cal S}^c_{dc}(-p_2) ]  & \\
    - & \Tr[ \ga_\mu \:{\cal S}^c_{ab}(p_1)\: \ga_\nu \:{\cal S}^s_{ba}(-x) ]\: \cdot
    \Tr[ {\cal S}^s_{cd}(x)\: {\cal S}^c_{dc}(-p_2) ]  & \\
    - & \Tr[ {\cal S}^c_{ab}(p_1) \:{\cal S}^s_{ba}(-x) ]\: \cdot
    \Tr[ \ga_\mu \:{\cal S}^s_{cd}(x)\: \ga_\nu \:{\cal S}^c_{dc}(-p_2) ]  & \\
    + & \Tr[ {\cal S}^c_{ab}(p_1)\: \ga_\nu \:{\cal S}^s_{ba}(-x) ]\: \cdot
    \Tr[ \ga_\mu \:{\cal S}^s_{cd}(x) \:{\cal S}^c_{dc}(-p_2) ]  &
  \end{array}
  \right\}
\end{eqnarray}
Since the current (\ref{Jx4350}) is not conserved, one must consider the $g_{\mu\nu}$ structure 
obtained from the correlation function as shown in Eq.(\ref{fc10}).
According to the Eq.(\ref{proj2}), it is necessary to calculate the invariant function $\Pi_1(q^2)$ 
since it gets contributions only from the vector $1^{-+}$ state.

The QCD sum rule calculation for this exotic meson, described by a $\DsxDso ~(1^{-+})$ 
molecular current, is done determining the invariant function $\Pi_1(q^2)$ and considering the 
OPE terms up to dimension-eight condensates, working at leading order in $\al_s$ in the 
operators and keeping terms which are linear in the strange quark mass. 
Finally, the contribution for each dimension in the OPE is calculated so that the expression 
for the spectral density $\rho^{_{OPE}}_{_{\DsxDso}}(s)$ is given by 
\begin{equation}	
  \rho^{_{OPE}}_{_{\DsxDso}}(s) = 
  \rho^{pert}_{_{\DsxDso}}(s) + \rho^{\qq[s]}_{_{\DsxDso}}(s) + \rho^{\GGi}_{_{\DsxDso}}(s) + 
  \rho^{\qGq[s]}_{_{\DsxDso}}(s) + \rho^{{\qq[s]}^2}_{_{\DsxDso}}(s) + 
  \rho^{\GGG}_{_{\DsxDso}} + \rho^{\qq[s] \qGq[s]}_{_{\DsxDso}}(s) ~~,
\end{equation}
where each term of this spectral density is shown below:
\begin{eqnarray}
  \rho^{pert}_{_{\DsxDso}}(s)&=&{1\over 2^{12} \pi^6}\int\limits_{\almin}^{\almax}{d\al \over \al^3}
    \int\limits_{\bemin}^{1-\al}{d\be\over\be^3}(1-\al-\be) {\cal F}^{\:3}_{(\al,\be)} 
    \bigg[ 2m_c^2(1-\al-\be)^2 \nno\\
    && +~ 3(1+\al+\be){\cal F}_{(\al,\be)} - 24 m_s m_c \:\beta (1-\al-\be) \bigg], \\
  \rho^{\qq[s]}_{_{\DsxDso}}(s) &=& {3\qq[s]\over2^{7} \pi^4} 
     \int\limits_{\almin}^{\almax} \!\!{d\al} \Bigg\{ \frac{m_s {\cal H}^{\:2}_{(\al)}}{\al(1-\al)} ~-~ 
     m_c \int\limits_{\bemin}^{1-\al} \frac{d\be}{\al^2\be} {\cal F}_{(\al,\be)} \bigg[ 
     2(1-\al-\be) {\cal F}_{(\al,\be)}  \nno \\     
    && +~ m_s m_c \:\al(3+\al+\be) \bigg] \Bigg\}, \\
  \rho^{\GGi}_{_{\DsxDso}}(s) &=& {\GG\over3\cdot 2^{12}\pi^6} 
  \int\limits_{\almin}^{\almax} \!\!{d\al\over\al^3} \int\limits_{\bemin}^{1-\al} \!\!{d\be\over\be} \Bigg\{ 
  m_{c}^{4} \:\be(1 \!-\! \al \!-\! \be)^{3} - 6\al(1 \!-\! 2\al \!-\! 2\be) {\cal F}^{\:2}_{(\al,\be)} \nno \\
  && +~ 3m_c^{2}(1-\al-\be) \Big[ 1+\al(1-2\al) +\be(\al+3\be) \Big] {\cal F}_{(\al,\be)} \Bigg\}, \\
  \rho^{\qGq[s]}_{_{\DsxDso}}(s) &=& -\frac{\qGq[s]}{2^{8} \pi^4} 
    \int\limits_{\almin}^{\almax} \!\!{d\al} \:\Bigg\{ \frac{m_s}{\al} 
    \Big( 8m_c^2 \:\al + (2-7\al) {\cal H}_{(\al,\be)} \Big) 
    - \int\limits_{\bemin}^{1-\al}\!\!\frac{d\be}{\al^2\be} \bigg[ 3m_c \:{\cal F}_{(\al,\be)} \nno \\
    && \times \Big( \al(1 \!-\! \al) -\be (5\al \!+\! 2\be) \Big)  + m_s \:\al\be 
    \Big( m_c^2(3 \!+\! 4\al \!+\! 3\be) + {\cal F}_{(\al,\be)} \Big) \bigg] \Bigg\}, ~~~~~\\
    \rho^{\qq[s]^2}_{_{\DsxDso}}(s) &=& -\frac{\rho \qq[s]^2}{3 \cdot 2^6 \pi^2} 
    \bigg[ s + 8m_c^2 + 6m_s m_c \bigg] v, \\
    \rho^{\GGGi}_{_{\DsxDso}}(s) &=& {\GGG \over 3\cdot 2^{13}\pi^6} \Bigg\{
    3 \int\limits_{\almin}^{\almax} \!\!d\al \int\limits_{\bemin}^{1-\al} \!\!\frac{d\be}{\be^3} \bigg[ 
    m_c^2 \bigg( 1-2\be+(\al+\be)(3\al+\be) \bigg) \nno\\
    && \hspace{-1cm} +~ (1 \!+\! \al \!+\! \be) {\cal F}_{(\al,\be)} \bigg]  - m_c^4
    \int\limits_{0}^{1} \!\!d\al \!\!\int\limits_{0}^{1-\al} \!\!\frac{d\be}{\be^4} (1 \!-\! \al \!-\! \be)^3 
    ~\delta\bigg( s - \frac{m_c^{2}(\al+\be)}{\al\be} \bigg) \Bigg\}, ~~\\
    \rho^{\qq[s] \qGq[s]}_{_{\DsxDso}}(s) &=& {\qq[s] \qGq[s] \over 3\cdot 2^{7}\pi^2} \Bigg\{
    \frac{6m_c^2 v}{s} + m_c \int\limits_{0}^{1} \frac{d\al}{\al(1-\al)^2} \Bigg[ 
    6m_c (1-\al) \bigg( \al(1 - 3\al) + 2 m_c^2 \:\tau \bigg) \nno\\
    && \hspace{-2cm}  -~ m_s \bigg( \al(1 \!-\! \al) (6 \!-\! 13\al \!+\! 20\al^2) - 
    2 m_c^2 \tau (3 \!-\! 2\al \!-\! 5\al^{2}) \bigg) \Bigg] 
    \delta\bigg( s - \frac{m_c^{2}}{\al(1-\al)} \bigg) \Bigg\}  ~.
\end{eqnarray}
Note that, one must consider the definitions (\ref{def:variables}) and the integration limits 
(\ref{def:limits}).

\subsubsection{Numerical Results}
To evaluate the mass of the $\DsxDso ~(1^{-+})$ molecular state, one uses Eq.(\ref{massa}) 
with the parameters given in Table (\ref{TabParam}).  
 
\begin{figure}[t]
    \hspace{-1.1cm}
    \subfloat[]{\includegraphics[width=7.5cm]{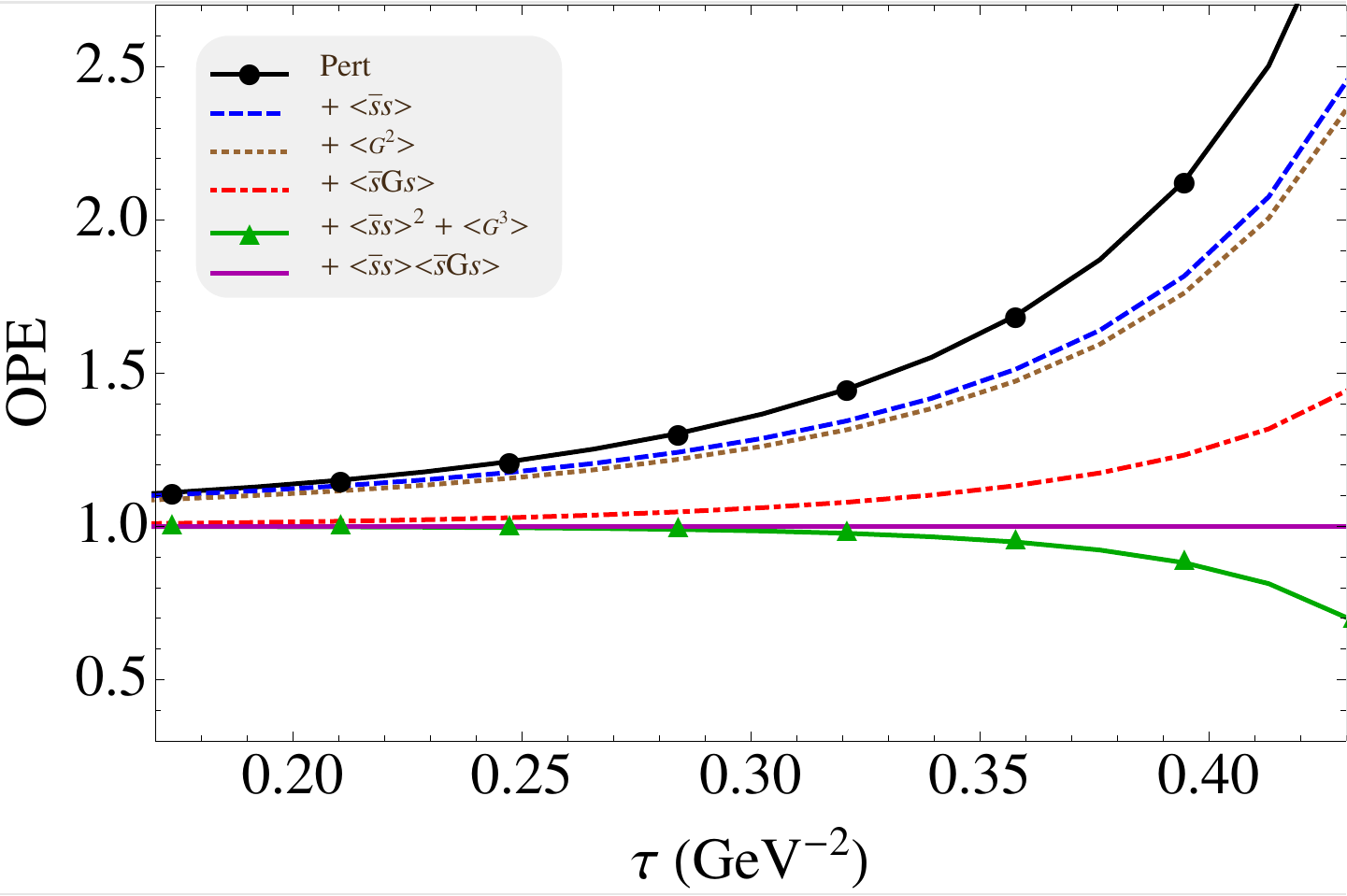}}\\ \vspace{-1.4cm}
    
    \hspace{-1.1cm}
    \subfloat[]{\includegraphics[width=7.5cm]{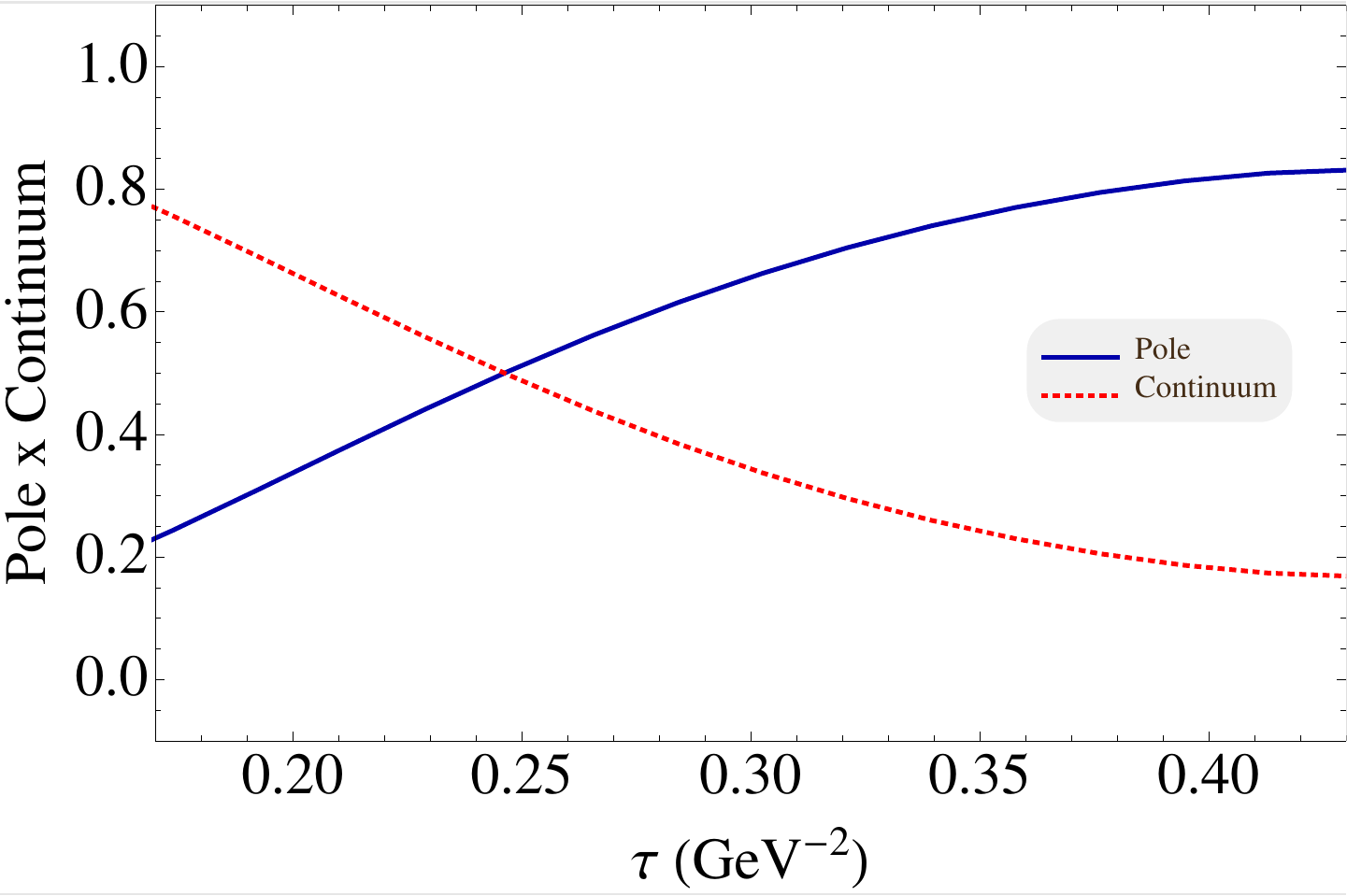}} \hspace{0.5cm}
    \subfloat[]{\includegraphics[width=10cm]{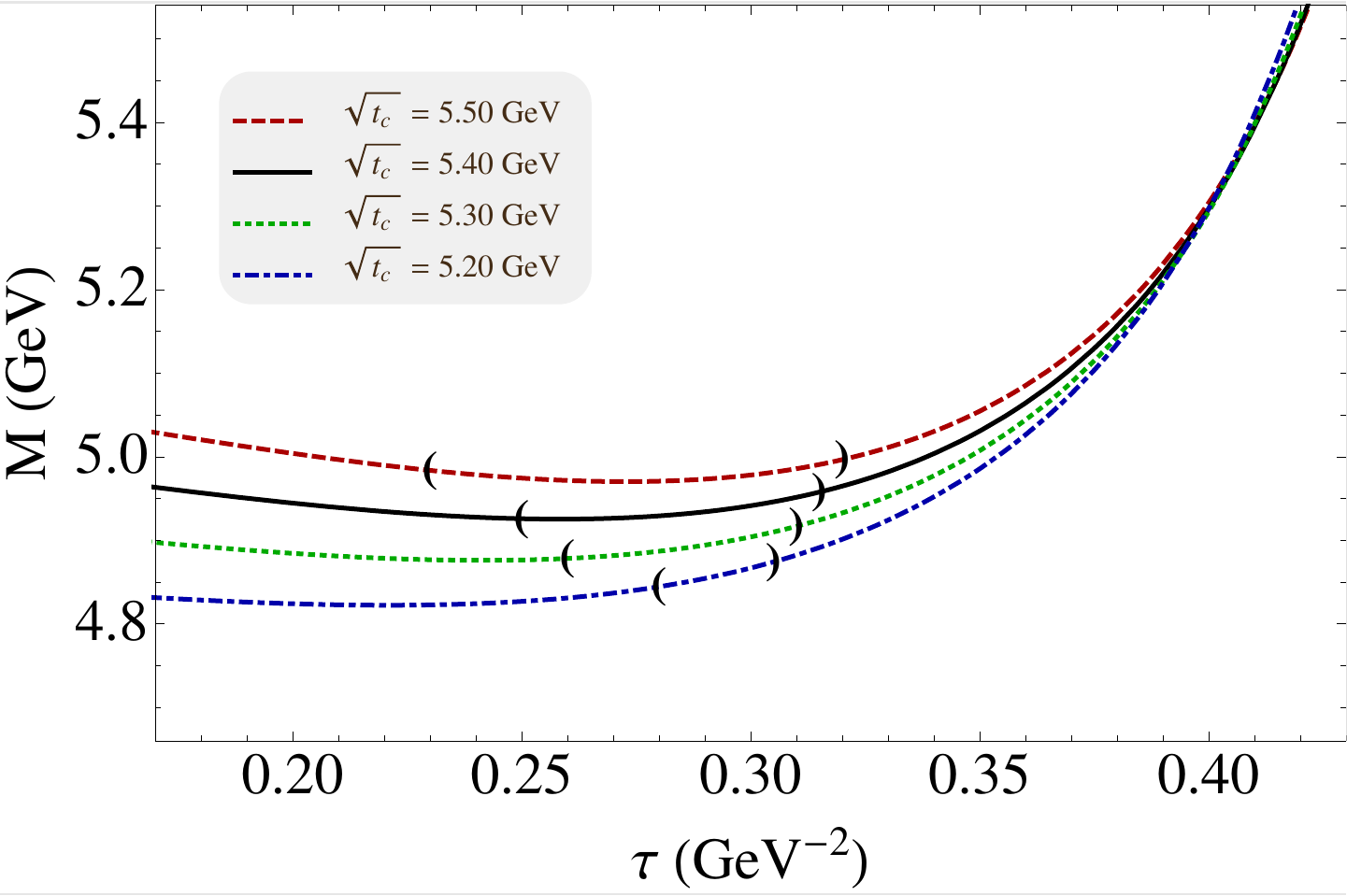}}    
\caption{\footnotesize $\DsxDso ~(1^{-+})$ molecule, considering the OPE contribution 
up to dimension-eight condensates and $m_c=1.23\GeV$.
{\bf (a)} OPE convergence in the region $0.17 \leq \tau \leq 0.43~\GeV^{-2}$ for
$\sqrt{t_c} = 5.40 \GeV$. The lines show the relative contribution after adding of one 
extra condensate in the expansion: $+ \qq[s]$, $+ \GG$, $+ \qGq[s]$, 
$+ \qq[s]^2 + \GGGi$ and $+ \qq[s] \qGq[s]$.
{\bf (b)} Pole vs$.$ Continuum contribution, for $\sqrt{t_c} = 5.40 \GeV$.
{\bf (c)} The mass as a function of the sum rule parameter $\tau$, for different values of 
$\sqrt{t_c}$. The parentheses indicate the upper and lower limits of a valid Borel window.}
\label{fig:DsxDs0x}
\end{figure}

In Fig.(\ref{fig:DsxDs0x}a), one can see the relative contribution of all the terms in the QCD side 
of the sum rule. From this figure, it is possible to check that, for $\tau \leq 0.32 \GeV^{-2}$, the 
contribution of the dimension-eight condensate is less than $5\%$ of the total contribution, 
which indicates a good OPE convergence. Therefore, one fixes the maximum $\tau$ value 
as $\tau_{max} = 0.32 \GeV^{-2}$.

From Fig.(\ref{fig:DsxDs0x}b), the pole contribution is bigger than the continuum contribution 
when $\tau \geq 0.25 \GeV^{-2}$, for $\sqrt{t_c} = 5.40 \GeV$. This minimum value for $\tau$ 
defines the lower limit of the Borel window as $\tau_{min} = 0.25 \GeV^{-2}$. Notice that, for 
$\sqrt{t_c} < 5.20 \GeV$, there is no allowed Borel window.

In Fig.(\ref{fig:DsxDs0x}c), the mass of the $\DsxDso ~(1^{-+})$ molecular state is exhibited 
for different values of the continuum threshold in the range $5.20 \leq \sqrt{t_c} \leq 5.50 \GeV$.
One can verify that the optimal choice for this parameter is given by $\sqrt{t_c} = 5.40 \GeV$.

Taking into account the uncertainties from the other parameters given in Table (\ref{TabParam}), 
the final value obtained is 
\begin{equation}
  M_{_{D^\ast_s D^\ast_{s0}}} = (5.05 \pm 0.15) \GeV ~~.
\end{equation}
Obviously this value is not compatible with the mass of the narrow structure $X(4350)$, 
observed by Belle. Also notice that this mass is well above the meson-meson threshold 
$E_{th}\left[D^\ast_s D^\ast_{s0}\right] = 4.43 \GeV$. Therefore, the $\DsxDso ~(1^{-+})$ molecular 
state is not a good candidate to describe the exotic state $X(4350)$ properly.

\subsection{$\DxDo ~(1^{-+})$}
From the above study, it is easy to get results for the $\DxDo ~(1^{-+})$ molecular 
state. For this, one simply takes the limit $m_s \rightarrow 0$ and does the exchanges 
of the quark condensates $\qq[s] \rightarrow \qq[q]$ and the mixed condensates 
$\qGq[s] \rightarrow \qGq[q]$ in the spectral density equations for the 
$\DsxDso ~(1^{-+})$ molecular state.

\begin{figure}[t]
    \hspace{-1.1cm}
    \subfloat[]{\includegraphics[width=7.5cm]{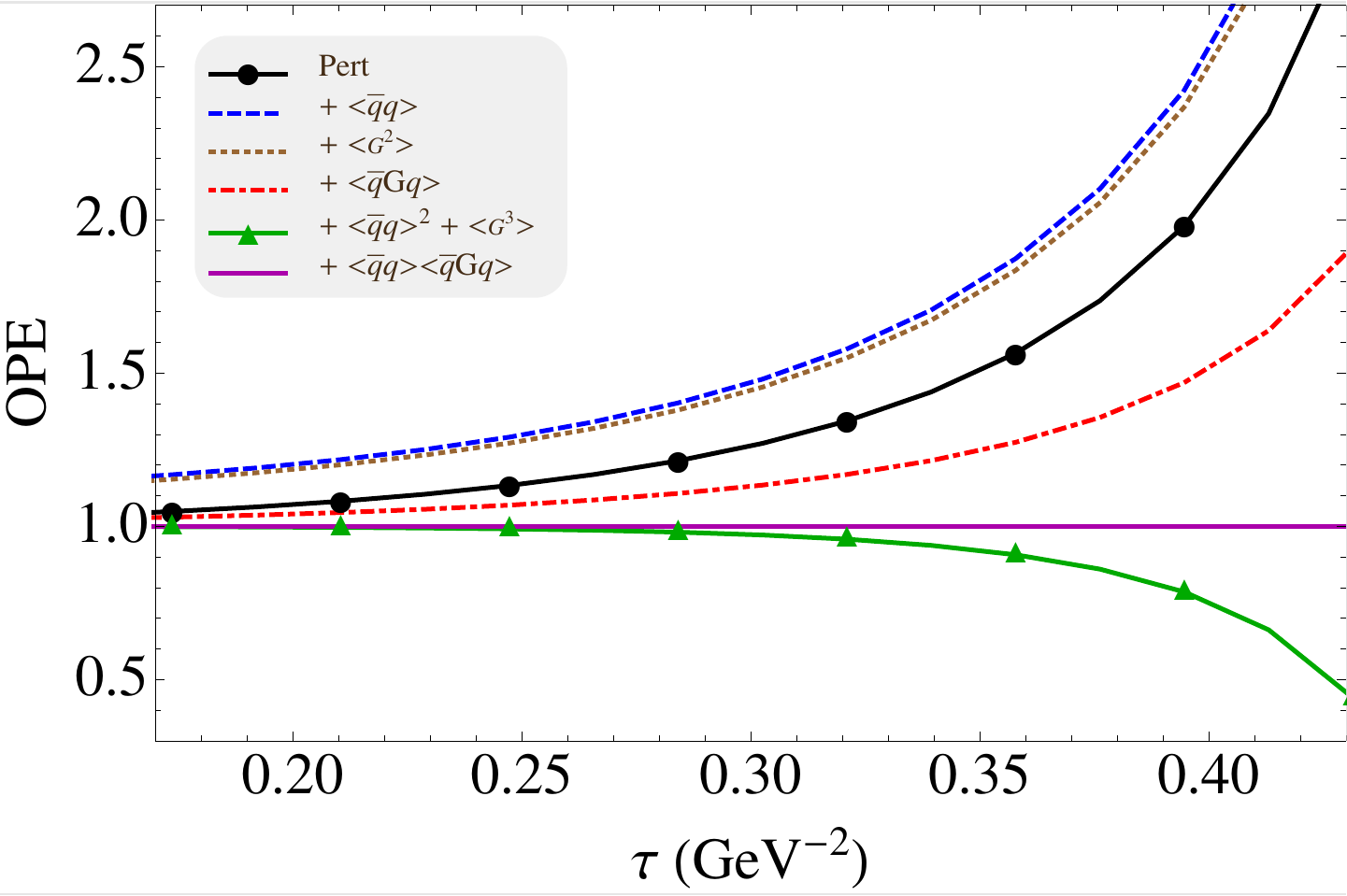}}\\ \vspace{-1.4cm}
    
    \hspace{-1.1cm}
    \subfloat[]{\includegraphics[width=7.5cm]{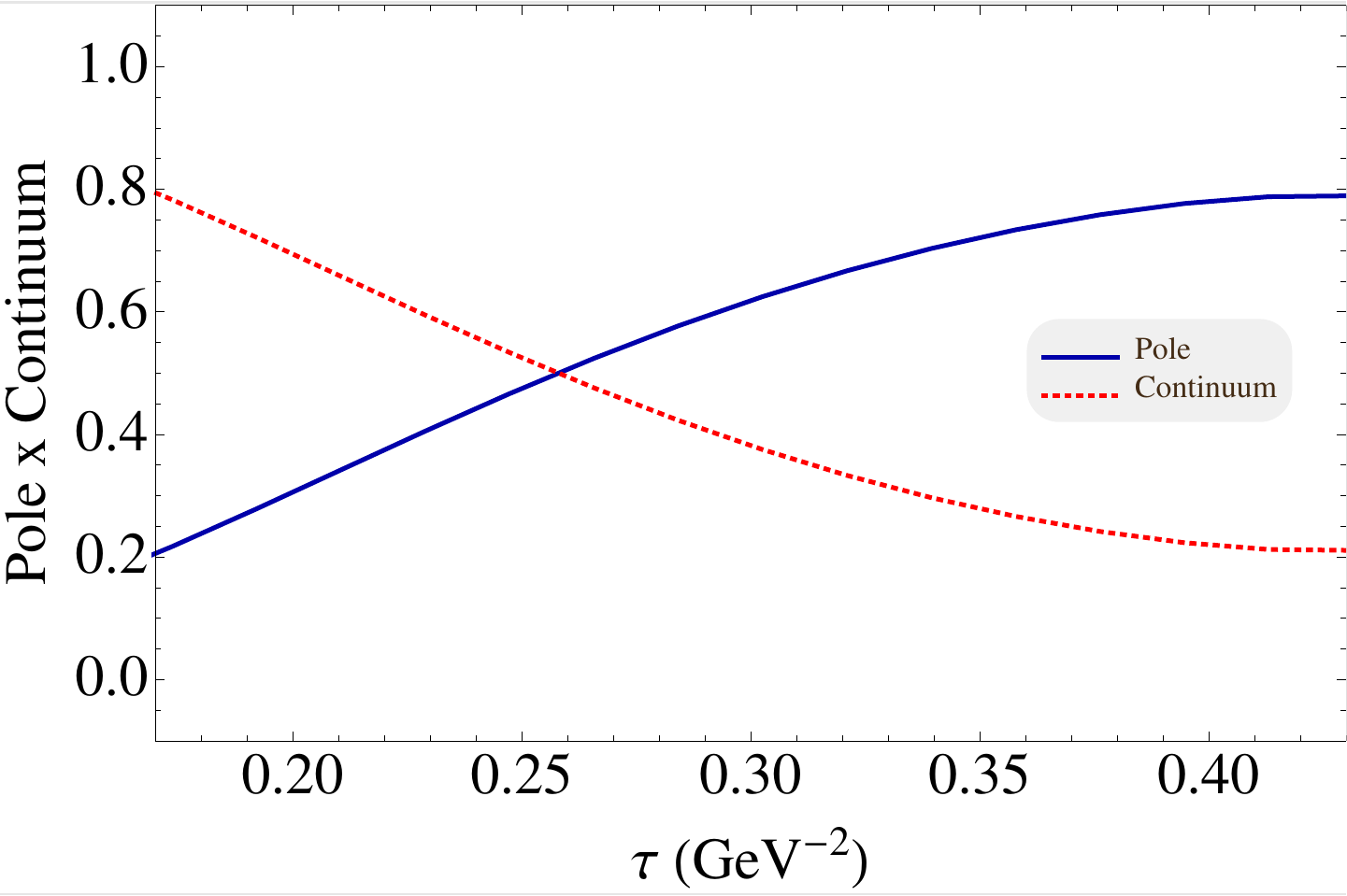}} \hspace{0.5cm}
    \subfloat[]{\includegraphics[width=10.0cm]{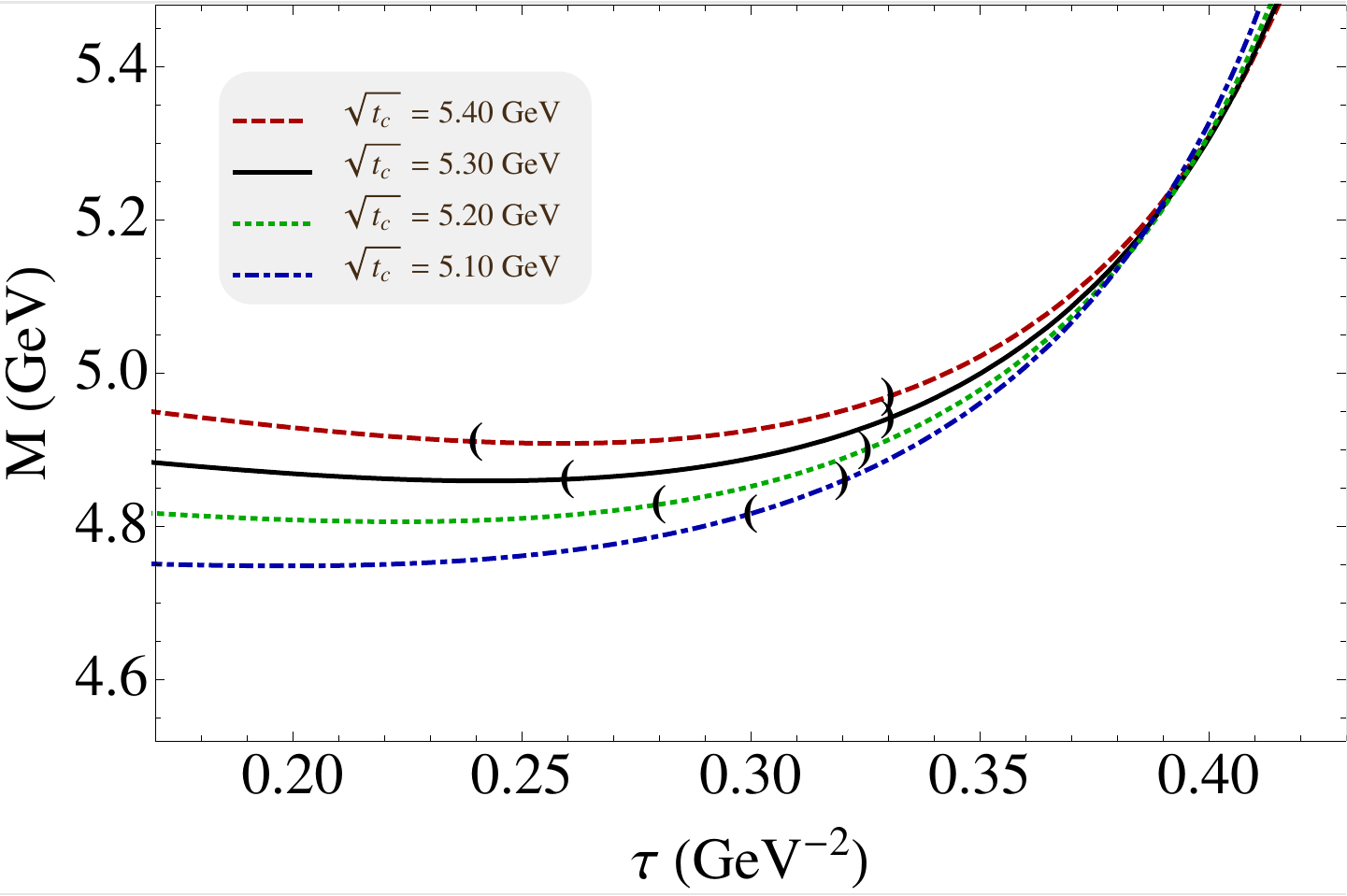}}
\caption{\footnotesize $\DxDo ~(1^{-+})$ molecule, considering the OPE contribution 
up to dimension-eight condensates and $m_c=1.23\GeV$.
{\bf (a)} OPE convergence in the region $0.17 \leq \tau \leq 0.43~\GeV^{-2}$ for
$\sqrt{t_c} = 5.30 \GeV$. The lines show the relative contribution after adding of one 
extra condensate in the expansion: $+ \qq[s]$, $+ \GG$, $+ \qGq[s]$, 
$+ \qq[s]^2 + \GGGi$ and $+ \qq[s] \qGq[s]$.
{\bf (b)} Pole vs$.$ Continuum contribution, for $\sqrt{t_c} = 5.30 \GeV$.
{\bf (c)} The mass as a function of the sum rule parameter $\tau$, for different values of 
$\sqrt{t_c}$. The parentheses indicate the upper and lower limits of a valid Borel window.}
\label{fig:DxD0x}
\end{figure}

\subsubsection{Numerical Results}
As in the previous case, one obtains the OPE convergence at the point where the relative 
contribution of the dimension-eight condensate is less than $5 \%$ of the total contribution - 
see Fig.(\ref{fig:DxD0x}a). For the $\DxDo ~(1^{-+})$ molecular state, the pole dominance 
and the mass results are shown in Figs.(\ref{fig:DxD0x}b, c). Using the values of the 
continuum threshold in the range $(5.10 \leq \sqrt {t_c} \leq 5.40) \GeV$, taking into account 
the uncertainties from the parameters given in Table (\ref{TabParam}) and evaluating the 
mass in a valid Borel window, one obtains the final value 
\begin{equation}
    M_{_{D^\ast D^\ast_0}} = (4.92 \pm 0.12) \GeV
\end{equation}
which is approximately $100 \MeV$ below the value obtained for the 
$\DsxDso ~(1^{-+})$ molecular state and also well above the meson-meson threshold
$E_{th}\left[D^\ast D^\ast_{0}\right] \simeq 4.41 \GeV$.

Therefore, from a QCDSR point of view, the $X(4350)$ exotic state observed by Belle collaboration 
is consistent neither $\DsxDso ~(1^{-+})$ nor $\DxDo ~(1^{-+})$ molecular states.
All of these discussions and results are published in Ref.\cite{rjm}.

\subsection{$\DsxDso ~(1^{--})$ and $\DxDo ~(1^{--})$}
A molecular state with a vector $D^\ast_s$ and a scalar $D^\ast_{s0}$ mesons, with negative 
parity and charge conjugation, was studied for the first time in Ref.\cite{ramar1} and the 
obtained mass was $(4.42 \pm 0.10) \GeV$. 
In this section, one presents the study of a $D^\ast_s D^\ast_{s0}$ molecular state, 
with $J^{PC} = 1^{--}$, combining the results from two distinct sum rule methods: 
QCDSR and FESR. 

A possible current for this molecular state can be constructed using the combination
$D^{\ast +}_{s}D^{\ast -}_{s0} + D^{\ast -}_{s}D^{\ast +}_{s0}$. Thus, 
\begin{eqnarray}
  j^\mu_{\DsxDso} &=& {1\over\sqrt{2}} \Big[
  (\bar{s}_a\gamma^\mu c_a)(\bar{c}_b s_b) + 
  (\bar{c}_a\gamma^\mu s_a)(\bar{s}_b c_b)\Big] ~.
  \label{Jmol1}
\end{eqnarray}
Inserting the current (\ref{Jmol1}) into the correlation function, one obtains the following 
expression in the QCD side:
\begin{eqnarray}
  \Pi^{_{OPE}}_{\mu\nu}(q) &=& \frac{i}{2^9\:\pi^8} 
  \int\! d^{4}x \:d^{4}p_1 \:d^{4}p_2 ~e^{ix \cdot (q-p_1-p_2)} \nno \\ && \times \left\{ \:
  \begin{array}{clc}
    \ & \Tr[ \ga_\mu \:{\cal S}^c_{ab}(p_1) \:{\cal S}^s_{ba}(-x) ]\: \cdot
    \Tr[ {\cal S}^s_{cd}(x)\: \ga_\nu \:{\cal S}^c_{dc}(-p_2) ]  & \\
    + & \Tr[ \ga_\mu \:{\cal S}^c_{ab}(p_1)\: \ga_\nu \:{\cal S}^s_{ba}(-x) ]\: \cdot
    \Tr[ {\cal S}^s_{cd}(x)\: {\cal S}^c_{dc}(-p_2) ]  & \\
    + & \Tr[ {\cal S}^c_{ab}(p_1) \:{\cal S}^s_{ba}(-x) ]\: \cdot
    \Tr[ \ga_\mu \:{\cal S}^s_{cd}(x)\: \ga_\nu \:{\cal S}^c_{dc}(-p_2) ]  & \\
    + & \Tr[ {\cal S}^c_{ab}(p_1)\: \ga_\nu \:{\cal S}^s_{ba}(-x) ]\: \cdot
    \Tr[ \ga_\mu \:{\cal S}^s_{cd}(x) \:{\cal S}^c_{dc}(-p_2) ]  &
  \end{array}
  \right\}
  \label{fcmol1}
\end{eqnarray}
In such a case the current (\ref{Jmol1}) is not conserved, then it is possible to write the correlation 
function in terms of two independent Lorentz structures, $\Pi_1(q^2)$ and $\Pi_0(q^2)$, where 
these functions contain the contributions from the spin 1 and 0 states, respectively. Therefore, 
for studying the $1^{--}$ state, one should use the projector (\ref{proj1}) to extract the invariant 
function $\Pi_1(q^2)$ from the correlation function (\ref{fcmol1}). 
The spectral densities are calculated up to dimension-six condensates, working at leading order 
terms in $\alpha_s$ and keeping terms which are linear in the strange quark mass. Then, 
one finally gets 
\vspace{-0.4cm}
\begin{eqnarray}
  \rho^{_{OPE}}_{_{\DsxDso}}(s) = 
  \rho^{pert}_{_{\DsxDso}}(s) + \rho^{\qq[s]}_{_{\DsxDso}}(s) + \rho^{\GGi}_{_{\DsxDso}}(s) + 
  \rho^{\qGq[s]}_{_{\DsxDso}}(s) + \rho^{{\qq[s]}^2}_{_{\DsxDso}}(s) + 
  \rho^{\GGG}_{_{\DsxDso}}(s) ,~~~~
\end{eqnarray}
where the expressions for the spectral densities are given by
\vspace{-0.4cm}
\begin{eqnarray}
\rho^{pert}_{_{\DsxDso}}(s) &=& -\frac{1}{2^{12} \:\pi^6} 
	\int\limits^{\alpha_{max}}_{\alpha_{min}} \!\frac{d\alpha}{\alpha^3} 
	\int\limits^{1 -\alpha}_{\beta_{min}} \!\frac{d\beta}{\beta^3}
	(1 \!-\! \al \!-\! \be) {\cal F}^3_{(\al,\be)}
	\left[ 2 m_Q^2 \left(1 \!-\! \al \!-\! \be \right)^2 -3(1 \!+\! \al \!+\! \be) {\cal F}_{(\al,\be)} \right] \\
\rho^{\qq[s]}_{_{\DsxDso}}(s) &=& \frac{3m_s \qq[s]}{2^{7} \:\pi^4} \Bigg{\{}
	\int\limits^{\alpha_{max}}_{\alpha_{min}} \!\frac{d\alpha \:{\cal H}_{(\al)}^2}{\al(1-\al)} - 
	\int\limits^{\al_{max}}_{\al_{min}} \!\frac{d\al}{\al} \int\limits^{1-\al}_{\be_{min}} \!\frac{d\beta}{\beta}
	{\cal F}_{(\al,\be)} \left[ m_Q^2(5 \!-\! \al \!-\! \be) \!+\! 2 {\cal F}_{(\al,\be)} \right] \Bigg{\}} \hspace{1.2cm}\\
\rho^{\GGi}_{_{\DsxDso}}(s) &=& -\frac{\GG}{3 \cdot 2^{12} \:\pi^6} 
	\int\limits^{\alpha_{max}}_{\alpha_{min}} \!\frac{d\alpha}{\al}
	\int\limits^{1-\alpha}_{\beta_{min}} \!\frac{d\beta}{\beta^3} \Bigg{\{}
	m_Q^4 \al(1-\al-\be)^3  +	3 m_Q^2 (1-\al-\be) \nno\\
	&& \times~ \big[ 1- \al(4+\al+\be) + \be(1-2\al-2\be) \big] {\cal F}_{(\al,\be)}
	+6\be \left( 1-2\al-2\be \right) {\cal F}_{(\al,\be)}^2 \Bigg{\}} \\
\rho^{\qGq[s]}_{_{\DsxDso}}(s) &=& \frac{\qGq[s]}{2^{8} \:\pi^4} \Bigg\{
	3 m_Q \!\!\int\limits^{\alpha_{max}}_{\alpha_{min}} \!\!\!\frac{d\alpha}{\al^2}
	\!\!\int\limits^{1-\al}_{\be_{min}} \!\!\!\frac{d\be}{\be} \bigg( \al^2 \!-\! \al(1 \!+\! \be) - 2\be^2 \bigg)  
	{\cal F}_{(\al,\be)} - m_s \Bigg[ \int\limits^{\alpha_{max}}_{\alpha_{min}} \!\!\!\frac{d\alpha}{\al} \nno\\
	&& \times~ \bigg( 8 m_Q^2 \al + (2 \!-\! \al) {\cal H}_{(\al)} \bigg) 
	-\int\limits^{\alpha_{max}}_{\alpha_{min}} \!\!\!\!d\alpha
	\!\!\!\int\limits^{1-\al}_{\be_{min}} \!\!\!\frac{d\be}{\be} 
	\bigg( m_Q^2 (9 \!-\! 3\al \!-\! 4\be) + 7 {\cal F}_{(\al,\be)} \!\bigg) \Bigg]  \Bigg\} \\	
\rho^{{\qq[s]}^2}_{_{\DsxDso}}(s) &=& -\frac{\rho\qq[s]^2}{2^{6} \:\pi^2} 
	\int\limits^{\alpha_{max}}_{\alpha_{min}} \!d\alpha
	\left[ 3m_Q^2 - \al(1-\al)s \right] \\
\rho^{\GGG}_{_{\DsxDso}}(s) &=& -\frac{\GGG}{5\cdot3 \cdot 2^{16} \:\pi^6} \Bigg{\{}
	5\int\limits^{\alpha_{max}}_{\alpha_{min}} \!d\alpha
	\int\limits^{1-\al}_{\be_{min}} \!\frac{d\be}{\be^3} (1-\al-\be) \Bigg[ 
	m_Q^2 \bigg(5 \alpha^2 - \alpha (37 - 19 \beta) \nno\\
	&& +~ 14 (1 - \beta)^2 \bigg) - 3 (7 + 9 \alpha + 9 \beta) {\cal F}_{(\al,\be)} \Bigg]
	+ m_Q^4 \int\limits^{1}_{0} \!d\alpha
	\int\limits^{1-\al}_{0} \!\frac{d\be}{\be^5} ~e^{-\frac{(\al+\be)}{\al\be} m_Q^2 \tau} \nno\\
	&& \times~ 
	(1 \!-\! \al \!-\! \be) \Bigg[ 2 m_Q^2 \tau (1 \!-\! \alpha \!-\! \beta)^2 -\beta \bigg( 50 \alpha^2 \!-\! 
	\alpha (61 \!-\! 85 \beta) \!+\! 35 (1 \!-\! \beta)^2 \!\bigg) \Bigg] \Bigg\} ~~~~~~~~~~	
\end{eqnarray}
Note that, one must consider the definitions (\ref{def:variables}) and the integration limits 
These expressions could also be applied for the following molecular states:
\begin{eqnarray}
  \DxDo ~(1^{--}): && \rho^{_{OPE}}_{\DxDo}(s) ~~=~~ 
  \lim_{m_s \rightarrow 0} ~~\rho^{_{OPE}}_{\DsxDso}(s) \\
  \BsxBso ~(1^{--}): && \rho^{_{OPE}}_{\BsxBso}(s) ~~=~~ 
  \lim_{m_c \rightarrow m_b} ~~\rho^{_{OPE}}_{\DsxDso}(s) \\
  \BxBo ~(1^{--}): && \rho^{_{OPE}}_{\BxBo}(s) ~~=~~ 
  \lim\limits_{\tiny \begin{matrix} m_c \rightarrow m_b \\ m_s \rightarrow 0 \end{matrix}} 
  ~~\rho^{_{OPE}}_{\DsxDso}(s) ~.
\end{eqnarray}

\subsubsection{Numerical Results} 
\begin{figure}[t]
\begin{center}
    \subfloat[]{\includegraphics[width=10cm]{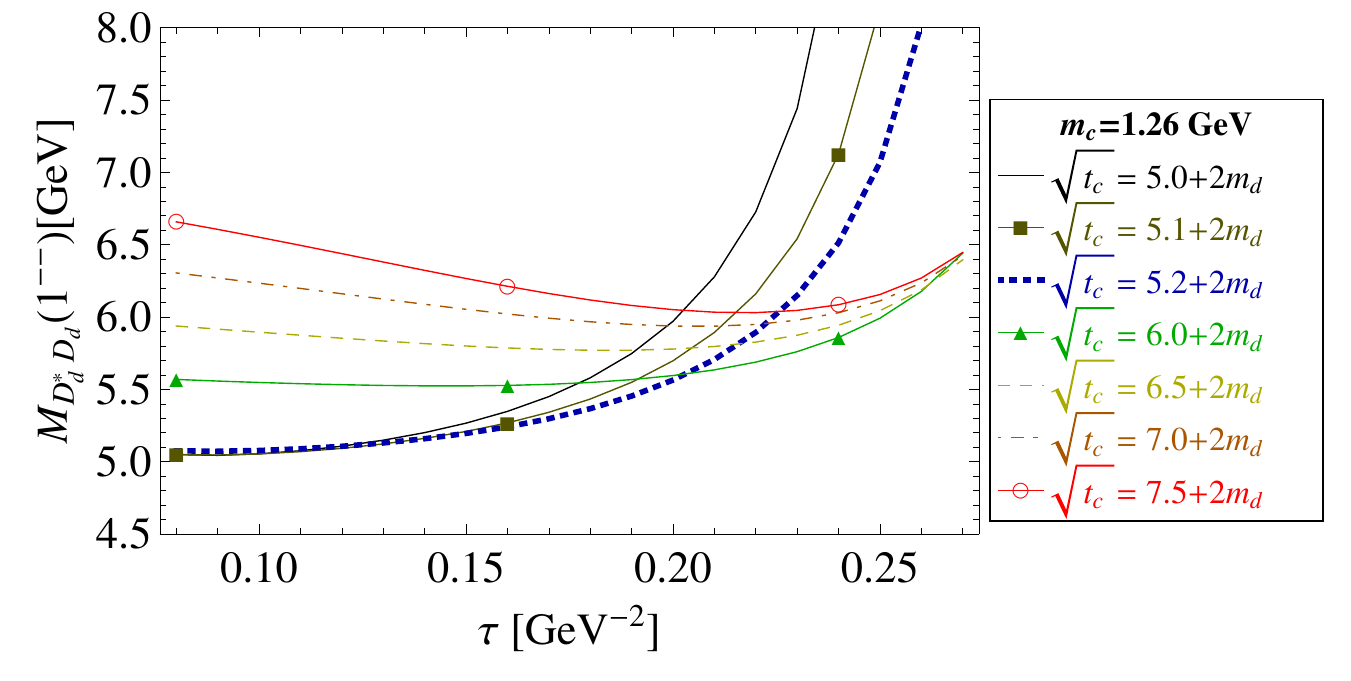}}\\
    \subfloat[]{\includegraphics[width=10cm]{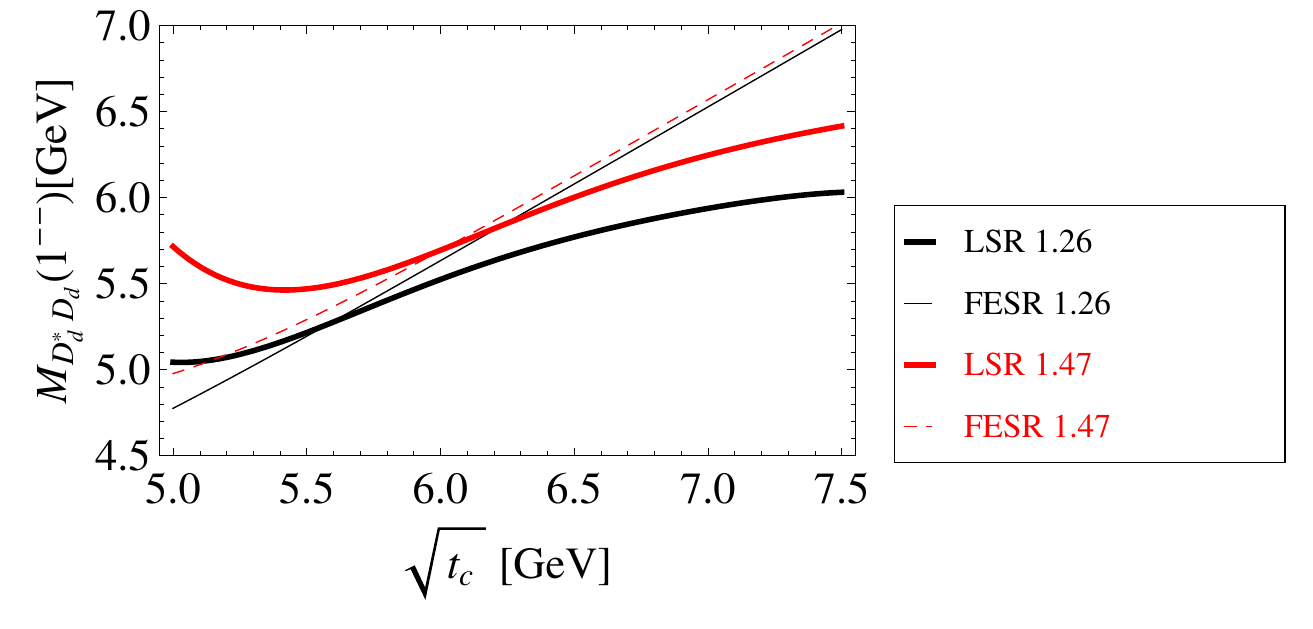}}    
\end{center}
\caption{ \footnotesize Mass of the $\DxDo ~(1^{--})$ molecular state, considering the OPE 
contributions up to dimension-six condensates:
{\bf (a)} as a function of $\tau$, for different values of the continuum threshold $\sqrt{t_c}$
and running $(m_c=1.26\GeV)$ $c$-quark mass .
{\bf (b)} as a function of $\sqrt{t_c}$, obtained from the $\tau$-stability points. The results 
from both methods - QCDSR and FESR - are presented, for running $(m_c=1.26\GeV)$ and 
on-shell $(m_c=1.47\GeV)$ $c$-quark mass.}
\label{fig:molD} 
\end{figure}

Using the QCD parameters given in Table (\ref{TabParam}), one calculates 
the mass of the $\DxDo ~(1^{--})$ molecular state. The result is presented in
Fig.(\ref{fig:molD}a), as a function of sum rule parameter $\tau$ and for different values 
of the continuum threshold $\sqrt{t_c}$. In this calculation, one considers the running mass 
for the $c$-quark $(m_c = 1.26 \GeV)$. Notice that the $\tau$-stability is reached only for 
$\sqrt{t_c} \geq 5.10 \GeV$. Evaluating the mass from the $\tau$-stability points, one obtains 
the $\sqrt{t_c}$-behavior shown in Fig.(\ref{fig:molD}b).
The same analysis is done considering the on-shell mass for $c$-quark 
$(m_c = 1.47 \GeV)$, where the result as a function of $\sqrt{t_c}$ is shown in 
Fig.(\ref{fig:molD}b), as well. Then, the FESR is calculated for this molecular state considering
$n=1$ in Eq.(\ref{mfesr}). The FESR results can be compared with the ones from QCDSR 
as shown in Fig.(\ref{fig:molD}b). 

Finally, combining the results obtained with the QCDSR and FESR, one can deduce the 
common solutions, considering the running $(m_c = 1.26 \GeV)$ and on-shell 
$(m_c = 1.47 \GeV)$ masses for $c$-quark:
\begin{eqnarray}
  M_{\DxDo} &=& 5.28 \GeV ~~~ \mbox{for } \sqrt{t_c} = 5.58 \GeV ~~\mbox{and } m_c =1.26 \GeV\nno\\
  &=& 5.70  \GeV ~~~ \mbox{for } \sqrt{t_c} = 6.10 \GeV ~~\mbox{and } m_c =1.47 \GeV.
\end{eqnarray}
\vfill

In order to fix the values of $M_{\DxDo}$, at leading order in $\al_s$, one must take a glance at 
the analysis made for the charmonium $J/\psi$ mass, which indicate that the on-shell $c$-quark 
mass value tends to overestimate $M_{J/\psi}$ \cite{NNN}. The same feature happens 
to evaluate of the $X ~(1^{++})$ four-quark state mass \cite{ricnar}. Therefore, in the 
present work, one considers the sum rule predictions using the running mass as the final 
result from a QCD sum rule calculation. Thus, including different sources of uncertainties 
from the parameters in Table (\ref{TabParam}), one obtains the final value:
\begin{eqnarray}
  M_{\DxDo} &=& 5268 ~(14)_{m_c} (3)_{\Lambda} (19)_{\qq[q]} (0)_{\GGi} 
  (0)_{m_0^2} (2)_{\GGGi} (5)_{\rho} ~\MeV~~,\nno\\
  &=& 5268 ~(24) \MeV~.
  \label{Mdxdo}
\end{eqnarray}
Using the fact that the FESR (with $n=1$) gives a more robust correlation between the mass of the 
lowest ground state and the onset of continuum threshold $\sqrt{t_c}$, one could assume that 
the mass of the first radial excitation is approximately equal to $\sqrt{t_c}$, then the mass-splitting 
can be estimated as 
\vspace{-0.2cm}
\begin{eqnarray}
  M^\prime_{\DxDo} - M_{\DxDo} &\simeq& 300 \MeV ~~.
\end{eqnarray}
This splitting is much lower than the one intuitively used in the current literature:
\begin{eqnarray}
  M_{\psi(2S)} - M_{J/\psi} &\simeq& 590 \MeV ~,
\end{eqnarray}
for fixing the arbitrary value of $t_c$ entering in different QCDSR of the molecular states. 
This difference may signal some new dynamics for the exotic states compared with the 
usual $c\bar{c}$ charmonium states and need to be tested from some other approaches 
such as potential models, heavy quark symmetry, AdS/QCD and/or lattice calculations.
\begin{figure}[t] 
\begin{center}
    \hspace{1.0cm} \subfloat[]{\includegraphics[width=10.5cm]{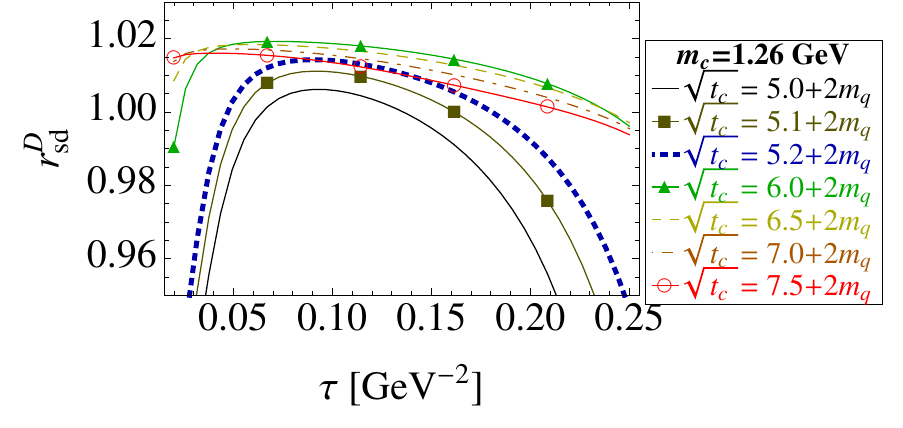}}\\
    \subfloat[]{\includegraphics[width=8cm]{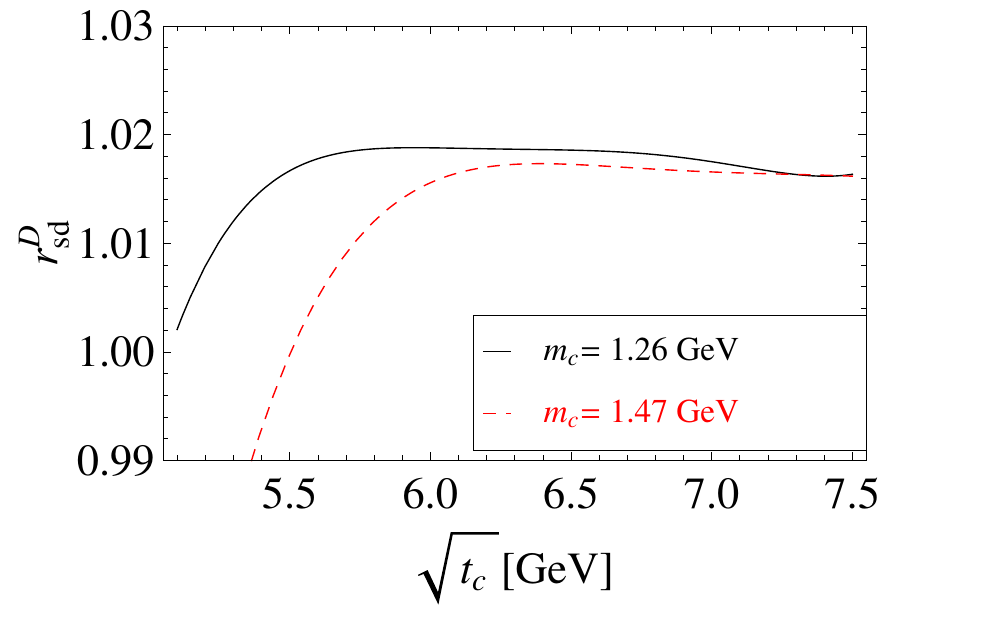}}    
\end{center}
\caption{ \footnotesize The mass ratio between the $\DsxDso ~(1^{--})$ and $\DxDo ~(1^{--})$ 
molecular states using the DRSR, considering the OPE contributions up to 
dimension-six condensates:
{\bf (a)} as a function of $\tau$, for different values of the continuum threshold $\sqrt{t_c}$
and running $(m_c=1.26\GeV)$ $c$-quark mass.
{\bf (b)} as a function of $\sqrt{t_c}$, obtained from the $\tau$-stability points, for running 
$(m_c=1.26\GeV)$ and on-shell $(m_c=1.47\GeV)$ $c$-quark mass.}
\label{fig:rmolD} 
\end{figure}
In Fig.(\ref{fig:su3mol0}), the DRSR calculation is done to estimate the mass ratio between the 
$\DsxDso ~(1^{--})$  and $\DxDo ~(1^{--})$ molecular states. From this figure, one gets the 
following result:
\begin{eqnarray}
  r^D_{sd} &=& \frac{M_{\DsxDso}}{M_{\DxDo}}  ~~=~~ 
  1.018 ~(1)_{m_c} (4)_{m_s} (0.8)_{\kappa} (0.5)_{\qq[q]} (0.2)_{\rho} (0.1)_{\GGGi} ~~.
\end{eqnarray}
Using the previous results in Eq.(\ref{Mdxdo}), one gets 
\begin{eqnarray}
  M_{\DsxDso} &=& 5363 (33) \MeV \\
  \Delta M_{sd}^{D} &\simeq& 95 \MeV.
\end{eqnarray}
where $\Delta M_{sd}^{D} \equiv M_{\DsxDso} - M_{\DxDo}$. These results indicate that 
the masses of the $\DsxDso ~(1^{--})$ and $\DxDo ~(1^{--})$ are well above the meson-meson 
thresholds $E_{th}\left[D^\ast_s D^\ast_{s0}\right] = 4.43 \GeV$ and 
$E_{th}\left[D^\ast_s D^\ast_{s0}\right] = 4.41 \GeV$, respectively. This could be an indication that 
these exotic states are weakly bounded or even cannot correspond to a bound state.
The masses obtained with this criterion for fixing the value of the continuum thresholds - 
through the intersection points of the QCDSR and FESR - are compatible with the ones 
in Ref.\cite{ramar1}.

\subsection{$\DsoDso ~(0^{++})$ and $\DoDo ~(0^{++})$}
Subsequently, one extends the previous analysis to the scalar $0^{++}$ molecular states. 
Extracting the longitudinal component of the correlation function (\ref{fcmol1}), one gets the 
invariant function $\Pi_0(q^2)$, which provides the spectral densities calculated up to 
dimension-six condensates in the OPE. Thus, 
$\rho^{_{OPE}}_{0}(s) = \frac{1}{\pi} ~\mbox{Im } \Pi_0(s)$.
The expressions for the $\DsoDso ~(0^{++})$ molecular state are given by\\
\begin{eqnarray}
\rho^{pert}_{\DsoDso} &=& -\frac{1}{2^{12} \:\pi^6} 
	\int\limits^{\alpha_{max}}_{\alpha_{min}} \!\!\!\frac{d\alpha}{\alpha^3} 
	\int\limits^{1-\alpha}_{\beta_{min}} \!\!\!\frac{d\beta}{\beta^3} (1-\al-\be)  
	\Bigg{[} 12 m_Q^4 (\al+\be) (1-\al-\be)^2 {\cal F}_{(\al,\be)}^{2} \nno\\
	&& -~ 2m_Q^2 (1-\al-\be)(7-19\al-19\be) {\cal F}_{(\al,\be)}^{3} 
	-\!\! 3(7-9\al-9\be) {\cal F}_{(\al,\be)}^{4} \Bigg{]} \\ && \nno\\	
\rho^{\qq[s]}_{\DsoDso} &=& -\frac{3m_s \qq[s]}{2^{7} \:\pi^4} \Bigg{\{}
	\int\limits^{\alpha_{max}}_{\alpha_{min}} \!d\alpha\frac{ {\cal H}_{(\al)}^{2}}{\al(1-\al)} + 
	\int\limits^{\al_{max}}_{\al_{min}} \!\!\!\frac{d\al}{\al} 
	\int\limits^{1-\al}_{\be_{min}} \!\!\!\frac{d\beta}{\beta}
	\Bigg[ 2m_Q^4(\al+\be) (1\!-\! \al \!-\! \be) \nno\\
	&& -~ m_Q^2 (7-11\al-11\be) {\cal F}_{(\al,\be)} -  10 {\cal F}_{(\al,\be)}^{2} \Bigg] \Bigg{\}} \\
	&& \nno\\
\rho^{\GGi}_{\DsoDso} &=& -\frac{\GG}{3 \cdot 2^{12} \:\pi^6} \Bigg{\{}
	\int\limits^{\alpha_{max}}_{\alpha_{min}} \!\frac{d\alpha}{\al}
	\int\limits^{1-\alpha}_{\beta_{min}} \!\frac{d\beta}{\beta^3}  \Bigg[ 
	m_Q^4 (1\!-\! \alpha \!-\! \beta)^2 \bigg( 7 \alpha^2 + \alpha (5 + 19 \beta) + 
	6 \beta (1 + 2 \beta) \bigg) \nno\\
	&& +~ 3 m_Q^2 (1 \!-\! \alpha \!-\! \beta)  \bigg( \alpha (4 \!+\! 3\al \!+\! 25 \beta) \!-\! 
	\beta (3 \!-\! 22 \beta) \!-\! 3 \bigg) {\cal F}_{(\al,\be)}  - \!\! 6 \beta (9 \!-\! 10 \alpha \!-\! 10 \beta) 
	{\cal F}_{(\al,\be)}^{2} \Bigg] \nno\\
	&& -~ 2 m_Q^6 \int\limits^{1}_{0} 
	\!\!d\alpha \!\!\int\limits^{1-\al}_{0} \!\!d\beta \bigg[\frac{(\al+\be)(1-\al-\be)^3}{\al \beta^4}\bigg] 
	~e^{-\frac{(\al+\be)}{\al\be} m_Q^2 \tau} \Bigg{\}}\\
	&& \nno\\
\rho^{\qGq[s]}_{\DsoDso} &=& -\frac{\qGq[s]}{2^{8} \:\pi^4} \Bigg{\{}
	3m_Q \!\!\int\limits^{\alpha_{max}}_{\alpha_{min}} \!\frac{d\alpha}{\al^2}
	\int\limits^{1-\al}_{\be_{min}} \!\frac{d\be}{\be} \Bigg{[} 2m_Q^2(1 \!-\! \al \!-\! \be)
	(\al \!+\! \be)(\al \!-\! 2\be) - \bigg(3\al(1 \!+\! \al \!+\! \be) \nno\\
	&&-~ 2\be(2 \!-\! 3\be) \bigg) {\cal F}_{(\al,\be)} \Bigg] + m_s 
	\int\limits^{\alpha_{max}}_{\alpha_{min}} \!\frac{d\alpha}{\al} 
	\left[ 2m_Q^2 \al  \!-\! (2 \!+\! 9\al) {\cal H}_{(\al)} \right] + m_s 
	\!\!\int\limits^{\alpha_{max}}_{\alpha_{min}} \!\!\!d\alpha
	\int\limits^{1-\al}_{\be_{min}} \!\!\!\frac{d\be}{\be} \bigg( 21 {\cal F}_{(\al,\be)} \nno\\
	&& +~ m_Q^2 (9 \!-\! 17\al \!-\! 18\be) \bigg) + 
	2m_s m_Q^4 \!\int\limits^{1}_{0} \!\!d\alpha \!\!\int\limits^{1-\al}_{0} \!\!d\beta 
	~e^{-\frac{(\al+\be)}{\al\be} m_Q^2 \tau} \:\frac{(\al \!+\! \be)(3 \!-\! 3\al \!-\! 4\be)}{\al \beta^2}  
	\Bigg{\}} 
\end{eqnarray}

\begin{eqnarray}
\rho^{\qq[s]^2}_{\DsoDso} &=& \frac{3\rho\qq[s]^2}{2^5 \:\pi^2} 
	\int\limits^{\alpha_{max}}_{\alpha_{min}} \!d\alpha
	\left[ m_Q^2 - \al(1-\al)s \right]\\
\rho^{\GGGi}_{\DsoDso} &=& \frac{\GGG}{5\cdot3\cdot 2^{16} \:\pi^6}  \Bigg\{
	5 \!\!\int\limits^{\alpha_{max}}_{\alpha_{min}} \!\!\!d\alpha
	\!\!\int\limits^{1-\al}_{\be_{min}} \!\!\!\frac{d\be}{\be^3} (1 \!-\! \al \!-\! \be) 
	\Bigg[ m_Q^2 \bigg(14 \!-\! \alpha (91 \!-\! 59\al \!-\! 127 \beta) 
	\!-\! 2\be(41 \!-\! 34 \beta) \bigg) \nno\\
	&& +~ (33 \!-\! 81 \alpha \!-\! 81 \beta) {\cal F}_{(\al,\be)} \Bigg]
	+~ m_Q^4 \int\limits^{1}_{0} \!\!\frac{d\al}{\al} 
	\!\!\int\limits^{1-\al}_{0} \!\!\frac{\:d\be}{\be^6} (1 \!-\! \al \!-\! \be) ~ 
	e^{-\frac{(\al+\be)}{\al\be} m_Q^2\tau} \Bigg{[} 4 m_Q^4 \tau^2 (\al \!+\! \be) \nno\\
	&& \times~ (1 \!-\! \al \!-\! \be)^2 + 2 m_Q^2 \tau \be  (1- \al- \be) \bigg( 
	53 \al^2 - \al (38 - 88 \be) - 35 \beta (1 \!-\! \beta) \!\bigg)  \nno\\
	&& +~ \beta^2 \bigg( \!100 \alpha^3 \!
	+\! 140 \beta (1 \!-\! \beta)^2 -\!13 \alpha^2 (23 \!-\! 25 \beta)
	+ 5 \alpha (1 \!-\! \beta) (35 \!-\! 73 \beta) \bigg) \Bigg] \Bigg\}~.  
\end{eqnarray}
These expressions can also be used for the molecular states as follows:
\begin{eqnarray}
  \DoDo ~(0^{++}): && \rho^{_{OPE}}_{\DoDo}(s) ~~=~~ 
  \lim_{m_s \rightarrow 0} ~~\rho^{_{OPE}}_{\DsoDso}(s) \\
  \BsoBso ~(0^{++}): && \rho^{_{OPE}}_{\BsoBso}(s) ~~=~~ 
  \lim_{m_c \rightarrow m_b} ~~\rho^{_{OPE}}_{\DsoDso}(s) \\
  \BoBo ~(0^{++}): && \rho^{_{OPE}}_{\BoBo}(s) ~~=~~ 
  \lim\limits_{\tiny \begin{matrix} m_c \rightarrow m_b \\ m_s \rightarrow 0 \end{matrix}} 
  ~~\rho^{_{OPE}}_{\DsoDso}(s) ~.
\end{eqnarray}

\subsubsection{Numerical Results}
\begin{figure}[t]
\begin{center}
  \subfloat[]{\includegraphics[width=10cm]{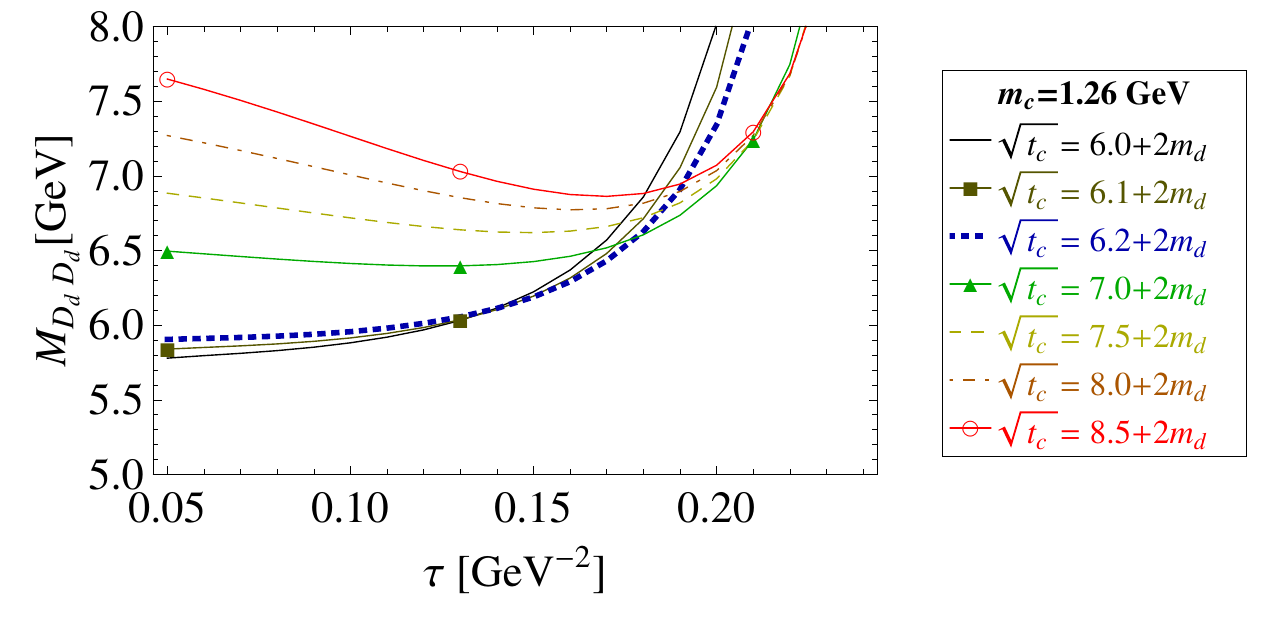}}\\
  \subfloat[]{\includegraphics[width=10cm]{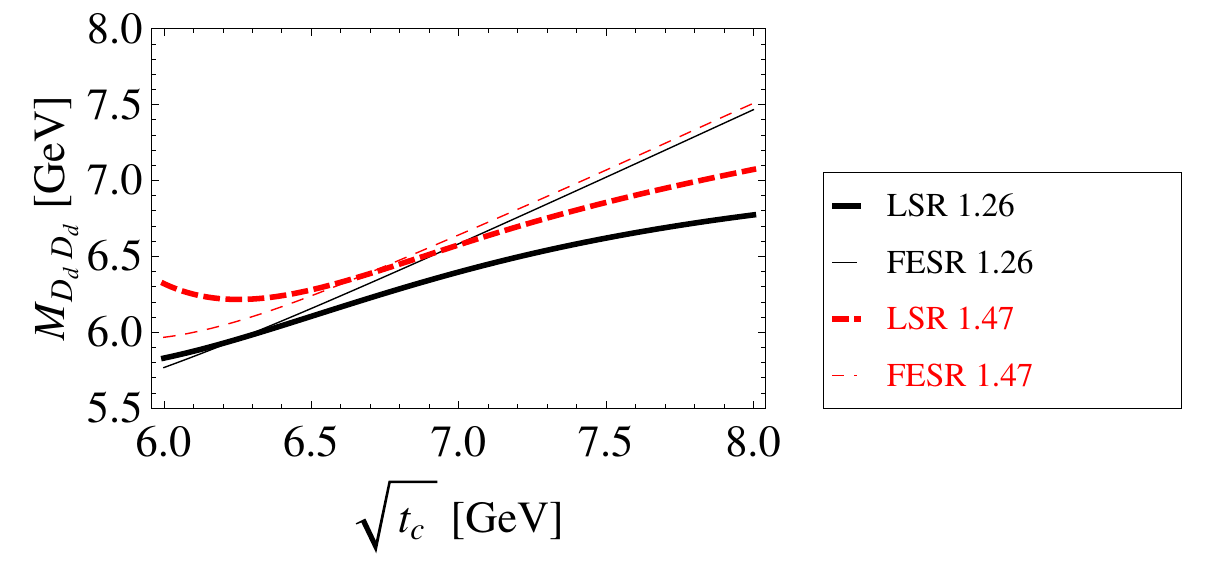}}    
\end{center}
\caption{ \footnotesize Mass of the $\DoDo ~(0^{++})$ molecular state, considering the OPE 
contributions up to dimension-six condensates:
{\bf (a)} as a function of $\tau$, for different values of the continuum threshold $\sqrt{t_c}$ and 
running $(m_c=1.26\GeV)$ $c$-quark mass.
{\bf (b)} as a function of $\sqrt{t_c}$, obtained from the $\tau$-stability points. The results 
from both methods - QCDSR and FESR - are presented, for running $(m_c=1.26\GeV)$ and 
on-shell $(m_c=1.47\GeV)$ $c$-quark mass.}
\label{fig:mol0} 
\end{figure}

To evaluate the mass of the $\DoDo ~(0^{++})$ molecular state, one uses the result obtained 
from the match between QCDSR and FESR. Therefore, using Eqs.(\ref{massa}) and (\ref{mfesr}), 
along with the spectral densities $\rho^{_{OPE}}_{\DsoDso}(s)$ and the parameters in Table 
(\ref{TabParam}), one obtains the results shown in Fig.(\ref{fig:mol0}).

\begin{figure}[t]
\begin{center}
  \subfloat[]{\includegraphics[width=10cm]{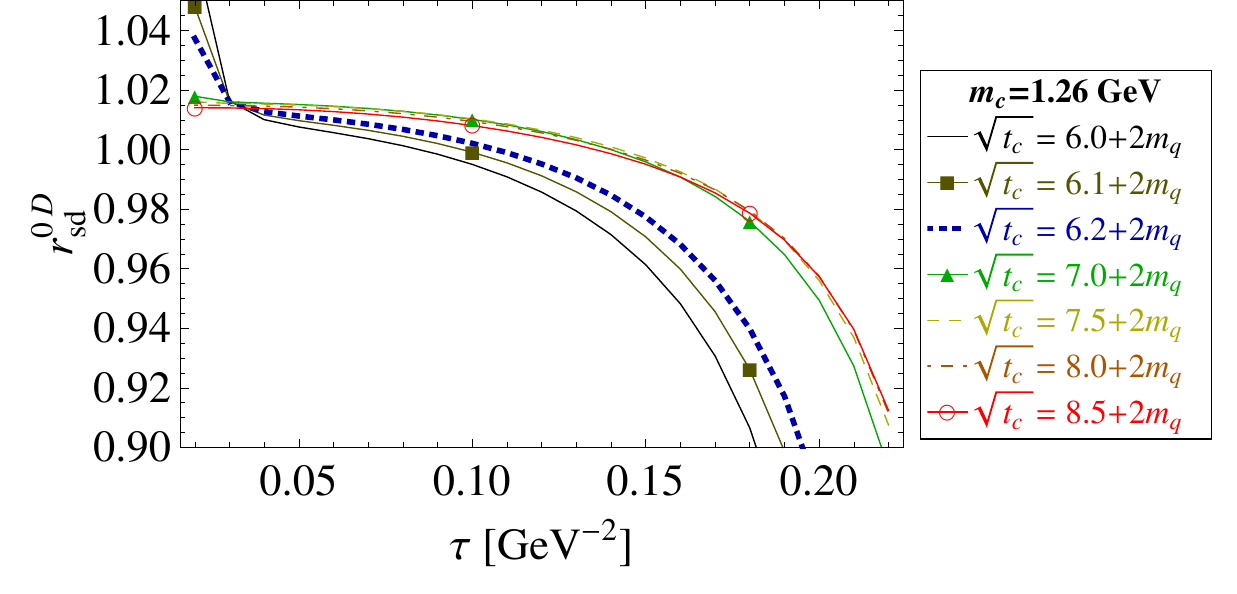}}\\
  \subfloat[]{\includegraphics[width=10cm]{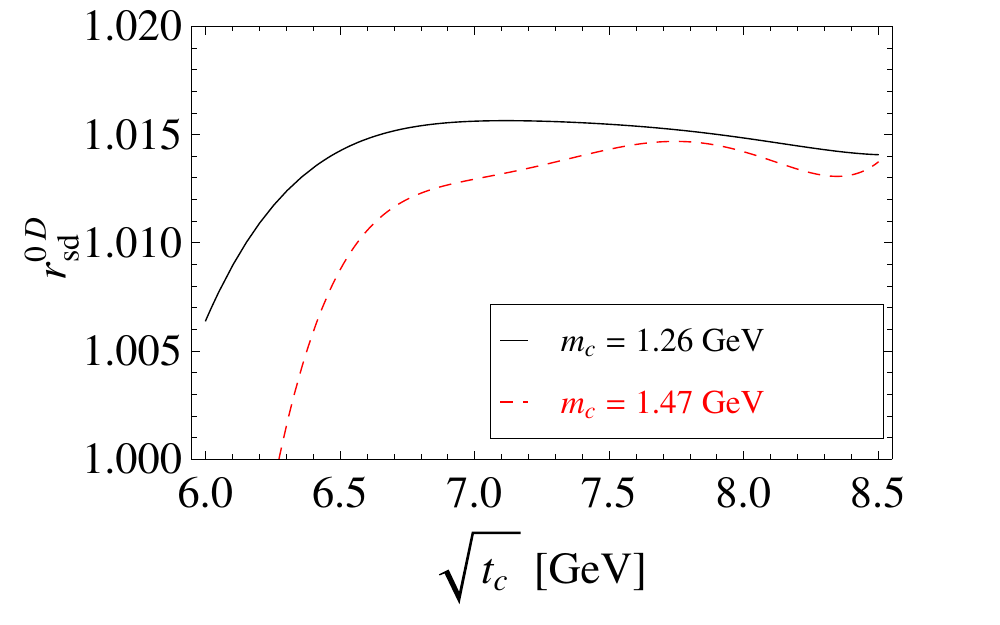}}    
\end{center}
\caption{ \footnotesize The mass ratio between the $\DsoDso ~(0^{++})$ and 
$\DoDo ~(0^{++})$ molecular states using the DRSR, considering the OPE 
contributions up to dimension-six condensates:
{\bf (a)} as a function of $\tau$, for different values of the continuum threshold $\sqrt{t_c}$
and running $(m_c=1.26\GeV)$ $c$-quark mass.
{\bf (b)} as a function of $\sqrt{t_c}$, obtained from the $\tau$-stability points, for running 
$(m_c=1.26\GeV)$ and on-shell $(m_c=1.47\GeV)$ $c$-quark mass.}
\vspace{-0.5cm}
\label{fig:su3mol0} 
\end{figure}

In Fig.(\ref{fig:mol0}a), the mass values are shown as a function of $\tau$, for different values of 
$\sqrt{t_c}$. As one can see the $\tau$-stability is obtained when $\sqrt{t_c} \geq 6.20 \GeV$. 
Evaluating the mass from the $\tau$-stability points, one obtains the $\sqrt{t_c}$-behavior shown 
in Fig.(\ref{fig:mol0}b). Using the FESR, with $n=1$ in Eq.(\ref{mfesr}), one can deduce 
\vspace{-0.5cm}
\begin{eqnarray}
  M_{\DoDo} &=& 5955 ~(24)_{t_c} (14)_{m_c} (5)_{\Lambda} (36)_{\qq[q]} (4)_{\GGi} 
  (4)_{\GGGi} (12)_{\rho} ~~,\nno\\
  &=& 5955 ~(48) ~\MeV~.
  \label{Mdodo}  
\end{eqnarray}
In Fig.(\ref{fig:su3mol0}), the DRSR calculation is done to estimate the mass ratio between the 
$\DsoDso ~(0^{++})$ and $\DoDo ~(0^{++})$ molecular states. From this figure, one gets the 
following result:
\vspace{-0.5cm}
\begin{eqnarray}
  r^{0D}_{sd} &=& \frac{M_{\DsoDso}}{M_{\DoDo}}  ~~=~~ 
  1.015 ~(1)_{m_c} (4)_{m_s} (2)_{\kappa} (1)_{\qq[q]} (0.5)_{\rho} ~~.
\end{eqnarray}
Using the previous values of $M_{\DoDo}$, estimated in Eq.(\ref{Mdodo}), one has 
\begin{eqnarray}
  M_{\DsoDso} &=& 6044 (56) \MeV \\
  \Delta M_{sd}^{0D} &\simeq& 89 \MeV.
\end{eqnarray}
These results indicate that the masses of the scalar molecular states are much higher than the 
vector ones. As an interesting example, the same occurs for the mass of the charmonium 
$J/\psi ~(1^{--})$ which is smaller than the one for the scalar $\chi_{c0} ~(0^{++})$ meson.
All of these results for the $\DsxDso ~(1^{--})$, $\DxDo ~(1^{--})$, $\DsoDso ~(0^{++})$ and 
$\DoDo ~(0^{++})$ molecular states are published in Ref.\cite{SNmad}.

\begin{figure}[t]
\begin{center}
    \subfloat[]{\includegraphics[width=8cm]{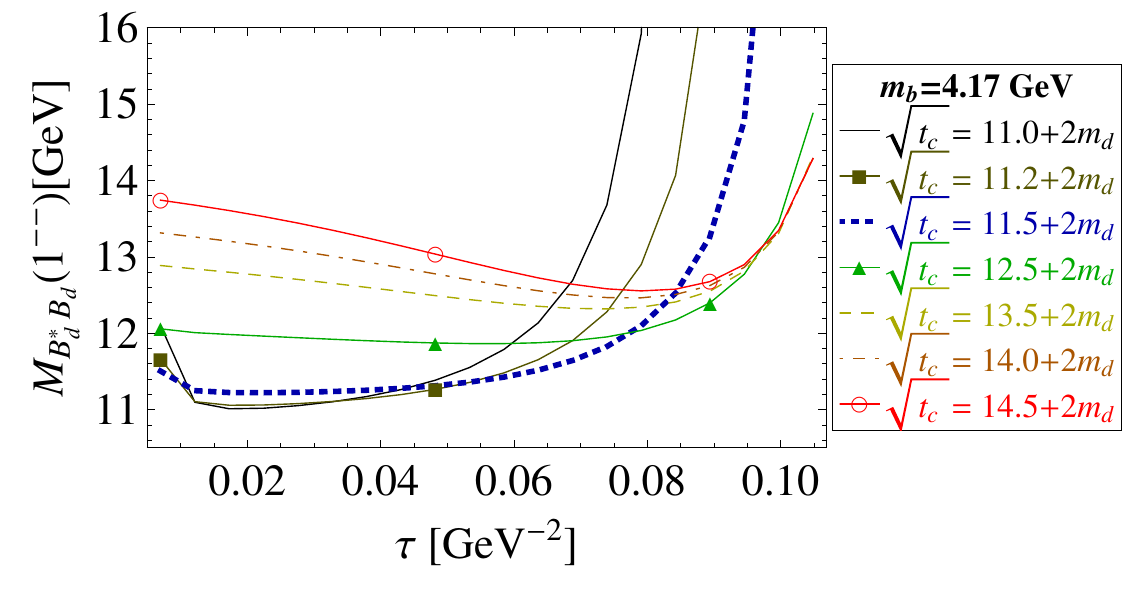}} ~
    \subfloat[]{\includegraphics[width=8cm]{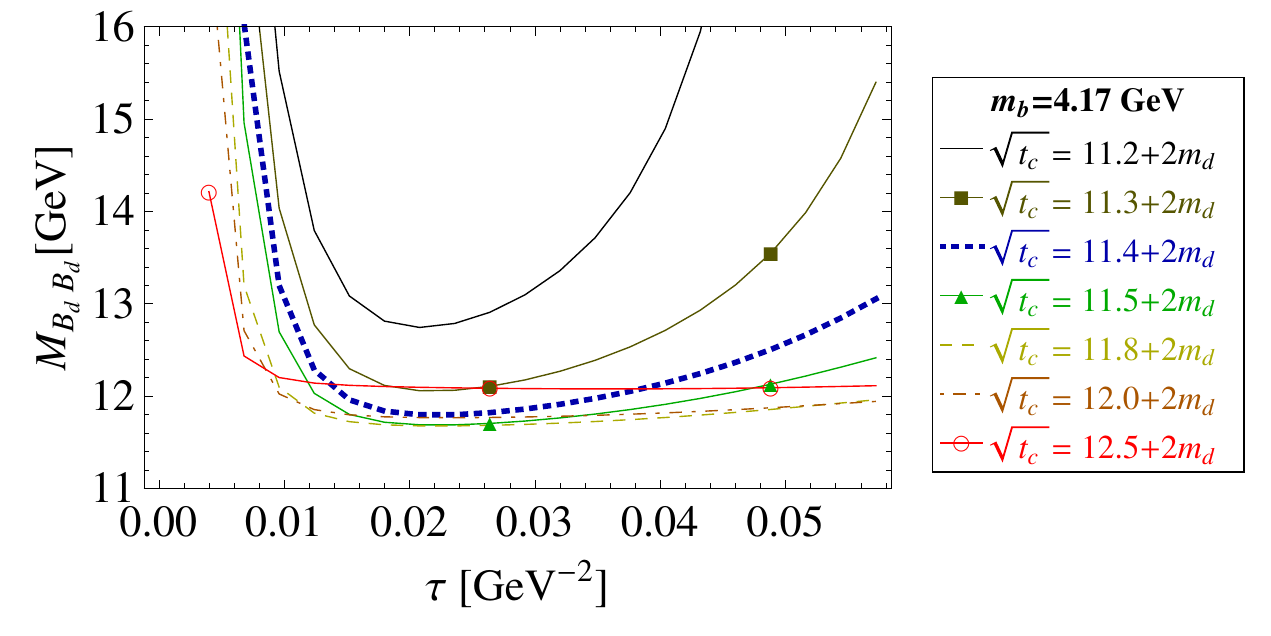}} \\
    \subfloat[]{\includegraphics[width=8cm]{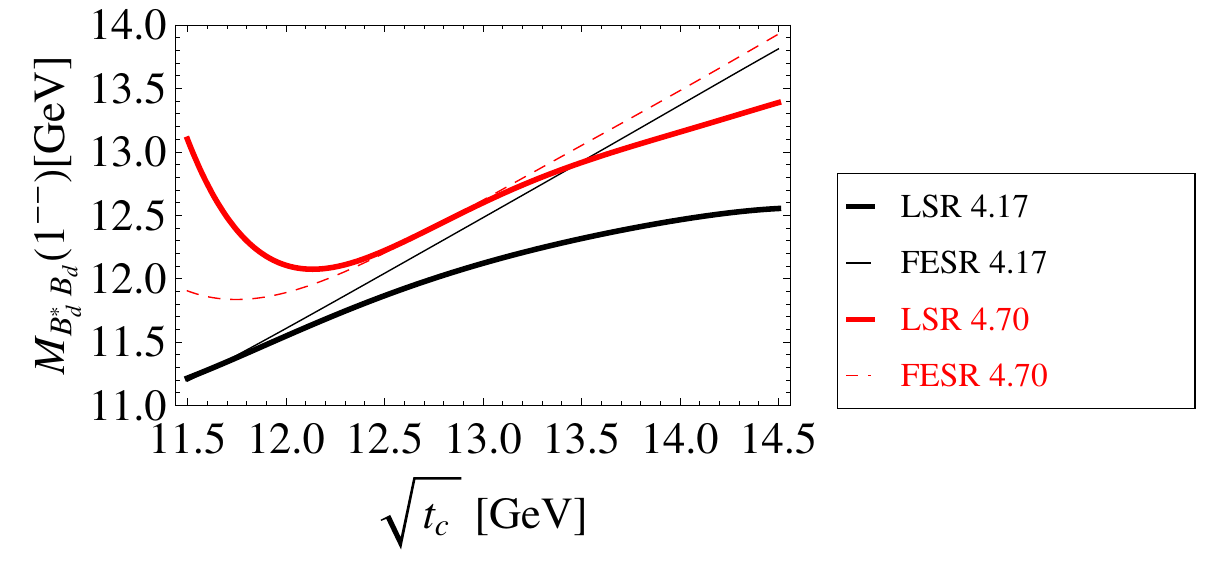}} ~
    \subfloat[]{\includegraphics[width=8cm]{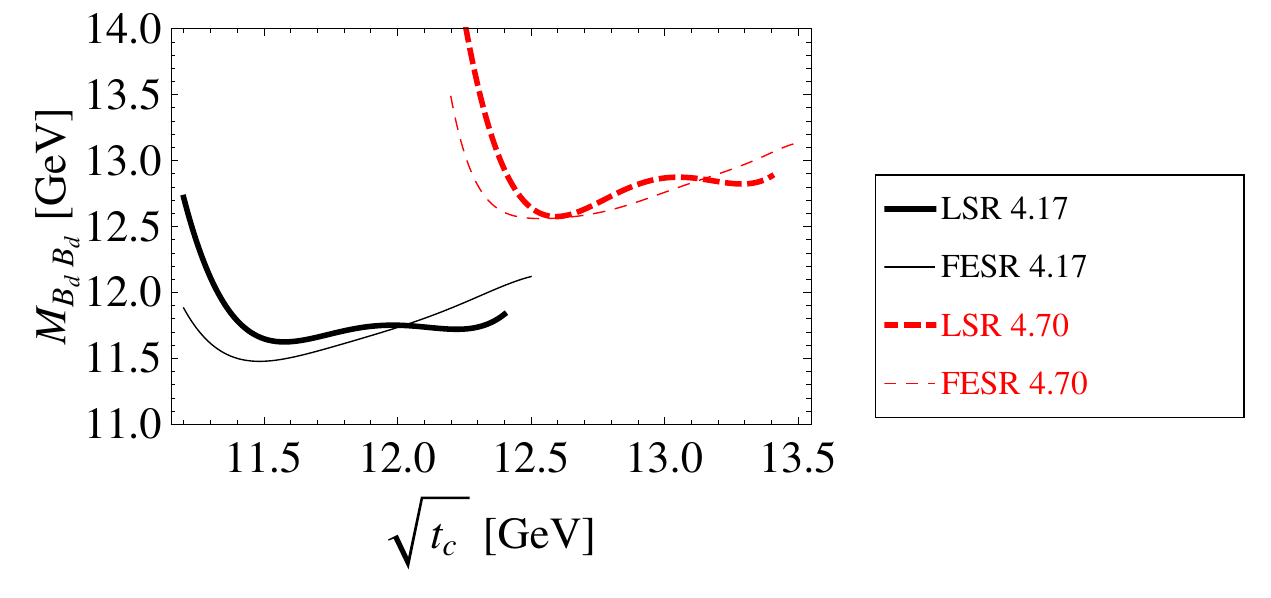}}    
\end{center}
\caption{ \footnotesize Masses of the $\BxBo ~(1^{--})$ and $\BoBo ~(1^{--})$ molecular states, 
considering the OPE contributions up to dimension-six condensates:
{\bf (a)} as a function of $\tau$, for different values of the continuum threshold $\sqrt{t_c}$ and 
running $(m_b=4.17\GeV)$ $b$-quark mass.
{\bf (b)} as a function of $\sqrt{t_c}$, obtained from the $\tau$-stability points. The results 
from both methods - QCDSR and FESR - are presented, for running $(m_b=4.17\GeV)$ and 
on-shell $(m_b=4.70\GeV)$ $c$-quark mass.}
\vspace{1.2cm}
\label{fig:molB}
\end{figure}
%

\subsection{$\BsxBso ~(1^{--})$, $\BxBo ~(1^{--})$, $\BsoBso ~(0^{++})$ and $\BoBo ~(0^{++})$}
The expressions of the spectral densities for $\DsxDso ~(1^{--})$ and $\DsoDso ~(0^{++})$ 
molecular states can be used to obtain the ones for $\BsxBso ~(1^{--})$ and $\BsoBso ~(0^{++})$
by exchanging the heavy quark mass: $m_c \to m_b$. Thus, analogous to the calculations 
that have been done in the charmonium sector, one obtains the results presented in 
Table (\ref{BB}).

The $\tau$-behavior is shown in Fig.(\ref{fig:molB}) for the $\BxBo ~(1^{--})$ and 
$\BoBo ~(0^{++})$ molecular states, as well as the figure comparing the results from QCDSR 
and FESR. In both cases, the values of the mass are extracted at the intersection point of the 
two methods and considering the value of the running $b$-quark mass $(m_b = 4.17 \GeV)$ - 
see Table (\ref{BB}).

\begin{figure}[t]
\begin{center}
  \subfloat[]{\includegraphics[width=8cm]{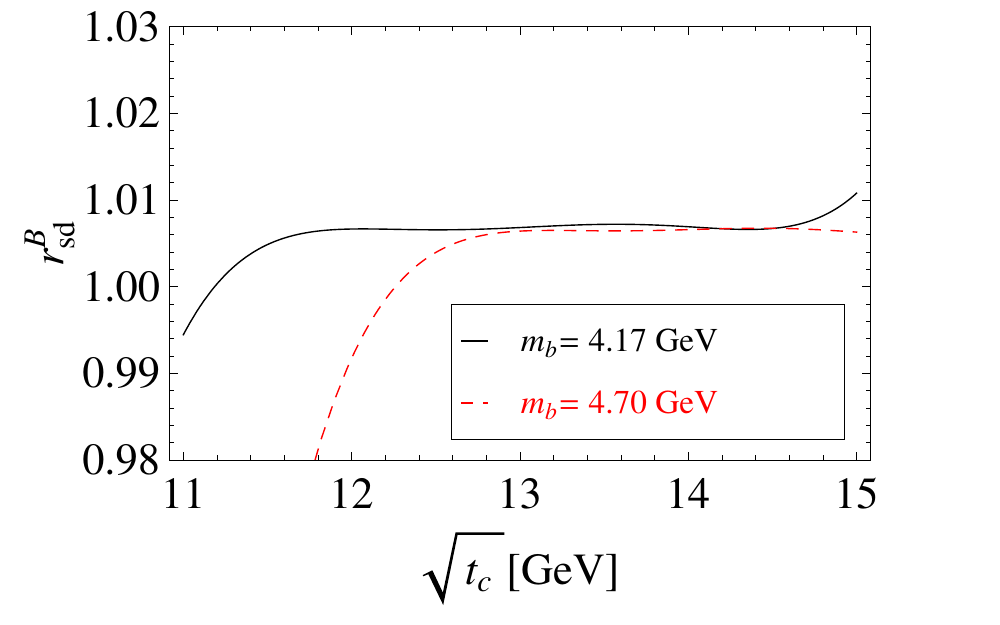}}
  \subfloat[]{\includegraphics[width=8cm]{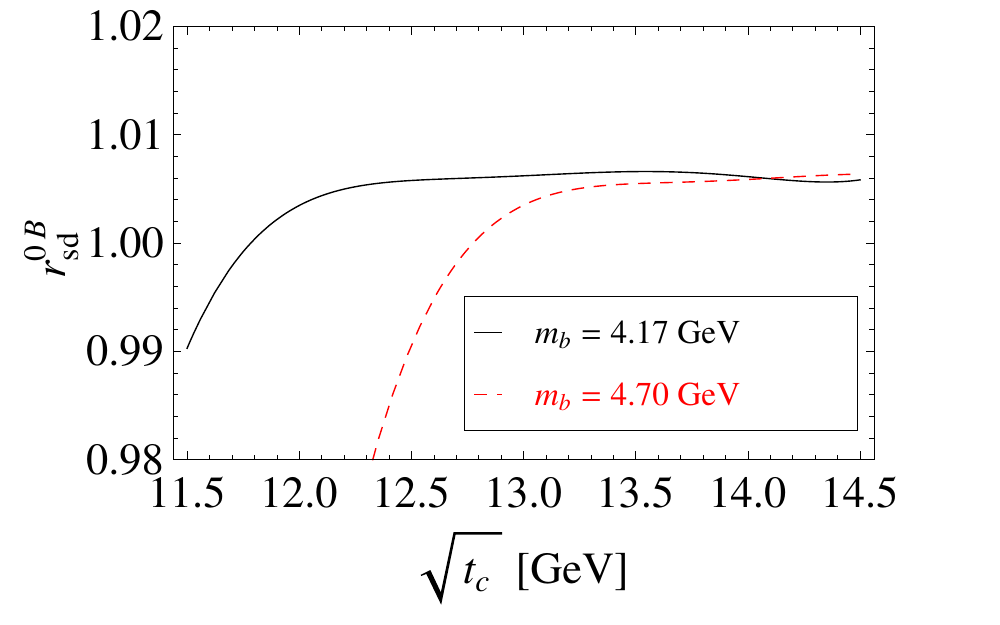}}    
\end{center}
\caption{\footnotesize The mass ratio between the $\BsoBso ~(0^{++})$ and 
$\BoBo ~(0^{++})$ molecular states using the DRSR, considering the OPE 
contributions up to dimension-six condensates, working with running $(m_b=4.17 \GeV)$ 
and on-shell $(m_b = 4.70 \GeV)$ $b$-quark mass and presented as a function 
of $\sqrt{t_c}$, obtained from the $\tau$-stability points:
{\bf (a)} $M_{\BsxBso} / M_{\BxBo}$ and {\bf (b)} $M_{\BsoBso}/ M_{\BoBo}$.}
\label{fig:su3B} 
\end{figure}

Subsequently, one uses the DRSR to estimate the mass ratio for the 
$\BsxBso$ and $\BxBo$ states and the results are shown in Fig.(\ref{fig:su3B}). 
The DRSR for $1^{--}$ vector states is given by
\begin{eqnarray}
  r^B_{sd} ~~=~~ \frac{M_{\BsxBso}}{M_{\BxBo}}  ~~=~~ 
  1.006 ~(1)_{m_b} (2)_{m_s} (1)_{\kappa} (0.5)_{\qq[q]} (0.2)_{\rho} (0.1)_{\GGGi} ~~.
\end{eqnarray}
while for $0^{++}$ scalar states:
\begin{eqnarray}
  r^{0B}_{sd} ~~=~~ \frac{M_{\BsoBso}}{M_{\BoBo}} ~~=~~
  1.008 ~(1)_{m_b} (4)_{m_s} (2)_{\kappa} (1)_{\qq[q]} (0.5)_{\rho} ~~.
\end{eqnarray}
These ratios lead to the masses as shown in Table (\ref{BB}).

As one can see, the masses obtained for all molecular states studied in this section are not in 
agreement with the ones observed for the new bottomonium states: $Y_b(10890)$ and $Y_b(11020)$.
All of these results for the $\BsxBso ~(1^{--})$, $\BxBo ~(1^{--})$, $\BsoBso ~(0^{++})$ and 
$\BoBo ~(0^{++})$ molecular states are also published in Ref.\cite{SNmad}.

{\small
\begin{table}[htbp]
\centering
\setlength{\tabcolsep}{1.25pc}
\caption{\small Masses of the bottomonium molecular states from the present analysis combining 
	QCDSR and FESR. These results correspond to the value of the running heavy quark 
	masses.}
\begin{tabular}{ccl}
&\\
\hline
State &$J^{PC}$ & Mass ($\MeV$)\\
\hline
$\BxBo$ & $1^{--}$ & $11302 (30)$ \\
$\BsxBso$ & $1^{--}$ & $11370 (40)$ \\
$\BoBo$ & $0^{++}$ & $11750 (40)$ \\
$\BsoBso$ & $0^{++}$ & $11844 (50)$ \\
\hline \\ \\
\end{tabular}
\label{BB}
\end{table}}

\subsection{$J/\psi \:f_0(980) ~(1^{--})$}
The first $1^{--}$ state observed in $e^+e^-$ annihilation through initial state radiation (ISR) 
was the $Y(4260)$ \cite{babar3}. Conducting a similar experiment which led to the 
observation of the $Y(4260)$ state, in the channel 
$e^+e^- \rightarrow \gamma_{_{ISR}} \psi(2S) \pi^+ \pi^-$, BaBar collaboration \cite{babary} 
has identified another broad peak at a mass around $4.32 \GeV$, which was confirmed by 
Belle collaboration \cite{belle}, and also announced another new resonance on charmonium 
spectroscopy, the $Y(4660)$ state.

There are many theoretical interpretations for these states \cite{nnl, olsen, brambilla}. In the 
case of the $Y(4260)$ state, although it seems not to fit into the conventional charmonium 
spectroscopy \cite{pdg}, the authors in Ref.\cite{estrada} describe it as a charmonium 
resonance state $\psi(4S)$. There are many other interpretations for this state, like: 
tetraquarks \cite{maiani1}, hadronic molecules $D_1D$, $D_0D^\ast$ \cite{ramar1, ding}, 
$\Xi_{c1} \omega$ \cite{ywm}, $\Xi_{c1} \rho$ \cite{lzl}, $J/\psi \:f_0(980)$ \cite{oset}, a 
hybrid charmonium \cite{zhu1} and cusp \cite{rupp}.

In the present work, one uses the QCDSR approach to study if a correlation function 
based on a $J/\psi \:f_0(980)$ molecular current, with $J^{{PC}} = 1^{--}$, could 
describe the new observed charmonium resonance $Y(4260)$. 
A possible current is given by
\begin{eqnarray}
  j_\mu &=& \big( \bar{c}_a \gamma_\mu c_a \big) \big(\bar{s}_b s_b \big) ~.
  \label{eq:SngJpsif0}
\end{eqnarray}
Although, there are conjectures that the $f_0(980)$ itself could be a tetraquark state \cite{jaffe}, 
in Ref.\cite{ricnav} it was discussed the difficulties in to explain the light scalars as tetraquark states 
from a QCDSR calculation. Therefore, one assumes that a single quark-antiquark current could 
describe the $f_0(980)$.

Another possibility for the current is considering the vector and scalar parts in a color octet 
configuration:
\begin{eqnarray}
  j^\lambda_\mu &=& \big( \bar{c}_a \lambda^A_{ab} \gamma_\mu c_b \big) 
  \big(\bar{s}_c \lambda^A_{cd} s_d \big) ~.
  \label{eq:OctJpsif0}
\end{eqnarray}
where $\lambda^A$ are the Gell-Mann matrices. The two currents can be related by the 
change: $j^\lambda_\mu \rightarrow j_\mu$ with $\lambda^A_{ab} \rightarrow \delta_{ab}$. 
Although, the current in Eq.(\ref{eq:OctJpsif0}) cannot be interpreted as a meson-meson 
current since the vector and scalar terms carry color, for simplicity, it will still be called as a 
molecular current. Since the currents in Eqs.(\ref{eq:SngJpsif0}) and (\ref{eq:OctJpsif0}) 
have the lowest dimension for a four-quark current with the $1^{--}$ quantum numbers, 
from the theory of composite-operator renormalization \cite{dixon} it is expected that 
these currents to be multiplicatively renormalizable.

Inserting the current (\ref{eq:SngJpsif0}) into the correlation function, one obtains the 
expression in terms of the full propagators in QCD:
\begin{eqnarray}
  \Pi^{_{OPE}}_{\mu\nu}(q) &=& \frac{i}{(2\pi)^8} \int\! d^4x \:d^4p_1 \:d^4p_2 ~
  e^{ix \cdot (q-p_1-p_2)} ~ \nno\\
  && ~\times ~
  \Tr \left[ {\cal S}^c_{ab}(p_1) \:\ga_\nu\: {\cal S}^c_{ba}(-p_2) \:\ga_\mu \right] ~
  \Tr \left[ {\cal S}^s_{cd}(x) {\cal S}^s_{dc}(-x) \right] ~.
  \label{eq:FCjpsif0}
\end{eqnarray}

With Eq.(\ref{eq:FCjpsif0}) one can determine the spectral density in the QCD 
side for the $J/\psi \:f_0(980)$ molecular state, with $J^{PC} = 1^{--}$. Considering the
OPE contributions up to dimension-six condensates, $\rho^{_{OPE}}(s)$ can be written as
\begin{eqnarray}
  \rho^{_{OPE}}(s) &=& \rho^{pert}(s) + \rho^{\qq[s]}(s) + 
  \rho^{\GGi}(s) + \rho^{\qGq[s]}(s) + \rho^{\qq[s]^2}(s) ~.
\end{eqnarray}
Notice that the difference between the currents (\ref{eq:SngJpsif0}) and (\ref{eq:OctJpsif0}) 
is only proportional to a color factor. Thus, the spectral density above can be expressed in 
terms of both currents, through insertion of the following definitions:
\[
{\footnotesize
\begin{array}{ccc}
&\\
\hline
\mbox{Color Factor} ~~&~~ \mbox{Mesons in} ~~&~~ \mbox{Mesons in}\\
&~~ \mbox{Color Singlet} ~~&~~ \mbox{Color Octet}\\
\hline
\mathcal{N} & 9 / 2^5 & 1 \\
\mathcal{N}^\ast & -9 / 2^2 & 1\\
\hline\\
\end{array}}\]
where the factors were normalized to the color octet configuration. Therefore, the 
expressions are given by
\begin{eqnarray}
\label{rhoope}
\rho^{pert}(s)&=&{\mathcal{N} \over 3\cdot 2^{7} \pi^6}\!\!
\int\limits_{\almin}^{\almax}\!\!\!\!{d\al\over\alpha^3} \!\!
\int\limits_{\bemin}^{1-\al}\!\!\!\!{d\be\over\be^3}(1 \!-\! \al \!-\! \be) {\cal F}^3_{(\al,\be)}
\bigg[ (1 \!+\! \al \!+\! \be) {\cal F}_{(\al,\be)} - 4m_c^2(1 \!-\! \al \!-\! \be) \bigg],  ~~~~~~~~\\
\nno \\
\rho^{\qq[s]}(s)&=&{\mathcal{N} m_s \qq[s] \over 4 \pi^4}
  \Bigg\{ \int\limits_{\almin}^{\almax}\!\!\!\! {d\al} {{\cal H}^2_{(\al)} \over \al(1-\al)} - 
  \!\!\int\limits_{\almin}^{\almax}\!\!\!\! {d\al \over \al} \!\!\int
  \limits_{\bemin}^{1-\al}\!\!\!\!{d\be\over\be} {\cal F}_{(\al,\be)}
  \bigg( {\cal F}_{(\al,\be)} + 2m_c^2 \bigg) \Bigg\}, \\
\nno \\
\rho^{\GGi}(s) &=& -{\GG \over 3^2 \cdot 2^{11}\pi^6} \!\!\int\limits_{\almin}^{\almax}\!\!\!\!{d\al}
\Bigg\{ \mathcal{N}^\ast \frac{ 6 {\cal H}^2_{(\al)} }{\al(1-\al)} - 
\mathcal{N}^\ast \!\!\int\limits_{\bemin}^{1-\al}\!\!\!\!{d\be\over \al^2 \be^2} {\cal F}_{(\al,\be)}
\bigg[ 12 m_c^2 \al\be + {\cal F}_{(\al,\be)} \nno\\
&& \times \Big( 3(1 \!-\! \al^2 \!-\! \be^2) -4 (\al \!+\! \be)(1 \!-\! \al \!-\! \be) \Big) \bigg] +
32\mathcal{N} m_c^2 \int\limits_{\bemin}^{1-\al}\!\!\!\!{d\be\over \al^3\be} (1 \!-\! \al \!-\! \be) \nno\\
&& \times \Bigg[ m_c^2 \:\be(1-\al-\be) + \Big( 3(1-\al-\be) - \be(1\!+\!\al\!+\!\be) \Big) 
{\cal F}_{(\al,\be)} \Bigg] \Bigg\}, \\
\nno \\
\rho^{\qGq[s]}(s)&=&-{m_s\qGq[s]\over 3^2\cdot 2^{4} \pi^4} (32\mathcal{N} + 
3\mathcal{N}^\ast)
\!\!\int\limits_{\almin}^{\almax}\!\! {d\al} ~\al(1-\al) s,\\
\nno \\
\rho^{\qq[s]^2}(s)&=&-{4\mathcal{N} \:\rho \qq[s]^2 \over 9\pi^2}\!\!\int
\limits_{\almin}^{\almax}\!\! {d\al} ~\al(1-\al) s
\end{eqnarray}

To extract reliable results from the sum rule it is necessary to establish that the relative 
contribution of the higher dimension condensate is smaller than $20\%$ of the total 
contribution, as well as imposing that the pole contribution is bigger than the continuum 
contribution. It is also noteworthy that there is $\tau$-stability inside the Borel window.

\begin{figure}[t]
    \hspace{-1.1cm}
    \subfloat[]{\includegraphics[width=7.5cm]{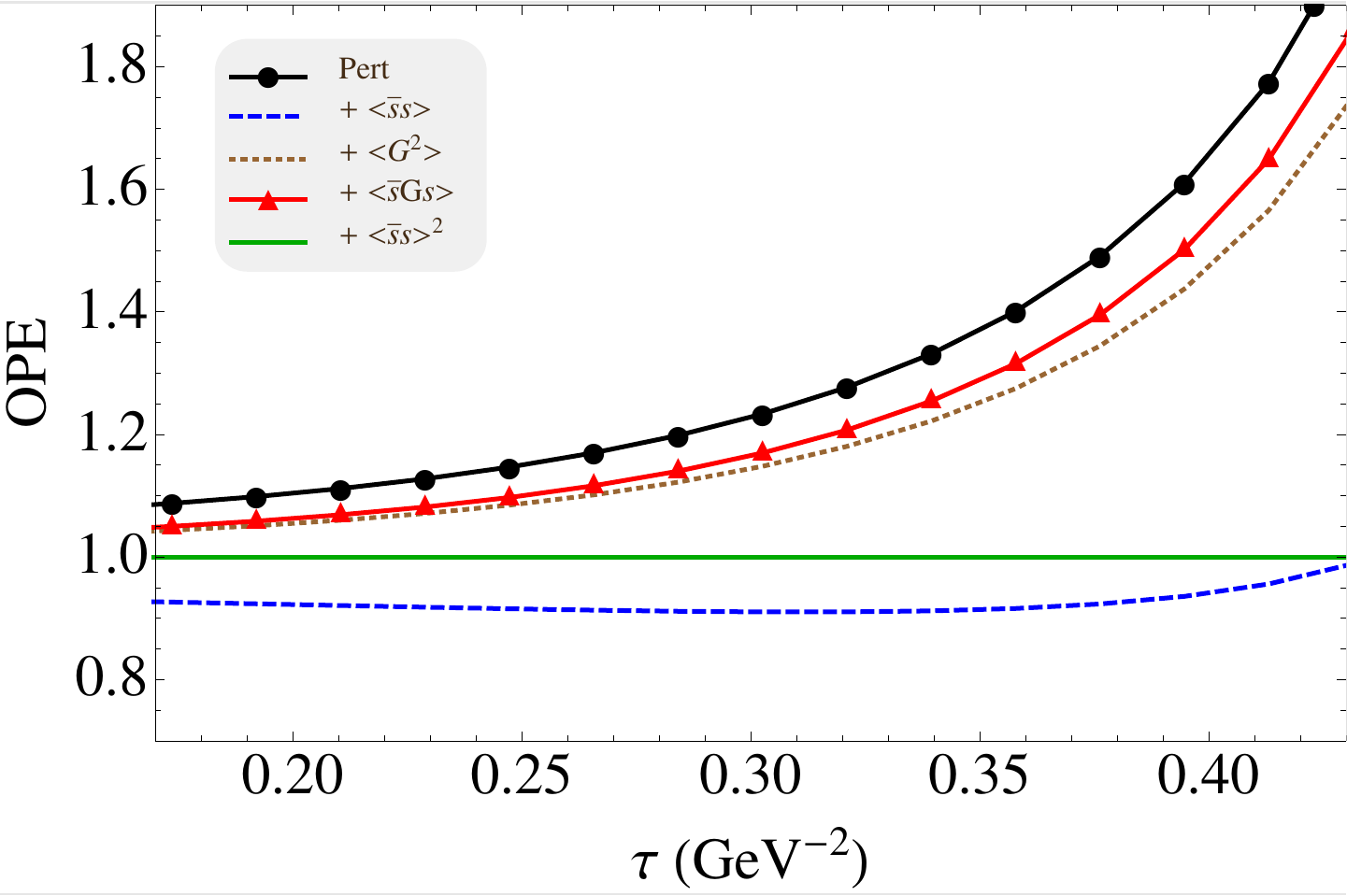}}\\ \vspace{-1.4cm}
    
    \hspace{-1.1cm}
    \subfloat[]{\includegraphics[width=7.5cm]{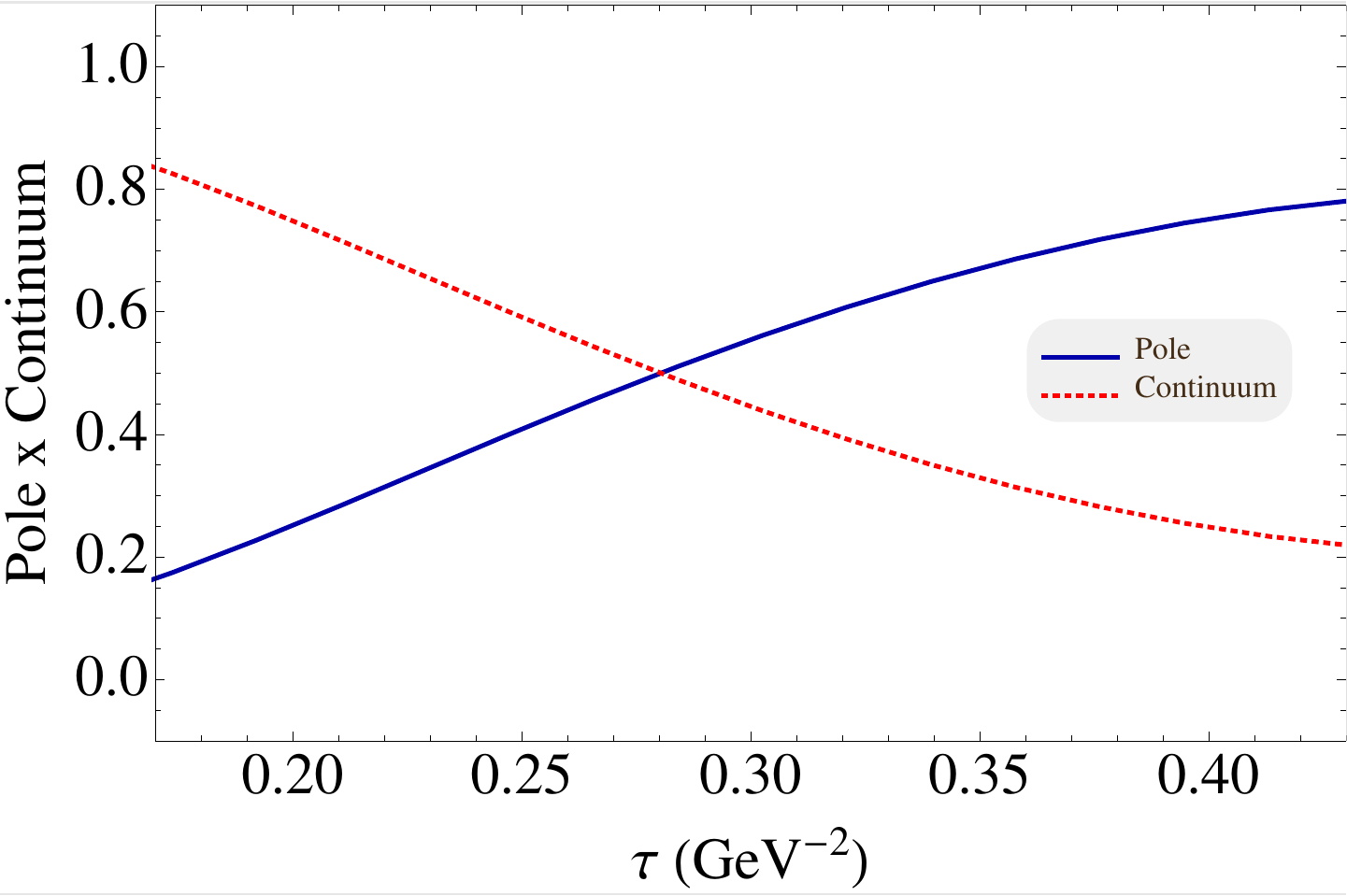}} \hspace{0.5cm}
    \subfloat[]{\includegraphics[width=10cm]{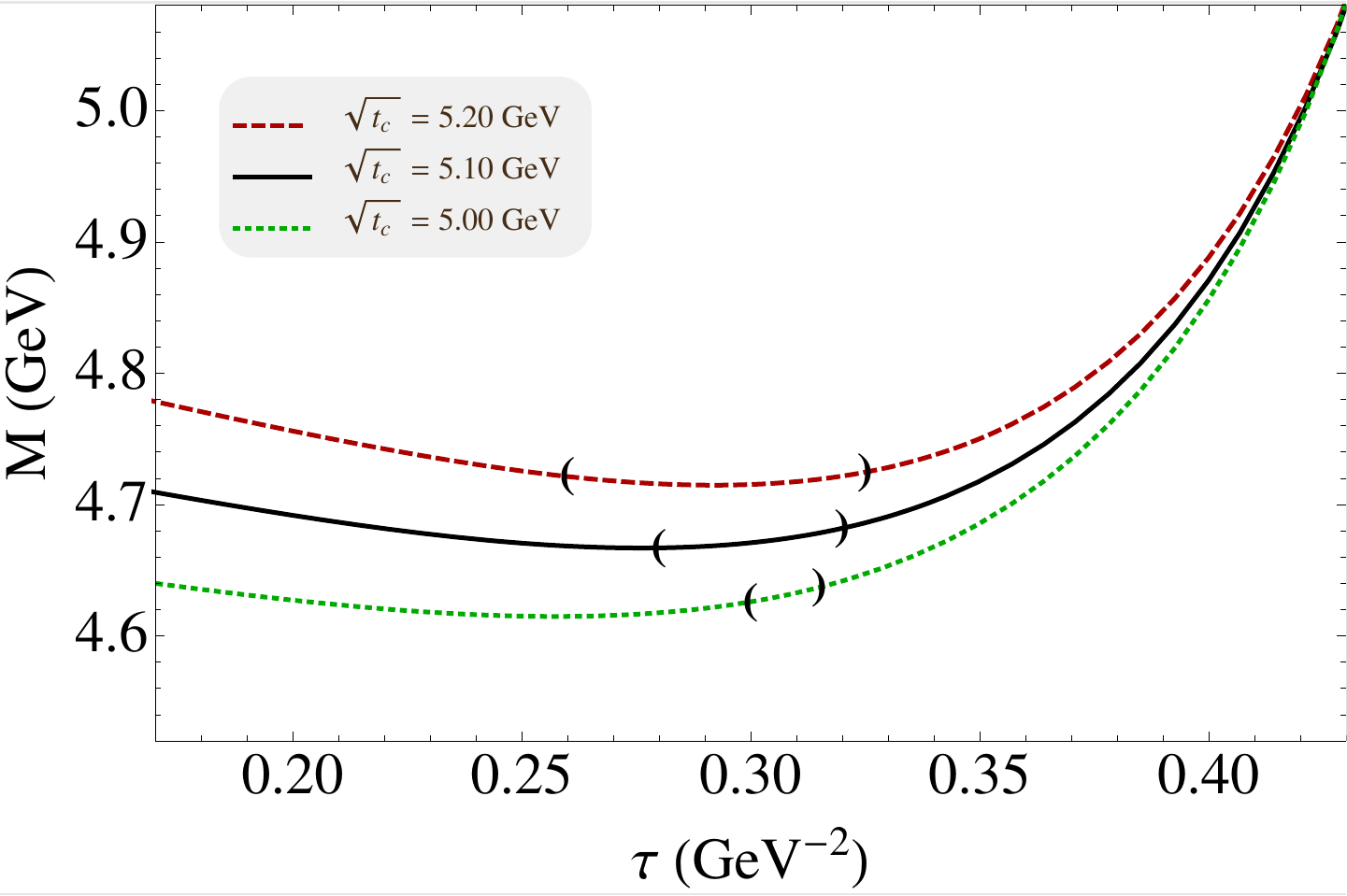}}    
\caption{\footnotesize $J/\psi \:f_0(980)$ molecule in a color singlet configuration, 
considering the OPE contributions up to dimension-six condensates and $m_c=1.23\GeV$.
{\bf (a)} OPE Convergence in the region $0.17 \leq \tau \leq 0.43~\GeV^{-2}$ for
$\sqrt{t_c} = 5.10 \GeV$. The lines show the relative contributions starting with the perturbative 
contribution and each other line represents the relative contribution after adding of one extra 
condensate in the expansion:  $+ \qq[s]$, $+ \GG$, $+ \qGq[s]$, $+ \qq[s]^2$.
{\bf (b)} Pole vs$.$ Continuum contribution, for $\sqrt{t_c} = 5.10 \GeV$.
{\bf (c)} The mass as a function of the sum rule parameter $\tau$, for different values of 
$\sqrt{t_c}$. The parentheses indicate the upper and lower limits of a valid Borel window.}
\label{fig:SngJpsif0}
\end{figure}

In Fig.(\ref{fig:SngJpsif0}a), one can see the relative contribution of all the terms in the OPE 
of the sum rule, for $\sqrt{t_c} = 5.10 \GeV$. From this figure, it is possible to verify that only for 
$\tau \leq 0.32 \GeV^{-2}$ the relative contribution of the dimension-six condensate is 
less than $20\%$ of the total contribution, which indicates a good OPE convergence. 
Therefore, one fixes the maximum value for $\tau$ in the Borel window as 
$\tau_{max} = 0.32 \GeV^{-2}$.
From Fig.(\ref{fig:SngJpsif0}b), the pole contribution is bigger than the continuum 
contribution only for $\tau \geq 0.28 \GeV^{-2}$. Then, the Borel window is fixed 
as $(0.28 \leq \tau \leq 0.32) \GeV^{-2}$.

The results for the mass are shown in Fig.(\ref{fig:SngJpsif0}c), for different values of 
$\sqrt{t_c}$. For each case, the valid Borel window is indicated through the parentheses. 
The allowed values for the continuum threshold are defined in the region 
$5.00 \leq \sqrt{t_c} \leq 5.20 \GeV$, where the selection of the central value for 
$\sqrt{t_c}$ is determined by the one that provides improved stability as a function of 
$\tau$, such that $\sqrt{t_c} = 5.10 \pm 0.10 \GeV$. 
Varying the other parameters as indicated in Table (\ref{TabParam}), and evaluating the 
mass in a valid Borel window, one finally obtains the result:
\begin{eqnarray}
  M_{_{J/\psi \:f_0}} = (4.67 \pm 0.12) \GeV ~.
  \label{eq:MassSngJpsif0}
\end{eqnarray}
This mass is not compatible with the proposition contained in Ref.\cite{oset}, which 
describes the $Y(4260)$ state as the $J/\psi \:f_0(980)$ molecular state. On the other hand, 
this result is in an excellent agreement with the mass of the $Y(4660)$ state. Notice that the 
obtained mass is largely above the meson-meson threshold 
$E_{th}\left[J/\psi \:f_0(980)\right] \simeq 4.09 \GeV$ and, therefore, such molecular state cannot 
be a bound state. However, one has to consider that the current in Eq.(\ref{eq:SngJpsif0}), 
besides coupling to the $J/\psi$ and $f_0(980)$ mesons, it also couples to all excited states 
of mesons having the same $1^{--}$ quantum numbers. 
This fact could lead to an interesting interpretation from the QCDSR calculation: the current 
(\ref{eq:SngJpsif0}) only could warrant that the mass in Eq.(\ref{eq:MassSngJpsif0}) is related 
to the low-lying state of the meson-meson molecule described by the current in 
Eq.(\ref{eq:SngJpsif0}), but not taking care of its meson constituents. Therefore, it would be 
possible that the mass obtained in Eq.(\ref{eq:MassSngJpsif0}) corresponds to the ground state 
of a molecule bounded by the first state excited of the $J/\psi$ meson with the $f_0(980)$ 
meson, the so-called $\psi^\prime \:f_0(980)$ molecular state. Considering that the new 
meson-meson threshold $E_{th}\left[\psi^\prime \:f_0(980)\right] \simeq 4.68 \GeV$ is slightly above the 
mass found in (\ref{eq:MassSngJpsif0}), there is an extra motivation to relate the current 
(\ref{eq:SngJpsif0}) with the $\psi^\prime \:f_0(980)$ molecular state.

It is also important to mention that the results found indicate that, from a QCDSR point of view, 
there is no formation of a $J/\psi \:f_0(980)$ bound state.

The interpretation of the $Y(4660)$ as a $\psi^\prime \:f_0(980)$ molecular state was first 
proposed in Ref.\cite{ghm}, which is also in agreement with the dominant decay channel for 
the $Y(4660)$ state:
$Y(4660) \to \psi^\prime \:\pi^+ \pi^-$.

\begin{figure}[t]
    \hspace{-1.1cm}
    \subfloat[]{\includegraphics[width=7.5cm]{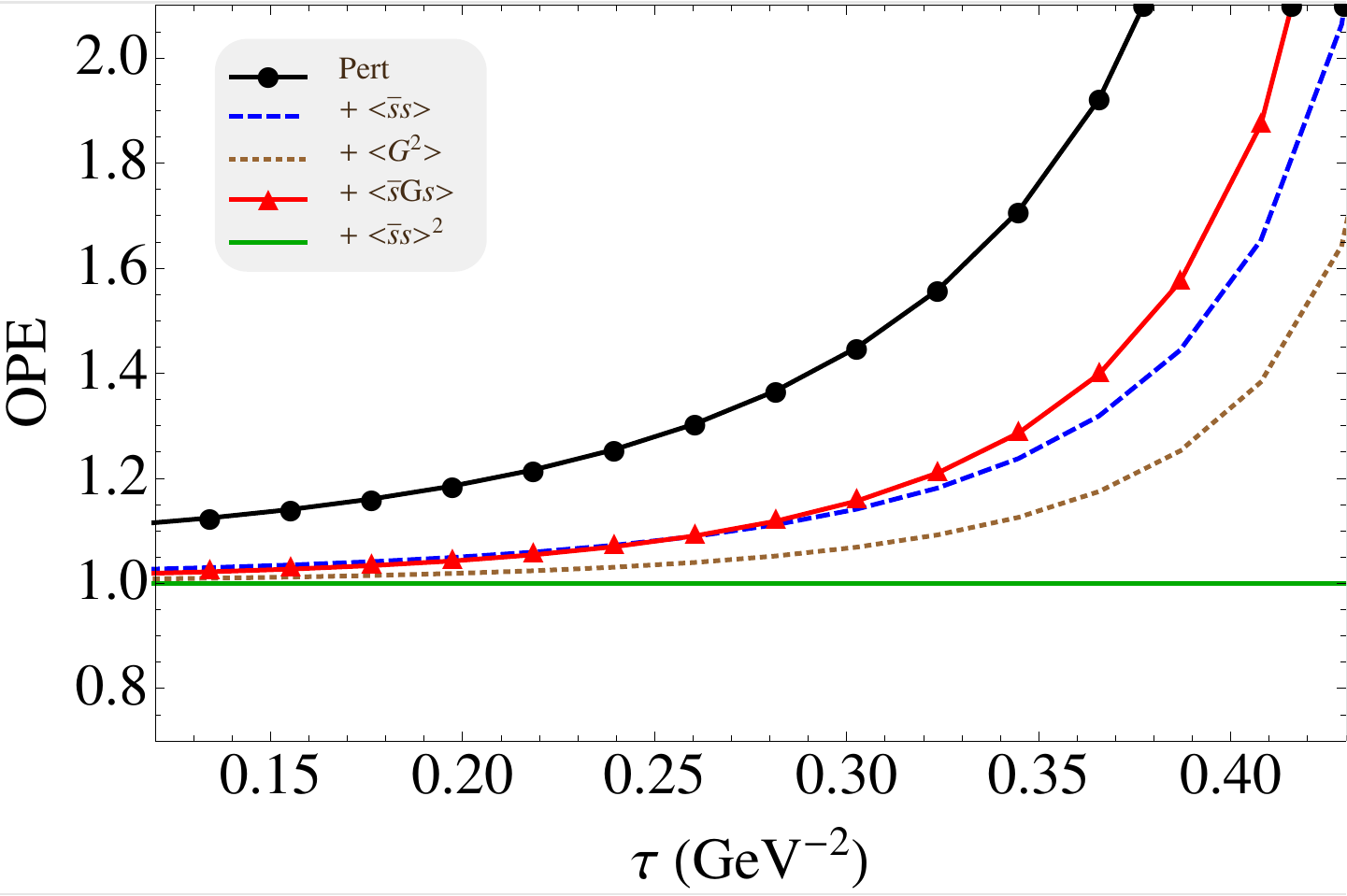}}\\ \vspace{-1.4cm}
    
    \hspace{-1.1cm}
    \subfloat[]{\includegraphics[width=7.5cm]{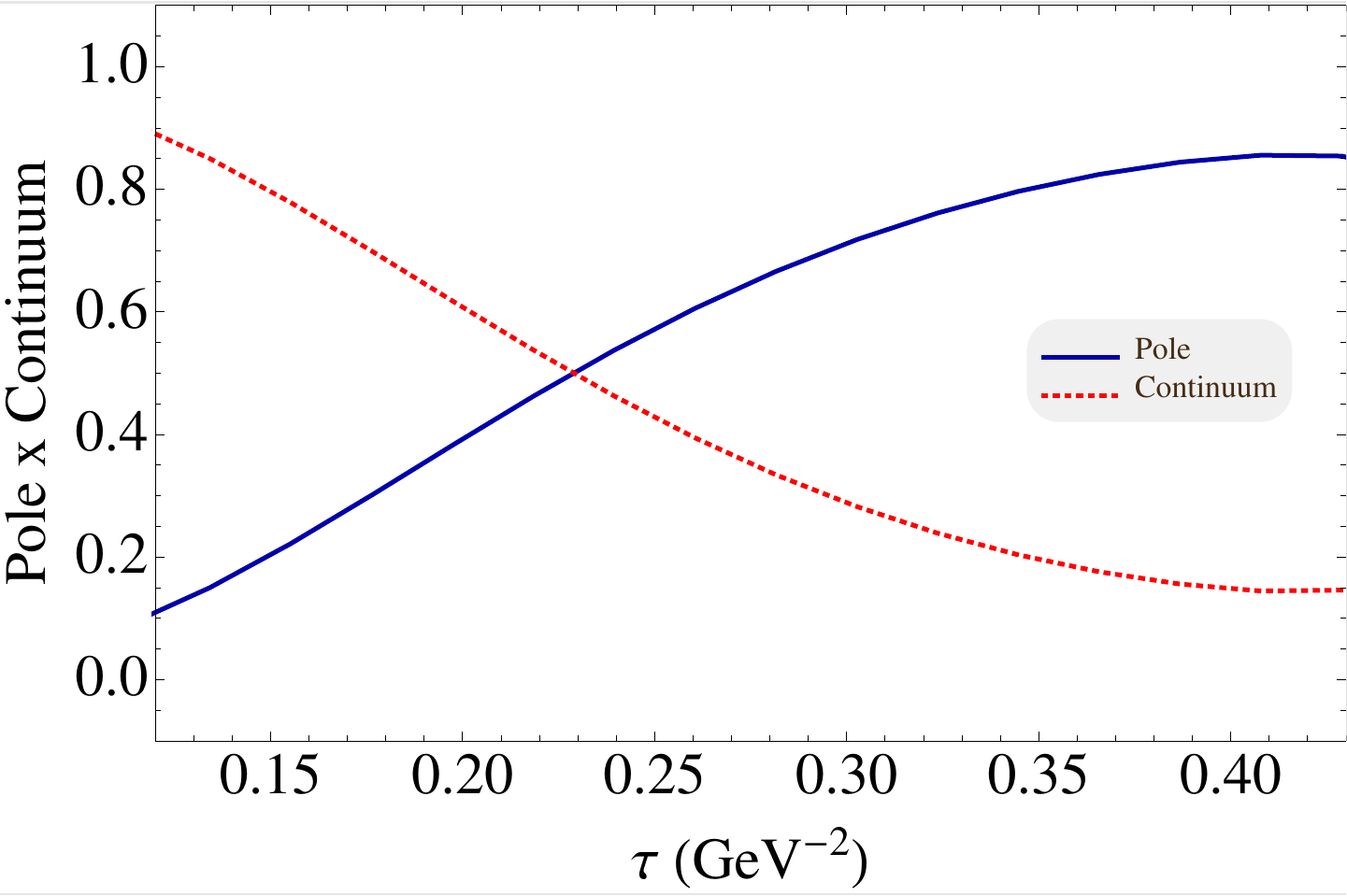}} \hspace{0.5cm}
    \subfloat[]{\includegraphics[width=10cm]{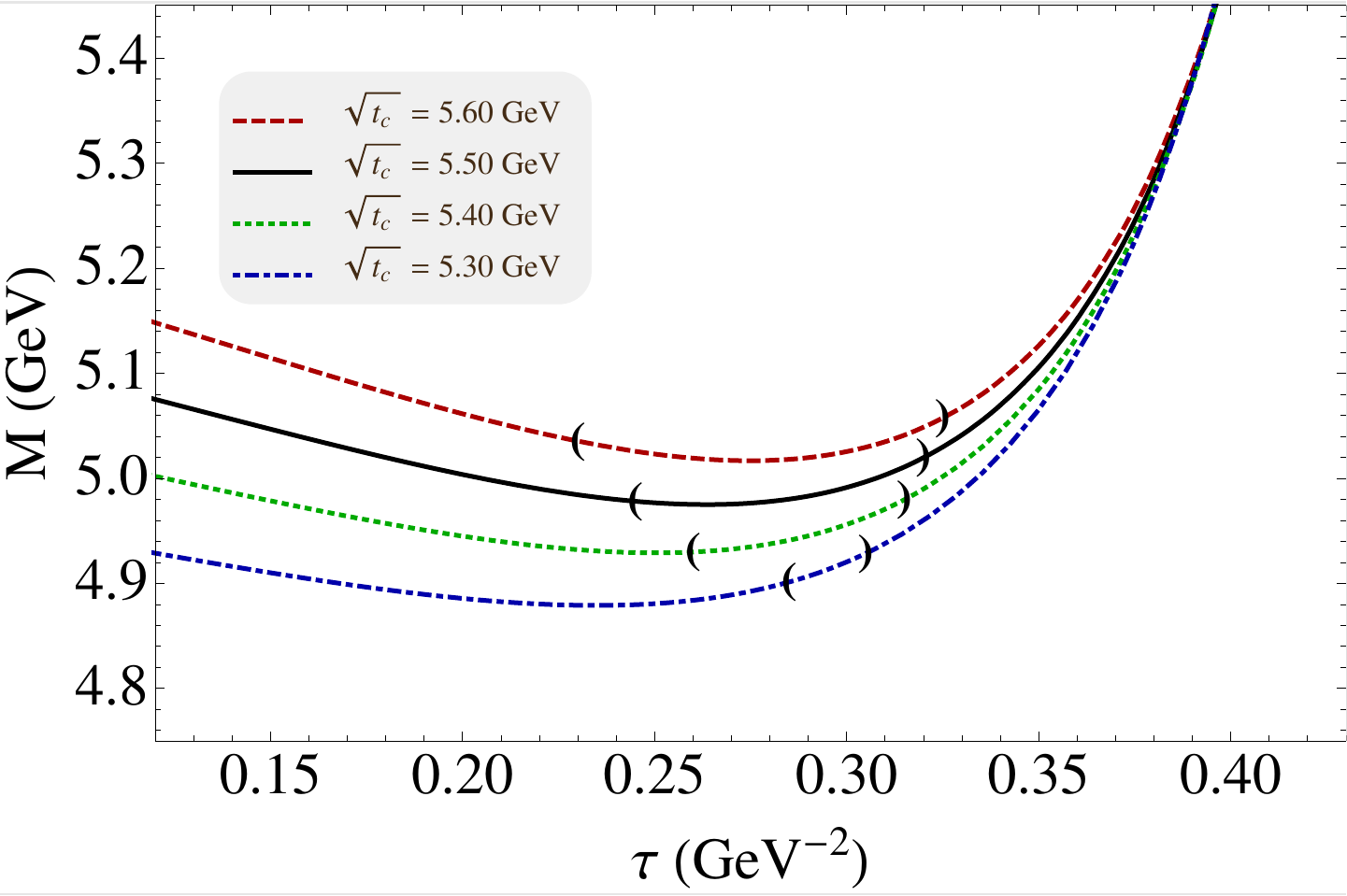}}    
\caption{\footnotesize $J/\psi \:f_0(980)$ molecule in a color octet configuration, 
considering the OPE contributions up to dimension-six condensates and $m_c=1.23\GeV$.
{\bf (a)} OPE Convergence in the region $0.12 \leq \tau \leq 0.43~\GeV^{-2}$ for
$\sqrt{t_c} = 5.50 \GeV$. The lines show the relative contributions starting with the perturbative 
contribution and each other line represents the relative contribution after adding of one extra 
condensate in the expansion:  $+ \qq[s]$, $+ \GG$, $+ \qGq[s]$, $+ \qq[s]^2$.
{\bf (b)} Pole vs$.$ Continuum contribution, for $\sqrt{t_c} = 5.50 \GeV$.
{\bf (c)} The mass as a function of the sum rule parameter $\tau$, for different values of 
$\sqrt{t_c}$. The parentheses indicate the upper and lower limits of a valid Borel window.}
\label{fig:OctJpsif0}
\end{figure}

Similarly, one can consider the case of the current (\ref{eq:OctJpsif0}) which describes the 
molecular state in a color octet configuration. In this case, the region considered for the 
continuum threshold is $(5.40 \leq \sqrt {t_c} \leq 5.60) \GeV$. The results are shown in 
Fig.(\ref{fig:OctJpsif0}), for $\sqrt{t_c} = 5.50 \GeV$, and the Borel window is fixed as 
$(0.23 \leq \tau \leq 0.32) \GeV^{-2}$.
Extracting the masses in a valid Borel window, varying the continuum threshold $\sqrt{t_c}$ 
and considering the uncertainties as indicated in Table (\ref{TabParam}), one obtains
\begin{eqnarray}
  M^\lambda_{_{J/\psi \:f_0}} = (5.02 \pm 0.12) \GeV ~.
  \label{eq:MassOctJpsif0}
\end{eqnarray}
This value for the mass is not compatible with any observed charmonium state. Besides, 
comparing the results in Eqs.(\ref{eq:MassSngJpsif0}) and (\ref{eq:MassOctJpsif0}), it is 
possible to conclude that a molecular state with $\bar{c} \gamma_\mu c$ and $\bar{s}s$ 
in a color octet configuration has a bigger mass than the similar state, with the same 
constituents, but in a color singlet configuration instead. This result is the opposite to 
that found in Ref.\cite{NNN} for a $J/\psi \:\pi$ molecular current. However, in 
Ref. \cite{NNN} one uses the same range for the continuum threshold for both currents. 
In the present work, if one considers the value $\sqrt{t_c} = (5.10 \pm 0.10) \GeV$ for the 
current (\ref{eq:OctJpsif0}) then it is not possible to establish a valid Borel window. As one can 
see in Fig.(\ref{fig:OctJpsif0}c), the lowest allowed value for the continuum threshold is 
$\sqrt{t_c} = 5.30 \GeV$.

\subsection{$J/\psi \:\sigma(600) ~(1^{--})$}
It is straightforward to extend the study presented in the above section for the non-strange case,
the $J/\psi \:\sigma(600)$ molecular state. To do that, one only has to use $m_s \rightarrow 0$, 
$\qq[s] \rightarrow \qq[q]$ and $\qGq[s] \rightarrow \qGq[q]$ in the spectral density expressions 
for the $J/\psi \:f_0(980)$ molecular state. In this case, one needs to define the maximum value 
for $\tau$ imposing that the dimension-six condensate could be at most $25\%$ of the 
total contribution. This indicates that the OPE convergence is worse as compared with the 
$J/\psi \:f_0(980)$ case. This fact is directly related to the absence of the dimension-three 
$\qq[q]$ and dimension-five $\qGq[q]$ condensates contributions once they are proportional 
to the strange quark mass. 

{\small
\begin{table}[htbp]
\centering
\vspace*{0.5cm}
\setlength{\tabcolsep}{1.0pc}
\caption{\small Results for the $J/\psi~\sigma$ currents.}
\vspace*{-0.5cm}
\begin{tabular}{ccc}
&\\
\hline
current in Eq.&$M(\GeV)$&$\sqrt{t_c}(\GeV)$\\
\hline
(\ref{eq:SngJpsif0}) & $4.65 \pm 0.13$& $5.10\pm0.10$ \\
(\ref{eq:OctJpsif0}) & $4.99 \pm 0.11$& $5.50\pm0.10$ \\
\hline \\ \\ 
\end{tabular}
\label{psi-sigma}
\end{table}}

In Table (\ref{psi-sigma}), one presents the masses for both currents, (\ref{eq:SngJpsif0}) and (\ref{eq:OctJpsif0}), including the respective range used for the continuum threshold.
As one can see, when the constituents mesons of the $J/\psi \:\sigma(600)$ molecular state 
are in a color octet configuration the mass is bigger than the one obtained for a color 
singlet configuration. Again, in both cases, the masses are largely above the meson-meson
threshold $E_{th}\left[ J/\psi \:\sigma(600) \right] \simeq 3.70 \GeV$. Therefore, there is no 
formation of the $J/\psi$ and $\sigma(600)$ bound state. Another possibility would be to 
consider the bound state formed by the $\sigma(600)$ meson and the excited meson state 
$\psi^\prime$. However, the obtained masses are still above of this new threshold 
$E_{th}\left[ \psi^\prime \:\sigma(600)\right] \simeq 4.32 \GeV$.

Therefore, one cannot interpret the $Y(4660)$ state neither as $J/\psi \:\sigma(600)$ 
nor $\psi^\prime \:\sigma(600)$ molecular state, despite the fact that obtained mass 
for a color singlet configuration is in agreement with the experimental mass for the 
$Y(4660)$ state.

\subsection{$\Upsilon \:f_0(980) ~(1^{--})$ and $\Upsilon \:\sigma(600) ~(1^{--})$}
It is also straightforward to extend the previous study to the b-sector. To do that, one only 
has to make the change $m_c \rightarrow m_b$ and for the non-strange case 
$m_s \rightarrow 0$ in the spectral density expressions given for the $J/\psi \:f_0(980)$ 
molecular state. These procedures allow to study the following molecular currents: 
$\Upsilon \:f_0(980)$ and $\Upsilon \:\sigma(600)$, in both, color singlet and color octet 
configuration. The study of these currents is an attempt to explain the new states which have 
been observed in the bottomonium sector. In particular, Belle collaboration has announced the 
observation of a new resonance \cite{chen}, called $Y_b(10890)$, with a mass: 
$M_{Y_b} = (10888.4 \pm2.7\pm1.2) \MeV$. 
The tetraquark structure was proposed for this new state in Ref.\cite{AliZhang}. 
Another possible explanation is that the resonance observed is related to the high production of 
$\Upsilon(nS) \pi^+\pi^-$ ($n=1,2,3$) in the decay channel \cite{simo}:
\begin{eqnarray*}
  \Upsilon(5S) \rightarrow B^{(\ast)}B^{(\ast)} \rightarrow \Upsilon(1S, 2S) \pi^+\pi^- ~.  
\end{eqnarray*}

\subsubsection{Numerical Results}
The numerical results for these currents present a good OPE convergence and a pole 
dominance similar to those found in the charmonium sector. In Table (\ref{upsilon}), 
the results obtained for the masses are presented for the states described by the currents 
$\Upsilon \:f_0(980)$ and $\Upsilon \:\sigma(600)$, including the respective values 
to the continuum threshold and the Borel window.

Considering the uncertainties, all the masses obtained with these two currents are 
compatible with the mass of the $Y_b(10890)$ state. However, analyzing the following 
meson-meson thresholds:
\begin{eqnarray*}
	E_{th}[\Upsilon(1S) \:f_0(980)] &\simeq& 10.44 \GeV \\
	E_{th}[\Upsilon(2S) \:f_0(980)] &\simeq& 11.00 \GeV
\end{eqnarray*}
and considering that the thresholds containing the $\sigma(600)$ meson are approximately 
$380 \MeV$ below these values, the only possible interpretation for the $Y_b(10890)$ as 
a molecule, among the molecular states studied, is that one can be described by the 
$\Upsilon(2S) \:f_0(980)$ molecular state.

All of these results obtained with QCDSR for the $J/\psi \:f_0(980)$, $J/\psi \:\sigma(600)$, 
$\Upsilon \:f_0(980)$ and $\Upsilon \:\sigma(600)$ molecular states are published in 
Ref.\cite{rnr}.

{\footnotesize
\begin{table}[htbp]
\centering
\vspace*{0.5cm}
\setlength{\tabcolsep}{.75pc}
\caption{\small Results for the currents $\Upsilon~f_0(980)$ and $\Upsilon~\sigma(600)$.}
\begin{tabular}{lccc}
&\\
\hline
States& $M_H$& Borel Window & $\sqrt{t_c}$\\
&($\GeV$)&($\GeV^{-2}$)&($\GeV$)\\
\hline
{\tiny Color Singlet}\\
$\Upsilon \:f_0(980)$ & $10.75 \pm 0.12$& $0.11 \leq \tau \leq 0.15$ &$11.3\pm0.1$ \\
$\Upsilon \:\sigma(600)$ & $10.74 \pm 0.09$& $0.11 \leq \tau \leq 0.13$& $11.3
\pm0.1$ \\
&&& \\
{\tiny Color Octet}\\
$\Upsilon \:f_0(980)$ &$11.08 \pm 0.11$& $0.11 \leq \tau \leq 0.14$& $11.7\pm0.1$ \\
$\Upsilon \:\sigma(600)$ & $11.09 \pm 0.10$& $0.10 \leq \tau \leq 0.13$& $11.7
\pm0.1$ \\
\hline \\ \\ \\ 
\end{tabular}
\label{upsilon}
\end{table}}

\subsection{$DB ~(0^+)$, $D^\ast B ~(1^+)$, $D B^\ast ~(1^+)$ and $D^\ast B^\ast ~(0^+)$}
In Ref.\cite{sunliu}, a one boson exchange model (OBE) was used to investigate hadronic 
molecules with both open charm and open bottom - they are called as $B_c$-like molecules. 
With OBE model, the authors categorized these molecules using a hand-waving notation,
with five-stars, four-stars, ... according to the probability of the charmed mesons 
${\cal D}^{(*)} = \big[ D^{(*)\:0}, D^{(*)\:+}, D^{(*)\:+}_s \big]$ and bottom mesons 
${\cal B}^{(*)} = \big[ B^{(*)\:0}, B^{(*)\:+}, B^{(*)\:+}_s \big]$ form bound states.
Thus, a five-star state would indicate that this molecular state probably exists, while a one-star 
state probably does not correspond to a bound state. In Ref.\cite{sunliu}, they found five 
five-stars molecular states, all of them isosinglets in the light sector, with no strange quarks. 

In this sense, it would be interesting to use the QCDSR approach to check if some of these five-star 
molecular states correspond to bound states. The molecular states considered in this study are 
shown in Table (\ref{mol}), with their respective molecular currents.

{\small
\begin{table}[t]
\centering
\setlength{\tabcolsep}{1.25pc}
\caption{\small Currents describing possible $B_c$-like molecules.}
\begin{tabular}{ccl}
&\\
\hline
State &$I(J^{P})$ & Current\\
\hline \\
$D\:B$ & $0(0^{+})$ & $ j_{_{DB}} = (\bar{q} \gamma_5 c)(\bar{b} \gamma_5 q)$ \\
$D^\ast B^\ast$ & $0(0^{+})$ & $j_{_{D^\ast B^\ast}}=(\bar{q} \gamma_\mu c)(\bar{b}\gamma^\mu q)$ \\
$D^\ast B$ & $0(1^{+})$ & $ j^\mu_{_{D^\ast B}} = i(\bar{q} \gamma^\mu c)(\bar{b}\gamma_5 q)$ \\
$D B^*$ & $0(1^{+})$ & $ j^\mu_{_{DB^\ast}} = i(\bar{q} \gamma_5 c)(\bar{b}\gamma^\mu q)$ \\ \\
\hline
\end{tabular}
\label{mol}
\end{table}}

In the Phenomenological side, the correlation function is calculated by inserting intermediate 
states for the hadronic state, $H = \big( DB, D^\ast B, D B^\ast,  D^\ast B^\ast \big)$, and 
parameterizing the coupling of these states to their respective current, in terms of a generic 
coupling parameter $\lambda_H$, so that for the $0^+$ scalar states one has 
\begin{eqnarray}
  \langle 0| j_H |0 \rangle &=& \lambda_H ~.
  \label{scDB}
\end{eqnarray}
For the $1^+$ axial states one has
\begin{eqnarray}
  \langle 0| j^\mu_H |0 \rangle &=& \lambda_H \:\epsilon^\mu~,
  \label{pvDB}
\end{eqnarray}
where $\epsilon^\mu$ is the polarization vector.
First consider the molecular current for the scalar $DB ~(0^+)$ state. Calculating the 
correlation function with the current $j_{_{DB}}(x)$, one obtains
\begin{eqnarray}
  \Pi^{_{OPE}}(q) &=& \frac{i}{(2\pi)^8} \int\! d^4x \:d^4p_1 \:d^4p_2 ~
  e^{ix \cdot (q-p_1-p_2)} ~ \nno\\
  && ~\times ~
  \Tr \left[ {\cal S}^b_{ab}(-p_1) \:\ga_5\: {\cal S}^q_{ba}(x) \:\ga_5 \right] ~
  \Tr \left[ {\cal S}^c_{cd}(p_2) {\cal S}^q_{dc}(-x) \right] ~.
  \label{eq:FCDB}
\vspace{-0.3cm}
\end{eqnarray}
For the QCDSR calculation of these molecular states, the OPE contributions were 
evaluated up to dimension-eight condensates, working at leading order in $\alpha_s$. 
Then, the spectral density obtained from Eq.(\ref{eq:FCDB}) is given by 
\begin{equation}	
  \rho^{_{OPE}}_{_{\DB}}(s) = 
  \rho^{pert}_{_{\DB}}(s) + \rho^{\qq[q]}_{_{\DB}}(s) + \rho^{\GGi}_{_{\DB}}(s) + 
  \rho^{\qGq[q]}_{_{\DB}} + \rho^{{\qq[q]}^2}_{_{\DB}}(s) + 
  \rho^{\GGG}_{_{\DB}}(s) + \rho^{\qq[q] \qGq[q]}_{_{\DB}}(s) ~~,
\end{equation}
where the explicit expression of each term above is presented as follows:

\begin{eqnarray}
  \rho^{pert}_{_{DB}}(s) &=& \frac{3}{2^{11} \pi^6}
  	\int\limits^{\almax}_{\almin} \!\!\frac{d\al}{\alpha^3}  
	\int\limits^{\bemax}_{\bemin} \!\!\frac{d\be}{\beta^3} \:
	(1-\alpha -\beta) \Fbc^4 , \\
  \nno \\
  \rho^{\qq[q]}_{_{DB}}(s) &=& -\frac{3 \qq[q]}{2^{7} \pi^4}
  	\int\limits^{\almax}_{\almin} \!\!\frac{d\al}{\alpha^2}  
	\int\limits^{\bemax}_{\bemin} \!\!\frac{d\be}{\beta^2} \:
	(\be \:m_c + \al \:m_b) \Fbc^2 ,\\
  \nno \\
  \rho^{\GGi}_{_{DB}}(s) &=& \frac{\GG}{2^{12} \pi^6}
  	\int\limits^{\almax}_{\almin} \!\!\frac{d\al}{\alpha^3}  
	\int\limits^{\bemax}_{\bemin} \!\!\frac{d\be}{\beta^3} \:
	\Fbc \bigg[ 3 \al \be (\alpha +\beta) \Fbc \nno\\
	&& +~ 2 (1-\al-\be) (\be^3 m_c^2 + \al^3 m_b^2 ) \bigg] ,\\
  \nno \\
  \rho^{\qGq[q]}_{_{DB}}(s) &=& - \frac{3\qGq[q]}{2^{8} \pi^4} \Bigg[
  	\int\limits^{\almax}_{\almin} \!\!\!\!\frac{d\al}{\al(1\!-\! \al)} \bigg( 
	m_c \!+\! \al(m_b \!-\! m_c) \bigg) \Hbc \nno\\
	&& -~ 2 \!\!\int\limits^{\almax}_{\almin} \!\!\frac{d\al}{\al^2} \!\!
	\int\limits^{\bemax}_{\bemin} \!\! \frac{d\be}{\be^2} 
	( \be^2 m_c \!+\! \al^2 m_b ) \Fbc \Bigg], \\
  \nno \\
  \rho^{\qq[q]^2}_{_{DB}}(s) &=& \frac{m_b m_c \:\rho \:\qq[q]^2}{16 \pi^2} \lambda_{bc} \:v_{bc},\\
  \nno \\
  \rho^{\GGGi}_{_{DB}}(s) &=& \frac{\GGG}{2^{13} \pi^6}
  	\!\!\int\limits^{\almax}_{\almin} \!\!\frac{d\al}{\al^3} \!\!
	\int\limits^{\bemax}_{\bemin} \!\! \frac{d\be}{\be^3} (1-\al-\be) \bigg[ 
	2 (\be^4 m_c^2 \!+\! \al^4 m_b^2 ) \nno\\
	&& +~ (\al^3 + \be^3) \Fbc ~\bigg], \\
  \nno \\
  \rho^{\qq[q]\qGq[q]}_{_{DB}}(s) &=& \frac{m_c m_b\rho \qq[q]\qGq[q]}{2^{5} \pi^2}
  	\int\limits^{1}_{0} \!\!\frac{d\al}{\al(1-\al)} \Bigg[ 1-\al+\al^2 \nno\\
	&& -~ \bigg( m_c^2 -\al (m_c^2 - m_b^2) \bigg) \tau \Bigg]
	\:\delta\bigg( s - \frac{m_c^2 -\al( m_c^2 - m_b^2)}{\al(1-\al)}\bigg) ~.
\label{eq:rhoDB}
\end{eqnarray}

For molecules containing both $c$- and $b$-quarks in their internal structures, one must 
define the new functions $\Fbc$, $\Hbc$ and $v_{bc}$, so that:
\begin{eqnarray}
\addtolength{\fboxsep}{10pt} 
\boxed{ \begin{gathered} 
  \begin{array}{rcl} 
    \Fbc &=& m_b^2 \al + m_c^2 \be - \al\be s, \\
    \Hbc &=& m_b^2 \al + m_c^2(1-\al) - \al(1-\al) s, \\ 
    v_{bc} &=& \displaystyle \sqrt{1 - \frac{4m_c^2 / s}{\lambda_{bc}^2}}, \\ 
    \lambda_{bc} &=& 1 + (m_c^2 - m_b^2) / s
  \end{array}
\end{gathered}} 
\end{eqnarray}
and also define the new integration limits: 
\begin{eqnarray}
\addtolength{\fboxsep}{10pt} 
\boxed{ \begin{gathered} 
  \begin{array}{rcl} 
    \al_{max} &=& \lambda_{bc} (1 + v_{bc})/2, \\
    \al_{min} &=& \lambda_{bc} (1 - v_{bc})/2, \\
    \be_{max} &=& 1-\al, \\
    \be_{min} &=& \displaystyle \frac{\al \:m_b^2}{\al s - m_c^2} \\ 
  \end{array}
\end{gathered}} 
\end{eqnarray}\\

\subsubsection{Numerical Results}
\begin{figure}[t]
    \hspace{-1.1cm}
    \subfloat[]{\includegraphics[width=7.5cm]{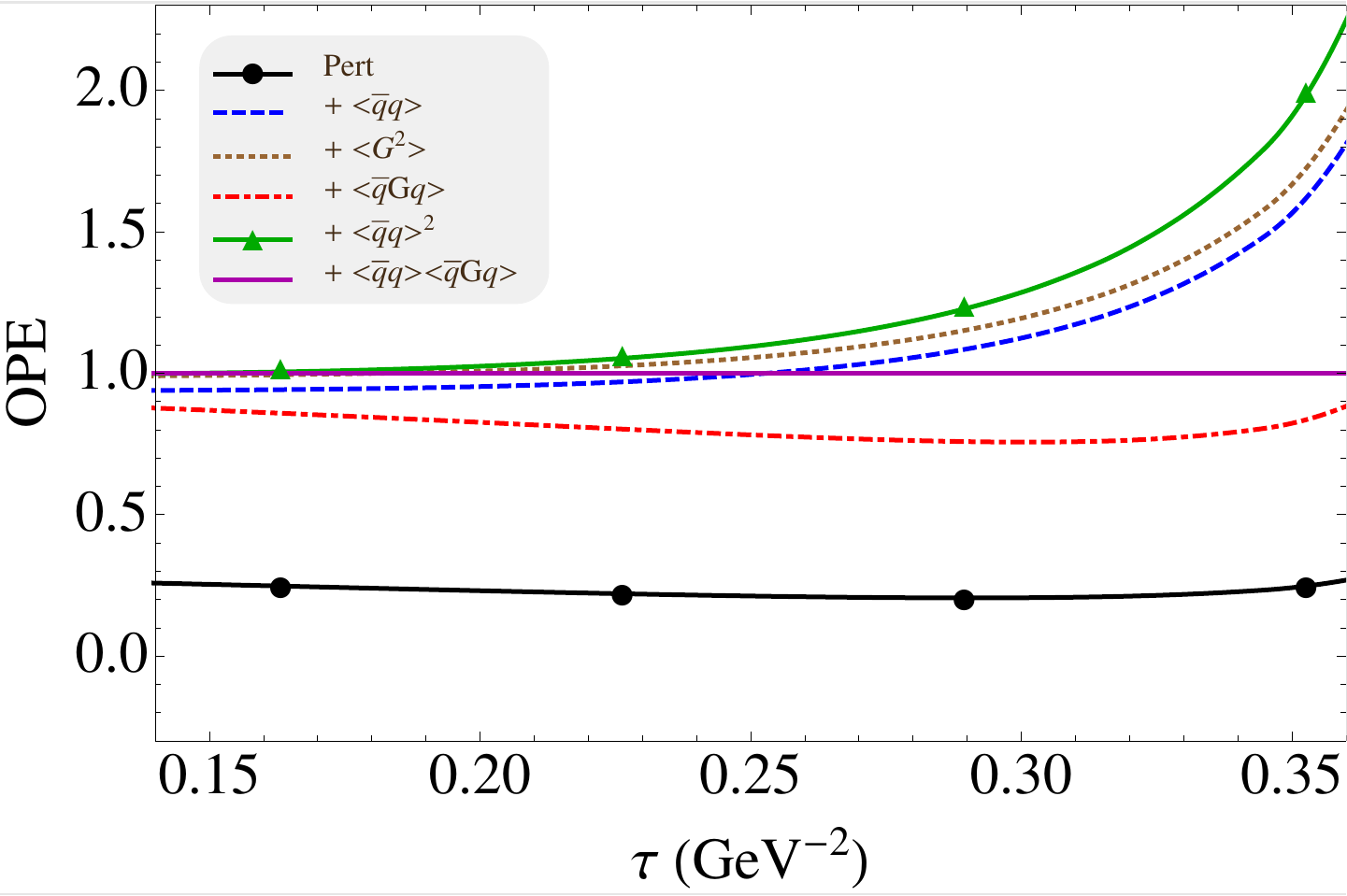}}\\ \vspace{-1.4cm}
    
    \hspace{-1.1cm}
    \subfloat[]{\includegraphics[width=7.5cm]{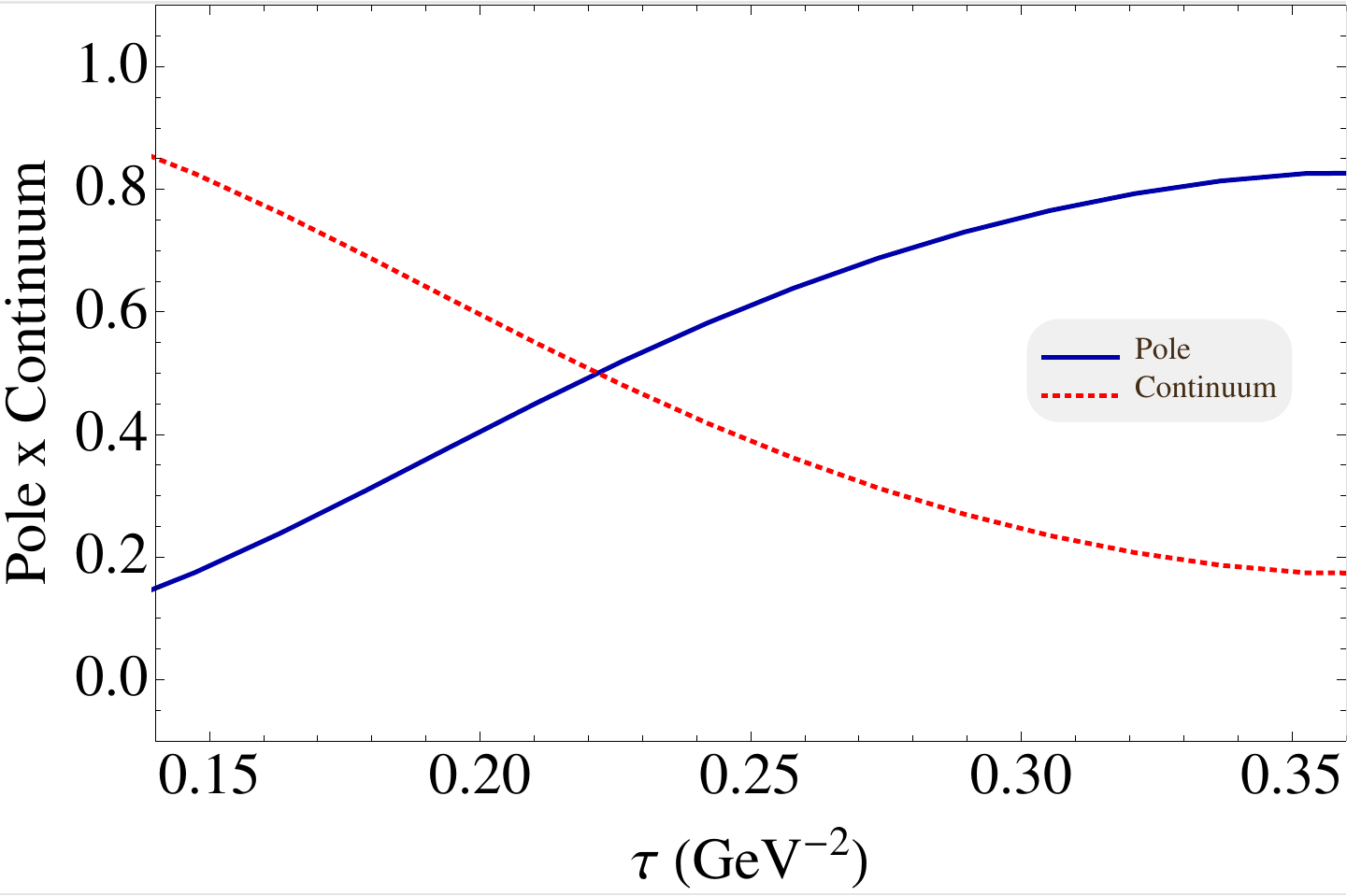}} \hspace{0.5cm}
    \subfloat[]{\includegraphics[width=10cm]{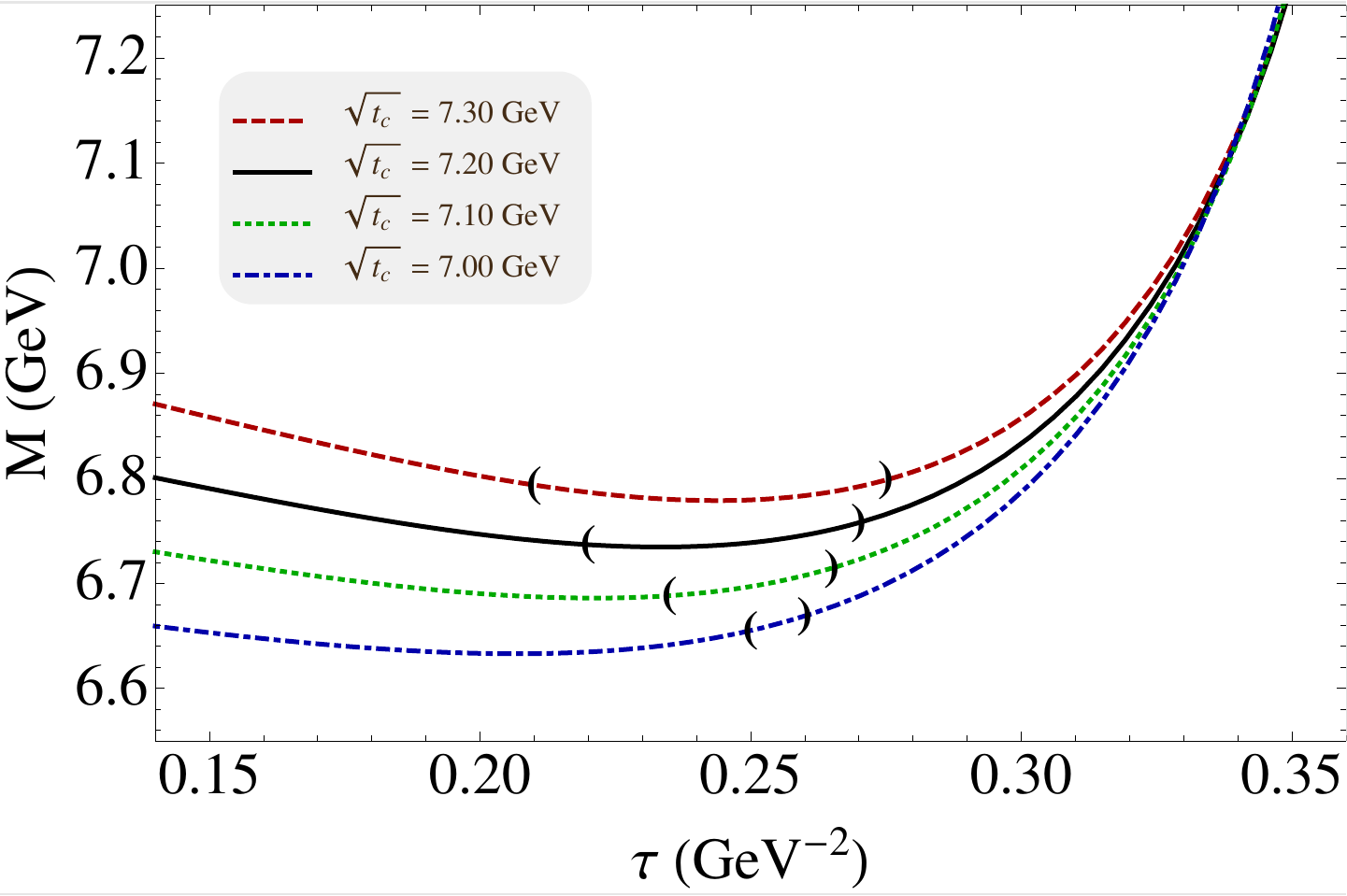}}    
\caption{\footnotesize $DB \:(0^+)$ molecule, considering the OPE contributions up to 
dimension-eight condensates, $m_c=1.23\GeV$ and $m_b=4.24\GeV$.
{\bf (a)} OPE Convergence in the region $0.14 \leq \tau \leq 0.36~\GeV^{-2}$ for
$\sqrt{t_c} = 7.20 \GeV$. The lines show the relative contributions starting with the perturbative 
contribution and each other line represents the relative contribution after adding of one extra 
condensate in the expansion:  $+ \qq[s]$, $+ \GG$, $+ \qGq[q]$, $+ \qq[s]^2 + \GGGi$ and 
$+ \qq[q] \qGq[q]$.
{\bf (b)} Pole vs$.$ Continuum contribution, for $\sqrt{t_c} = 7.20 \GeV$.
{\bf (c)} The mass as a function of the sum rule parameter $\tau$, for different values of 
$\sqrt{t_c}$. The parentheses indicate the upper and lower limits of a valid Borel window.}
\label{fig:DB}
\end{figure}

To evaluate the mass of the $DB (0^+)$ molecular state, one uses Eq.(\ref{massa}) 
considering the expressions for the spectral density $\rho^{_{OPE}}_{_{\DB}}(s)$ and 
the parameters in Table (\ref{TabParam}). The results obtained are shown in 
Fig.(\ref{fig:DB}).

In Fig.(\ref{fig:DB}a), the relative contribution of the OPE terms is presented, for 
$\sqrt{t_c} = 7.20 \GeV$. From this figure, one verifies that the contribution of the 
dimension-eight is smaller than $15\%$ of the total contribution, for values of 
$\tau \leq 0.27 \GeV^{-2}$, which indicates a good OPE convergence and fixes
the maximum value of $\tau$ in the Borel window as $\tau_{max} = 0.27 \GeV^{-2}$.

In Fig.({\ref{fig:DB}b), one obtains the minimum value for the Borel window 
$\tau_{min} = 0.22 \GeV^{-2}$, considering that at this point the pole 
contribution is bigger than the continuum contribution.

The results for the mass are shown in Fig.(\ref{fig:DB}c), as a function of $\tau$, for 
different values of $\sqrt{t_c}$. As one can see, the Borel window (indicated through 
the parentheses) gets smaller as the value of $\sqrt{t_c}$ decreases. So, the minimum 
value allowed for the continuum threshold is given by $\sqrt{t_c} = 7.00 \GeV$. 
It is also possible to observe that the optimal choice for the continuum threshold is 
$\sqrt{t_c} = 7.20 \GeV$, because it provides the best $\tau$-stability inside of the Borel 
window, including the existence of a minimum point for the value of the mass.

Therefore, varying the value of the continuum threshold in the range 
$\sqrt{t_c} = (7.00 - 7.30) \GeV$, and the other parameters as indicated in Table
(\ref{TabParam}), one obtains
\begin{eqnarray}
    M^{\lag8\rag}_{DB} &=& (6.77 \pm 0.11) \GeV.
  \label{eq:massDB8}
\end{eqnarray}

As widely observed in other sum rule calculations, the most significant sources of uncertainty 
are the values of the heavy quark masses \cite{SNB, rry, svz}. In this sense, one could refer to 
the quoted uncertainty in Eq.(\ref{eq:massDB8}) as the OPE uncertainty. 
As discussed in Ref.\cite{Lucha}, there is another kind of uncertainty, called systematic 
uncertainty, related to the intrinsic limited accuracy of the method. The systematic uncertainty 
of the physical quantity extracted from the QCDSR represents, perhaps, the most subtle point 
in the application of the method. Without an estimate of the systematic uncertainty, the numerical 
value of the physical quantity one reads off from the Borel window might differ significantly from 
its true value. In Ref.\cite{Lucha} it was shown that the use of the Borel dependent continuum 
threshold allows to estimate the systematic uncertainty. In particular, for the case of the 
$D$ and $D_s$ mesons studied in \cite{Lucha}, the systematic uncertainty turns out to be of 
the same order of the OPE uncertainty. 
In an attempt to obtain some information about the systematic uncertainty, one repeats the 
analysis considering only terms up to dimension 6 in the OPE. These new results
are shown in Fig.(\ref{fig:DB6}). 

\begin{figure}[t]
    \hspace{-1.1cm}
    \subfloat[]{\includegraphics[width=7.5cm]{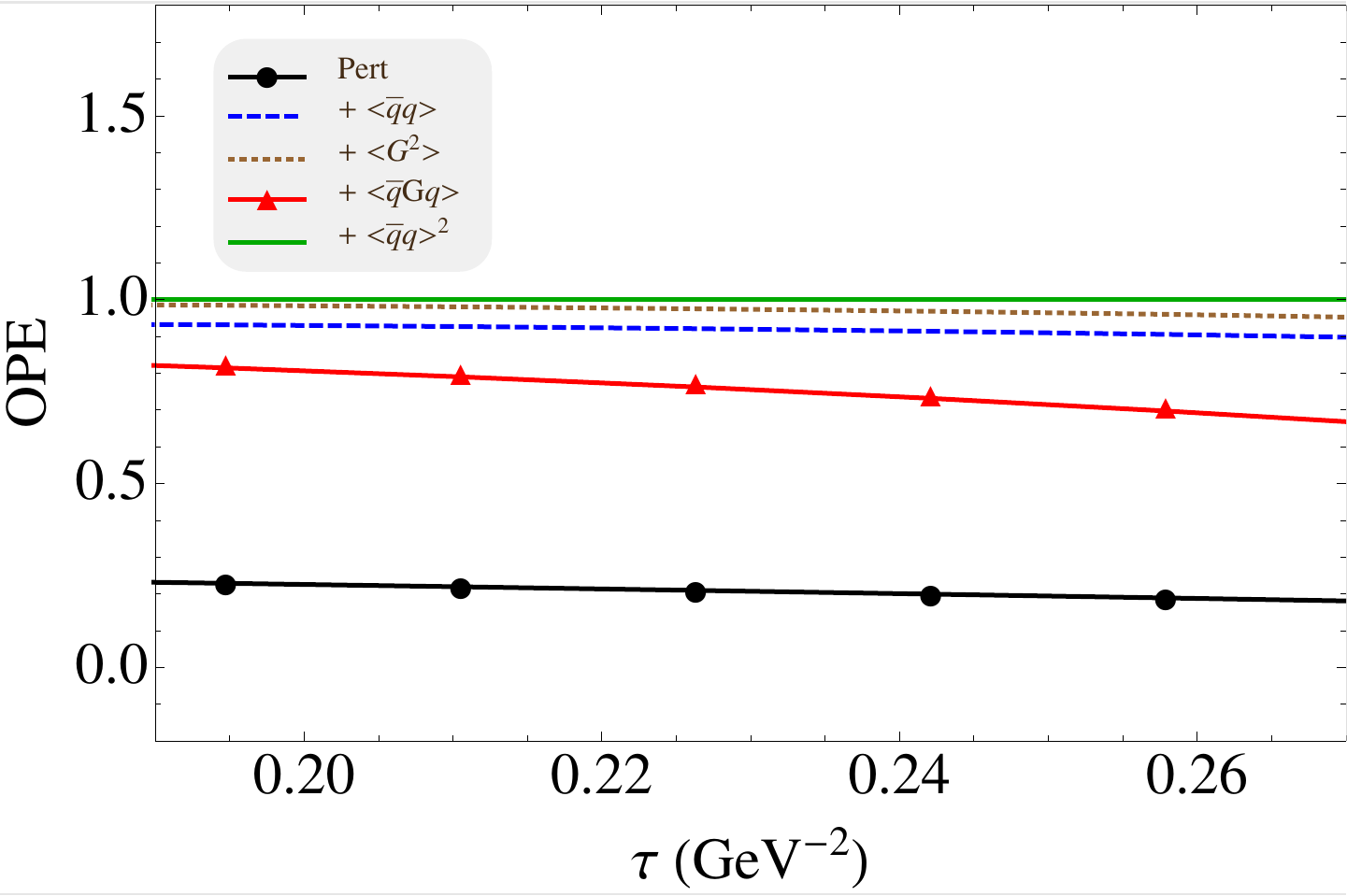}}\\ \vspace{-1.4cm}
    
    \hspace{-1.1cm}
    \subfloat[]{\includegraphics[width=7.5cm]{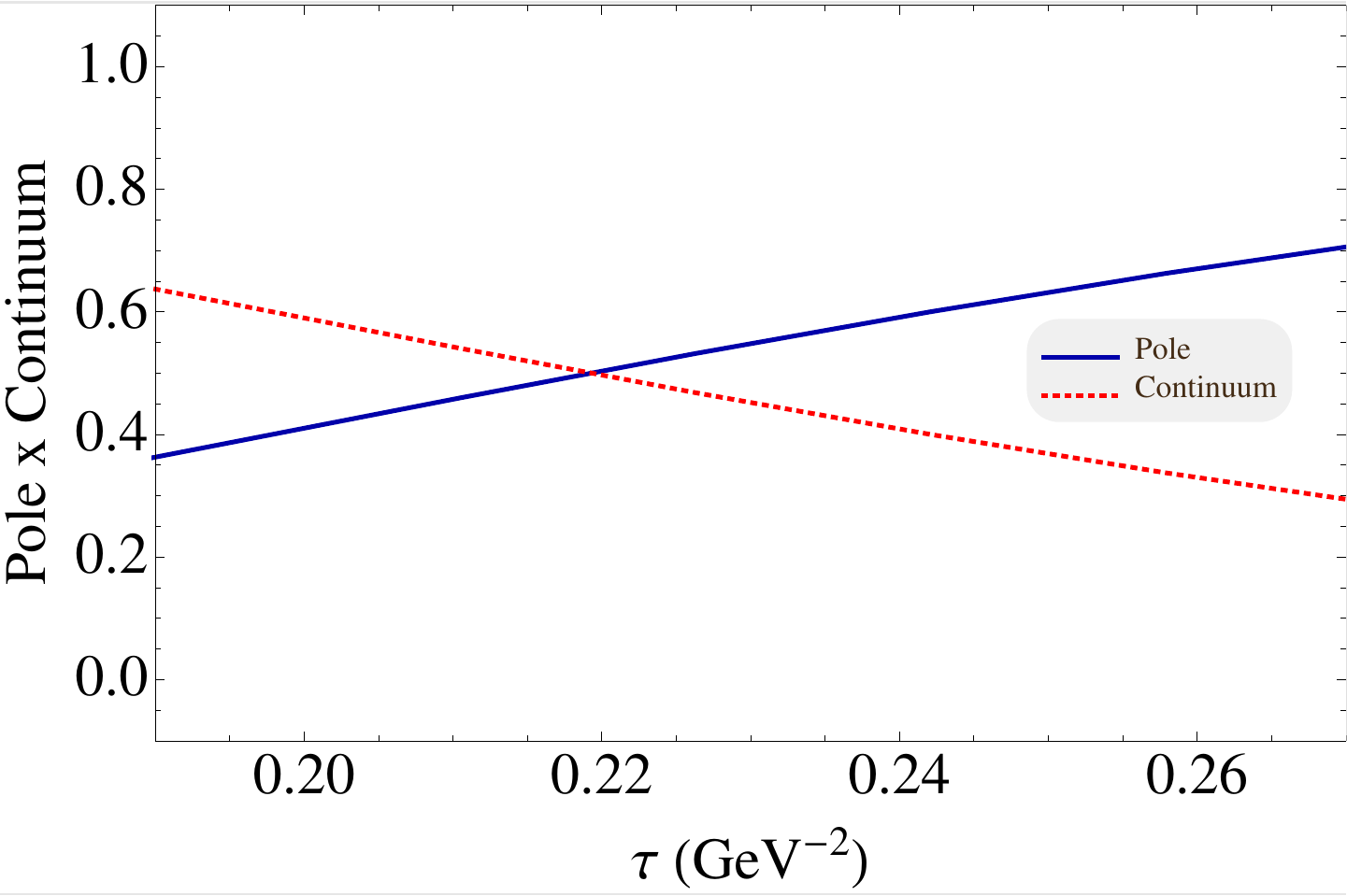}} \hspace{0.5cm}
    \subfloat[]{\includegraphics[width=10cm]{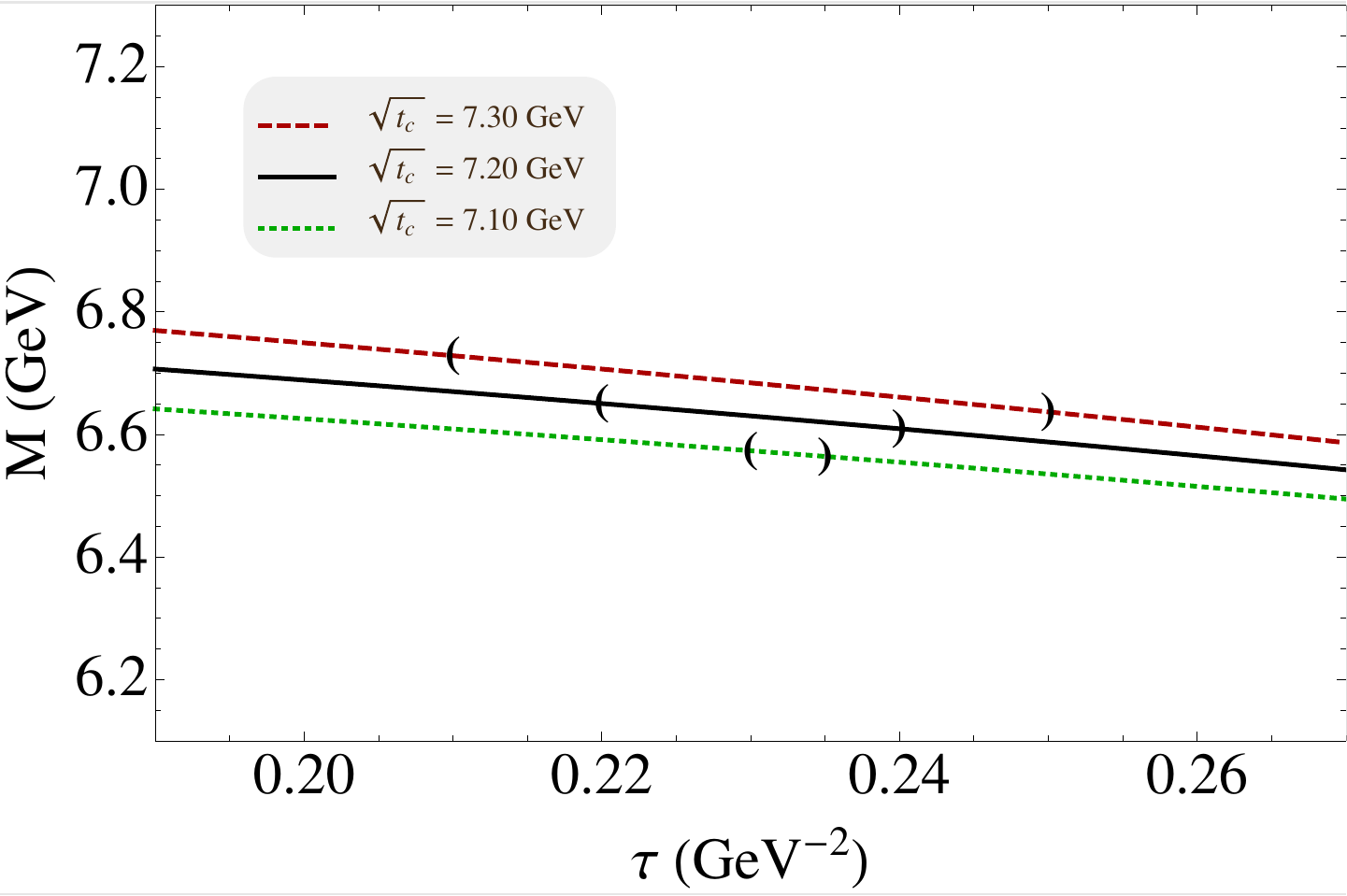}}    
\caption{\footnotesize $DB \:(0^+)$ molecule, considering the OPE contributions up to 
dimension-six condensates, $m_c=1.23\GeV$ and $m_b=4.24\GeV$.
{\bf (a)} OPE Convergence in the region $0.19 \leq \tau \leq 0.27~\GeV^{-2}$ for
$\sqrt{t_c} = 7.20 \GeV$. The lines show the relative contributions starting with the perturbative 
contribution and each other line represents the relative contribution after adding of one extra 
condensate in the expansion:  $+ \qq[s]$, $+ \GG$, $+ \qGq[q]$, $+ \qq[s]^2 + \GGGi$.
{\bf (b)} Pole vs$.$ Continuum contribution, for $\sqrt{t_c} = 7.20 \GeV$.
{\bf (c)} The mass as a function of the sum rule parameter $\tau$, for different values of 
$\sqrt{t_c}$. The parentheses indicate the upper and lower limits of a valid Borel window.}
\label{fig:DB6}
\end{figure}

As one can see in Fig.(\ref{fig:DB6}a), the OPE convergence gets worse after removing the 
dimension-eight condensate contributions, since the most important contributions to the OPE 
come from $\qq[q]$ and $\rho \:\qq[q]^2$ condensates. Thus, in order to be able to extract 
some results from this new analysis, one determines the maximum value of $\tau$ parameter 
imposing that the contribution of the dimension-6 condensate is smaller than $25\%$ of the 
total contribution, otherwise it is not possible to fix a valid Borel window for this sum rule. 
The minimum value of $\tau$ is not changed since the pole dominance behavior remains 
the same. Finally, one obtains the results shown in Fig.(\ref{fig:DB6}c) and given by 
\begin{eqnarray}
  M^{\lag6\rag}_{DB} &=& (6.63 \pm 0.09) \GeV.
  \label{eq:massDB6}
\end{eqnarray}
Notice that the value in Eq.(\ref{eq:massDB6}) differs at maximum only $\sim ~~5.0\%$ to 
that in Eq.(\ref{eq:massDB8}). Besides, the inclusion of dimension-8 condensate provides a 
better OPE convergence, $\tau$-stability and improved Borel window. Therefore, 
the contribution from the dimension-eight condensates for the $D B$, $D^\ast D$, 
$D B^\ast$ and $D^\ast B^\ast$ molecular states in a QCDSR calculation will be 
considered hereafter. 
Then, the final value for the $D B \:(0^+)$ molecular state is given by 
\begin{eqnarray}
  M_{DB} &=& (6.75 \pm 0.14) \GeV.
  \label{eq:massDB}
\end{eqnarray}
This mass is $\sim 400 \MeV$ below the $E_{th} [DB]$ threshold indicating that such 
molecular state would be tightly bound. This result, for the binding energy, is very different 
than the obtained in Ref.\cite{sunliu} for the $D B \:(0^+)$ molecular state. The authors 
of Ref.\cite{sunliu} found that the $D B \:(0^+)$ molecular state is loosely bound with a 
binding energy smaller than $14 \MeV$. However, it is very important to notice that since 
the molecular currents given in Table (\ref{mol}) are local, they do not represent extended 
objects, with two mesons separated in space, but rather a very compact object with two 
singlet quark-antiquark pairs. Therefore, the result obtained here may suggest that, although 
a loosely bound $D B \:(0^+)$ molecular state can exist, it may not be the ground state for 
a four-quark exotic state with the same quantum numbers and quark content.
Having the hadron mass, it is also possible evaluate the coupling parameter, $\lambda_{DB}$:
\begin{eqnarray}
  \lambda_{DB} &=& (0.029 \pm 0.008) \GeV^5 ~~.
  \label{lDB}
\end{eqnarray}
As discussed before, the parameter $\lambda_{DB}$ gives a measure of the strength 
of the coupling between the current and the state. The result in Eq.(\ref{lDB}) has the same 
order of magnitude as the coupling obtained for the $X(3872)$ \cite{ricnar}, for example. 
This indicates that such state could be very well represented by the respective current in 
Table (\ref{mol}).

The following step is to extend the same analysis to study other molecular 
states presented in Table (\ref{mol}). Inserting the respective currents into the 
correlation function and considering the OPE contributions up to dimension-eight 
condensates, the spectral densities for $D^\ast B^\ast \:(0^+)$ molecular state
are given by 
\begin{eqnarray}
  \rho^{pert}_{_{D^\ast B^\ast}}(s) &=& \frac{3}{2^{9} \pi^6}
  	\int\limits^{\almax}_{\almin} \!\!\frac{d\al}{\alpha^3}  
	\int\limits^{\bemax}_{\bemin} \!\!\frac{d\be}{\beta^3} \:
	(1-\alpha -\beta) \Fbc^4, \\
  \nno\\
  \rho^{\qq[q]}_{_{D^\ast B^\ast}}(s) &=& -\frac{3 \qq[q]}{2^{6} \pi^4}
  	\int\limits^{\almax}_{\almin} \!\!\frac{d\al}{\alpha^2}  
	\int\limits^{\bemax}_{\bemin} \!\!\frac{d\be}{\beta^2} \:
	(\be \:m_c + \al \:m_b) \Fbc^2, \\
  \nno\\
  \rho^{\GGi}_{_{D^\ast B^\ast}}(s) &=& \frac{\GG}{2^{9} \pi^6}
  	\int\limits^{\almax}_{\almin} \!\!\frac{d\al}{\alpha^3}  
	\int\limits^{\bemax}_{\bemin} \!\!\frac{d\be}{\beta^3} \:
	(1-\al-\be) ( \be^3 m_c^2 + \al^3 m_b^2 ) \Fbc,\\
  \nno\\
  \rho^{\qGq[q]}_{_{D^\ast B^\ast}}(s) &=& - \frac{3 \qGq[q]}{2^{7} \pi^4}
  	\int\limits^{\almax}_{\almin} \!\!\frac{d\al}{\al(1-\al)} \:
	\bigg( m_c - \al(m_c-m_b) \bigg) \Hbc, \\
  \nno\\
  \rho^{\qq[q]^2}_{_{D^\ast B^\ast}}(s) &=& \frac{m_c m_b \:\rho \qq[q]^2}{4 \pi^2} 
  \lambda_{bc} \:v_{bc},\\
  \nno\\
  \rho^{\GGGi}_{_{D^\ast B^\ast}}(s) &=& \frac{\GGG}{2^{11} \pi^6}
  	\!\!\int\limits^{\almax}_{\almin} \!\!\frac{d\al}{\al^3} \!\!
	\int\limits^{\bemax}_{\bemin} \!\! \frac{d\be}{\be^3} (1-\al-\be) \bigg[ 
	2 (\be^4 m_c^2 \!+\! \al^4 m_b^2 )\nno\\
	&& +~ (\al^3 + \be^3) \Fbc ~\bigg], \\
  \rho^{\qq[q]\qGq[q]}_{_{D^\ast B^\ast}}(s) &=& - \frac{m_c m_b \:\rho \qq[q]\qGq[q]}{8 \pi^2} \!\!\!
  	\int\limits^{1}_{0} \!\!\frac{d\al}{\al(1 \!-\! \al)} 
	\delta\bigg( s - \frac{m_c^2 \!-\! \al( m_c^2 \!-\! m_b^2)}{\al(1-\al)}\bigg) \nno\\
	&\times&\Bigg[ \al(1-\al) + \bigg( m_c^2 -\al (m_c^2 - m_b^2) \bigg) \tau \Bigg]. 
\end{eqnarray}

For the $D^\ast B \:(1^+)$ molecular state one gets
\begin{eqnarray}
  \rho^{pert}_{_{D^*B}}(s) &=& \frac{3}{2^{12} \pi^6}
  	\int\limits^{\almax}_{\almin} \!\!\frac{d\al}{\alpha^3}  
	\!\!\int\limits^{\bemax}_{\bemin} \!\!\frac{d\be}{\beta^3} \:
	(1 \!-\! \alpha \!-\! \beta)(1\!+\! \al \!+\! \be) \Fbc^4,\\
  \nno\\
  \rho^{\qq[q]}_{_{D^*B}}(s) &=& -\frac{3 \qq[q]}{2^{7} \pi^4}
  	\!\!\int\limits^{\almax}_{\almin} \!\!\frac{d\al}{\alpha^2}  
	\!\!\int\limits^{\bemax}_{\bemin} \!\!\frac{d\be}{\beta^2} \:
	\bigg[ \be \:m_c + \al(\al+\be) \:m_b \bigg] \Fbc^2, 
\end{eqnarray}

\begin{eqnarray}
  \rho^{\GGi}_{_{D^*B}}(s) &=& \frac{\GGG}{2^{12} \pi^6}
  	\!\!\int\limits^{\almax}_{\almin} \!\!\frac{d\al}{\alpha^3}  
	\!\!\int\limits^{\bemax}_{\bemin} \!\!\frac{d\be}{\beta^3} \: \Fbc \Bigg[
	\al \be \bigg( 3\al (\al +\be) - \be(2-\al -\be) \bigg) \Fbc \nno\\
	&& + ( \be^3 m_c^2 + \al^3 m_b^2 ) (1-\al-\be)(1+\al+\be) \Bigg], \\
  \nno\\
  \rho^{\qGq[q]}_{_{D^*B}}(s) &=& - \frac{3\qGq[q]}{2^{8} \pi^4} \Bigg[
  	\int\limits^{\almax}_{\almin} \!\!\!\!\frac{d\al}{\al(1\!-\! \al)} 
	\bigg( m_c - \al(m_c-m_b) \bigg) \Hbc \nno\\
	&& -~ m_b \!\!\int\limits^{\almax}_{\almin} \!\!d\al \!\!
	\int\limits^{\bemax}_{\bemin} \!\!\frac{d\be}{\be^2} \: (2\al+3\be) \Fbc \Bigg], ~~~~\\
  \nno\\
  \rho^{\qq[q]^2}_{_{D^*B}}(s) &=& \frac{m_c m_b \:\rho \qq[q]^2}{16 \pi^2} \lambda_{bc} \:v_{bc}, \\
  \nno\\
  \rho^{\GGGi}_{_{D^*B}}(s) &=& \frac{\GGG}{2^{14} \pi^6}
  	\!\!\int\limits^{\almax}_{\almin} \!\!\frac{d\al}{\al^3} \!\!
	\int\limits^{\bemax}_{\bemin} \!\! \frac{d\be}{\be^3} (1-\al-\be)(1+\al+\be) \bigg[ \nno\\
	&& \times~ 2 (\be^4 m_c^2 \!+\! \al^4 m_b^2 ) + (\al^3 + \be^3) \Fbc ~\bigg], \\
  \nno\\
  \rho^{\qq[q]qGq[q]}_{_{D^*B}}(s) &=& \frac{m_c m_b \:\rho \qq[q]\qGq[q]}{2^{5} \pi^2}
  	\int\limits^{1}_{0} \!\!\frac{d\al}{\al(1 \!-\! \al)} 
	\delta\bigg( s - \frac{m_c^2 -\al( m_c^2 \!-\! m_b^2)}{\al(1-\al)}\bigg) \nno\\
	&& \times\bigg[ \al^2 - \Big( m_c^2 -\al (m_c^2 - m_b^2) \Big) \tau \bigg].
\end{eqnarray}
Finally, the expressions for the $D B^\ast \:(1^+)$ molecular state are given by 
\begin{eqnarray}
  \rho^{pert}_{_{DB^*}}(s) &=& \frac{3}{2^{12} \pi^6}
  	\int\limits^{\almax}_{\almin} \!\!\frac{d\al}{\alpha^3}  
	\int\limits^{\bemax}_{\bemin} \!\!\frac{d\be}{\beta^3} \:
	(1 \!-\! \alpha \!-\! \beta)(1\!+\! \al \!+\! \be) \Fbc^4, \\
  \nno\\
  \rho^{\qq[q]}_{_{DB^*}}(s) &=& -\frac{3 \qq[q]}{2^{7} \pi^4}
  	\int\limits^{\almax}_{\almin} \!\!\frac{d\al}{\alpha^2}  
	\int\limits^{\bemax}_{\bemin} \!\!\frac{d\be}{\beta^2} \:
	 \bigg[ \be(\al+\be) \:m_c + \al \:m_b \bigg] \Fbc^2, 
\end{eqnarray}

\begin{eqnarray}  
  \rho^{\GGi}_{_{DB^*}}(s) &=& \frac{\GG}{2^{12} \pi^6}
  	\int\limits^{\almax}_{\almin} \!\!\frac{d\al}{\alpha^3}  
	\int\limits^{\bemax}_{\bemin} \!\!\frac{d\be}{\beta^3} \: \Fbc \Bigg[
	\al \be \bigg( 3\be (\al +\be) -\al(2-\al -\be) \bigg) \Fbc \nno\\
	&& + ( \be^3 m_c^2 + \al^3 m_b^2 ) (1-\al-\be)(1+\al+\be) \Bigg], \\
  \rho^{\qGq[q]}_{_{DB^*}}(s) &=& - \frac{3\qGq[q]}{2^{8} \pi^4} \Bigg[
  	\int\limits^{\almax}_{\almin} \!\!\frac{d\al}{\al(1 \!-\! \al)} 
	\bigg( m_c - \al(m_c - m_b) \bigg) \Hbc \nno\\
	&& -~ m_c \!\!\int\limits^{\almax}_{\almin} \!\!\frac{d\al}{\al^2} \!\!
	\int\limits^{\bemax}_{\bemin} \!\! d\be \: (3\al+2\be) \Fbc \Bigg], ~~~~\\
  \rho^{\qq[q]^2}_{_{DB^*}}(s) &=& \frac{m_c m_b \:\rho \qq[q]^2}{16 \pi^2} \lambda_{bc} \:v_{bc}, \\
  \rho^{\GGGi}_{_{DB^*}}(s) &=& \frac{\GGG}{2^{14} \pi^6}
  	\!\!\int\limits^{\almax}_{\almin} \!\!\frac{d\al}{\al^3} \!\!
	\int\limits^{\bemax}_{\bemin} \!\! \frac{d\be}{\be^3} (1-\al-\be)(1+\al+\be) \bigg[ \nno\\
	&& \times~ 2 (\be^4 m_c^2 \!+\! \al^4 m_b^2 ) + (\al^3 + \be^3) \Fbc ~\bigg], \\
  \rho^{\qq[q]\qGq[q]}_{_{DB^*}}(s) &=& \frac{m_c m_b \:\rho \qq[q]\qGq[q]}{2^{5} \pi^2} \!\!\!
  	\int\limits^{1}_{0} \!\!\!\!\frac{d\al}{\al(1 \!-\! \al)} 
	\delta\bigg( s - \frac{m_c^2 \!-\! \al( m_c^2 \!-\! m_b^2)}{\al(1-\al)}\bigg) \nno\\
	&& \times \Bigg[ (1-\al)^2 - \bigg( m_c^2 -\al (m_c^2 - m_b^2) \bigg) \tau \Bigg].
\end{eqnarray}

For all of them, one obtains a good OPE convergence in a region where the pole contribution is 
bigger than the continuum contribution. The results are shown in Fig.(\ref{fig:DxBx}).

\begin{figure}[t]
    \centering
    \subfloat[]{\includegraphics[width=7.4cm]{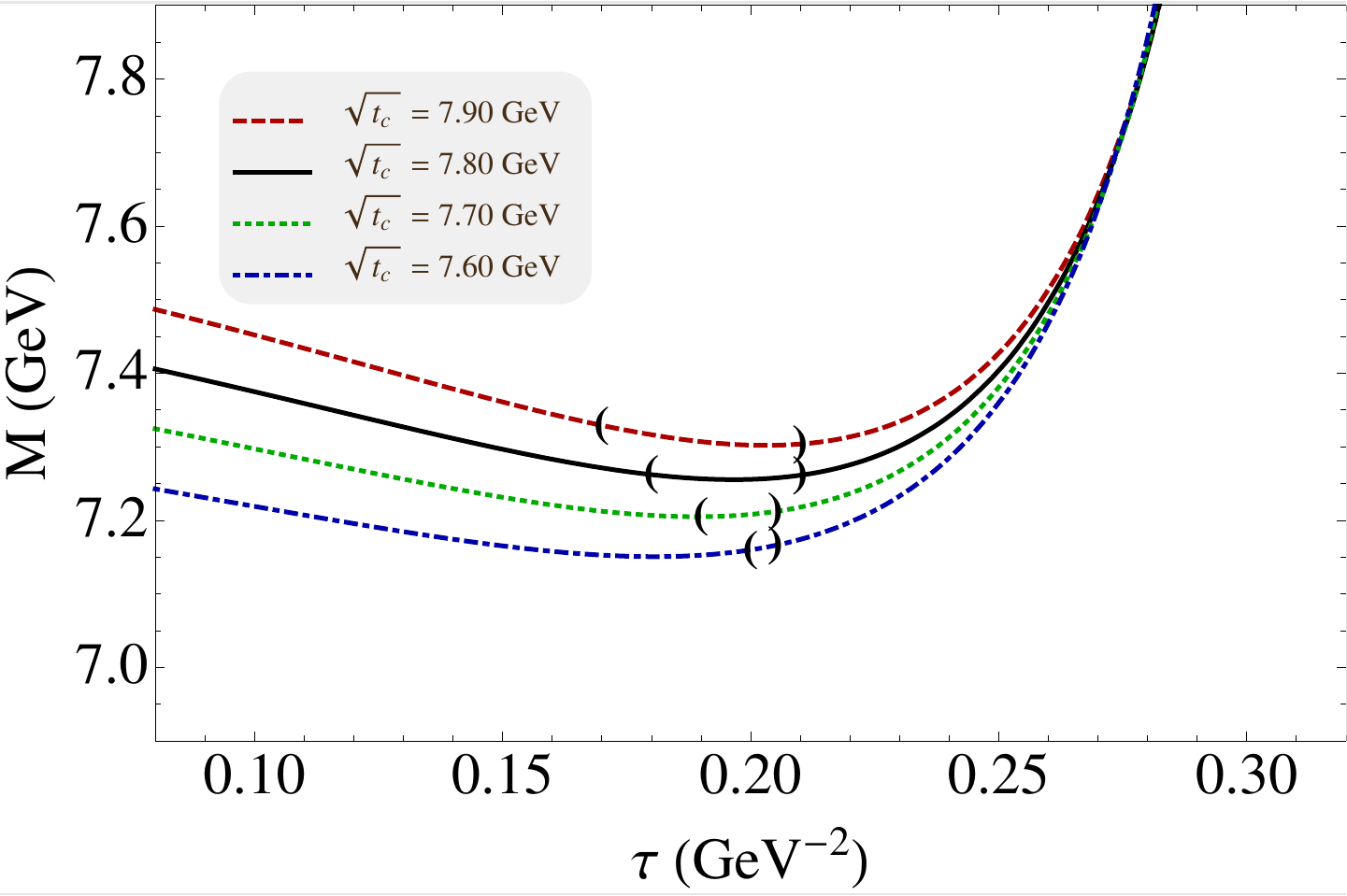}} \hspace{1cm}
    \subfloat[]{\includegraphics[width=7.4cm]{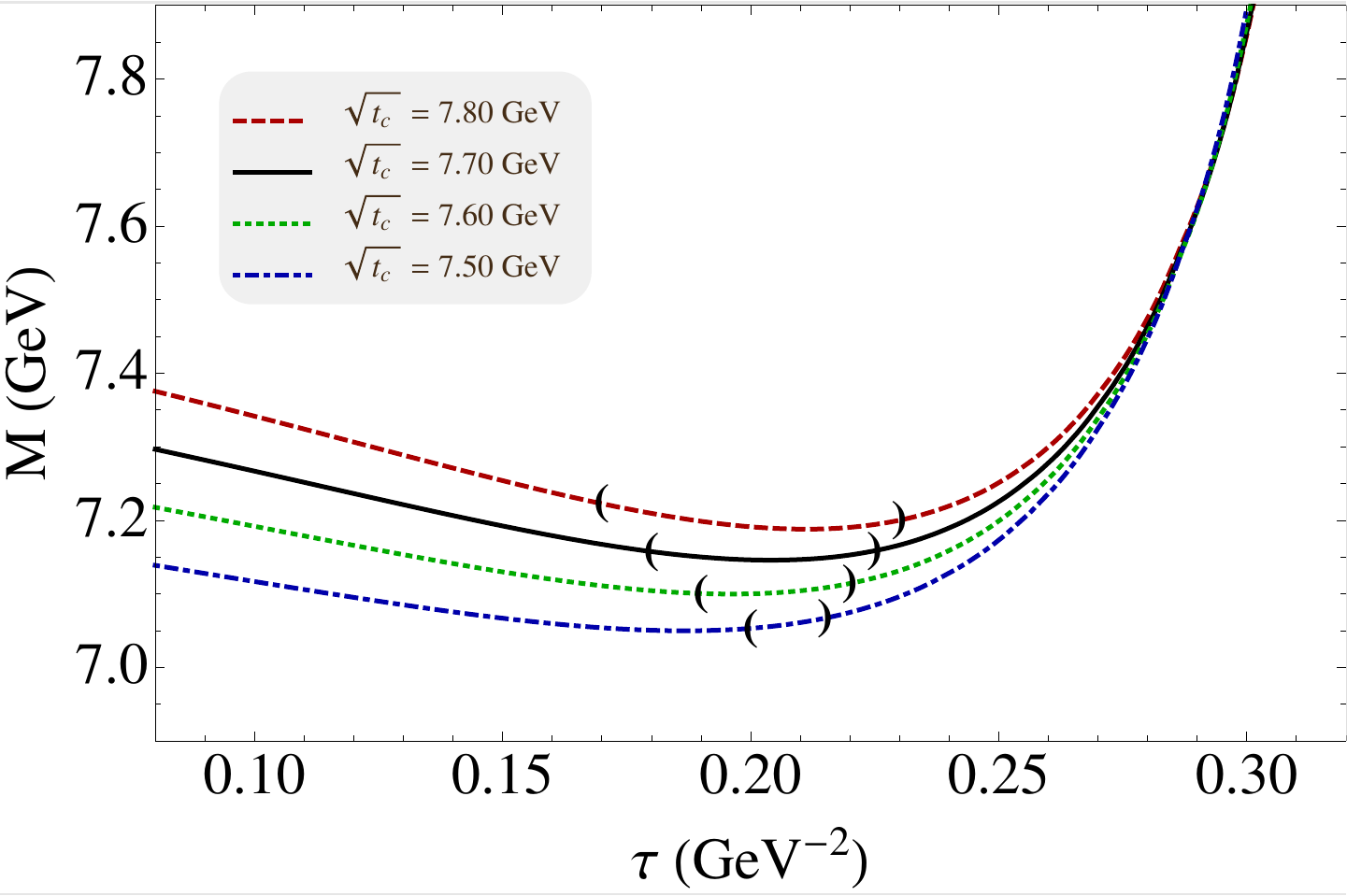}}\\
    \subfloat[]{\includegraphics[width=7.4cm]{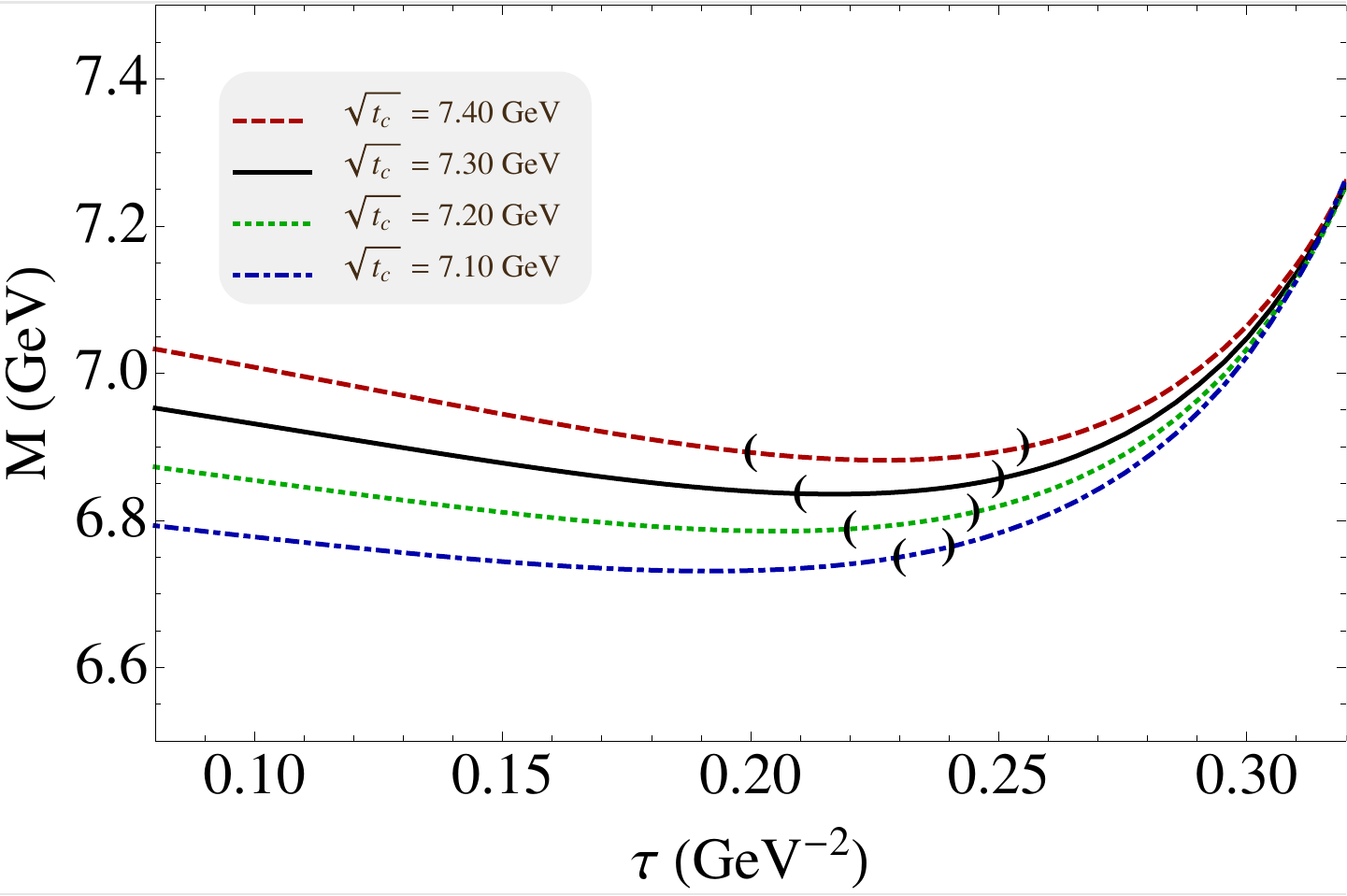}}    
\caption{\footnotesize The mass as a function of $\tau$, considering the OPE contributions 
up to dimension-eight, $m_c=1.23\GeV$, $m_b=4.24\GeV$ and different values of $\sqrt{t_c}$
for the:
{\bf (a)} $D^\ast B^\ast \:(0^+)$ molecular state;
{\bf (b)} $D^\ast B \:(1^+)$ molecular state;
{\bf (c)} $D B^\ast \:(1^+)$ molecular state.
For each line, the parentheses indicate the upper and lower limits of a valid Borel window.}
\label{fig:DxBx}
\end{figure}

In Fig.(\ref{fig:DxBx}a), the ground state mass is presented for the $D^\ast B^\ast \: (0^+)$ 
molecular state, as a function of $\tau$. For $\sqrt{t_c} = 7.80~\GeV$, the Borel window is fixed as 
$(0.18 \leq \tau \leq 0.21) \GeV^{-2}$. From this figure, it is also possible to verify that there is 
a good $\tau$-stability in the determined Borel window. Varying the value continuum threshold 
in the range $\sqrt{t_c} = (7.60 - 7.90) \GeV$, the other parameters as indicated in Table 
(\ref{TabParam}) and also estimating the uncertainty by neglecting the dimension-eight 
condensate contribution, one obtains 
\begin{eqnarray}
 M_{_{D^\ast B^\ast}} &=& (7.27\pm 0.12)~\GeV, \\
 \lambda_{_{D^\ast B^\ast}} &=& (0.115\pm 0.021)~\GeV^5  ~~.
 \label{eq:massDxBx}
\end{eqnarray}
This mass value indicates a binding energy of the order of $\sim \! 50\MeV$ below the 
threshold $E_{th} [D^\ast B^\ast]$. Considering the uncertainties, it is even possible that this 
state cannot be related to a bounded state. In such a case, our central result is in a good 
agreement with the result obtained in Ref.\cite{sunliu} for the $D^\ast B^\ast \:(0^+)$ 
molecular state.

Now considering the $D^\ast B \:(1^+)$ molecular state, the low-lying mass is shown in 
Fig.(\ref{fig:DxBx} b), as a function of $\tau$. For $\sqrt{t_c} = 7.70~\GeV$, the Borel window 
is fixed as $(0.18 \leq \tau \leq 0.22) \GeV^{-2}$.

Varying the value of continuum threshold in the range $\sqrt{t_c} = (7.50 - 7.80) \GeV$, the 
other parameters as indicated in Table (\ref{TabParam}) and also estimating the uncertainty 
by neglecting the dimension-eight condensate contribution, one obtains 
\begin{eqnarray}
 M_{_{D^\ast B}} &=& (7.16\pm0.12)~\GeV, \\
 \lambda_{_{D^\ast B}} &=& (0.058\pm 0.013)~\GeV^5  ~~.
 \label{eq:massDxB}
\end{eqnarray}
This mass value indicates a central binding energy of the order of $\sim \! 130 \MeV$, for the 
$D^\ast B \:(1^+)$ molecular state. Considering the uncertainty, this result might be 
compatible with the one obtained by the authors in Ref.\cite{sunliu}, where this molecule
would correspond to a loosely bound state between its constituents mesons.

Finally, it is analyzed the molecular current for $D B^\ast \:(1^+)$ state. As one can see 
from Fig.(\ref{fig:DxBx}c), there is a very good $\tau$-stability inside of the Borel window: 
$(0.21 \leq \tau \leq 0.25) \GeV^{-2}$, for $\sqrt{t_c} = 7.30~\GeV$. Estimating the uncertainties 
in the range $\sqrt{t_c} = (7.10 - 7.40) \GeV$, one obtains the following results
\begin{eqnarray}
 M_{_{DB^\ast}} &=& (6.85\pm0.15)~\GeV,
 \label{eq:massDBx}\\
 \lambda_{_{DB^\ast}} &=& (0.036\pm 0.011)~\GeV^5
 \label{lDBx}
\end{eqnarray}
which indicate a binding energy of the order of $\sim \! 330\MeV$, much bigger than that 
obtained in Ref.\cite{sunliu}. 

All of these results obtained with QCDSR for the $DB ~(0^+)$, $D^\ast B ~(1^+)$,
$DB^\ast ~(1^+)$ and $D^\ast B^\ast ~(0^+)$ molecular states are published in Ref.\cite{DB}.
\cleardoublepage	
\chapter{Heavy Baryons in QCD}
The heavy baryons are composed by at least one heavy quark ($c$ or $b$). 
Their spectroscopy is an area of considerable interest for hadronic physics 
since several of these hadrons were observed through the last past 
decades in the particle accelerators, see Tables (\ref{BarCharm}) and (\ref{BarBottom}). 
Then, it is extremely encouraging to use the available data - such as mass, quantum 
numbers, decay width - to develop physical models capable of predicting, with an improved 
confidence level, properties of the heavy baryons that have not yet been observed 
experimentally.

\begin{table}[bht]
  \setlength{\tabcolsep}{0.5pc}
  \caption{\small Spin $1/2^+$ and $3/2^+$ Baryons with $c$-quark observed experimentally.
  The data contained in this table are presented as follows: the updated masses of the 
  heavy baryons, according to the Particle Data Group \cite{pdg}, the year of the first experimental 
  evidence and the working group that carried out such experiment.}
  \begin{center}\vspace*{-1cm}
  \begin{tabular}{llcccr}
  &\\
  \hline \hline
  & Baryons & Quarks & Isospin & Mass (MeV) & $1^{st}$ Experimental Evidence\\
  \hline \hline
  $J^P = 1/2^+$\\
  &$\Lambda_c^+$ & $(cud)$ & $0$ & $2286.46 \pm 0.14$ & BNL, 1975 \:\cite{CLSpp}\\
  &$\Sigma^{++}_c$ & $(cuu)$ & $1$ & $2453.98 \pm 0.16$ & BNL, 1975 \:\cite{CLSpp}\\
  &$\Sigma^{+}_c$ & $(cud)$ & $1$ & $2452.9 \pm 0.4$ & BEBc, 1980 \:\cite{CSp}\\
  &$\Sigma^{0}_c$ & $(cdd)$ & $1$ & $2453.74 \pm 0.16$ & ITEP, 1986 \:\cite{CS0}\\
  &$\Xi^{+}_c$ & $(csu)$ & $1/2$ & $2467.8^{+0.4}_{-0.6}$ & CERN, 1983 \:\cite{CXp}\\
  &$\Xi^{0}_c$ & $(csd)$ & $1/2$ & $2470.88^{+0.34}_{-0.80}$ & CLEO, 1989 \:\cite{CX0}\\
  &$\Xi'^{+}_c$ & $(csu)$ & $1/2$ & $2575.6 \pm 3.1$ & CLEO, 1999 \:\cite{CXprime}\\
  &$\Xi'^{0}_c$ & $(csd)$ & $1/2$ & $2577.9 \pm 2.9$ & CLEO, 1999 \:\cite{CXprime}\\
  &$\Omega^{0}_c$ & $(css)$ & $0$ & $2695.2 \pm 1.7$ & CERN, 1985 \:\cite{CO0}\\
  $J^P = 3/2^+$\\
  &$\Sigma^{* \:++}_c$ & $(cuu)$ & $1$ & $2517.9 \pm 0.6$ & SERP, 1993 \:\cite{CS*pp}\\
  &$\Sigma^{* \:+}_c$ & $(cud)$ & $1$ & $2517.5 \pm 2.3$ & CLEO, 2001 \:\cite{CS*p}\\
  &$\Sigma^{* \:0}_c$ & $(cdd)$ & $1$ & $2518.8 \pm 0.6$ & CLEO, 1997 \:\cite{CS*0}\\
  &$\Xi^{* \:+}_c$ & $(csu)$ & $1/2$ & $2645.9^{+0.5}_{-0.6}$ & CLEO, 1996 \:\cite{CX*p}\\
  &$\Xi^{* \:0}_c$ & $(csd)$ & $1/2$ & $2645.9 \pm 0.5$ & CLEO, 1995 \:\cite{CX*0}\\
  &$\Omega^{* \:0}_c$ & $(css)$ & $0$ & $2765.9 \pm 2.0$ & BABAR, 2006 \:\cite{CO*0}\\
  \hline \hline
  \end{tabular}
  \end{center}
  \label{BarCharm}
\end{table}

A good review on the heavy baryons can be found in Ref.\cite{RICHARD}.
Among the recent discoveries involving the baryons with a $b$-quark, the ones most 
prominent are: $\Lambda_b^0$ baryon, observed by LHCb collaboration \cite{BLnew} 
in the final states including the exclusive decay mode $J/\psi \rightarrow \mu^+\mu^-$; 
$\Sigma_b$, $\Sigma_b^\ast$, $\Xi_b^-$ and $\Omega_b^-$ baryons observed in $\bar{p}p$ 
collisions by CDF and D0 collaborations. The most recent acquisition for this list, the 
$\Xi_b^{\ast \:-}$ baryon, was observed by CMS collaboration at LHC, through the strong 
decay $\Xi_b^{\ast \:-} \rightarrow \Xi_b^- \:\pi^+$ produced in $pp$ collisions.

Nowadays, with the recent progress in the experiments carried out by CDF, D0, Belle, BaBar 
and LHC collaborations there is a strong expectation for further information on heavy 
baryons spectroscopy, especially to those related to the baryons with two and three heavy 
quarks, called as Doubly and Triply Heavy Baryons respectively. These results could 
support (or not) the theoretical predictions made by Potential Models, Lattice QCD, QCDSR 
and several other models. In this sense, the present work provides the studies on heavy 
baryons spectroscopy, using the QCDSR formalism and also the Double Ratio of sum rules 
to predict the masses of these new particles. The results here found for the Singly Heavy Baryons 
(baryons with one heavy quark) are published in Ref.\cite{rnm}, while the results for the 
Doubly Heavy Baryons (baryons with two heavy quarks) are published in Ref.\cite{rnar}.

The QCDSR calculations for heavy baryons have been done using the QCD parameters listed 
in Table (\ref{TabParam}).

\begin{table}[thp]
  \setlength{\tabcolsep}{0.5pc}
  \caption{\small Spin $1/2^+$ and $3/2^+$ baryons with one $b$-quark observed experimentally.}
  \begin{center}\vspace*{-1cm}
  \begin{tabular}{llcccr}
  &\\
  \hline \hline
  & Baryons & Quarks & Isospin & Mass (MeV) & $1^{st}$ Experimental Evidence\\
  \hline \hline
  $J^P = 1/2^+$\\
  &$\Lambda_b^0$ & $(bud)$ & $0$ & $5619.4 \pm 0.7$ & CERN, 1981 \:\cite{BL}\\
  &$\Sigma^{+}_b$ & $(buu)$ & $1$ & $5811.3 \pm 1.9$ & CDF, 2007 \:\cite{BSpm}\\
  &$\Sigma^{-}_b$ & $(bdd)$ & $1$ & $5815.5 \pm 1.8$ & CDF, 2007 \:\cite{BSpm}\\
  &$\Xi^{0}_b$ & $(bsu)$ & $1/2$ & $5788 \pm 5$ & DELPHI, 1995 \:\cite{BX0}\\
  &$\Xi^{-}_b$ & $(bsd)$ & $1/2$ & $5791.1 \pm 2.2$ & CDF, 2007 \:\cite{BXm}\\
  &$\Omega^{-}_b$ & $(bss)$ & $0$ & $6071 \pm 40$ & D0, 2008 \:\cite{BOm}\\
  $J^P = 3/2^+$\\
  &$\Sigma^{* \:+}_b$ & $(buu)$ & $1$ & $5832.1 \pm 1.9$ & CDF, 2007 \:\cite{BSpm}\\
  &$\Sigma^{* \:-}_b$ & $(bdd)$ & $1$ & $5835.1 \pm 1.9$ & CDF, 2007 \:\cite{BSpm}\\
  &$\Xi^{* \:-}_b$ & $(bsd)$ & $1/2$ & $5945.0 \pm 0.7$ & CMS, 2012 \:\cite{BX*m}\\
  \hline \hline
  \end{tabular}
  \end{center}
  \label{BarBottom}
\end{table}
%

\section{Singly Heavy Baryons $(Q\MakeLowercase{qq})$}
In Ref.\cite{BAGAN}, there is a complete study in sum rules of the baryons with 
one heavy quark, $(Q=c,b)$, and two light quarks, $(q=u, d)$. 
Therefore, a natural complement to this work would be the inclusion of the strange 
$s$-quark into the baryonic currents used in Ref.\cite{BAGAN}, which allows to 
evaluate the SU(3) mass-splittings of the heavy baryons $(Qqq)$, $(Qsq)$ and $(Qss)$.

\subsubsection{\boldmath Spin $1/2^+$ Baryons $(Qsq)$}
The possible currents to describe these baryons can be obtained modifying the ones from the 
Ref.\cite{BAGAN}, in such a way the new expressions are given by:
\begin{eqnarray}
  \eta_{\Xi_Q}&=&\epsilon_{abc}\left[(q_a^TC\gamma_5s_b)+
    b(q_a^TCs_b) \gamma_5\right] Q_c~, \\
  \eta_{\Lambda_Q}&=&\eta_{\Xi_Q}~~~~(s\rightarrow q^\prime)~,\\
  \eta_{\Omega_Q}&=&\epsilon_{abc}\left[(s_a^TC\gamma_5Q_b)+
    b(s_a^TCQ_b)\gamma_5\right] s_c~, \\
  \eta_{\Sigma_Q}&=&\eta_{\Omega_Q}~~~~(s\rightarrow q)~, \\
  \eta_{\Xi^\prime_Q}&=&{\epsilon_{abc}\over\sqrt{2}}\Big{[}(s_a^TC\gamma_5Q_b)q_c +
  (q_a^TC\gamma_5Q_b)s_c + 
  b\big( (s_a^TCQ_b)\gamma_5q_c + (q_a^TCQ_b)\gamma_5 s_c\big) \Big] ~~~~
  \label{JQqq1}  
\end{eqnarray}
where the usual indices and notations are applied and $q^\prime \neq q$. The $b$
is an arbitrary mixing parameter and its value has been found to be \cite{JAMI2}:
\begin{equation}
  b= -1/5 ~,
\end{equation}
in the case of light baryons. For non-strange heavy baryons, its value is defined in the range:
\begin{equation}
  -0.5 \leq b \leq 0.5 ~.
  \label{bvalue}
\end{equation}
For both cases, the allowed values for $b$ parameter do not favor the Ioffe 
choice \cite{HEID}: 
\begin{eqnarray}
  b=-1  ~.
\end{eqnarray}
One should be careful in not confusing the $b$ parameter with the bottom quark field $b_a$.

\subsubsection{\boldmath Spin $3/2^+$ Baryons $(Qqq)$}
Using again the suggestions proposed in Ref.\cite{BAGAN}, the currents for the spin $3/2^+$ 
baryons with one heavy quark are given by:
\begin{eqnarray}
  \eta^\mu_{\Xi^*_Q} &=& \sqrt{2\over3} \epsilon_{abc} \big{[} (q_a^TC\gamma_\mu Q_b) s_c + 
  (s_a^TC\gamma_\mu Q_b)q_c+(q_a^TC\gamma_\mu s_b)Q_c\big{]} \\
  \eta^\mu_{\Omega^*_Q}&=&{1\over\sqrt{2}}\eta^\mu_{\Xi^*_Q} ~~~~(q\rightarrow s) ~.
  \label{JQqq2}
\end{eqnarray}
As in Ref.\cite{BAGAN}, one obtains the current for $\Sigma^*_Q$ baryon doing the 
following change:
\begin{eqnarray}
  \eta^\mu_{\Sigma^*_Q}&=&{1\over\sqrt{2}}\eta^\mu_{\Xi^*_Q}~~~~(s\rightarrow q)~.
\end{eqnarray}

\subsection{$\Xi_Q \:(Qsq)$ and $\Lambda_Q \:(Qqq')$}
The spectral density expressions for $\Lambda_Q$ baryon have been obtained in the chiral 
limit, $m_q = 0$, in Ref.\cite{BAGAN2}. The expressions for $\Xi_Q$ baryon, including the 
SU(3) breaking corrections, were estimated in Ref.\cite{MARINA}. Inserting the current 
$\eta_{\Xi_Q}$ into the correlation function (\ref{2point}), one obtains:
\begin{eqnarray}
  \Pi^{\Xi_Q}(q) &=& -\frac{i \:\epsilon_{ijk} \epsilon_{lmn}}{2^4 \pi^4} 
  \int\limits\! d^4x \:d^4p ~e^{ix \cdot (q-p)} \Bigg\{
  {\cal S}^Q_{il}(p) \:\Tr \left[ \ga_5 \:{\cal S}^s_{jm}(x) \:\ga_5 \:C \:{\cal S}^q_{kn}(x) \:C \right] \nno\\
  && +~ b^2 \:\ga_5 \:{\cal S}^Q_{il}(p) \:\ga_5 
  \:\Tr \left[{\cal S}^s_{jm}(x) \:C \:{\cal S}^q_{kn}(x) \:C \right] \Bigg\} ~~.
\end{eqnarray}
The expressions for $\Lambda_Q$ baryon are obtained directly from the expression above, 
by exchanging the $s$-quark with the $q^\prime$-quark, so that: 
$\Pi^{\Lambda_Q}(q) = \Pi^{\Xi_Q}(q) ~~~~ (s \rightarrow q^\prime)$.
The spectral density expressions are obtained using the full propagator of the light/heavy quarks 
and considering the OPE contributions up to dimension-six condensate. 
According to the Eq.(\ref{fcbarion}), in the case of baryons, there are two structures to be 
considered: $F_1$ and $F_2$.

\subsubsection{Spectral Densities for \boldmath $\Xi_Q$ and \boldmath $\Lambda_Q$ baryons}
\noindent $F_1$ structure:
\begin{eqnarray}
  \Img F_1^{pert} &=& \frac{m_{_Q}^4 (1+b^2)}{2^9 \pi^3} \Bigg( {1\over x^2} - {8 \over x} + 
  8x - x^2 - 12 \Log(x) \Bigg) \\
  \Img F_1^{\qq[q]} &=& \frac{m_s}{2^5 \pi} \Bigg( (1+b^2)\qq[s] - 2(1-b^2)\qq[q] \Bigg) (1-x^2)\\
  \Img F_1^{\GG} &=& \frac{(1+b^2)\GG}{3 \cdot 2^{10} \pi^3} \Bigg( 1+ 4x - 5x^2 \Bigg) \\
  \Img F_1^{\qGq[q]} &=& \frac{m_s}{3 \cdot 2^6 \pi} \Bigg( 
  (1+b^2)\qGq[s] + 6(1-b^2) \qGq[q] \Bigg) \:\de \!\left[ s - m_{_Q}^2 \right] \\
  \Img F_1^{\qq[q]^2} &=& \frac{\pi (1-b^2)}{6} \:\rho \qq[q] \qq[s]  
  \:\de\!\left[ s - m_{_Q}^2 \right]
\end{eqnarray}

\noindent $F_2$ structure:
\begin{eqnarray}  
  \Img F_2^{pert} &=& \frac{m_{_Q}^5 (1-b^2)}{2^7 \pi^3} \Bigg( 
  {1 \over x^2} + {9 \over x} - 9 - x + 6(1+1/x) \Log(x) \Bigg) \\
  \Img F_2^{\qq[q]} &=& \frac{m_{_Q} m_s}{2^4 \pi} \Bigg( (1-b^2)\qq[s] - 2(1+b^2)\qq[q] \Bigg) (1 - x) \\
  \Img F_2^{\GG} &=& \frac{m_{_Q} (1-b^2) \GG}{3 \cdot 2^9 \pi^3} \Bigg( {2\over x} + 
  5 - 7x + 6\Log(x) \Bigg) \\
  \Img F_2^{\qGq[q]} &=& \frac{m_{_Q} m_s}{3 \cdot 2^6 \pi} \Bigg( 
  (1-b^2)\qGq[s] + 6(1+b^2)\qGq[q] \Bigg)  \:\de \!\left[ s - m_{_Q}^2 \right] \\
  \Img F_2^{\qq[q]^2} &=& \frac{\pi \:m_{_Q} (1+b^2)}{6} \:\rho \qq[q]\qq[s] 
  \:\de \!\left[ s - m_{_Q}^2 \right]
\end{eqnarray}
where $x \equiv m_{_Q}^2 / s$. The spectral densities for $\Lambda_Q$ baryons are
obtained directly from the expressions above by doing the following changes:
\begin{eqnarray}
  m_s &\rightarrow& m_q \nno \\
  \qq[s] &\rightarrow& \qq[q] \nno \\
  \qGq[s] &\rightarrow& \qGq[q] ~~.
\end{eqnarray}

\subsubsection{Mass Ratio \boldmath $\Xi_c (csq) / \Lambda_c (cud)$}
The DRSR calculation is done imposing that the sum rule satisfies three stability criteria: 
1) the optimal value for the $b$ parameter is extracted from a $b$-stability point; 
2) the sum rule should present a good $\tau$-stability and 3) a good $t_c$-stability.
According to the Eqs.(\ref{DRSR}), the DRSR approach contains three equations to evaluate the 
baryon mass: $r^{sq}_{21}$, $r^{sq}_{1}$ and $r^{sq}_{2}$. It is appropriate to choose the 
sum rule which satisfies all three stability criteria previously defined.

\begin{figure}[t]
\begin{center}
{\footnotesize a)}
\epsfig{figure=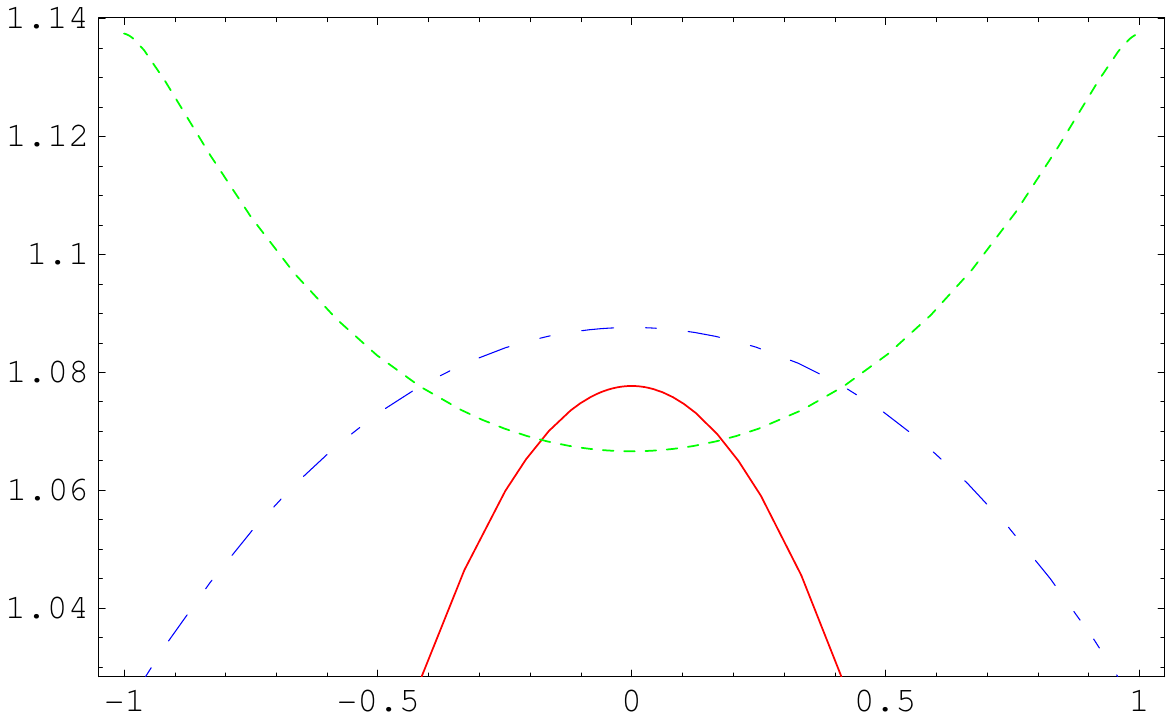,height=45mm}\hfill
{\footnotesize b)}
\epsfig{figure=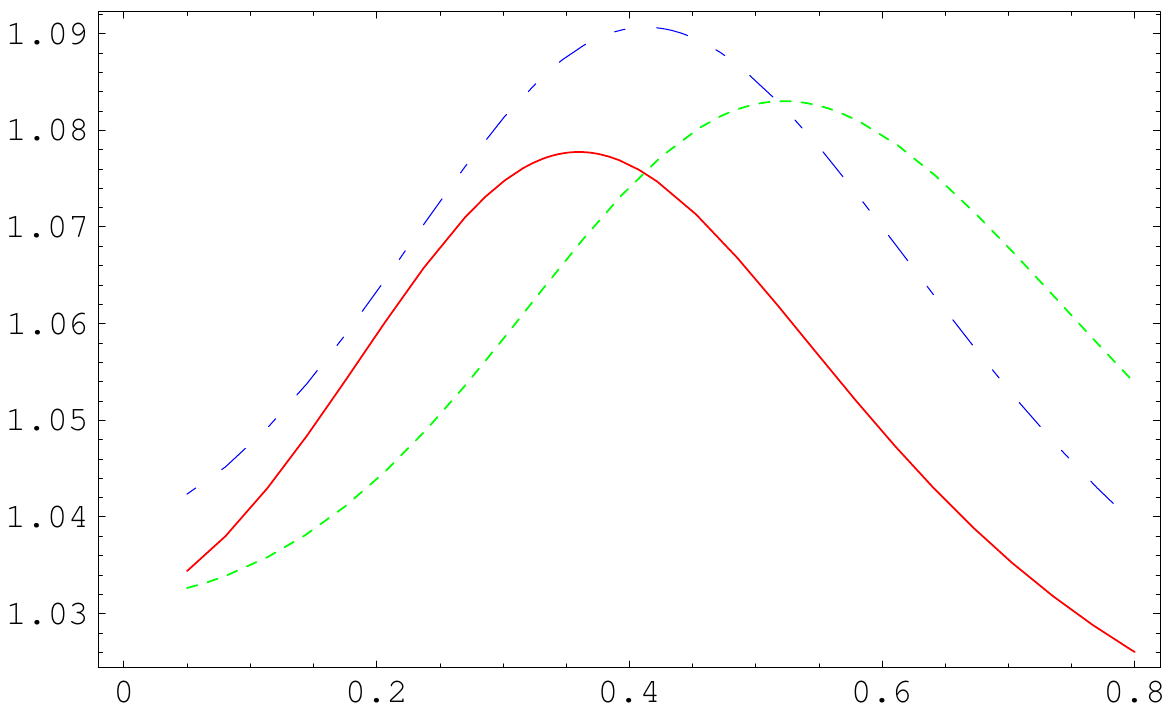,height=45mm}\hfill\\ \vspace{0.5cm}
{\footnotesize c)}
\epsfig{figure=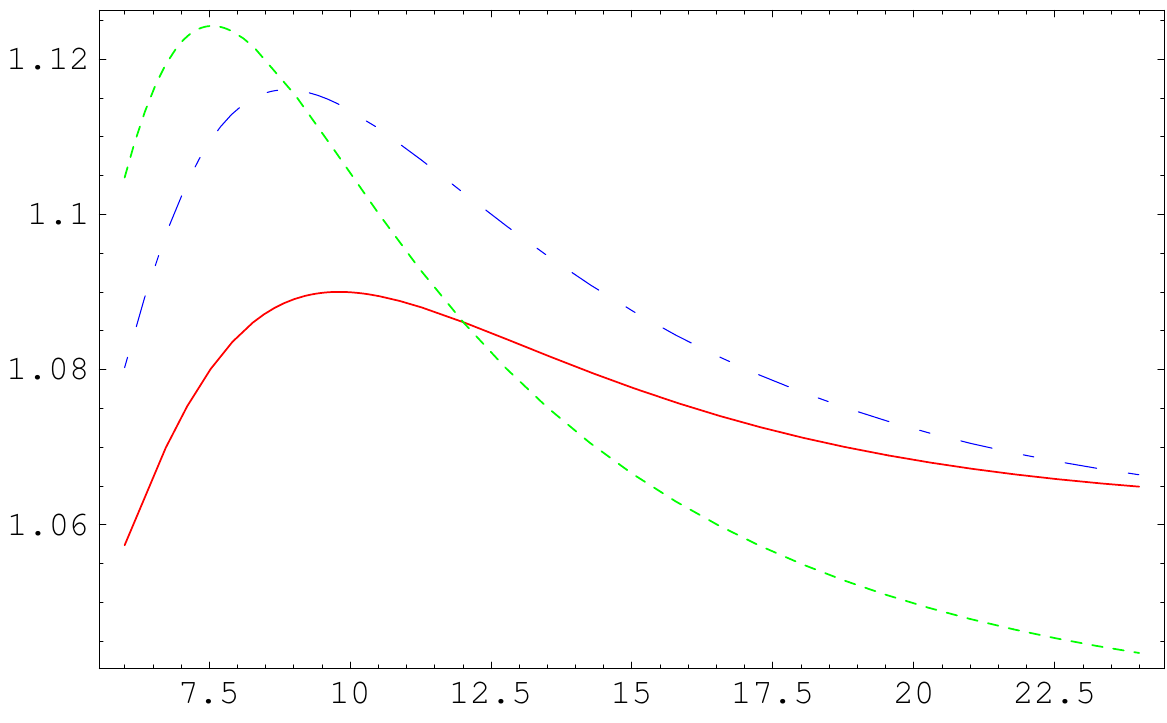,height=45mm}
\caption{\footnotesize DRSR $\Xi_c / \Lambda_c$: {\bf a)} $b$-behavior, 
for $\tau = 0.35 \GeV^{-2}$ and $t_c = 15.0 \GeV^2$, where $r^{sq}_{1}$ dot-dashed line 
(blue), $r^{sq}_{2}$ dotted line (green) and $r^{sq}_{21}$ continuous line (red);
{\bf b)} $\tau$-behavior, for $b=0$ and $t_c = 15.0 \GeV^2$; {\bf c)} the continuum 
threshold $t_c$-behavior of $r^{sq}_{21}$, for $b=0$ and $\tau = 0.35 \GeV^{-2}$.}
\label{FigChic}
\end{center}
\end{figure}

The $b$-behavior of the DRSR, presented in Fig.(\ref{FigChic}a), is obtained by 
fixing $\tau = 0.35 \GeV^{-2}$ and $t_c = 15.0 \GeV^2$. The result presents a good stability 
around the point $b = 0$, which is in the range given in Eq.(\ref{bvalue}). So, for the 
$\Xi_c$ and $\Lambda_c$ baryons, this is the optimal choice for the $b$ parameter:
\begin{equation}
  b \simeq 0 ~~.
\end{equation}

The $\tau$-behavior of different DRSR is presented in Fig.(\ref{FigChic}b), for
$t_c = 15.0 \GeV^2$ and $b=0$. Then, in Fig.(\ref{FigChic}c), fixing $b=0$ and 
$\tau = 0.35 \GeV^{-2}$ from the previous analysis, the DRSR results are shown as a 
function of the continuum threshold. Among the three sum rules, one can see that only 
$r^{sq}_{21}$ sum rule is the most stable in $\sqrt{t_c}$ and less affected by the higher 
state contributions. From the $r^{sq}_{21}$ sum rule, one can deduce the mass ratio:
\begin{equation}
  r^{sq}_{21} \equiv \frac{M_{\Xi_c}}{M_{\Lambda_c}} = 
  1.080 ~(10)_{t_c} (2)_{\tau} (6)_{m_c} (2)_{m_s} (2)_{\rho}
  \label{SRChic}
\end{equation}
where $t_c = (15.0 \pm 5.0) \GeV^2$ and $\tau = (0.35 \pm 0.05) \GeV^{-2}$. 
The most relevant uncertainties are shown explicitly in Eq.(\ref{SRChic}) and they are 
estimated using the values in Table (\ref{TabParam}). The uncertainties due to $b$ 
and some other QCD parameters are negligible. Using as input the experimental 
$\Lambda_c$ baryon mass \cite{pdg}
\begin{equation}
  M^{exp}_{\Lambda_c} = (2286.46 \pm 0.14) \MeV ~,
  \label{eq:expmassLc}
\end{equation}
and adding the different errors quadratically, one can deduce:
\begin{equation}
  M_{\Xi_c} = (2469.4 \pm 26.6) \MeV ~,
  \label{eq:massXic}
\end{equation}
which agrees nicely with the data \cite{pdg}:
\begin{equation}
  M^{exp}_{\Xi_c} = (2469.3 \pm 0.5) \MeV ~.  
  \label{eq:expmassXic}
\end{equation}

\subsubsection{Mass Ratio \boldmath $\Xi_b (bsq) / \Lambda_b (bud)$}
One repeats the previous analysis in the b-quark sector. The DRSR calculations
for $\Xi_b (bsq)$ and $\Lambda_b (bud)$ baryons show similar curves when compared to
the charm case except the obvious change of scale on the parameters $\tau$ and $t_c$. 
Using $r^{sq}_{21}$ for extracting the results, considering $\kappa = 0.74$, one obtains:
\begin{equation}
  r^{sq}_{21} \equiv \frac{M_{\Xi_b}}{M_{\Lambda_b}} = 
  1.030 ~(2.5)_{t_c} (0.5)_{\tau} (1.5)_{m_b} (0.5)_{m_s} (0.5)_{\rho}
  \label{SRChib}  
\end{equation}
for the values $t_c = (62.5 \pm 17.5) \GeV^2$ and $\tau = (0.18 \pm 0.05) \GeV^{-2}$. 
Using as input the experimental $\Lambda_b$ baryon mass \cite{pdg}
\begin{equation}
  M^{exp}_{\Lambda_b} = (5619.4 \pm 0.7) \MeV ~,
\end{equation}
and adding the different errors quadratically, one can deduce:
\vspace{-0.3cm}
\begin{equation}
  M_{\Xi_b} = (5789 \pm 16) \MeV ~,
\end{equation}
which also agrees quite well with the data \cite{pdg}:
\vspace{-0.3cm}
\begin{equation}
  M^{exp}_{\Xi_b} = (5792.4 \pm 3.0) \MeV ~.  
\end{equation}

\subsection{$\Omega_Q \:(Qss)$ and $\Sigma_Q \:(Qqq)$}
Inserting the current $\eta_{\Omega_Q}$ into the correlation function (\ref{2point}), one obtains:
\vspace{-0.2cm}
{\normalsize
\begin{eqnarray}
  \Pi^{\Omega_Q}(q) &=& -\frac{i \:\epsilon_{ijk} \epsilon_{lmn}}{2^4 \pi^4} 
  \int\limits\! d^4x \:d^4p ~e^{ix \cdot (q-p)} \nno \\ && \hspace{-1cm} \times \left\{ \:
  \begin{array}{l}
    \Tr [{\cal S}^s_{il}(x) \:\ga_5 C \:{\cal S}^{c \:T}_{jm}(p) \:C \ga_5 ] \:{\cal S}^s_{kn}(x) +
    {\cal S}^s_{il}(x) \:\ga_5 C \:{\cal S}^{c \:T}_{jm}(p) \:C \ga_5 \:{\cal S}^s_{kn}(x) \\
    \!\!+~ b \left[ {\cal S}^s_{il}(x) \:C \:{\cal S}^{c \:T}_{jm}(p) \:C \ga_5 \:{\cal S}^s_{kn}(x) \:\ga_5 +
    \ga_5 \:{\cal S}^s_{il}(x) \:\ga_5 C \:{\cal S}^{c \:T}_{jm}(p) \:C \:{\cal S}^s_{kn}(x) \right] \\
    \!\!+~ b^2 \left[ \ga_5 \:{\cal S}^s_{il}(x) \:\ga_5 \:\Tr[ C \:{\cal S}^{c \:T}_{jm}(p) \:C \:{\cal S}^s_{kn}(x) ] +
    \ga_5 \:{\cal S}^s_{il}(x) \:C \:{\cal S}^{c \:T}_{jm}(p) \:C \:{\cal S}^s_{kn}(x) \:\ga_5 \right]  \\
  \end{array}
  \right\} \nno \\ &&
\end{eqnarray}}
One uses this expression to calculate the spectral densities of $\Omega_Q$ baryons up to 
dimension-six in the OPE, working at leading order in $\alpha_s$ and keeping terms
which are linear in the strange quark mass. The correlation function for $\Sigma_Q$ baryons 
is obtained directly from the expression above. For this, one only needs to do the exchange of the 
$s$-quark propagator by the one for the light quark propagator. Notice that the spectral 
densities for $\Sigma_Q$ have already been calculated in Ref.\cite{BAGAN}.

\subsubsection{Spectral Densities for \boldmath $\Omega_Q$ and $\Sigma_Q$ baryons}
\noindent $F_1$ structure:}
\begin{eqnarray}
  \Img F_1^{pert} &=& \frac{m_{_Q}^4 (5b^2+2b+5)}{2^{11} \pi^3} \Bigg( 
  {1\over x^2} - {8\over x} + 8x - x^2 - 12\Log(x) \Bigg) \nno \\
  &&+ \frac{3m_s m_{_Q}^3 (1-b^2)}{2^8 \pi^3} \Bigg({2\over x} + 3 - 6x + x^2 + 6\Log(x) \Bigg)
\end{eqnarray}

\begin{eqnarray}  
  \Img F_1^{\qq[s]} &=&-\frac{3 \qq[s]}{2^6 \pi} \Bigg( 2m_{_Q}(1-b^2) (1-x)^2 - m_s (1+b)^2 (1-x^2) \Bigg) \\
  \Img F_1^{\GG} &=& \frac{\GG}{3 \cdot 2^{12} \pi^3} \Bigg( (1+b^2)(5-11x)+2b (1-7x) \Bigg) (1-x) \\
  \Img F_1^{\qGq[s]} &=& -\frac{\qGq[s]}{3 \cdot2^7 \pi \:s} \Bigg( 3m_{_Q} (1-b^2)(6-13x) - 6m_s(1+b)^2 x \nno\\
  && + m_s (7 + 22b + 7b^2) s \:\de \!\left[ s - m_{_Q}^2 \right] \Bigg) \hspace{1cm} \\
  \Img F_1^{\qq[s]^2} &=& \frac{\pi \rho \qq[s]^2}{3 \cdot 2^3} \Bigg( 
  (1-b)^2 - 3m_s m_{_Q} \tau (1-b^2) \Bigg)  \:\de \!\left[ s - m_{_Q}^2 \right] \\ && \nno \\
  \mbox{\noindent $F_2$ structure:}&&\nno \\
  \Img F_2^{pert} &=& \frac{m_{_Q}^5 (1-b)^2}{2^9 \pi^3} \Bigg( 
  {1\over x^2} + {9\over x} - 9 - x + 6(1+1/x) \Log(x) \Bigg) \nno \\
  && + \frac{3m_s m_{_Q}^4 (1-b^2)}{2^8 \pi^3} \Bigg( {1\over x^2} - {6\over x} + 3 + 2x - 6 \Log(x) \Bigg) \\
  \Img F_2^{\qq[q]} &=& -\frac{3m_{_Q} \qq[s]}{2^5 \pi} \Bigg( m_{_Q}(1-b^2) (1/x - 2 + x) 
  -m_s (3+2b+3b^2)(1-x) \Bigg) \\
  \Img F_2^{\GG} &=& \frac{m_{_Q} (1-b)^2 \GG}{3 \cdot 2^{11} \pi^3} \Bigg( {2\over x} - 7 + 5x \Bigg)\\
  \Img F_2^{\qGq[s]} &=& \frac{m_{_Q} \qGq[s]}{3 \cdot2^7 \pi \:s} \Bigg( 
  3m_{_Q} (1-b^2)(1+6/x) - 3m_s(5+6b+5b^2) \nno\\
  && + m_s (25 + 22b + 25b^2)s  \:\de \!\left[ s - m_{_Q}^2 \right]  \Bigg) \\
  \Img F_2^{\qq[s]^2} &=& \frac{\pi \:\rho \qq[s]^2}{3 \cdot 2^3} \Bigg( 
  m_{_Q}(5+2b+5b^2) - 3m_s (1 + m_{_Q}^2\tau) (1-b^2) \Bigg) \:\de \!\left[ s - m_{_Q}^2 \right] ~~.
  \hspace{1cm}
\end{eqnarray}\\
The spectral densities for $\Sigma_Q$ baryons are calculated from the above expressions 
by doing the following changes : $m_s \to m_q$, $\qq[s] \to \qq[q]$ and $\qGq[s] \to \qGq[q]$.

\begin{figure}[t]
\begin{center}
{\footnotesize a)}
\epsfig{figure=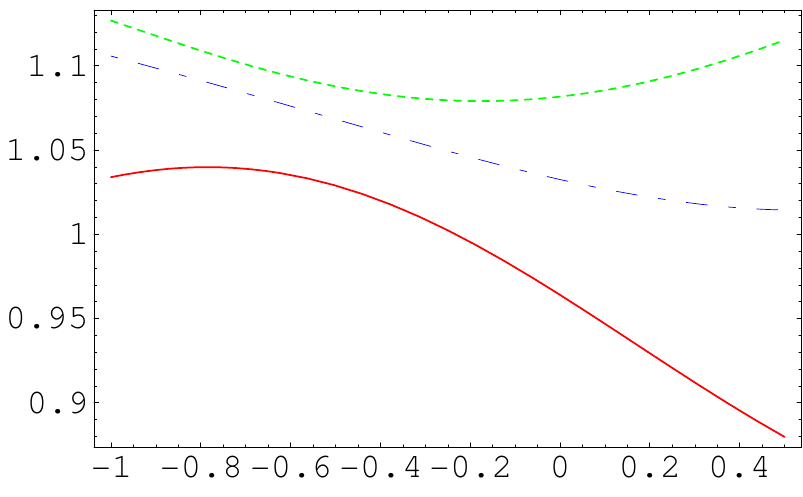,height=40mm}\hfill
{\footnotesize b)}
\epsfig{figure=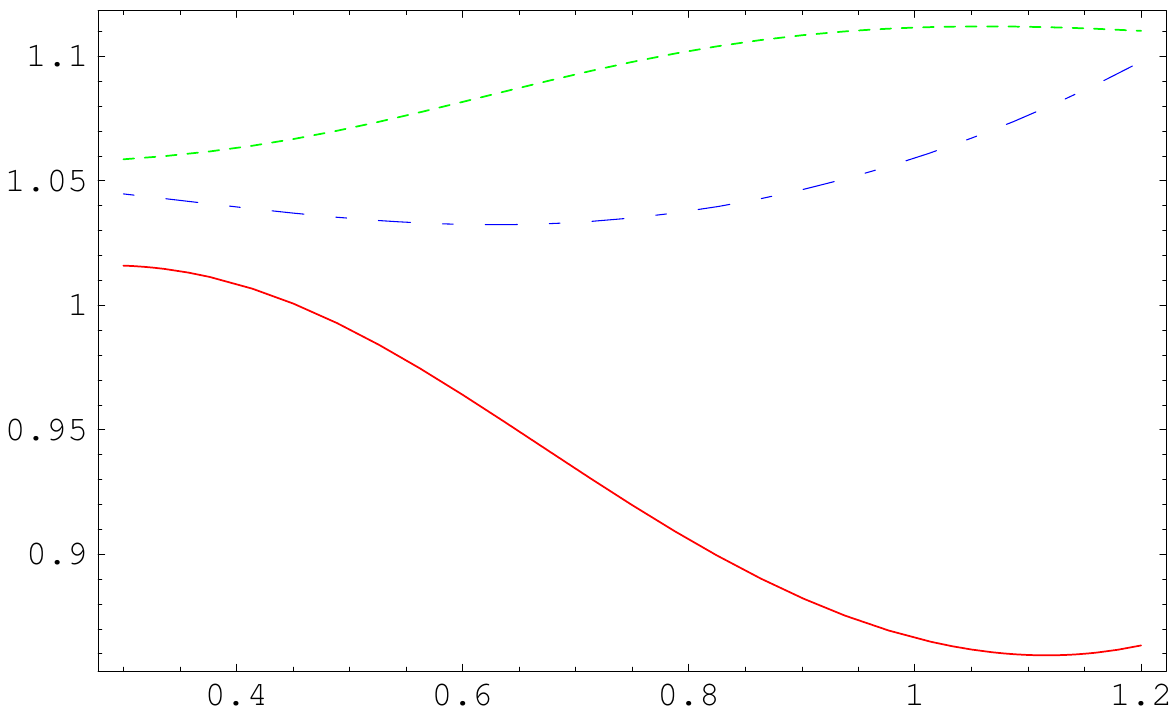,height=40mm}\hfill\\ \vspace{0.5cm}
{\footnotesize c)}
\epsfig{figure=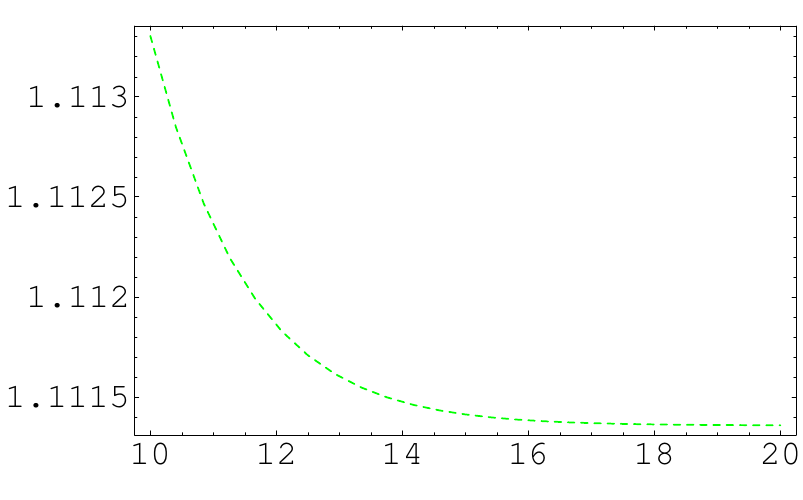,height=50mm}
\caption{\footnotesize DRSR $\Omega_c / \Sigma_c$: {\bf a)} $b$-behavior, 
for $\tau = 0.60 \GeV^{-2}$ and $t_c = 14.0 \GeV^2$, where $r^{sq}_{1}$ dot-dashed line 
(blue), $r^{sq}_{2}$ dotted line (green) and $r^{sq}_{21}$ continuous line (red);
{\bf b)} $\tau$-behavior, for $b=0$ and $t_c = 15.0 \GeV^2$; {\bf c)} the continuum 
threshold $t_c$-behavior of $r^{sq}_{2}$, for $b=0$ and $\tau = 1.0 \GeV^{-2}$.}
\label{FigOmegac}
\end{center}
\end{figure}

\subsubsection{Mass Ratio \boldmath $\Omega_c (css) / \Sigma_c (cqq)$}
In Fig.(\ref{FigOmegac}a), the $b$-behavior of various DRSR is analysed. One can notice 
that the best choice of the $b$ parameter is the same as before: $b=0$. 
Notice that only $r^{sq}_{2}$ presents simultaneously $b$- and $\tau$-stabilities from 
which the optimal result, for the mass ratio between $\Omega_c$ and $\Sigma_c$, is evaluated.
Thus, from the Fig.(\ref{FigOmegac}c), one can deduce:
\begin{equation}
  r^{sq}_{2} \equiv \frac{M_{\Omega_c}}{M_{\Sigma_c}} = 
  1.111 ~(1.4)_{\tau} (1.3)_{m_c} (16.4)_{m_s} (0.2)_{\rho}
  \label{SROmegac}    
\end{equation}
for the values $t_c = 14.0 \GeV^2$ and $\tau = (1.1 \pm 0.1) \GeV^{-2}$. 
In such a case, the uncertainty from the variations in $t_c$ is negligible. Using as input 
the experimental data for $\Sigma_c$ baryon mass \cite{pdg}
\begin{equation}
  M^{exp}_{\Sigma_c} = (2453.6 \pm 2.5) \MeV ~,
\end{equation}
and adding the different errors quadratically, one can estimate:
\begin{equation}
  M_{\Omega_c} = (2726.9 \pm 40.5) \MeV ~,
\end{equation}
which is an excelent agreement with the data observed \cite{pdg}
\begin{equation}
  M^{exp}_{\Omega_c} = (2697.5 \pm 2.6) \MeV ~.  
\end{equation}

\subsubsection{Mass Ratio \boldmath $\Omega_b (bss) / \Sigma_b (bqq)$}

\begin{figure}[t]
\begin{center}
{\footnotesize a)}
\epsfig{figure=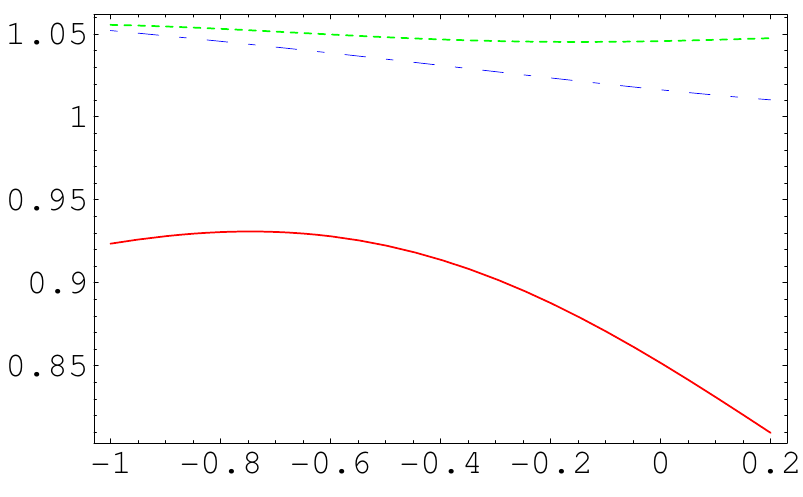,height=40mm}\hfill
{\footnotesize b)}
\epsfig{figure=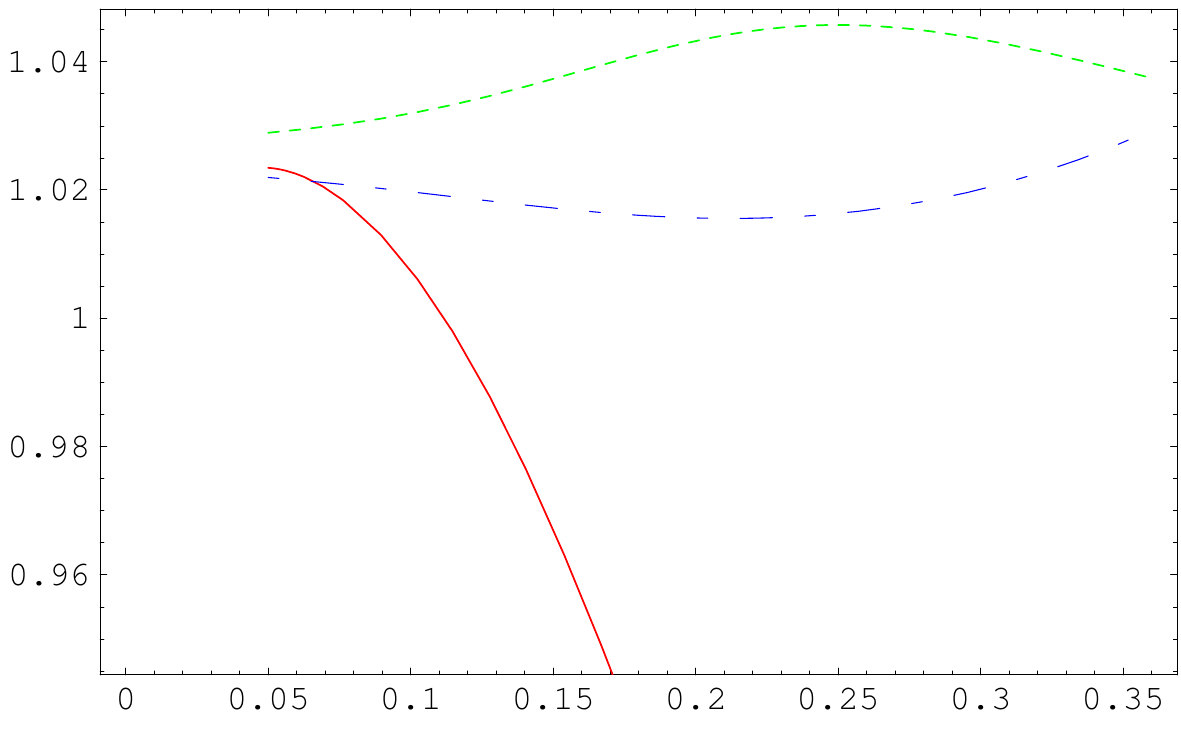,height=40mm}\hfill\\ \vspace{0.5cm}
{\footnotesize c)}
\epsfig{figure=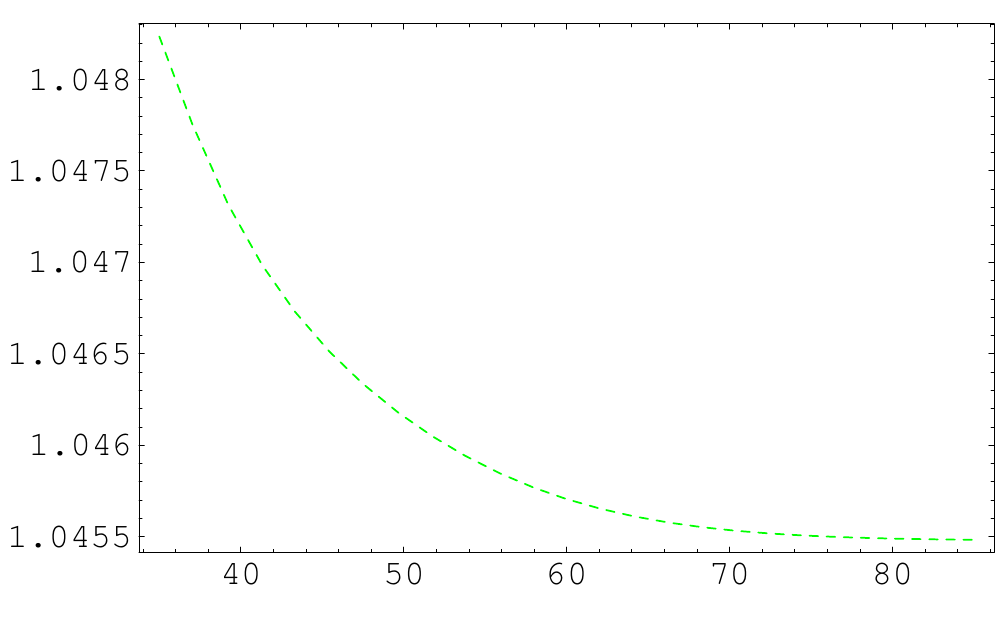,height=50mm}
\caption{\footnotesize DRSR $\Omega_b / \Sigma_b$: {\bf a)} $b$-behavior, 
for $\tau = 0.25 \GeV^{-2}$ and $t_c = 60.0 \GeV^2$, where $r^{sq}_{1}$ dot-dashed line 
(blue), $r^{sq}_{2}$ dotted line (green) and $r^{sq}_{21}$ continuous line (red);
{\bf b)} $\tau$-behavior, for $b=0$ and $t_c = 60.0 \GeV^2$; {\bf c)} the continuum 
threshold $t_c$-behavior of $r^{sq}_{2}$, for $b=0$ and $\tau = 0.25 \GeV^{-2}$.}
\label{FigOmegab}
\end{center}
\end{figure}

Repeating the previous analysis in the case of the $b$-quark, 
one obtains the results for the $\Omega_b$ baryon, only by exchanging the heavy 
quark mass: $m_c \to m_b$. Therefore, in Fig.(\ref{FigOmegab}), it is analysed the $b$-, 
$\tau$- and $t_c$-behaviors, from where it is possible to conclude that only $r^{sq}_{2}$ 
satisfies all stability criteria. Then, the mass ratio is estimated in
\begin{equation}
  r^{sq}_{2} \equiv \frac{M_{\Omega_b}}{M_{\Sigma_b}} = 
  1.0455 ~ (20)_{\tau} (22)_{m_b} (41)_{m_s} (13)_{\rho}
  \label{SROmegab}  
\end{equation}
for the values $t_c = 60.0 \GeV^2$ and $\tau = (0.25 \pm 0.05) \GeV^{-2}$. 
Using as input the experimental $\Sigma_b$ baryon mass \cite{pdg}
\begin{equation}
  M^{exp}_{\Sigma_b} = 5811.2 \MeV ~,
\end{equation}
and adding the different errors quadratically, one can deduce:
\begin{equation}
  M_{\Omega_b} = (6075.6 \pm 37.2) \MeV ~,
\end{equation}
which, considering the uncertainties, is in agreement with the experimental data observed by
CDF collaboration \cite{cdf3}: $M_{\Omega_b}^{_{CDF}} = (6054.4 \pm 6.9) \MeV$, but it 
is not compatible with the mass observed by D0 collaboration \cite{d02}: 
$M_{\Omega_b}^{_{D0}} = (6165.0 \pm 13.0) \MeV$.

\subsection{Estimation for $\kappa$}
Since the masses of the $\Xi_c$ and $\Lambda_c$ baryons are already known from the
experiments, one could control the uncertainty in the determination of $\kappa$, calculating 
the DRSR with the experimental masses of these baryons and assuming that the uncertainty 
in $\kappa$ produces a variation of $1\sigma$ on the expected value. Thus, one finally can 
estimate the following value for $\kappa$:
\begin{equation}
  \kappa = 0.74 \pm 0.06 ~,
\end{equation}
which could be considered as an improved ratio between the condensates 
$\qq[s]$ and $\qq[q]$, when compared with the existing values used in the current sum rules.
In the present work, this value will be used in the sum rule for the baryons that have not yet 
been observed experimentally.

\subsection{$\Xi^\prime_Q \:(Qsq)$}
Finally, one estimates the DRSR for $\Xi^\prime_c$ and $\Xi^\prime_b$ baryons to determine 
their masses. Inserting the current $\eta_{\Xi^\prime_Q}$ into the correlation function (\ref{2point}), 
one obtains:
\begin{eqnarray}
  \Pi^{\Xi^\prime_Q}(q) &=& -\frac{i \:\epsilon_{ijk} \epsilon_{lmn}}{2^5 \pi^4} 
  \int\limits\! d^4x \:d^4p ~e^{ix \cdot (q-p)} \nno \\ && \hspace{-1.0cm}\times \left\{ \:
  \begin{array}{l}
    \Tr [{\cal S}^q_{il}(x) \:\ga_5 C \:{\cal S}^{c \:T}_{jm}(p) \:C \ga_5 ] \:{\cal S}^s_{kn}(x) +
    \Tr [{\cal S}^s_{kn}(x) \:\ga_5 C \:{\cal S}^{c \:T}_{jm}(p) \:C \ga_5 ] \:{\cal S}^q_{il}(x) \\
    \!\!+~ {\cal S}^s_{kn}(x) \:\ga_5 C \:{\cal S}^{c \:T}_{jm}(p) \:C \ga_5 \:{\cal S}^q_{il}(x) +
    {\cal S}^q_{il}(x) \:\ga_5 C \:{\cal S}^{c \:T}_{jm}(p) \:C \ga_5 \:{\cal S}^s_{kn}(x) \\
    \!\!+~ b \left[ {\cal S}^s_{kn}(x) \:C \:{\cal S}^{c \:T}_{jm}(p) \:C \ga_5 \:{\cal S}^q_{il}(x) \:\ga_5 +
    \ga_5 \:{\cal S}^s_{kn}(x) \:\ga_5 C \:{\cal S}^{c \:T}_{jm}(p) \:C \:{\cal S}^q_{il}(x) \right] \\
    \!\!+~ b \left[ {\cal S}^q_{il}(x) \:C \:{\cal S}^{c \:T}_{jm}(p) \:C \ga_5 \:{\cal S}^s_{kn}(x) \:\ga_5 +
    \ga_5 \:{\cal S}^q_{il}(x) \:\ga_5 C \:{\cal S}^{c \:T}_{jm}(p) \:C \:{\cal S}^s_{kn}(x) \right] \\
    \!\!+~ b^2 \left[ \ga_5 \:{\cal S}^s_{kn}(x) \:\ga_5 \:\Tr[ C \:{\cal S}^{c \:T}_{jm}(p) \:C {\cal S}^q_{il}(x) ] +
    \ga_5 \:{\cal S}^s_{kn}(x) \:C \:{\cal S}^{c \:T}_{jm}(p) \:C \:{\cal S}^q_{il}(x) \:\ga_5 \right]  \\
    \!\!+~ b^2 \left[ \ga_5 \:{\cal S}^q_{il}(x) \:\ga_5 \:\Tr[ C \:{\cal S}^{c \:T}_{jm}(p) \:C \:{\cal S}^s_{kn}(x) ] +
    \ga_5 \:{\cal S}^q_{il}(x) \:C \:{\cal S}^{c \:T}_{jm}(p) \:C \:{\cal S}^s_{kn}(x) \:\ga_5 \right]  \\
  \end{array}
  \right\} \nno \\ &&
\end{eqnarray}
That is the expression used for calculating the spectral densities of $\Xi^\prime_Q$ baryon 
up to dimension-six in the OPE, working at leading order in $\alpha_s$ and keeping terms 
which are linear in the strange quark mass.

\subsubsection{Spectral Densities for \boldmath $\Xi^\prime_Q$ baryon}
\noindent $F_1$ structure:
\begin{eqnarray}
  \Img F_1^{pert} &=& \frac{m_{_Q}^4 (5b^2+2b+5)}{2^{11} \pi^3} \Bigg( 
  {1\over x^2} - {8\over x} + 8x - x^2 - 12\Log(x) \Bigg) \nno \\
  &&+ \frac{3m_s m_{_Q}^3 (1-b^2)}{2^9 \pi^3} \Bigg({2\over x} + 3 - 6x + x^2 + 6\Log(x) \Bigg) \\
  \Img F_1^{\qq[s]} &=&-\frac{3 m_{_Q}(\qq[s] + \qq[q])}{2^6 \pi} (1-b^2)(1-x)^2 \nno\\
  && + \frac{m_s}{2^7 \pi} \Bigg[ (5b^2 + 2b + 5)\qq[s] - 2(1-b)^2 \qq[q] \Bigg] (1-x^2)
\end{eqnarray}

\begin{eqnarray}
  \Img F_1^{\GG} &=& \frac{\GG}{3 \cdot 2^{12} \pi^3} \Bigg( (1+b^2)(5-11x)+2b (1-7x) \Bigg) (1-x) \\
  \Img F_1^{\qGq[s]} &=& -\frac{m_{_Q} (\qGq[s] + \qGq[q])}{2^8 \pi \:s} (1-b^2)(6-13x) \nno\\
  && + \frac{m_s \:x}{2^8 \pi \:s} \Bigg( (3+2b+3b^2) \qGq[s] - (1-b)^2 \qGq[q] \Bigg) \nno\\
  && - \frac{m_s}{3 \cdot 2^8 \pi} \Bigg( (13 + 10b + 13b^2)\qGq[s] - 6(1-b)^2\qGq[q] \Bigg)
  \:\de \!\left[ s - m_{_Q}^2 \right] \hspace{1cm} \\
  \Img F_1^{\qq[s]^2} &=& \frac{\pi \rho \qq[q]\qq[s]}{3 \cdot 2^4} \Bigg( 
  2(1-b)^2 - 3m_s m_{_Q} \tau (1-b^2) \Bigg) \:\de \!\left[ s - m_{_Q}^2 \right] ~~\\
  &&\nno\\ \mbox{\noindent $F_2$ structure:} &&\nno\\
  \Img F_2^{pert} &=& \frac{m_{_Q}^5 (1-b)^2}{2^9 \pi^3} \Bigg( 
  {1\over x^2} + {9\over x} - 9 - x + 6(1+1/x) \Log(x) \Bigg) \nno \\
  && + \frac{3m_s m_{_Q}^4 (1-b^2)}{2^9 \pi^3} 
  \Bigg( {1\over x^2} - {6\over x} + 3 + 2x - 6 \Log(x) \Bigg) \\
  \Img F_2^{\qq[q]} &=& -\frac{3m_{_Q} (\qq[s]+\qq[q])}{2^6 \pi} (1-b^2) (1/x - 2 + x) \nno \\
  && + \frac{m_s m_{_Q}}{2^6 \pi} \Bigg( (1-b)^2 \qq[s] - 2(5+2b+5b^2) \qq[q] \Bigg) (1-x) \\
  \Img F_2^{\GG} &=& \frac{m_{_Q} (1-b)^2 \GG}{3 \cdot 2^{11} \pi^3} \Bigg( {2\over x} - 7 + 5x \Bigg) \\
  \Img F_2^{\qGq[s]} &=& \frac{\qGq[s] + \qGq[q]}{2^8 \pi} (1-b^2)(6+x) \nno\\
  && + \frac{m_s m_{_Q}}{2^8 \pi \:s} \Bigg( (1-b)^2 \qGq[s] - 2(3+2b+3b^2) \qGq[q] \Bigg) \nno \\
  && - \frac{m_s m_{_Q}}{3 \cdot 2^8 \pi} \Bigg( 5(1-b)^2 \qGq[s] - 6(5 + 2b + 5b^2)\qGq[q] \Bigg) 
  \:\de \!\left[ s - m_{_Q}^2 \right] \hspace{1cm} \\
  \Img F_2^{\qq[s]^2} &=& \frac{\pi \:\rho \qq[q]\qq[s]}{3 \cdot 2^4} \Bigg( 
  2m_{_Q}(5 \!+\! 2b \!+\! 5b^2) - 3m_s (1 \!+\! m_{_Q}^2\tau) (1 \!-\! b^2) \Bigg)
   \:\de \!\left[ s \!-\! m_{_Q}^2 \right] ~~~~~~~~
\end{eqnarray}

\subsubsection{\boldmath Mass Ratio $\Xi^\prime_c (csq) / \Sigma_c (cqq)$}
From the Fig.(\ref{FigXiprimec}), one can see that only $r^{sq}_{2}$ satisfies all 
stability criteria and provides the following result:
\vspace{-0.3cm}
\begin{equation}
  r^{sq}_{2} \equiv \frac{M_{\Xi^\prime_c}}{M_{\Sigma_c}} = 
  1.043 ~ (1)_{\tau} (2)_{m_c} (6)_{m_s} (2)_\rho (3)_b (7)_\kappa
  \label{SRXiprimec}  
\end{equation}
for the values $t_c = 14.0 \GeV^2$, $\tau = (0.9 \pm 0.1) \GeV^{-2}$ and 
$b=-(0.4\pm0.2)$. Using as input the experimental $\Sigma_c$ baryon mass \cite{pdg}, 
the estimative for the mass is given by:
\begin{figure}[t]
\begin{center}
{\footnotesize a)}
\epsfig{figure=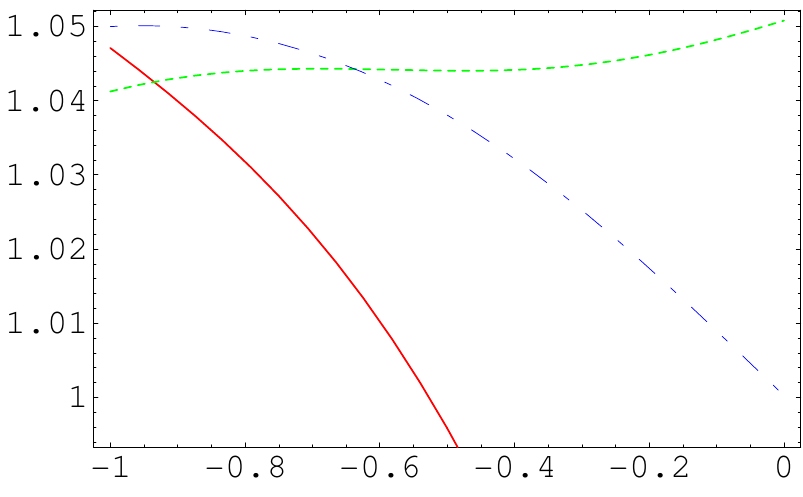,height=40mm}\hfill
{\footnotesize b)}
\epsfig{figure=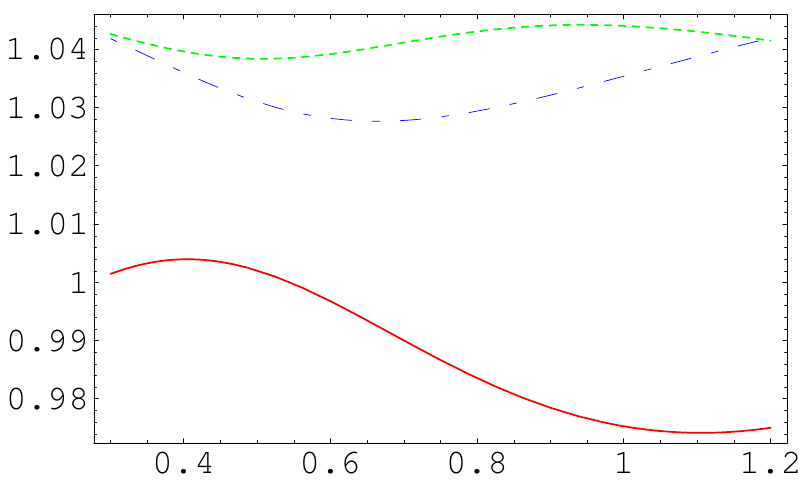,height=40mm}\hfill\\ \vspace{0.5cm}
{\footnotesize c)}
\epsfig{figure=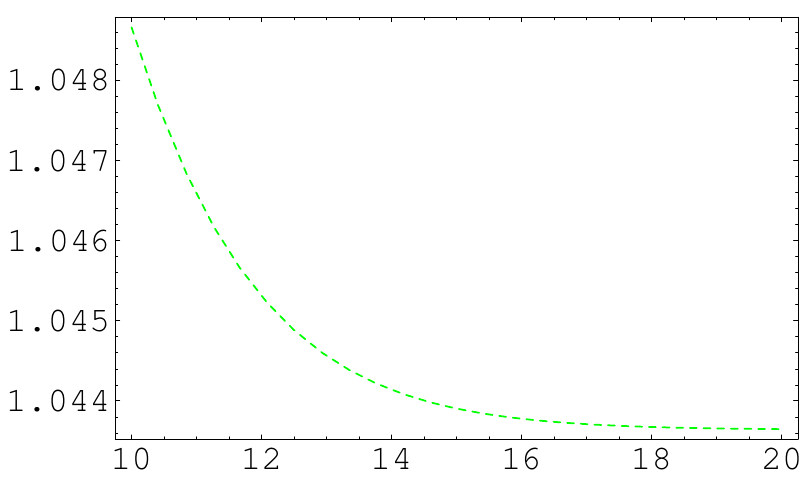,height=50mm}
\caption{\footnotesize DRSR $\Xi^\prime_c / \Sigma_c$: {\bf a)} $b$-behavior, 
for $\tau = 0.90 \GeV^{-2}$ and $t_c = 14.0 \GeV^2$, where $r^{sq}_{1}$ dot-dashed line 
(blue), $r^{sq}_{2}$ dotted line (green) and $r^{sq}_{21}$ continuous line (red);
{\bf b)} $\tau$-behavior, for $b=-0.4$ and $t_c = 14.0 \GeV^2$; {\bf c)} the continuum 
threshold $t_c$-behavior of $r^{sq}_{2}$, for $b=-0.4$ and $\tau = 1.0 \GeV^{-2}$.}
\label{FigXiprimec}
\end{center}
\end{figure}
\vspace{-0.3cm}
\begin{equation}
  M_{\Xi^\prime_c} = (2559 \pm 25) \MeV ~,
\vspace{-0.3cm}
\end{equation}
which is in an excellent agreement with experimental data observed \cite{pdg}:
\vspace{-0.3cm}
\begin{equation}
  M^{exp}_{\Xi^\prime_c} = (2576.8 \pm 3.0) \MeV ~.
\end{equation}

\begin{figure}[t]
\begin{center}
{\footnotesize a)}
\epsfig{figure=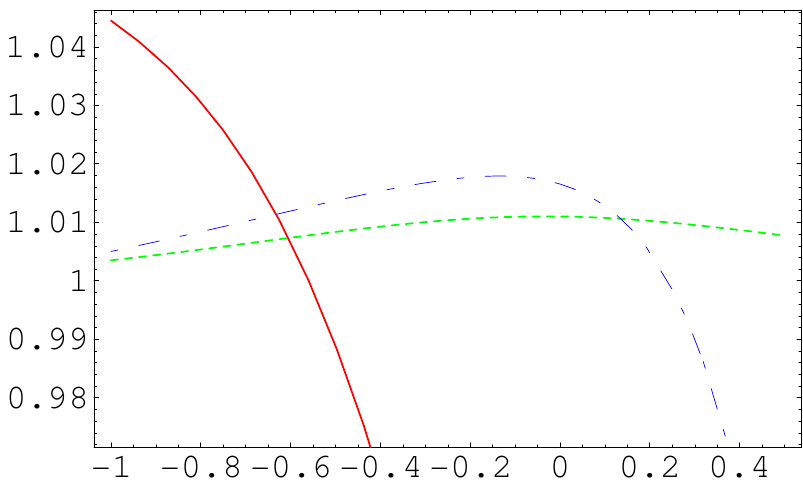,height=40mm}\hfill
{\footnotesize b)}
\epsfig{figure=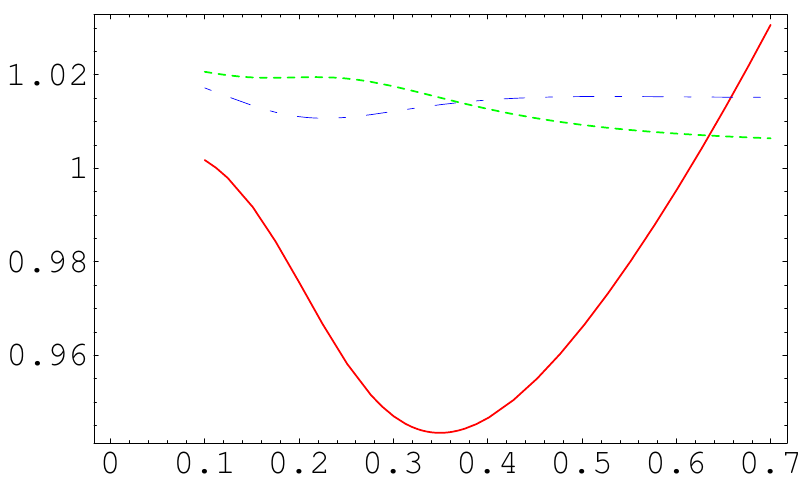,height=40mm}\hfill\\ \vspace{0.5cm}
{\footnotesize c)}
\epsfig{figure=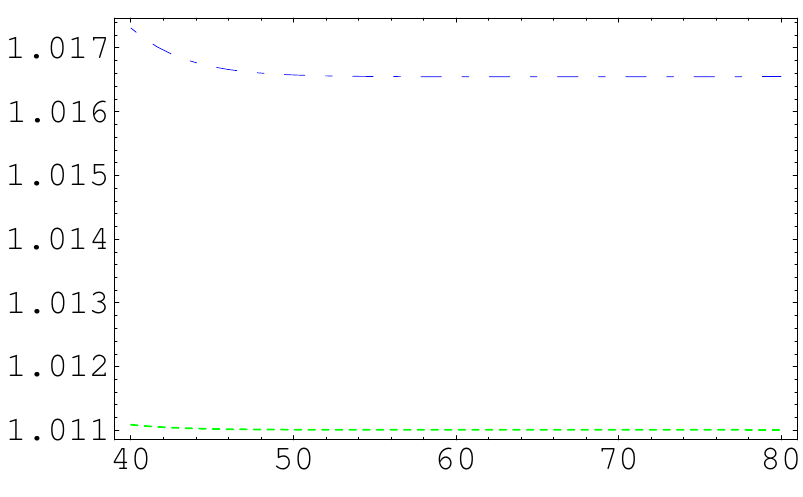,height=50mm}
\caption{\footnotesize DRSR $\Xi^\prime_b / \Sigma_b$: {\bf a)} $b$-behavior, 
for $\tau = 0.50 \GeV^{-2}$ and $t_c = 60.0 \GeV^2$, where $r^{sq}_{1}$ dot-dashed line 
(blue), $r^{sq}_{2}$ dotted line (green) and $r^{sq}_{21}$ continuous line (red);
{\bf b)} $\tau$ parameter, for $b=0$ and $t_c = 60.0 \GeV^2$; {\bf c)} the continuum 
threshold $t_c$-behavior of $r^{sq}_1$ and $r^{sq}_2$, for $b=0$ and $\tau = 0.50 \GeV^{-2}$.}
\label{FigXiprimeb}
\end{center}
\end{figure}

\subsubsection{\boldmath Mass Ratio $\Xi^\prime_b (bsq) / \Sigma_b (bqq)$}
A similar analysis is done for $\Xi^\prime_b$ baryon. In this case, one uses $m_{_Q} = m_b$ 
into the spectral density expressions to obtain the DRSR for the $\Xi^\prime_c$ baryon. The results 
are shown in Fig.(\ref{FigXiprimeb}), where it is possible to verify that both $r^{sq}_1$ and 
$r^{sq}_2$ sum rules satisfy the stability criteria. Therefore, one can estimate as the final result the 
mean value from these two sum rules:
\begin{equation}
  \bar{r}^{sq} \equiv \frac{M_{\Xi^\prime_b}}{M_{\Sigma_b}} = 
  1.0140 ~ (34)_{\tau} (50)_{m_b} (17)_{m_s} (5)_\rho (20)_b (5)_\kappa (30)_{_{SR}}
  \label{SRXiprime}  
\end{equation}
considering the values $t_c = 60 \GeV^2$, $\tau = (0.5 \pm 0.1) \GeV^{-2}$ and $b=(0.0\pm0.2)$. 
Using as input the experimental $\Sigma_b$ baryon mass \cite{pdg}:

\begin{equation}
  M^{exp}_{\Sigma_b} = 5811.2 \MeV ~,
\end{equation}
the $\Xi^\prime_b$ mass now can be estimated in:
\begin{equation}
  M_{\Xi^\prime_b} = (5893 \pm 42) \MeV ~.
\end{equation}	
This mass prediction for the non-observed $\Xi^\prime$ baryon could be compared in a 
near future with the results from the experiments carried out by LHC, CDF and D0 
collaborations.

\subsection{$\Xi^\ast_Q \:(Qsq)$ and $\Sigma^\ast_Q \:(Qqq)$}
The next list of particles to be studied from a DRSR calculation contains the spin $3/2^+$ baryons.
Inserting the current $\eta^\mu_{\Xi^\ast_Q}$ into the correlation function (\ref{2point}), one 
obtains:
{\small
\begin{eqnarray}
  \Pi^{\Xi^\ast_Q}_{\mu\nu}(q) &=& -\frac{2i \:\epsilon_{ijk} \epsilon_{lmn}}{3 \cdot 2^4 \pi^4} 
  \int\limits\! d^4x \:d^4p ~e^{ix \cdot (q-p)} \nno \\ && \hspace{-2.0cm}\times \left\{ \:
  \begin{array}{l}
    {\cal S}^s_{il}(x) \Bigg( \Tr [ {\cal S}^{c}_{jm}(p) \:\ga_\nu C \:{\cal S}^{q \:T}_{kn}(x) \:C\ga_\mu ] 
    + \ga_\nu C \:{\cal S}^{q \:T}_{kn}(x) \:C\ga_\mu \:{\cal S}^{c}_{jm}(p) 
    + \ga_\nu C \:{\cal S}^{c \:T}_{jm}(p) \:C\ga_\mu \:{\cal S}^{q}_{kn}(x) \Bigg) \\
    \!\!+ {\cal S}^c_{jm}(p) \Bigg( \Tr [ {\cal S}^{s}_{il}(x) \:\ga_\nu C \:{\cal S}^{q \:T}_{kn}(x) \:C\ga_\mu ] 
    + \ga_\nu C \:{\cal S}^{q \:T}_{kn}(x) \:C\ga_\mu \:{\cal S}^{s}_{il}(x) 
    + \ga_\nu C \:{\cal S}^{s \:T}_{il}(x) \:C\ga_\mu \:{\cal S}^{q}_{kn}(x) \Bigg) \\
    \!\!+ {\cal S}^q_{kn}(x) \Bigg( \Tr [ {\cal S}^{c}_{jm}(p) \:\ga_\nu C \:{\cal S}^{s \:T}_{il}(x) \:C\ga_\mu ] 
    + \ga_\nu C \:{\cal S}^{s \:T}_{il}(x) \:C\ga_\mu \:{\cal S}^{c}_{jm}(p) 
    + \ga_\nu C \:{\cal S}^{c \:T}_{jm}(p) \:C\ga_\mu \:{\cal S}^{s}_{il}(x) \Bigg) \\
  \end{array}
  \right\} \nno \\ &&
\end{eqnarray}}
This expression is used to calculate the spectral densities of $\Xi^\ast_Q$ baryons up to 
dimension-six in the OPE, working at leading order in $\alpha_s$ and keeping terms
which are linear in the strange quark mass. Notice that, according to the Eq.(\ref{fcbarion*}), 
only terms proportional to the structure $g_{\mu\nu}$ are important to calculate the invariant 
functions $F_1$ and $F_2$.

\subsubsection{Spectral Densities for \boldmath $\Xi^\ast_Q$ and $\Sigma^\ast_Q$ baryons}
\noindent $F_1$ structure:
\begin{eqnarray}
  \Img F_1^{pert} &=& \frac{m_{_Q}^4}{5 \cdot 3 \cdot 2^6 \pi^3} \Bigg( 
  {6\over x^2} - {45\over x} + 10 + 30x - x^3 - 60\Log(x) \Bigg) \nno \\
  &&+ \frac{m_s m_{_Q}^3}{3 \cdot 2^4 \pi^3} \Bigg({2\over x} + 3 - 6x + x^2 + 6\Log(x) \Bigg) \\
  \Img F_1^{\qq[s]} &=&-\frac{m_{_Q}}{6 \pi} \Big( \!\qq[s]\! + \!\qq[q]\! \Big) (1-x)^2 
  - \frac{m_s}{12 \pi} \Big( 2(1 \!-\! x^2)\qq[q] - (1 \!-\! x^3)\qq[s] \Big) ~~~\\
  \Img F_1^{\GGi} &=& -\frac{\GG}{3^2 \cdot 2^7 \pi^3} (7-3x-x^2)(1-x) \\
  \Img F_1^{\qGq[s]} &=& \frac{7m_{_Q} \:x}{3^2 \cdot 2^3 \pi \:s} \Big(\qGq[s] + \qGq[q] \Big)
  - \frac{m_s \:x}{3^2 \cdot 2^3 \pi \:s} \Big( (2-3x) \qGq[s] - \qGq[q] \Big) \nno\\
  && - \frac{m_s}{3 \cdot 2^4 \pi} \Big( 3\qGq[s] - 4\qGq[q] \Big)
  \:\de \!\left[ s - m_{_Q}^2 \right] \hspace{1cm} \\
  \Img F_1^{\qq[s]^2} &=& \frac{2\pi \rho \qq[q]\qq[s]}{9} \Big( 
  2 - m_s m_{_Q} \tau \Big) \:\de \!\left[ s - m_{_Q}^2 \right]
\end{eqnarray}

\noindent $F_2$ structure:
\begin{eqnarray}
  \Img F_2^{pert} &=& \frac{m_{_Q}^5}{2^9 \pi^3} \Bigg( 
  {9\over x^2} + {64\over x} - 72 - x^2 + 12(3+4/x) \Log(x) \Bigg) \nno \\
  && + \frac{m_s m_{_Q}^4}{3 \cdot 2^6 \pi^3} \Bigg( {3\over x^2} - {16\over x} + 
  12 + x^2 - 12 \Log(x) \Bigg)\\
  \Img F_2^{\qq[s]} &=& -\frac{m_{_Q}^2}{18 \pi} \Big( \!\qq[s]\! + \!\qq[q]\! \Big) 
  \Bigg( {2\over x} -3 + x^2 \Bigg) \nno\\
  && + \frac{m_s m_{_Q}}{12 \pi} \Bigg( (1 \!+\! x)\qq[s] - 6\qq[q] \Bigg) (1-x) ~~~~\\
  \Img F_2^{\GGi} &=& \frac{m_{_Q} \GG}{3^2 \cdot 2^7 \pi^3} \Bigg( 
  {8\over x} - 9 + 12x +11x^2 + 15 \Log(x) \Bigg) \\
  \Img F_2^{\qGq[s]} &=& \frac{m_{_Q}}{3^2 \cdot2^3 \pi} 
  \Bigg( \qGq[s] + \qGq[q] \Bigg) (4 + 3x^2) + \frac{m_s m_{_Q} \qGq[q]}{3^2 \cdot 2^3 \pi \:s} \nno \\
  &&- \frac{m_s m_{_Q}}{3 \cdot 2^3 \pi} \Bigg( \qGq[s] - 3\qGq[q] \Bigg) 
  \:\de \!\left[ s - m_{_Q}^2 \right]
\end{eqnarray}

\begin{eqnarray}
  \Img F_2^{\qq[s]^2} &=& \frac{2\pi \:m_{_Q} \:\rho \qq[q]\qq[s]}{9} \Bigg( 
  3 - m_s m_{_Q} \tau \Bigg) \:\de \!\left[ s - m_{_Q}^2 \right]
  \hspace{1cm}
\end{eqnarray}
The spectral densities for $\Sigma^\ast_Q$ baryons are obtained from the above 
expressions by doing the following changes: $m_s \rightarrow m_q$, $\qq[s] \rightarrow \qq[q]$ 
and $\qGq[s] \rightarrow \qGq[q]$.

\subsubsection{\boldmath Mass Ratio $\Xi^\ast_c (csq) / \Sigma^\ast_c (cqq)$}
\begin{figure}[!t]
\begin{center}
{\footnotesize a)}
\epsfig{figure=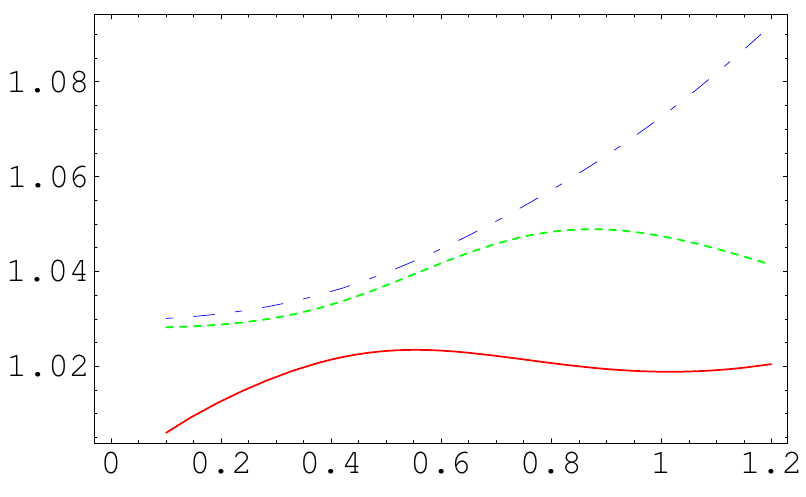,height=45mm}\hfill
{\footnotesize b)}
\epsfig{figure=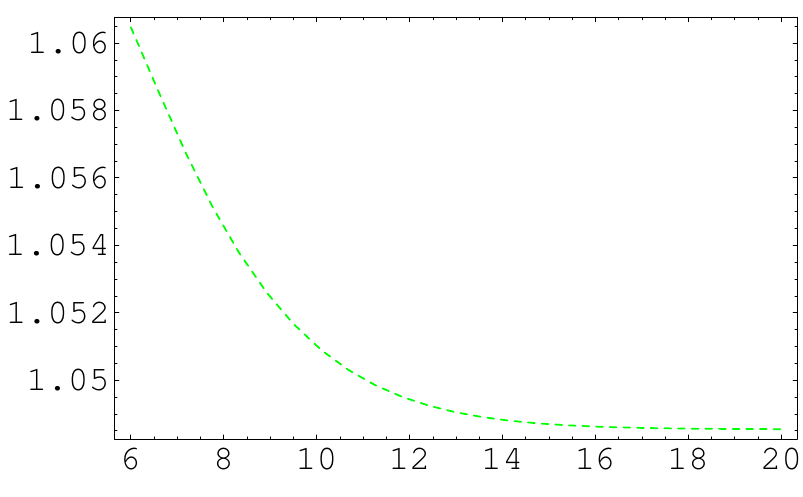,height=45mm}\hfill\\ \vspace{0.5cm}
\caption{\footnotesize DRSR $\Xi^\ast_c / \Sigma^\ast_c$: {\bf a)} $\tau$-behavior, 
for $t_c = 14.0 \GeV^2$, where $r^{sq}_{1}$ dot-dashed line (blue), $r^{sq}_{2}$ dotted line (green) 
and $r^{sq}_{21}$ continuous line (red);
{\bf b)} the continuum threshold $t_c$-behavior of $r^{sq}_2$, for $\tau = 0.90 \GeV^{-2}$.}
\label{FigXicstar}
\end{center}
\end{figure}

One repeats the previous DRSR analysis in the case of $\Xi^\ast_c$ baryon. 
In Fig.(\ref{FigXicstar}a), the $\tau$-behavior of the mass predictions is shown for 
$t_c = 14 \GeV^2$. From this figure, one does not retain $r^{sq}_{21}$ sum rule, which differs 
completely from the results found in $r^{sq}_1$ and $r^{sq}_2$ sum rules. One considers 
$r^{sq}_2$ sum rule since it is the most stable in $\tau$ and provides the following result:
\begin{equation}
  r^{sq}_2 \equiv \frac{M_{\Xi^\ast_c}}{M_{\Sigma^\ast_c}} = 
  1.049 ~ (1)_{\tau} (10)_{m_c} (4)_{m_s} (4)_\rho (20)_b (18)_\kappa
\end{equation}
for the values $\tau = (0.9 \pm 0.1) \GeV^{-2}$ and $t_c = 14.0 \GeV^2$. Using as input the 
experimental mass \cite{pdg}
\begin{equation}
  M^{exp}_{\Sigma^\ast_c} = (2518.1 \pm 1.2) \MeV ~,
  \label{MSigmacstar}
\end{equation}
and adding the different errors quadratically, one can deduce:
\begin{equation}
  M_{\Xi^\ast_c} = (2641 \pm 21) \MeV ~,
\end{equation}
which is in an excellent agreement with the value expected experimentally \cite{pdg}:
\begin{equation}
  M^{exp}_{\Xi^\ast_c} = (2645.9 \pm 0.5) \MeV ~.
\end{equation}

\subsubsection{\boldmath Mass Ratio $\Xi^\ast_b (bsq) /  \Sigma^\ast_b (bqq)$}
Considering now the $b$-quark into the current $\eta^\mu_{\Xi^*_Q}$, one obtains
analogous curves which are similar to the charm case. Then, the mass ratio can be 
evaluated and its result is given by:
\begin{equation}
  r^{sq}_2 \equiv \frac{M_{\Xi^\ast_b}}{M_{\Sigma^\ast_b}} = 
  1.022 ~ (2)_{\tau} (2)_{m_b} (0.5)_{m_s} (1)_\rho (2)_b
\end{equation}
for the values $\tau = (0.25 \pm 0.05) \GeV^{-2}$ and $t_c = 60.0 \GeV^2$. The uncertainties 
sources are the same as for $\Xi^\ast_c$ baryon. The ones due to some other parameters
are negligible. Using the averaged data \cite{pdg}
\begin{equation}
  M^{exp}_{\Sigma^\ast_b} = (5832.7 \pm 6.5) \MeV ~,
  \label{MSigmabstar}
\end{equation}
and adding the different errors quadratically, one can deduce:
\begin{equation}
  M_{\Xi^\ast_b} = (5961 \pm 21) \MeV ~.
\end{equation}
This is an estimative for $\Xi^\ast_b$ baryon mass and could be tested in a near future through
experiments carried out by LHC, CDF and D0 collaborations.

\subsection{$\Omega^\ast_Q \:(Qss)$}
Evaluating other currents for spin $3/2^+$ baryons, one can calculate
the correlation function (\ref{2point}) for the current $\eta^\mu_{\Omega^\ast_Q}$, so that:
{\footnotesize
\begin{eqnarray}
  \Pi^{\Omega^\ast_Q}_{\mu\nu}(q) &=& -\frac{2i \:\epsilon_{ijk} \epsilon_{lmn}}{3 \cdot 2^4 \pi^4} 
  \int\limits\! d^4x \:d^4p ~e^{ix \cdot (q-p)} \nno \\ && \hspace{-2cm}\times \left\{ \:
  \begin{array}{l}
    2{\cal S}^s_{il}(x) \Bigg( \Tr [ {\cal S}^{c}_{jm}(p) \:\ga_\nu C \:{\cal S}^{s \:T}_{kn}(x) \:C\ga_\mu ] 
    + \ga_\nu C \:{\cal S}^{s \:T}_{kn}(x) \:C\ga_\mu \:{\cal S}^{c}_{jm}(p) 
    + \ga_\nu C \:{\cal S}^{c \:T}_{jm}(p) \:C\ga_\mu \:{\cal S}^{s}_{kn}(x) \Bigg) \\
    \!\!+ {\cal S}^c_{jm}(p) \Bigg( \Tr [ {\cal S}^{s}_{il}(x) \:\ga_\nu C \:{\cal S}^{s \:T}_{kn}(x) \:C\ga_\mu ] 
    + \ga_\nu C \:{\cal S}^{s \:T}_{kn}(x) \:C\ga_\mu \:{\cal S}^{s}_{il}(x) 
    + \ga_\nu C \:{\cal S}^{s \:T}_{il}(x) \:C\ga_\mu \:{\cal S}^{s}_{kn}(x) \Bigg) \\
  \end{array}
  \right\} \nno \\ &&
\end{eqnarray}}
\vspace{-1.5cm}

\subsubsection{Spectral Densities for \boldmath $\Omega^\ast_Q$ baryon}
\noindent $F_1$ structure:
\begin{eqnarray}
  \Img F_1^{pert} &=& \frac{m_{_Q}^4}{5 \cdot 3 \cdot 2^5 \pi^3} \Bigg( 
  {6\over x^2} - {45\over x} + 10 + 30x - x^3 - 60\Log(x) \Bigg) \nno \\
  &&+ \frac{m_s m_{_Q}^3}{3 \cdot 2^3 \pi^3} \Bigg(
  {2\over x} + 3 - 6x + x^2 + 6\Log(x) \Bigg) \\
  \Img F_1^{\qq[s]} &=&- \frac{\qq[s]}{6\pi} \Bigg( 2m_{_Q} (1-x)^2 + 
  m_s ( 1- 2x^2 + x^3 ) \Bigg)\\
  \Img F_1^{\GGi} &=& -\frac{\GG}{3^2 \cdot 2^7 \pi^3} (7-3x-x^2)(1-x) \\
  \Img F_1^{\qGq[s]} &=& \frac{\qGq[s]}{3^2 \cdot 2^3 \pi \:s} \Bigg(
   28m_{_Q} \:x - 2m_s \:x(1-3x) + 3m_s \:s 
   \:\de \!\left[ s - m_{_Q}^2 \right] \Bigg)\hspace{1cm} \\  
  \Img F_1^{\qq[s]^2} &=& \frac{4\pi \rho \qq[s]^2}{9} \Bigg( 
  2 - m_s m_{_Q} \tau \Bigg) \:\de \!\left[ s - m_{_Q}^2 \right]
\end{eqnarray}
  
\noindent $F_2$ structure:
\begin{eqnarray}
  \Img F_2^{pert} &=& \frac{m_{_Q}^5}{3 \cdot 2^6 \pi^3} \Bigg( 
  {9\over x^2} + {64\over x} - 72 - x^2 + 12(3+4/x) \Log(x) \Bigg) \nno \\
  && + \frac{m_s m_{_Q}^4}{3 \cdot 2^5 \pi^3} \Bigg( {3\over x^2} - 
  {16\over x} + 12 + x^2 - 12 \Log(x) \Bigg)
\end{eqnarray}

\begin{eqnarray}
  \Img F_2^{\qq[s]} &=& -\frac{m_{_Q} \qq[s]}{18 \pi} \Bigg( 
  2m_{_Q}(2/ x -3 + x^2) + 3m_s (5-6x+x^2) \Bigg) \\
  \Img F_2^{\GGi} &=& \frac{m_{_Q} \GG}{3^3 \cdot 2^6 \pi^3} \Bigg( 
  {8\over x} - 9 + 12x -11x^2 + 24 \Log(x) \Bigg) \\
  \Img F_2^{\qGq[s]} &=& -\frac{\qGq[s]}{3^2\cdot 2^2 \pi \:s} \Bigg( 
  2s(4 + 3x^2) - m_s m_{_Q}  + 
  6m_s m_{_Q} s\tau \:\de \!\left[ s - m_{_Q}^2 \right]  \Bigg) ~~\\
  \Img F_2^{\qq[s]^2} &=& \frac{4\pi \:m_{_Q} \:\rho \qq[s]^2}{9} \Bigg( 
  3 - m_s m_{_Q} \tau \Bigg) \:\de \!\left[ s - m_{_Q}^2 \right]
  \hspace{1cm}
\end{eqnarray}\\

\subsubsection{\boldmath Mass Ratio $\Omega^\ast_c (css) / \Sigma^\ast_c (cqq)$}
\begin{figure}[!t]
\begin{center}
{\footnotesize a)}
\epsfig{figure=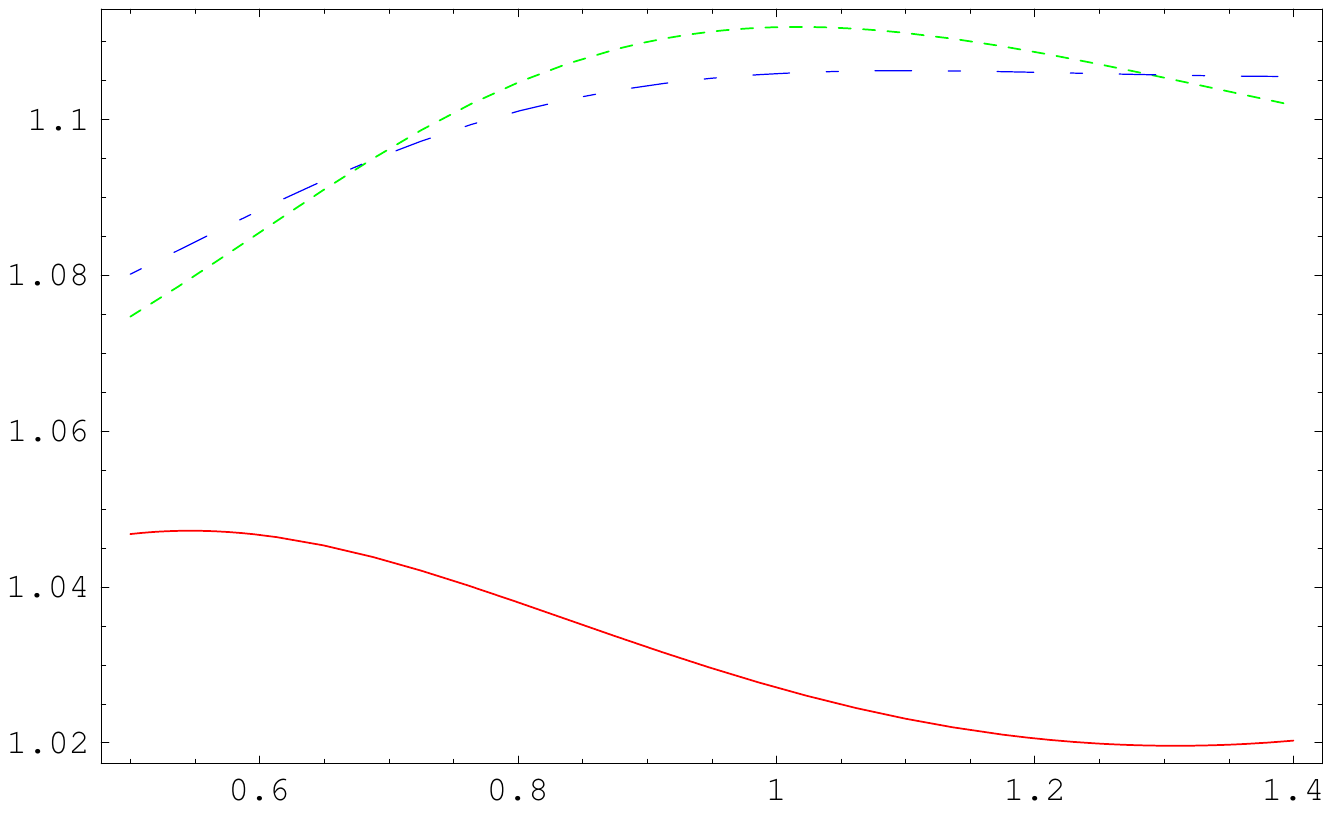,height=45mm}\hfill
{\footnotesize b)}
\epsfig{figure=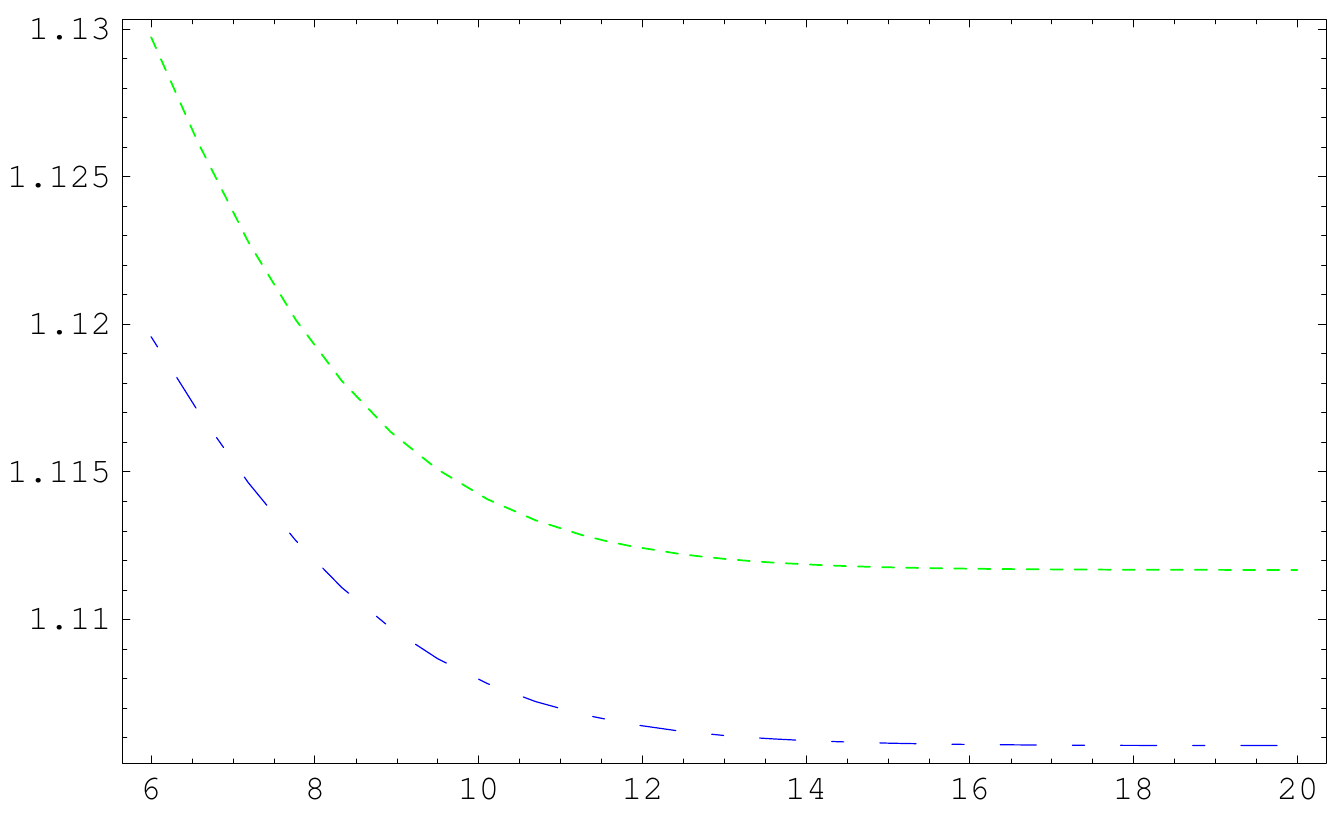,height=45mm}\hfill\\ \vspace{0.5cm}
\caption{\footnotesize DRSR $\Omega^\ast_c / \Sigma^\ast_c$: 
{\bf a)} $\tau$-behavior, for $t_c = 14.0 \GeV^2$, where $r^{sq}_{1}$ 
dot-dashed line (blue), $r^{sq}_{2}$ dotted line (green) and $r^{sq}_{21}$ continuous line (red);
{\bf b)} the continuum threshold $t_c$-behavior of $r^{sq}_1$ and $r^{sq}_2$, 
for $\tau = 1.0 \GeV^{-2}$.}
\label{FigOmegacstar}
\end{center}
\end{figure}

The DRSR calculation for $\Omega^\ast_c$ baryon provides the results shown in 
Fig.(\ref{FigOmegacstar}a), as a function of $\tau$. From this analysis, one can neglect
the $r^{sq}_{21}$ sum rule and extract the result directly from the mean value of the 
$r^{sq}_1$ and $r^{sq}_2$ sum rules. Then, 
\begin{equation}
  \bar{r}^{sq} \equiv \frac{M_{\Omega^\ast_c}}{M_{\Sigma^\ast_c}} = 
  1.109 ~ (10)_{\tau} (10)_{m_c} (4)_{m_s} (0.5)_\rho (8)_\kappa (3)_{_{SR}}
\end{equation}
where $\tau = (1.0 \pm 0.2) \GeV^{-2}$ and $t_c = 14.0 \GeV^2$. 
Using as input the experimental data (\ref{MSigmacstar}), one can estimate the mass:
\begin{equation}
  M_{\Omega^\ast_c} = (2792 \pm 38) \MeV ~.
\end{equation}
Considering the uncertainties, this result is in an good agreement with the 
experimental data \cite{pdg}:
\begin{equation}
  M^{exp}_{\Omega^\ast_c} = (2765.9 \pm 2.0) \MeV ~.
\end{equation}
\vfill

\subsubsection{\boldmath Mass Ratio $\Omega^\ast_b (bss) / \Sigma^\ast_b (bqq)$}
Finally, one considers the last heavy baryon: $\Omega^\ast_b$. The mass ratio between the 
$\Omega^\ast_b$ and $\Sigma^\ast_b$ baryons is obtained from the mean value of the
$r^{sq}_1$ and $r^{sq}_2$, so that:
\begin{equation}
  \bar{r}^{sq} \equiv \frac{M_{\Omega^\ast_b}}{M_{\Sigma^\ast_b}} = 
  1.040 ~ (4)_{\tau} (2)_{m_b} (4.6)_{m_s} (0.2)_\rho (6)_{_{SR}}
\end{equation}
where $\tau = (0.30 \pm 0.05) \GeV^{-2}$ and $t_c = 60.0 \GeV^2$. Using (\ref{MSigmabstar}) 
and adding the different errors quadratically, one can deduce:
\begin{equation}
  M_{\Omega^\ast_b} = (6066 \pm 49) \MeV ~.
\end{equation}
All the results and predictions for the Singly Heavy Baryons masses are summarized in Table (\ref{TabBarionsQqq}).

{\scriptsize
\begin{table}[hbt]
  \setlength{\tabcolsep}{1.0pc}
  \begin{center}
  \begin{minipage}[c]{14cm}
  \caption{\small Mass predictions for the heavy baryons $(Qsq)$ and $(Qss)$, from a DRSR 
  calculation with the QCD parameters as indicated in Table (\ref{TabParam}) and using as 
  input the baryon masses which have already been observed \cite{pdg}.}
  \begin{tabular}{ccccc}
  &\\
  \hline \hline
  Baryons &$I$& $r^{sq}$ & Mass (\MeV)& Exp. Data (\MeV)   \\
  \hline \hline 
  \\
  { $J^P={1\over 2}^+$}&&&&\\
  &\\
  $\Xi_c(csq)$&${1/2}$&&input&$2469.3\pm 0.5$\\
  $\Omega_c(css)$&${0}$&&input&$2697.5\pm 2.6$\\
  $\Xi_b(bsq)$&${1/2}$&&input&$5792.4\pm 3.0$\\
  $\Xi'_c(csq)$&${1/2}$&1.043(10)&2559(25)&$2576.8\pm 3.0$\\
  $\Xi'_b(bsq)$&${1/2}$&1.014(7)&5893(42)&$-$\\
  $\Omega_b(bss)$&${0}$&1.0455(64)&6076(37)&$6071 \pm 40$\\ \\
  %
  { $J^P={3\over 2}^+$}&&&& \\
  &\\
  $\Xi^*_c(csq)$&${1/2}$&1.049(8)&2641(21)&$2645.9\pm 0.5$\\
  $\Omega^*_c(css)$&0&1.109(17)&2792(38)&$2765.9\pm 2.0$\\
  $\Xi^*_b(bsq)$&${1/2}$&1.024(8)&5961(21)&$-$ \\
  $\Omega^*_b(bss)$&0&1.040(9)&6066(49)&$-$\\
  \hline \hline
  \label{TabBarionsQqq}
  \end{tabular}
  \end{minipage}
  \end{center}
\end{table}}

\subsection{Hyperfine Mass-splittings}
Combining the results for spin $1/2^+$ and $3/2^+$, given in Table (\ref{TabBarionsQqq}), 
one can deduce the values of the hyperfine mass-splittings, see Table (\ref{TabHyperQqq}). 
From this analysis, one expects that $\Omega^\ast_Q$ baryon could only decay 
electromagnetically as: $\Omega^\ast_Q \rightarrow \Omega_Q + \ga$ since
\begin{eqnarray}
  M_{\Omega^\ast_c} - M_{\Omega_c} &=& 95 (38) \MeV \\
  M_{\Omega^\ast_b} - M_{\Omega_b} &=& -10 (61) \MeV
\end{eqnarray}
and, therefore, there is not enough energy to produce more hadrons in the final state. 
On the other hand, $\Xi^\ast_Q$ baryon could, in addition, decay hadronically as: 
$\Xi^\ast_Q \rightarrow \Xi_Q + \pi$ since 
\begin{eqnarray}
  M_{\Xi^\ast_c} - M_{\Xi_c} &=& 173 (21) \MeV \\
  M_{\Xi^\ast_b} - M_{\Xi_b} &=& 169 (21) \MeV  ~~.
\end{eqnarray}
Notice that the predictions to the mass-splittings $M_{\Omega^\ast_b} - M_{\Omega_b}$ and
$M_{\Xi^\ast_b} - M_{\Xi^\prime_b}$ are in a agreement, considering the uncertainties, with the ones 
obtained from other theoretical models - see Table (\ref{TabHyperQqq}) - like Potential Models 
\cite{gisgur1, RICHARD} and $1/N_c$ Expansion \cite{JENKINS}. However, only with the 
sum rules it is possible to determine with a better precision the following result:
\begin{equation}
  M_{\Xi^\ast_b} - M_{\Xi_b} \simeq M_{\Xi^\ast_c} - M_{\Xi_c} ~~.
\end{equation}
Future precise measurements of $\Xi^\prime_b$, $\Xi^\ast_b$ and $\Omega^\ast_b$ shed 
light on the quark mass behavior of these mass-differences and will test the DRSR 
predictions for the hyperfine mass-splittings of the Singly Heavy Baryons.\\

{\scriptsize
\begin{table}[t]
  \setlength{\tabcolsep}{0.9pc}
  \begin{center} 
  \begin{minipage}[c]{15.5cm}
  \caption{\small Preditions for the hyperfine mass-splittings for the heavy baryons from a 
  DRSR calculation and from other theoretical models (in $\MeV$).}
  \vspace{-0.5cm}
  \begin{tabular}{crrrr}
    &\\
    \hline \hline
    Hyperfine Mass-splittings &  Exp. Data \cite{pdg}  & DRSR & 
    PM \cite{RICHARD} & $1/N_c$ \cite{JENKINS}\\
    \hline \hline
    $M_{\Xi^*_c}-M_{\Xi_c}$ & $179 (1)$ & 173 (21) & $-$ & $-$\\
    $M_{\Xi^*_c}-M_{\Xi'_c}$ & $70 (3)$ & 82 (33) & $-$ & 63.2 (2.6)\\
    $M_{\Omega^*_c} - M_{\Omega_c}$ & 70 (3) & 95 (38) & 70.8 (1.5) & 60.6 (5.7) \\
    $M_{\Xi^*_b}-M_{\Xi_b}$ & $-$ & 169 (21) & 164 (6) & $-$\\
    $M_{\Xi^*_b}-M_{\Xi'_b}$ & $-$ & 68 (47) & 29 (6) & 20.6 (1.9) \\
    $M_{\Omega^*_b}-M_{\Omega_b}$ & $-$ & -10 (61) & 30.7 (1.3) & 19.8 (3.1)\\
    \hline \hline
    \label{TabHyperQqq}
  \end{tabular}
  \end{minipage}
  \end{center}
  \vspace*{-0.5cm}
\end{table}}

\section{Doubly Heavy Baryons $(QQ\MakeLowercase{q})$}
Following the studies on the heavy baryons, in this section one uses the DRSR approach 
to calculate the masses of the baryons composed by two heavy quarks $(QQq)$ and $(QQs)$. 
The absolute values of the doubly heavy baryon masses of spin $1/2^+$ and $3/2^+$ have 
been obtained using QCDSR for the first time in Ref.\cite{BAGAN3}. The results found are 
given by
\begin{eqnarray}
  M_{\Xi^*_{cc}}(3/2^+)&=& 3.58 (5) ~\GeV,~~~~M_{\Xi^*_{bb}}(3/2^+)=10.33 (1.09) ~\GeV~,\nno\\
  M_{\Xi_{cc}}(1/2^+)&=& 3.48 (6) ~\GeV,~~~~M_{\Xi_{bb}}(3/2^+)=9.94 (91) ~\GeV~,
  \label{bagan}
\end{eqnarray}
and in Ref.\cite{Dosch}:
\begin{eqnarray}
  M_{\Xi_{bc}}&=&6.86(28)~ \GeV.
\end{eqnarray}

In fact, only one experiment carried out by SELEX collaboration supports the discovery of 
the first doubly heavy baryon observed in nature. The $\Xi^+_{cc} \:(ccd)$ baryon was 
observed in the decay channel $\Xi^+_{cc} \rightarrow \Lambda^+_c \:K^- \pi^+$,
with a mass:
\begin{eqnarray}
  M_{\Xi^+_{cc}} &=& 3518.9 \pm 0.9 \MeV ~.
  \label{eq:expmassXicc}
\end{eqnarray}
Since this decay channel is consistent with a particle having isospin $I = 1/2$, it could be 
useful searching for the $\Xi^{++}_{cc} \:(ccu)$ baryon as well. 

However, BaBar \cite{babarcc} and Belle \cite{bellecc} collaborations have not found any 
experimental evidence either for $\Xi^+_{cc}$ baryon in the decay channels 
$\Xi^+_{cc} \rightarrow \Lambda^+_c \:K^- \pi^+$ and $\Xi^+_{cc} \rightarrow \Xi^0_c \:\pi^+$, 
or for $\Xi^{++}_{cc}$ baryon in the decay channels 
$\Xi^{++}_{cc} \rightarrow \Lambda^+_c \:K^- \pi^+\pi^+$ and
$\Xi^{++}_{cc} \rightarrow \Xi^0_c \:\pi^+\pi^+$. Therefore, it is still needed more experimental 
data to establish the observation of Doubly Heavy Baryons.

There are some works for studying the Doubly Heavy Baryons in sum rules \cite{SR}. 
Their predictions for $M_{\Xi^\ast_{cc}}$ and $M_{\Xi_{cc}}$ masses are in an 
agreement with the ones expected experimentally for $\Xi_{cc}$ baryon as indicated in 
Eq.(\ref{eq:expmassXicc}).
In this sense, there is a strong motivation to improve the results of the mass ratio of the spin 
$3/2^+$ and $1/2^+$ baryons, using the DRSR approach.
Besides, it would be possible to extend this analysis to predict the masses of the new heavy 
baryons with two heavy quarks. These predictions could be tested in a near future through 
experiments carried out by LHC, CDF and/or D0 collaborations.

The Lorentz structures are obtained from the correlation function for the spin $1/2^+$ and 
$3/2^+$ baryons by using (\ref{fcbarion}) and (\ref{fcbarion*}) respectively.
The baryonic currents are given by:  
\begin{eqnarray}
  \mbox{\bf spin $1/2^+$}&&\nno\\
  \label{eq:currOmegaQQ}
  \eta_{\Omega_{QQ}} &=& \epsilon_{ijk} \left[(Q_i^TC\gamma_5 s_j) + 
    b(Q_i^T C s_j)\gamma_5\right] Q_k~, \\
  \label{eq:currXiQQ}  
  \eta_{\Xi_{QQ}} &=& \eta_{\Omega_{QQ}}~~~~(s \rightarrow q)~, \\ \nno\\
  \mbox{\bf spin $3/2^+$}&&\nno\\
  \label{eq:currOmega*QQ}  
  \eta^\mu_{\Omega^\ast_{QQ}} &=& {1\over\sqrt{3}}\epsilon_{ijk}\Big{[} 
  2(Q_i^T C\gamma^\mu s_j) Q_k + (Q_i^T C\gamma^\mu Q_j) s_k \Big{]}~, \\
  \label{eq:currXi*QQ}  
  \eta^\mu_{\Xi^\ast_{QQ}}&=&\eta^\mu_{\Omega^\ast_{QQ}}~~~~(s \rightarrow q)~,
\end{eqnarray}
where the usual notation is applied.
The spectral densities for $\Xi_{QQ}$ and $\Xi^\ast_{QQ}$ baryons have already been 
calculated in Ref.\cite{BAGAN3}, in the chiral limit $m_q = 0$, considering up to 
dimension-five condensates in the OPE. In the present work, the analysis are 
extended by including the linear strange quark mass corrections to the perturbative 
and quark condensate contributions, which allow the study of the $\Omega^\ast_{QQ}$ 
and $\Omega_{QQ}$ baryons. Then, the DRSR approach is used to estimate a most 
accurate value for $\Xi^\ast_{QQ}$ baryon mass calculated in \cite{BAGAN3}.
All the results for the Doubly Heavy Baryons: $\Xi_{QQ}$, $\Xi^\ast_{QQ}$, $\Omega_{QQ}$ 
and $\Omega^\ast_{QQ}$ obtained in the present work are published in Ref.\cite{rnar}.

\subsection{$\Omega_{QQ} \:(QQs)$ and $\Xi_{QQ} \:(QQq)$}
Inserting the current $\eta_{\Omega_{QQ}}$ into the correlation function (\ref{2point}), one obtains:

\begin{eqnarray}
  \Pi^{\Omega_{QQ}}(q) &=& -\frac{i \:\epsilon_{ijk} \:\epsilon_{i'j'k'}}{2^8 \pi^8} 
  \int\limits\! d^4x \:d^4p_1 \:d^4p_2 ~e^{ix \cdot (q-p_1-p_2)} \nno \\ && \hspace{-2.0cm}\times \left\{ \:
  \begin{array}{l}
    {\cal S}^Q_{kk'}(p_1) \Tr [ \ga_5 C \:{\cal S}^{s \:T}_{ii'}(x) \:C \ga_5 \:{\cal S}^Q_{jj'}(p_2)] +
    {\cal S}^Q_{kk'}(p_1) \:\ga_5 C \:{\cal S}^{s \:T}_{ii'}(x) \:C \ga_5 \:{\cal S}^Q_{jj'}(p_1) \\
    \!\!+~ b \left[ {\cal S}^Q_{kk'}(p_1) \:C \:{\cal S}^{s \:T}_{ii'}(x) \:C \ga_5 \:{\cal S}^Q_{jj'}(p_2) \:\ga_5 +
    \ga_5 \:{\cal S}^Q_{kk'}(p_1) \:\ga_5 C \:{\cal S}^{s \:T}_{ii'}(x) \:C \:{\cal S}^Q_{jj'}(p_2) \right] \\
    \!\!+~ b^2 \left[ \ga_5 \:{\cal S}^Q_{kk'}(p_1) \:\ga_5 
    \:\Tr[ C \:{\cal S}^{s \:T}_{ii'}(x) \:C {\cal S}^Q_{jj'}(p_2) ] +
    \ga_5 \:{\cal S}^Q_{kk'}(p_1) \:C \:{\cal S}^{s \:T}_{ii'}(x) \:C \:{\cal S}^Q_{jj'}(p_2) \:\ga_5 \right]  \\
  \end{array}
  \right\} \nno \\ &&
\end{eqnarray}
That is the expression used for calculating the spectral densities of the $\Omega_{QQ}$ baryon 
up to dimension-five in the OPE, working at leading order in $\alpha_s$ and keeping terms 
which are linear in the strange quark mass.

\subsubsection{Spectral Densities for \boldmath $\Omega_{QQ}$ and $\Xi_{QQ}$ baryon}
\noindent $F_1$ structure:
\begin{eqnarray}
  \Img F_1^{pert} &=& \frac{m_Q^4}{2^{11} \pi^3} \Bigg\{ 24 \Big[ 3(1+b)^2 + 4(1-b)^2 x - 
  	3(3 -2b + 3b^2)x^2 \Big] \:\mathcal{L}_v \nno\\
	&& + \:v \bigg[ 2(15\!-\! 42b \!+\! 15b^2) - 36(3 \!-\! 2b \!+\! 3b^2)x - 
	\frac{2}{x}(31 \!+\! 22b \!+\! 31b^2) \nno\\
	&& + \frac{1}{x^2}(5 \!+\! 2b \!+\! 5b^2)\bigg] - \frac{24m_s}{m_Q} (1 \!-\! b^2) \bigg[ 6(1 \!-\! 2x^2)
	\mathcal{L}_v - v \left( \frac{2}{x} \!+\! 1 \!+\! 6x \right) \bigg] \Bigg\} ~~~~~~~~~\\
  \Img F_1^{\qq[s]} &=& - \frac{m_Q \qq[s]}{2^8\pi} \Bigg\{ 24(1-b^2)v -
  	\frac{m_s}{m_Q} \bigg[ v \Big( 7 + 4b + 7b^2 + 8x(1+b+b^2) \Big) \nno\\
	&& + {3\over v}(1+b^2) \bigg] \Bigg\} \\
  \Img F_1^{\GGi} &=& \frac{\GG}{2^{12} \pi^3} (1 - 6b + b^2) \bigg[ (1 - 2x) v - 4x^2 \mathcal{L}_v\bigg] \\ 
  \Img F_1^{\qGq[s]} &=& -\frac{\qGq[s]}{2^6 \pi \:m_Q} (1-b^2) \frac{x}{v^3}(3-20x+38x^2)
\end{eqnarray}

\noindent $F_2$ structure:
\begin{eqnarray}
  \Img F_2^{pert} &=& \frac{m_Q^5}{2^9 \pi^3} \Bigg\{ 6(1-b^2) \bigg[
	6 \left( 2 - \frac{1}{x} - 2x \right)\mathcal{L}_v - v \left(6 - \frac{5}{x} - \frac{1}{x^2} \right) \bigg] \nno\\
	&& +\frac{m_s}{m_Q} \bigg[ 12 \Big( 2x (5 + 2b + 5b^2) - 3(3 + 2b + 3b^2) \Big) \mathcal{L}_v \nno \\
	&& + v \Big( 12(5 + 2b + 5b^2) + \frac{4}{x}(5 + 8b + 5b^2) + \frac{1}{x^2}(1-b)^2 \Big) \bigg] \Bigg\} \\
  \Img F_2^{\qq[s]} &=& -\frac{m_Q^2 \qq[s]}{2^7 \pi} \Bigg\{ 
  	2v \Big( 8(1 + b + b^2) + \frac{1}{x}(1-b)^2 \Big) \nno \\
	&& + \frac{3m_s}{m_Q} (1-b^2) \bigg(3v+{1\over v} \bigg) \Bigg\} \\
  \Img F_2^{\GGi} &=& \frac{m_Q \GG}{2^{10} \pi^3} (1-b^2) \bigg[ 
  \Big(3 + \frac{2}{x} \Big) v - 2(3 - x)\mathcal{L}_v \bigg] \\
  \Img F_2^{\qGq[s]} &=& -\frac{\qGq[s]}{2^7 \pi v^3} x \bigg[ 5 \!+\! 6b \!+\! 5b^2 - 
  2(9 \!+\! 14b \!+\! 9b^2) x + 16(1 \!+\! b \!+\! b^2) x^2 \bigg] ~~~
\end{eqnarray}
where the variables $x$, $v$ and ${\cal L}_v$ are defined as follows:
\begin{equation}
  x\equiv{m^2_Q\over s}~,~~v\equiv\sqrt{1-4x}~, ~~{\cal L}_v\equiv\Log {\Bigg({1+v\over 1-v}\Bigg)}   ~.
\end{equation}
Notice that the spectral densities for the $\Xi_{QQ}$ baryons are obtained from the above 
expressions by doing the following changes: $m_s \rightarrow m_q$, $\qq[s] \rightarrow \qq[q]$ 
and $\qGq[s] \rightarrow \qGq[q]$. These expressions have already been estimated 
in Ref.\cite{BAGAN3}.

According to the Ref.\cite{BAGAN3}, the mixed quark $\qGq[q]$ condensate contribution has  
terms which behave like $1/v^3$ (where $v$ is related to the heavy quark velocity). The presence 
of these terms could indicate Coulombic-like corrections and, therefore, would require a 
complete treatment of the non-relativistic Coulombic corrections which is beyond the aim of 
the present work. 
Therefore, for simplicity, the authors in Ref.\cite{BAGAN3} suggest the truncation of the OPE 
at the dimension-four condensates. The inclusion of the dimension-five condensate contribution 
will be considered only for controlling the accuracy of the approach or for improving the $\tau$ 
and/or $t_c$-stabilitiy of the analysis. 
As discussed previously, the DRSR can be evaluated using the three equations given by 
Eqs.(\ref{DRSR}). 

\subsubsection{\boldmath Mass Ratio $\Omega_{QQ}  / \Xi_{QQ}$}
\begin{figure}[!t]
\begin{center}
{\footnotesize a)}
\epsfig{figure=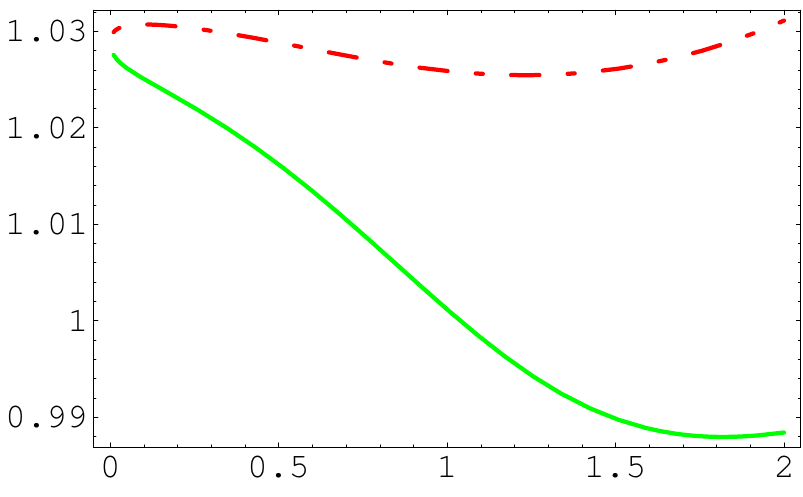,height=40mm}\hfill
{\footnotesize b)}
\epsfig{figure=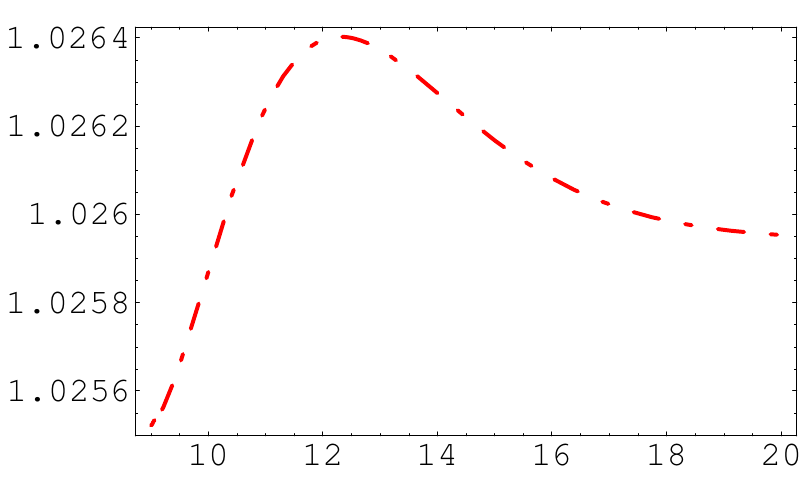,height=43mm}\hfill\\ \vspace{0.5cm}
\caption{\footnotesize DRSR $\Omega_{cc} / \Xi_{cc}$: 
{\bf a)} $\tau$-behavior, for $b =- 0.35$, $t_c =12.0 \GeV^2$ and $m_c = 1.26 \GeV$, 
where $r^{sq}_{1}(cc)$ dot-dashed line (red) and $r^{sq}_{2}(cc)$ continuous line (green);
{\bf b)} the continuum threshold $t_c$-behavior of $r^{sq}_{1}(cc)$ dot-dashed line (red), 
for $b=-0.35$ and $\tau = 1.0 \GeV^{-2}$.}
\label{FigOmegacc}
\end{center}
\end{figure}

In Fig.(\ref{FigOmegacc}a), the $\tau$-behavior is shown for the $r^{sq}_1(cc)$ and 
$r^{sq}_2(cc)$ sum rules, which are related to the mass ratio $M_{\Omega_{cc}}\over M_{\Xi_{cc}}$. 
From this figure, it is possible to check that only $r^{sq}_1(cc)$ satisfies the stability criteria 
for $\tau$. Thus, the most stable result is given by $r^{sq}_{1}(cc)$, whose $t_c$-behavior 
is presented in Fig.(\ref{FigOmegacc}b), for $\tau = 1.0 \GeV^{-2}$. 
At the $t_c$-stability point, one can deduce:
\begin{equation}
   r^{sq}_{1}(cc)\equiv {M_{\Omega_{cc}}\over M_{\Xi_{cc}}}=1.026(5)_{m_c}(2)_{\bar ss}(4)_{m_s}~,
\end{equation}
the indices indicate the different sources of uncertainties. Considering the experimental value
$M_{\Xi_{cc}} \simeq 3.52 \GeV$ \cite{pdg}, this mass ratio provides 
\begin{equation}
  M_{\Omega_{cc}}- M_{\Xi_{cc}}= 92(24) \MeV ~.
  \label{omegacc}
\end{equation}

\begin{figure}[!t]
\begin{center}
{\footnotesize a)}
\epsfig{figure=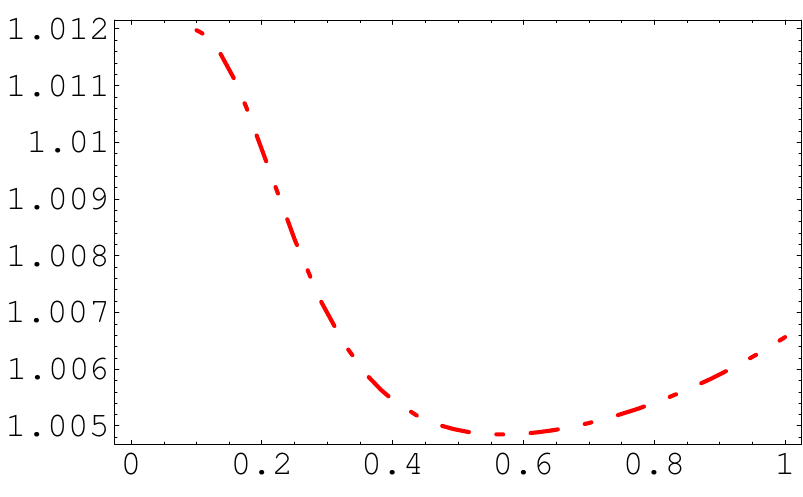,height=40mm}\hfill
{\footnotesize b)}
\epsfig{figure=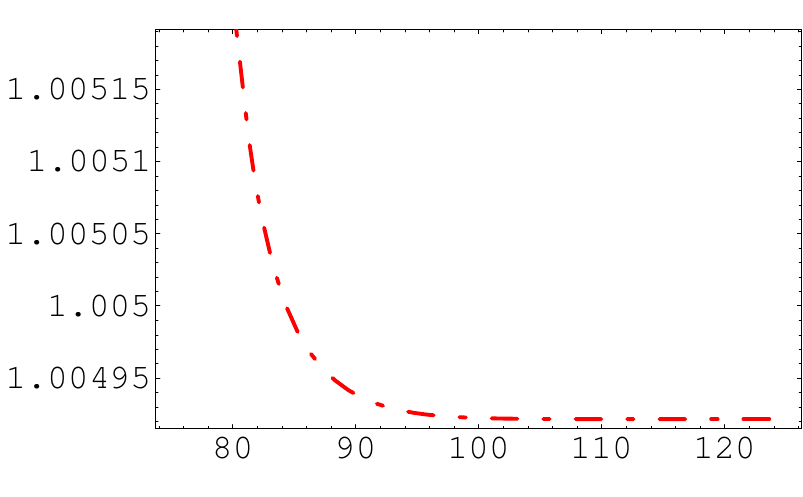,height=43mm}\hfill\\ \vspace{0.5cm}
\caption{\footnotesize DRSR $\Omega_{bb} / \Xi_{bb}$: 
{\bf a)} $\tau$-behavior, for $b =- 0.35$, $t_c =100.0 \GeV^2$ and $m_b = 4.22 \GeV$, 
where $r^{sq}_{1}(bb)$ dot-dashed line (red);
{\bf b)} the continuum threshold $t_c$-behavior of $r^{sq}_{1}(cc)$ dot-dashed line (red), 
for $b=-0.35$ and $\tau = 0.5 \GeV^{-2}$.}
\label{FigOmegabb}
\end{center}
\end{figure}

Performing a similar DRSR analysis in the $b$-channel, one obtains the results shown in 
Fig.(\ref{FigOmegabb}a). Only the DRSR $r^{sq}_1(bb)$ presents a good $\tau$-behavior. 
Thus, from the Fig.(\ref{FigOmegabb}b), one can extract the value for the mass ratio from the 
$t_c$-stability point:
\begin{equation}
   r^{sq}_1(bb)\simeq 1.0049(7)_{m_b}(3)_{\bar ss}(10)_{m_s}~,
\end{equation}
which corresponds to the following mass-splitting between the $\Omega_{bb}$ and 
$\Xi_{bb}$ baryons:
\begin{equation}
  M_{\Omega_{bb}} - M_{\Xi_{bb}}=49(13) \MeV ~.
  \label{omegabb}
\end{equation}
For above calculation, the value estimated in Ref.\cite{BAGAN3} was used: 
$M_{\Xi_{bb}} \simeq 9.94 \GeV$.\\

\subsection{$\Omega^\ast_{QQ} \:(QQs)$ and $\Xi^\ast_{QQ} \:(QQq)$}
For studying the spin $3/2^+$ baryons, it is necessary to calculate the correlation function 
using the currents (\ref{eq:currOmega*QQ}) and (\ref{eq:currXi*QQ}). 
Then, inserting the current $\eta_{\Omega^\ast_{QQ}}^\mu$ into the correlation function
(\ref{2point}) one obtains:

\begin{eqnarray}
  \Pi^{\Omega^\ast_{QQ}}_{\mu\nu}(q) &=& -\frac{i \:\epsilon_{ijk} \:\epsilon_{i'j'k'}}{3 \cdot 2^8 \pi^8} 
  \int\limits\! d^4x \:d^4p_1 \:d^4p_2 ~e^{ix \cdot (q-p_1-p_2)} \nno \\ && \hspace{-2.0cm}\times \left\{ \:
  \begin{array}{l}
    4 \:{\cal S}^Q_{kk'}(p_2) \:\Tr [ {\cal S}^{s}_{ii'}(x) \:\ga_\nu C\: {\cal S}^{Q \:T}_{jj'}(p_1) \:C \ga_\mu] +
    4 \:{\cal S}^Q_{kk'}(p_2) \:\ga_\nu C\: {\cal S}^{s \:T}_{ii'}(x) \:C \ga_\mu\: {\cal S}^Q_{jj'}(p_1) \\
    \!\!+ 2 \:{\cal S}^Q_{kk'}(p_2) \:\ga_\nu C\: {\cal S}^{Q \:T}_{jj'}(p_1) \:C \ga_\mu\: {\cal S}^s_{ii'}(x)
    + 2 \:{\cal S}^Q_{jj'}(p_1) \:\ga_\nu C\: {\cal S}^{Q \:T}_{kk'}(p_2) \:C \ga_\mu\: {\cal S}^s_{ii'}(x) \\
    \!\!+ 2 \:{\cal S}^s_{ii'}(x) \:\ga_\nu C\: {\cal S}^{Q \:T}_{jj'}(p_1) \:C \ga_\mu\: {\cal S}^Q_{kk'}(p_2)
    + 2 \:{\cal S}^s_{ii'}(x) \:\ga_\nu C\: {\cal S}^{Q \:T}_{kk'}(p_2) \:C \ga_\mu\: {\cal S}^Q_{jj'}(p_1) \\
    \!\!+ \:{\cal S}^s_{ii'}(x) \:\Tr [ {\cal S}^Q_{jj'}(p_1) \:\ga_\nu C\: {\cal S}^{Q \:T}_{kk'}(p_2) \:C \ga_\mu] +
    \:{\cal S}^s_{ii'}(x) \:\Tr [ {\cal S}^Q_{kk'}(p_2) \:\ga_\nu C\: {\cal S}^{Q \:T}_{jj'}(p_1) \:C \ga_\mu] \\
  \end{array}
  \right\} \nno \\ &&
\end{eqnarray}
The analogous expression for the current $\eta_{\Xi^\ast_{QQ}}^\mu$ is obtained from the 
above expression by doing the change of the strange quark propagators with the ones 
for light quark propagators. 
Therefore, the expression used to calculate the spectral densities for $\Omega^\ast_{QQ}$ and 
$\Xi^\ast_{QQ}$ baryons, considering the OPE contributions up to dimension-five condensates.
The density expressions for $\Xi^\ast_{QQ}$ baryons have already been calculated in 
Ref.\cite{BAGAN3}.

\subsubsection{Spectral Densities for $\Omega^\ast_{QQ}$ and $\Xi^\ast_{QQ}$ baryons}
\noindent $F_1$ structure:
\begin{eqnarray}
  \Img F_1^{pert} &=& -\frac{m_Q^4}{5 \cdot 3 \cdot 2^5 \pi^3}\Bigg\{ 
  60(1 \!-\! 4x \!+\! 4x^2 \!+\! 2x^3)\mathcal{L}_v - 
  v \bigg( \frac{3}{x^2} \!-\! \frac{19}{x} \!+\! 98 \!-\! 130x \!-\! 60x^2 \bigg) \nno \\
  && + \frac{20m_s}{m_Q} \bigg[ 6(1- 2 x^2) {\cal L}_v
  - v \bigg( {2\over x} +1 + 6x \bigg) \bigg] \Bigg\} \\
  \Img F_1^{\qq[s]} &=& - \frac{m_Q \qq[s]}{3 \cdot 2^5 \pi} \Bigg\{ 32 v
  - \frac{m_s}{m_Q} \bigg( v(3-4x) + \frac{5}{v} \bigg) \Bigg\} \\
  \Img F_1^{\GGi} &=& -\frac{\GG}{3^2 \cdot 2^7 \pi^3} \Bigg\{
  24x^2(2+x)\mathcal{L}_v + v (1 + 26x +12x^2) \Bigg\} \\ 
  \Img F_1^{\qGq[s]} &=& \frac{\qGq[s]}{3^2 \pi \:m_Q \:v^3} x^2(2-11x)
\end{eqnarray}

\noindent $F_2$ structure:
\begin{eqnarray}
  \Img F_2^{pert} &=& -\frac{m_Q^5}{3^2 \cdot 2^6 \pi^3}\Bigg\{ 
  48 \bigg( \frac{2}{x} - 3 + 5x^2 \bigg) \mathcal{L}_v -
  2v \bigg( \frac{9}{x^2} \!+\! \frac{34}{x} \!-\! 10 \!-\! 60x \bigg) \nno \\
  && + \frac{3m_s}{m_Q} \bigg[ 24(5 - 6x - x^2) {\cal L}_v
  - v \bigg( {3\over x^2} + {10\over x} + 74 + 12x \bigg) \bigg] \Bigg\} \\
  \Img F_2^{\qq[s]} &=& - \frac{m_Q^2 \qq[s]}{3^2 \cdot 2 \pi} \Bigg\{ 
  v \bigg( \frac{2}{x} + 7 \bigg) - \frac{3m_s}{m_Q} (1-2x)\frac{1}{v} \Bigg\} \\
  \Img F_2^{\GGi} &=& -\frac{m_Q \GG}{3^3 \cdot 2^6 \pi^3} \Bigg\{ 
  12(2+3x^2)\mathcal{L}_v - v \bigg( \frac{8}{x} - 11 - 18x \bigg) \Bigg\} \\ 
  \Img F_2^{\qGq[s]} &=& \frac{\qGq[s]}{3^2 \cdot 2^4 \:\pi \:v^3}(2-11x+12x^2-30x^3)
\end{eqnarray}
The spectral densities for $\Xi^\ast_{QQ}$ baryons are obtained from the above expressions 
by doing the changes: $m_s \rightarrow m_q$, $\qq[s] \rightarrow \qq[q]$ and 
$\qGq[s] \rightarrow \qGq[q]$.

\subsubsection{Mass Ratio $\Xi^\ast_{QQ} / \Xi_{QQ}$}
The DRSR approach is largely used to calculate the mass ratios for heavy baryons 
with or without $s$-quark. However, with the probable existence of the $\Xi^\ast_{cc}$ 
baryon, it is highly recommended introducing the DRSR for estimating the spin 
$1/2^+$ and $3/2^+$ baryon mass ratios, for instance, the ratio $\Xi^\ast_{QQ} / \Xi_{QQ}$.
For this, one uses the following DRSR equations:
\begin{equation}
  r^{3/1}_i\equiv \sqrt{{\cal R}^3_i\over {\cal R}^1_i}~:~~i=1,2~~;~~~~~~~~
  r^{3/1}_{21}\equiv {{\cal R}^3_{21}\over {\cal R}^1_{21}}~,
  \label{DR1}
\end{equation}
where the upper indices $3$ and $1$ corresponds to the spin $3/2^+$ and 
$1/2^+$ baryons, respectively.

\begin{figure}[!t]
\begin{center}
{\footnotesize a)}
\epsfig{figure=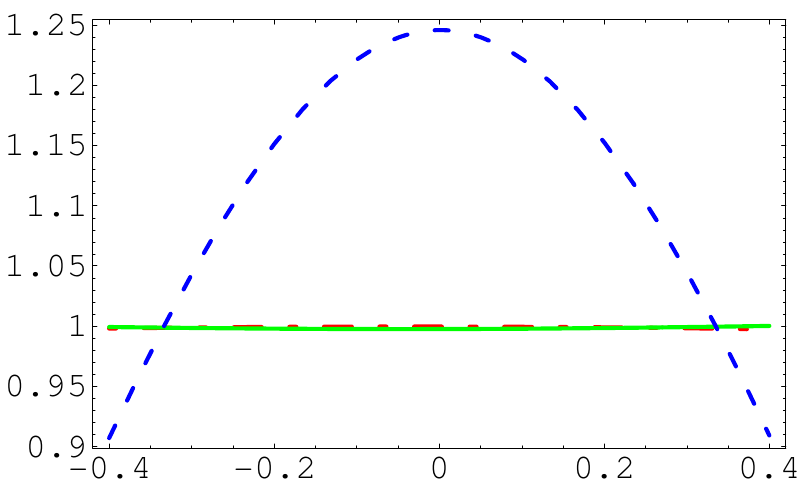,height=40mm}\hfill
{\footnotesize b)}
\epsfig{figure=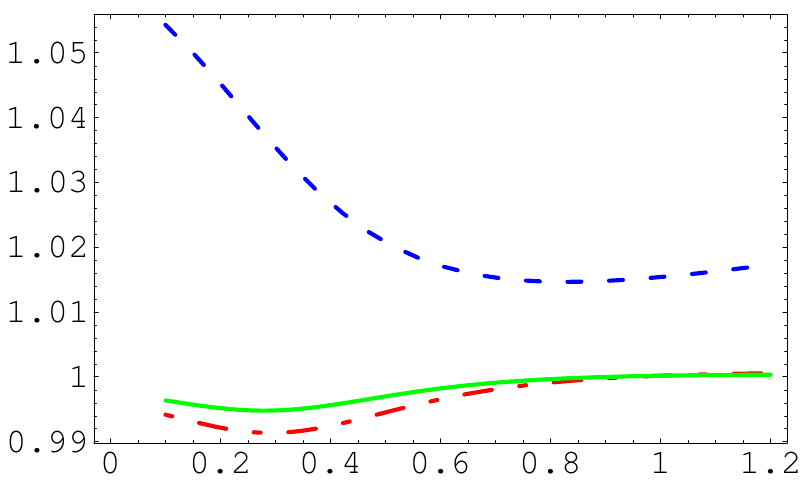,height=40mm}\hfill\\ \vspace{0.5cm}
{\footnotesize c)}
\epsfig{figure=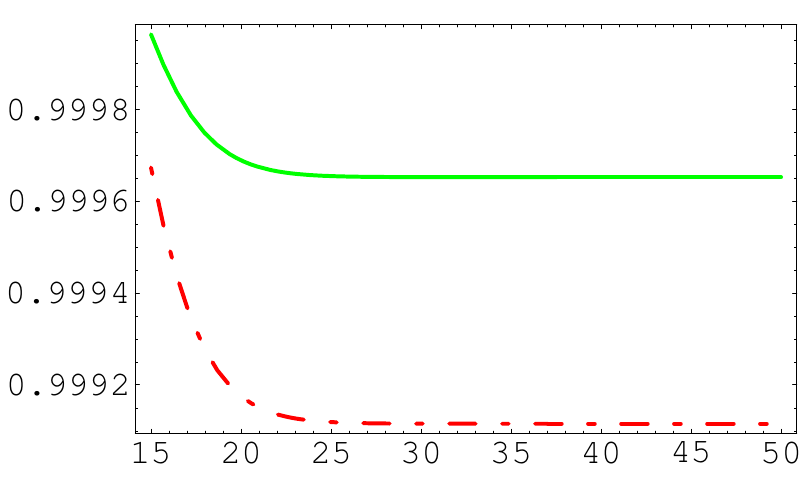,height=50mm}
\caption{\footnotesize DRSR $\Xi^\ast_{cc} / \Xi_{cc}$: 
{\bf a)} $b$-behavior, for $\tau =0.8 \GeV^{-2}$, $t_c =25.0 \GeV^2$ and $m_c = 1.26 \GeV$;
{\bf b)} $\tau$-behavior, for $b =- 0.35$, $t_c =25.0 \GeV^2$ and $m_c = 1.26 \GeV$, 
where $r^{3/1}_{1}$ dot-dashed line (red), $r^{3/1}_{2}$ continuous line (green) and
$r^{3/1}_{21}$ dotted line (blue); 
{\bf c)} the continuum threshold $t_c$-behavior of $r^{3/1}_{2}$ and $r^{3/1}_{21}$, 
for $b=-0.35$ and $\tau = 0.8 \GeV^{-2}$.}
\label{FigXistarcc}
\end{center}
\end{figure}

The $b$-behavior for these DRSR is presented in Fig.(\ref{FigXistarcc}a), by fixing 
$\tau = 0.8 \GeV^{-2}$ and $t_c = 25.0 \GeV^2$, which are inside the $\tau$- and 
$t_c$-stability regions. From this figure, only the $r^{3/1}_1$ and $r^{3/1}_2$ sum rules
present a good $b$-stability. Besides, some common solutions are obtained for
\begin{equation}
  b \simeq -0.35 \hspace{1cm} and \hspace{1cm} b \simeq 0.2 ~.
\end{equation}
For definiteness, the value $b=-0.35$ is fixed and used in the analysis of the 
$t_c$-behavior. The results obtained are presented in Figs.(\ref{FigXistarcc}b) and 
(\ref{FigXistarcc}c). In these figures, one has used the running mass $m_c = 1.26 \GeV$. 
One has also checked that the results are insensitive to the change of the charm mass 
to $m_c = 1.47 \GeV$.
Finally, one can deduce:
\begin{equation}
  \frac{M_{\Xi^\ast_{cc}}}{M_{\Xi_{cc}}} = 0.9994 (3)
\end{equation}
where the uncertainty is the quadratic sum due to $m_c$, $\alpha_s$ and $\GG$.

\begin{figure}[!t]
\begin{center}
{\footnotesize a)}
\epsfig{figure=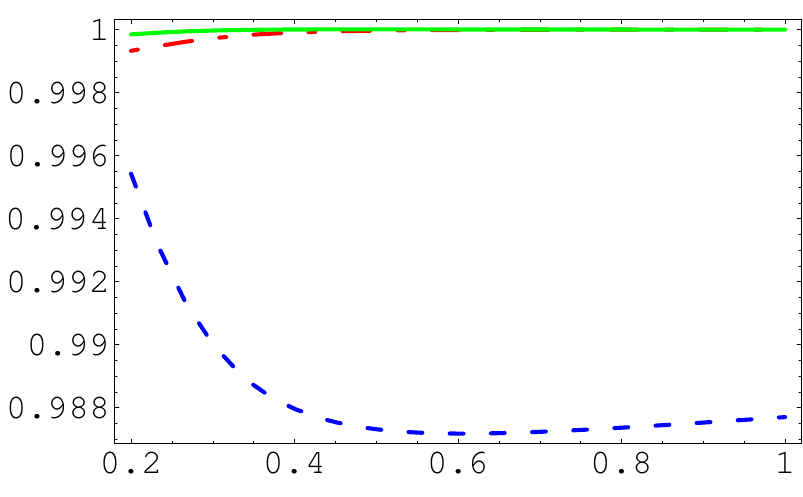,height=40mm}\hfill
{\footnotesize b)}
\epsfig{figure=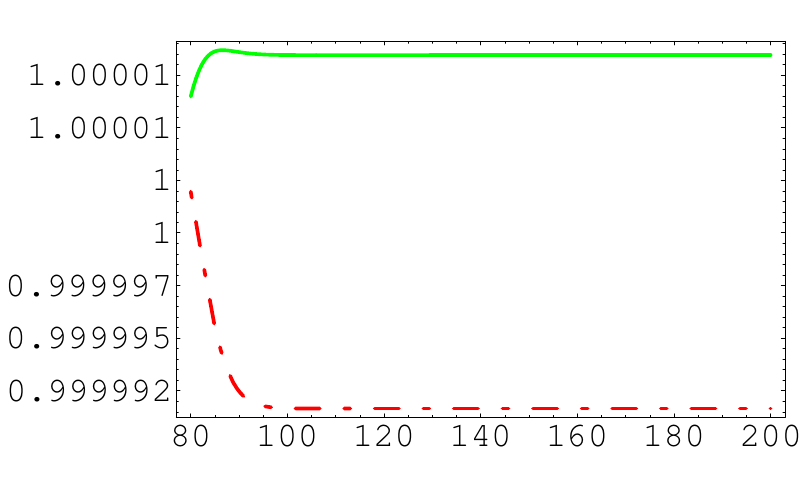,height=42mm}\hfill\\ \vspace{0.5cm}
\caption{\footnotesize DRSR $\Xi^\ast_{bb} / \Xi_{bb}$: 
{\bf a)} $\tau$-behavior, for $b =- 0.35$ and $t_c =100.0 \GeV^2$,  
where $r^{3/1}_{1}$ dot-dashed line (red), $r^{3/1}_{2}$ continuous line (green) and
$r^{3/1}_{21}$ dotted line (blue); 
{\bf b)} the continuum threshold $t_c$-behavior of $r^{3/1}_{1}$ and $r^{3/1}_{2}$, 
for $b=-0.35$ and $\tau = 0.6 \GeV^{-2}$.}
\label{FigXistarbb}
\end{center}
\end{figure}

Extending this analysis to the baryons with two bottom quarks, the correspondig curves are 
qualitatively similar to the charm case. Considering for this case $b=-0.35$, 
one can verify in Fig.(\ref{FigXistarbb}a) that the $\tau$-stability is reached for 
$\tau \geq 0.6 \GeV^{-2}$ and the most trustable DRSR result comes from 
$r^{3/1}_1$ and $r^{3/1}_2$ sum rules. In Fig.(\ref{FigXistarbb}b), both curves present a 
good $t_c$-stability, from one obtains the following result:
\begin{equation}
  \frac{M_{\Xi^\ast_{bb}}}{M_{\Xi_{bb}}} = 1.0000 ~.
\end{equation}

\subsubsection{\boldmath $\al_s$ Corrections}
Radiative corrections due to $\al_s$ are known to be large in the baryon two-point 
correlation functions \cite{KORNER,JAMI2}. However, one can easily inspect that in the simple 
ratios, ${\cal R}^3_i$ and ${\cal R}^1_i$, these huge corrections cancel out, and the 
only remain is the one induced by the anomalous dimension of the baryon operators.
Including the anomalous dimension $\ga = 2$ (resp. -2/3) for the spin $1/2^+$ 
(resp. $3/2^+$) baryons \cite{JAMI2}, one can generically write a crude approximation 
but very informative to the perturbative expressions of the sum rule:
\begin{equation}
  {\cal F}_i(\tau)|_{pert}\approx \left( \alpha_s(\tau) \right)^{-{\ga\over \beta_1} } 
  A_i~\tau^{-3}\Bigg( 1+K_i{\alpha_s\over\pi} \Bigg),
  \label{nlo}  
\end{equation}
where $\beta_1$ is the first coefficient of the $\beta$-function; $A_i$ is a known leading 
order expression; $K_i$ is the radiative correction which is known in some cases of light 
and heavy baryons \cite{KORNER,JAMI2}. From the previous expression in Eq.(\ref{nlo}), 
one can derive a new expression for the DRSR which take into account the effects from 
the radiate corrections (NLO) in the OPE:
\begin{equation}
  r_i^{3/1}|^{NLO}_{pert}\simeq  r_i^{3/1}|^{LO}_{pert}\times \Bigg{[}
  1+{2\over 9}{\alpha_s\over\pi}+{\cal O}\left( \alpha_s^2,  m^2_{_Q}\tau\right) \Bigg{]}~.
  \end{equation}
\vspace*{-0.5cm}

\begin{figure}[!t]
\begin{center}
{\footnotesize a)}
\epsfig{figure=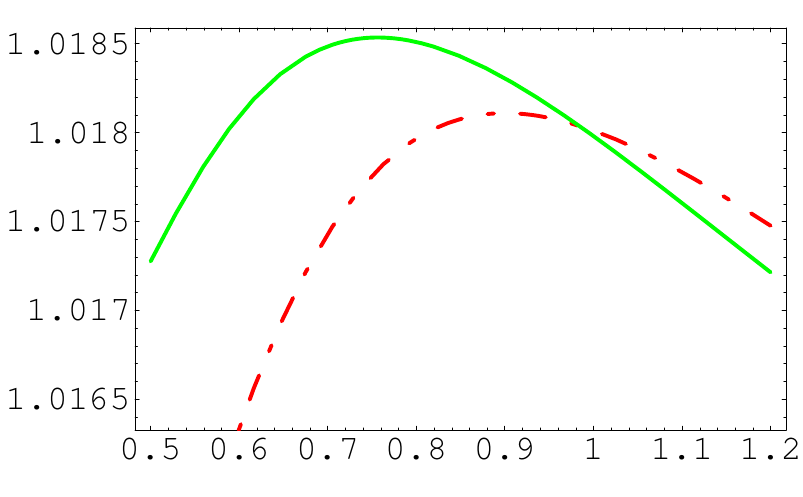,height=40mm}\hfill
{\footnotesize b)}
\epsfig{figure=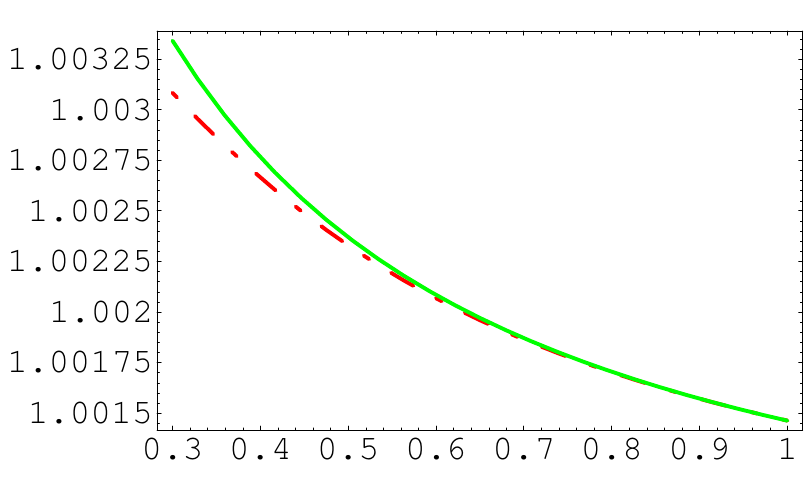,height=42mm}\hfill\\ \vspace{0.5cm}
\caption{\footnotesize NLO DRSR: {\bf a)} $\Xi^\ast_{cc} / \Xi_{cc}$ as a function of $\tau$, for 
$b =- 0.35$ and $t_c =25.0 \GeV^2$, where $r^{3/1}_{1}$ dot-dashed line (red) and 
$r^{3/1}_{2}$ continuous line (green);
{\bf b)} $\Xi^\ast_{bb} / \Xi_{bb}$ as a function of $\tau$, for $b=-0.35$ and 
$t_c = 100.0 \GeV^2$.}
\label{FigXistarQQ}
\end{center}
\end{figure}

It is important to notice for $r_i^{3/1}$ that the radiative correction has been only 
induced by the ones due to the anomalous dimensions, while the one due to $K_i$ 
cancels out to this order. This is not the case of $r_{21}^{3/1}$ where the radiative 
correction is only due to $K_2-K_1$ and needs to be evaluated which is beyond the 
aim of this work. Therefore, in the following, one only considers the results from 
$r_i^{3/1}$. The $\tau$-dependence of the DRSR is shown in Fig.(\ref{FigXistarQQ}). 
One shall take the range of $\tau$-values where the leading order terms have $\tau$-stability, 
which is $(0.7 \leq \tau \leq 1.0) \GeV^{-2}$ for charm and $(0.5 \leq \tau \leq 0.8) \GeV^{-2}$ 
for bottom. One can also notice that the NLO DRSR for charm presents a $\tau$-extremum 
in the above range rendering its prediction more reliable than for the bottom channel 
case. Thus, one can deduce:
\begin{equation}
  {M_{\Xi^*_{cc}}\over M_{\Xi_{cc}}}= 1.0167(10)_{\alpha_s}(16)_{m_c},~~~
  {M_{\Xi^*_{bb }}\over M_{\Xi_{bb}}}=1.0019(3)_{\alpha_s}(2)_{m_b} ~.
\end{equation}
These ratios would correspond to the mass-splittings:
\begin{eqnarray}
  {M_{\Xi^*_{cc }}- M_{\Xi_{cc}}}&=& 59(7) \MeV \\
  {M_{\Xi^*_{bb }}- M_{\Xi_{bb}}} &=& 19(3) \MeV~,
\label{chibcmass}
\end{eqnarray}
where the experimental value $M_{\Xi_{cc}} = 3518.9 \pm 0.9 \MeV$ is considered. 
For the evaluation of $\Xi_{bb}$ mass, one uses the estimated value in Ref.\cite{BAGAN3}: 
$M_{\Xi_{bb}} = 9.94 \GeV$.
The mass-splitting between $\Xi^*_{cc}$ and $\Xi_{cc}$ baryons is comparable with the 
one of about $70 \MeV$ from potential models \cite{RICHARD, Dosch}, but larger than the 
one of about $24 \MeV$ obtained in \cite{vijande}. The mass-splitting between $\Xi^*_{bb}$ 
and $\Xi_{bb}$ baryons also agrees with potential models \cite{RICHARD, Dosch}.
Another noteworthy observation is that the mass-splittings (\ref{chibcmass}) 
do not favor a hadronic decay channel like $\Xi^*_{QQ} \rightarrow \Xi_{QQ} +\pi's$ 
since there is not enough energy to produce pion pairs. Therefore, from a DRSR 
point of view, $\Xi^\ast_{QQ}$ baryon can only decay predominantly in an 
electromagnetically channel
\begin{eqnarray}
  \Xi^*_{QQ} &\rightarrow& \Xi_{QQ} + \gamma ~.
\end{eqnarray}
A future discovery of ${\Xi^*_{cc}}$ and ${\Xi^*_{bb}}$ baryons could support  
(or not) all of these predictions, including the existence of such decay channels.

Notice that, with the results obtained in Eq.(\ref{chibcmass}), it is also possible to 
estimate the $\Xi^\ast_{cc}$ and $\Xi^\ast_{bb}$ masses as
\begin{eqnarray}
  M_{\Xi^\ast_{cc}} \simeq 3.58 \GeV ~~~& \mbox{and} & ~~~  
  M_{\Xi^\ast_{bb}} \simeq 9.96 \GeV ~.  
\end{eqnarray}

\begin{figure}[t]
\begin{center}
{\footnotesize a)}
\epsfig{figure=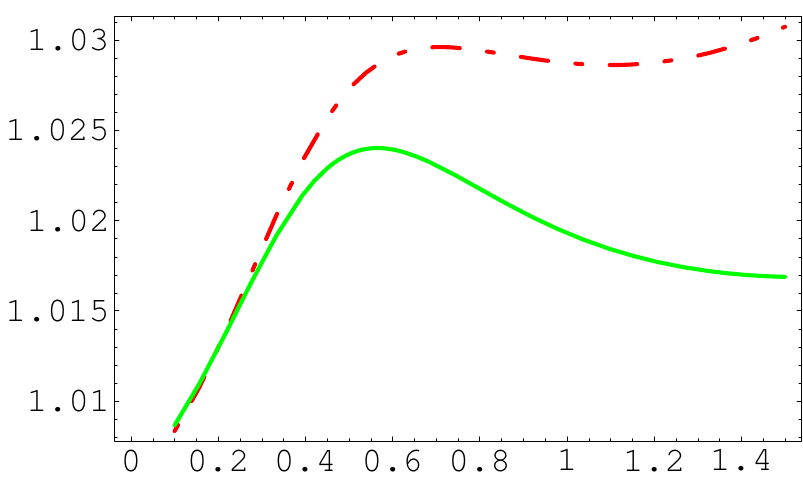,height=40mm}\hfill
{\footnotesize b)}
\epsfig{figure=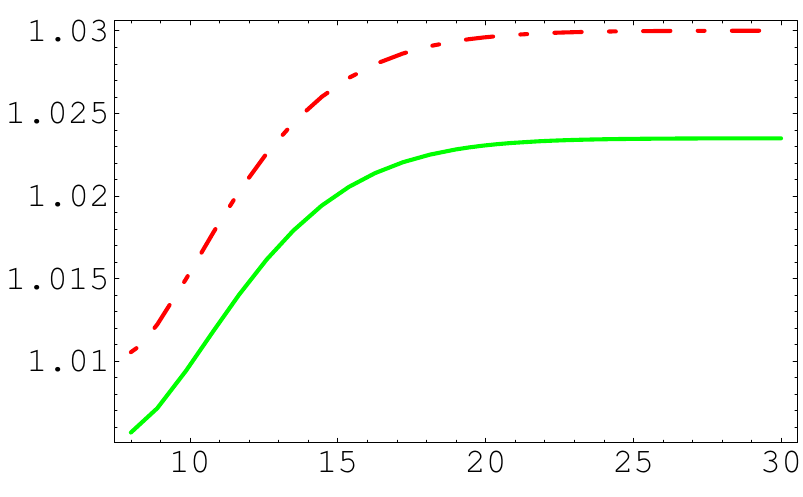,height=43mm}\hfill\\ \vspace{0.5cm}
\caption{\footnotesize DRSR $\Omega^\ast_{cc} / \Xi^\ast_{cc}$: {\bf a)} $\tau$-behavior, for
$t_c =25.0 \GeV^2$, where $r^{sq}_{1}(cc)^\ast$ dot-dashed (red) and $r^{sq}_{2}(cc)^\ast$ 
continuous (green);
{\bf b)} the continuum threshold $t_c$-behavior, for $\tau = 0.7 \GeV^{-2}$.}
\label{FigOmegastarcc}
\end{center}
\end{figure}

\subsubsection{Mass Ratio $\Omega^\ast_{QQ} / \Xi^\ast_{QQ}$}
Calculating the DRSR for $\Omega^\ast_{cc}$ and $\Xi^\ast_{cc}$ baryons, one obtains
the results presented in Fig.(\ref{FigOmegastarcc}). Both sum rules, $r^{sq}_1$ and 
$r^{sq}_2$, have a good $\tau$- and $t_c$-stability. These stabilities could be observed 
at the point $\tau \simeq 0.7 \GeV^{-2}$, in Fig.(\ref{FigOmegastarcc}a), and when 
$t_c \geq 20 \GeV^2$ in Fig.(\ref{FigOmegastarcc}b). Then, considering the mean value from 
the $r^{sq}_1$ and $r^{sq}_2$ sum rules inside the stability regions, one can deduce:
\begin{equation}
   r^{sq}(cc)^*\equiv {M_{\Omega^*_{cc}}\over M_{\Xi^*_{cc}}} = 
   1.026(4)_{\qq[s]}(4)_{m_s}(6)_{m_c}(1)_{t_c}~,
\end{equation}
which allows to estimate the mass-splitting between $\Omega^\ast_{cc}$ and $\Xi^\ast_{cc}$ 
baryons, as follows:
\begin{equation}
  M_{\Omega^*_{cc}} - M_{\Xi^*_{cc}}= 94(27) \MeV~,
  \label{omegastarcc}
\end{equation}
where the experimental data is given by \cite{pdg}:
$M_{\Xi^\ast_{cc}} = 3518.9 \pm 0.9 \MeV$.

\begin{figure}[!t]
\begin{center}
{\footnotesize a)}
\epsfig{figure=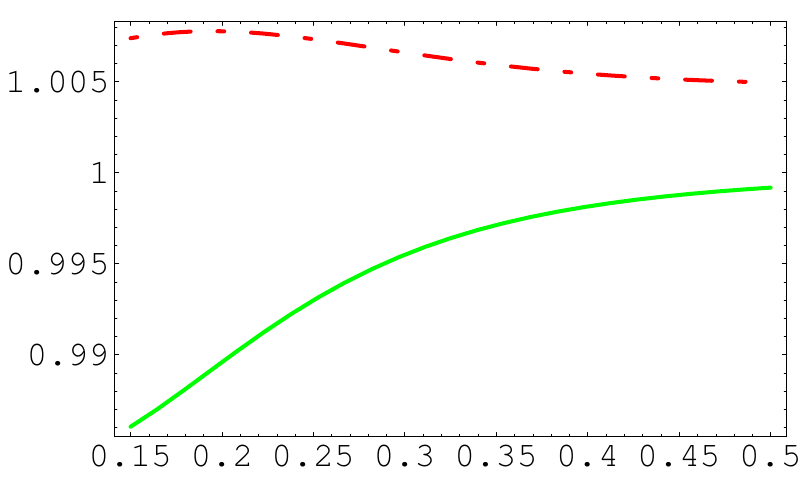,height=40mm}\hfill
{\footnotesize b)}
\epsfig{figure=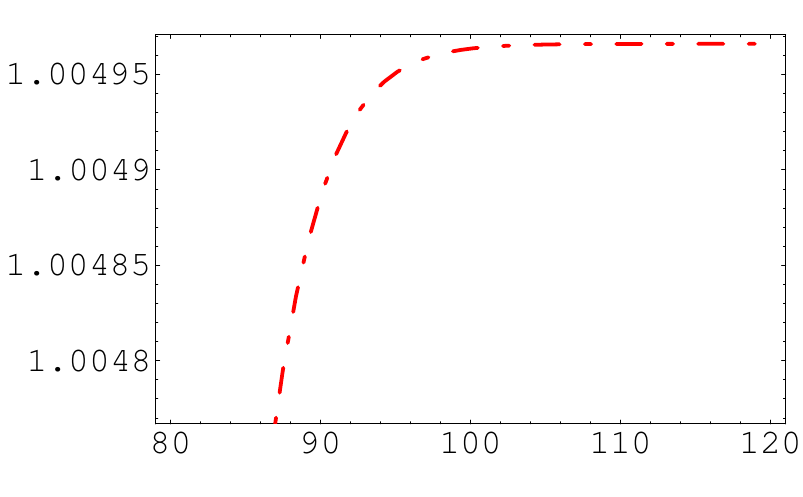,height=43mm}\hfill\\ \vspace{0.5cm}
\caption{\footnotesize DRSR $\Omega^\ast_{bb} / \Xi^\ast_{bb}$: {\bf a)} $\tau$-behavior, for
$t_c =100.0 \GeV^2$, where $r^{sq}_{1}(bb)^\ast$ dot-dashed (red) and $r^{sq}_{2}(bb)^\ast$ 
continuous (green);
{\bf b)} the continuum threshold $t_c$-behavior, for $\tau = 0.5 \GeV^{-2}$.}
\label{FigOmegastarbb}
\end{center}
\end{figure}

The analogous analysis for the spin $3/2^+$ baryons with two bottom quarks is presented in 
Fig.(\ref{FigOmegastarbb}). Considering the stability criteria, one can deduce:
\begin{equation}
   r^{sq}(bb)^*\equiv {M_{\Omega^*_{bb}}\over M_{\Xi^*_{bb}}} = 
   1.0050(3)_{\qq[s]}(10)_{m_s}(4)_{\tau}(10)_{m_b}~.
\end{equation}
Using the value estimated in the previous section: $M_{\Xi^\ast_{bb}} \simeq 9.96 \GeV$, 
it is possible to estimate the mass-splitting
\begin{equation}
  M_{\Omega^*_{bb}} - M_{\Xi^*_{bb}} = 50(15) \MeV ~.
  \label{omegastarbb}
\end{equation}
This value agrees with the one from the potential model \cite{Dosch}:
\begin{equation}
  M_{\Omega^*_{bb}}- M_{\Xi^*_{bb}} \:\vert_{PM} \simeq 60 \MeV ~.
\end{equation}
With the results obtained in Eqs.(\ref{omegastarcc}) and (\ref{omegastarbb}), it is 
also possible to estimate the $\Omega^\ast_{cc}$ and $\Omega^\ast_{bb}$ masses:
\vspace{-0.5cm}
\begin{eqnarray}
  M_{\Omega^\ast_{cc}} \simeq 3.61 \GeV ~~~& \mbox{and} & ~~~  
  M_{\Omega^\ast_{bb}} \simeq 10.01 \GeV ~.  
\end{eqnarray}
The mass-splittings given by Eqs.(\ref{omegastarcc}) and (\ref{omegastarbb}) 
are summarized in Table (\ref{TabQQq}).

\subsection{$\Omega_{bc} \:(bcs)$ and $\Xi_{bc} \:(bcq)$}
It is also possible to study the baryons with two different heavy quarks. The 
$\Xi_{bc} \:(bcq)$ and $\Omega_{bc} \:(bcs)$ spin $1/2^+$ baryons can be 
described by the corresponding currents:

\begin{eqnarray}
  J_{\Omega_{bc}} &=& \epsilon_{ijk} \left[(c_i^T C\gamma_5 s_j)+
  \beta (c_i^T C s_j) \gamma_5 \right] b_k ~,\\
  J_{\Xi_{bc}}&=&J_{\Omega_{bc}}  ~~~(s \rightarrow q)~,
\end{eqnarray}
where $q$ and $s$ are the light quark fields, $c$ and $b$ are the heavy quark fields and 
$\beta$ is an arbitrary mixing parameter. The expression of the corresponding two-point 
correlation function has been obtained in the chiral limit, $m_q = m_s = 0$, in 
Refs.\cite{BAGAN3, Dosch}. 
In the present work, one has checked all of these expressions and calculated the ones 
related to the linear terms in the strange quark mass for the perturbative and quark 
condensate contributions.

\subsubsection{Spectral Densities for $\Omega_{bc}$ and $\Xi_{bc}$ Baryons}
\noindent $F_1$ structure:
\begin{eqnarray}
  \Img F_1^{pert} &=& \frac{(1+\beta^2)}{2^9 \pi^3 \:s^2}\Bigg\{ 
  6 \Big[ (m_c^4 + m_b^4) s^2 - 2m_c^4 m_b^4 \Big] \mathcal{L}_v +
  12 (m_c^4 - m_b^4)s^2 \:\mathcal{L}_m \nno \\ 
  && +~ \omega_{bc} \Big[s^3 -7(m_c^2+m_b^2)s^2 
  - (7m_c^4+7m_b^4-12m_b^2m_c^2) s \nno\\
  && -~ 7 m_b^2m_c^2(m_c^2+m_b^2) + m_c^6+m_b^6 \Big] \Bigg\}
  -\frac{m_s m_c}{2^7 \:\pi^3 \:s^2} (1 \!-\! \beta^2) \Bigg\{ 3m_c^2 ( s^2 - 2 m_b^4) \:{\cal L}_v ~\nno \\
  && +~ 6 m_c^2 s^2 \:{\cal L}_m - \omega_{bc}  \Big[ 2s^2 + 
  (5m_c^2 - 4m_b^2) s - m_c^4 + 5m_b^2 m_c^2 + 2m_ b^4 \Big] \Bigg\}\\
  \nno\\
  \Img F_1^{\qq[s]} &=& \frac{m_c \qq[s]}{2^4 \pi \:s^2}
  \:(1-\beta^2) \:\omega_{bc} (m_c^2 - m_b^2- s) + \frac{m_s \qq[s]}{2^5 \:\pi \:s^3} 
  \:(1+\beta^2) \nno\\
  && \times~ \Bigg\{ \frac{2}{\omega_{bc}}  \Big[ (m_c^4+m_b^4) s^2 - 
  2(m_c^2+m_b^2)(m_c^2-m_b^2)^2 s + (m_c^2 - m_b^2)^4 \Big] \nno\\
  && +~ \omega_{bc} \Big[ s^2 + (m_c^2 + m_b^2) s - 2(m_c^2-m_b^2)^2 \Big]  \Bigg\}
\end{eqnarray}

\begin{eqnarray}
  \Img F_1^{\GGi} &=& -\frac{\GG}{3 \cdot 2^{10} \pi^3 \:s^2} \:(1+\beta^2)\:\omega_{bc} 
  \Big( s + 3m_c^2 -3 m_b^2 \Big) \\
  \nno\\  
  \Img F_1^{\qGq[s]} &=& -\frac{ m_c \qGq[s]}{2^5 \pi \:s^2} 
  \:\frac{(1-\beta^2)}{\omega_{bc}^3} \Big[ s^4 - (5m_c^2+2m_b^2) s^3 + 
  (9m_c^4+ 3m_c^2m_b^2 + 2m_b^4) s^2 \nno\\
  && -~ (m_c^2-m_b^2)(7m_c^4+ m_c^2m_b^2 - 2m_b^4)s + 
  (m_c^2-m_b^2)^3(2m_c^2-m_b^2) \Big]
\end{eqnarray}

\noindent $F_2$ structure:
\begin{eqnarray}
  \Img F_2^{pert} &=& -\frac{m_b}{2^7 \pi^3 \:s} (1-\beta^2)\Bigg\{ 
  3 \Big[ m_b^2 s^2 - (m_c^4 - m_b^4 + 2m_c^2 m_b^2) s + 
  2m_c^4 m_b^2 \Big] \mathcal{L}_v \nno \\
  && -~ 6(m_b^2 s + m_c^4 - 2m_c^2 m_b^2 + m_b^4)s \: \mathcal{L}_m 
  -~ \omega_{bc} \Big[s^2 - 5(m_c^2 - 2m_b^2)s \nno \\
  && -~ 2m_c^4 - 5m_c^2m_b^2 + m_b^4 \Big]  
  - \frac{3 m_s m_c m_b}{2^6 \:\pi^3 \:s} \:(1+\beta^2) \Bigg\{ \Big[ (m_c^2+m_b^2) s \nno\\
  && -~ 2m_c^2 m_b^2 \Big] {\cal L}_v + 2(m_c^2 - m_b^2) s \:{\cal L}_m 
  - \omega_{bc}(s + m_ c^2 + m_ b^2) \Bigg\} \\
  \nno\\  
  \Img F_2^{\qq[s]} &=& -\frac{m_c m_b \qq[s]}{2^3 \pi \:s} \:(1+\beta^2)
  \: \omega_{bc} + \frac{m_s m_b \qq[s]}{2^4 \:\pi s^2} \:(1-\beta^2) \Bigg\{ 
  \omega_{bc} (m_c^2 - m_b^2 + s) \nno\\
  && +~ \frac{1}{\omega_{bc}} \Big[ m_b^2 s^2+ ( m_c^4+m_b^2 m_c^2-2m_b^4) s 
  - ( m_c^2 - m_b^2)^3 \Big]  \Bigg\} \\
  \nno\\  
  \Img F_2^{\GGi} &=& -\frac{\GG \:(1 \!-\! \beta^2)}{3 \cdot 2^{9} \pi^3 \:m_b \:s} \Bigg\{ 
  \omega_{bc} (2 s \!-\! 2m_c^2 \!+\! 5m_b^2) - 6m_b^2 s ({\cal L}_v - {\cal L}_m) \Bigg\} \\
  \nno\\  
  \Img F_2^{\qGq[s]} &=& -\frac{ m_c m_b \qGq[s]}{2^5 \pi \:s} 
  \:\frac{(1+\beta^2)}{\omega_{bc}^3} \Bigg\{ s^3 - (3m_c^2+m_b^2) s^2 
  + (m_c^2 + m_b^2)(3m_c^2 - m_b^2) s ~~~\nno\\
  && -~ (m_c^2 - m_b^2)^3 \Bigg\} 
\end{eqnarray}
where the new definitions are introduced through

\begin{eqnarray}
  \lambda_{bc} &=& 1+ (m_c^2 - m_b^2)/s \\
  \omega_{bc} &=& (s + m_c^2 - m_b^2) v_{bc} \\ 
  v_{bc} &=& \sqrt{1- \frac{4m_c^2 /s}{\lambda_{bc}^2}} \\
  {\cal L}_v &=& \Log{\bigg( \frac{1+v_{bc}}{1-v_{bc}} \bigg)} \\
  {\cal L}_m &=& \Log{\bigg[ \frac{(m_c^2 + m_b^2) s - 
    (m_c^2 - m_b^2)(\omega_{bc} + m_c^2 - m_b^2)}{2 \:m_c \:m_b \:s} \bigg]} ~.
\end{eqnarray}

\begin{figure}[!t]
\begin{center}
{\footnotesize a)}
\epsfig{figure=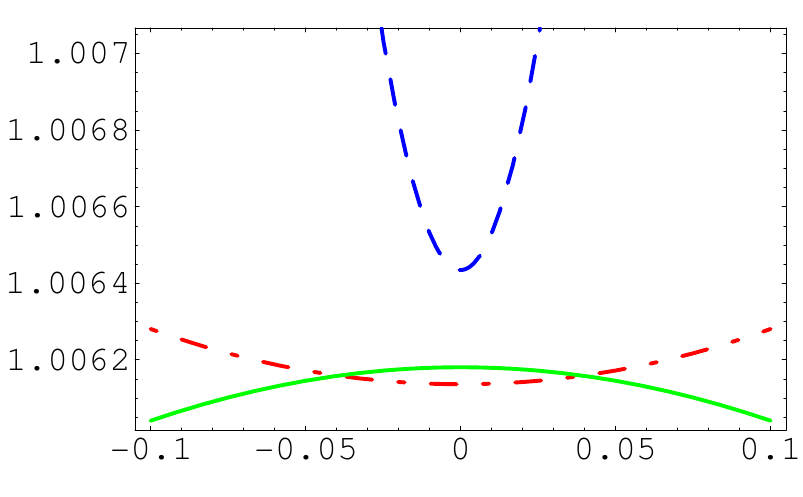,height=40mm}\hfill
{\footnotesize b)}
\epsfig{figure=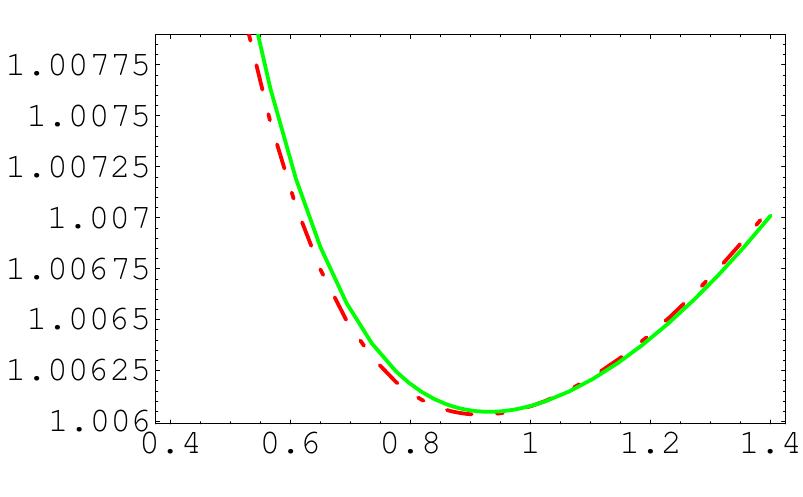,height=40mm}\hfill\\ \vspace{0.5cm}
{\footnotesize c)}
\epsfig{figure=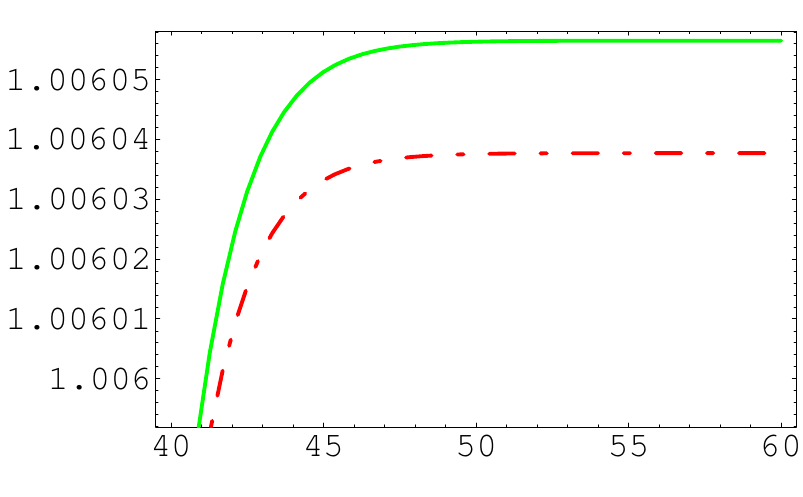,height=50mm}
\caption{\footnotesize DRSR $\Omega_{bc} / \Xi_{bc}$: 
{\bf a)} $\beta$-behavior, for $\tau =0.8 \GeV^{-2}$ and $t_c = 50.0 \GeV^2$;
{\bf b)} $\tau$-behavior, for $\beta =- 0.05$ and $t_c =50.0 \GeV^2$, 
where $r^{sq}_{1}(bc)$ dot-dashed line (red), $r^{sq}_{2}(bc)$ continuous line (green) 
and $r^{sq}_{21}(bc)$ dotted line (blue); 
{\bf c)} the continuum threshold $t_c$-behavior of $r^{sq}_1(bc)$ and $r^{sq}_2(bc)$ , 
for $\beta=-0.05$ and $\tau = 0.9 \GeV^{-2}$.}
\label{FigOmegabc}
\end{center}
\end{figure}

Like in previous sections, one studies the differents DRSR and the result are presented in 
Fig.(\ref{FigOmegabc}). As one can see in Fig.(\ref{FigOmegabc}a), $r^{sq}_1(bc)$ and 
$r^{sq}_2(bc)$ are quite stable in $\beta$ and present common solutions when
\begin{equation}
	\beta = \pm 0.05~.
\end{equation}
Since $r^{sq}_{21}(bc)$ does not intersect with the other ones DRSR, its results will not be 
considered hereafter. In Figs.(\ref{FigOmegabc}b) and (\ref{FigOmegabc}c), both DRSR 
present excellent $\tau$- and $t_c$-stabilities. From these figures, it is possible to obtain 
the result:
\begin{equation}
 r^{sq}(bc)\equiv {M_{\Omega_{bc}}\over M_{\Xi_{bc}}} = 
   1.006(0.2)_{\bar ss}(1.4)_{m_s}(1)_{m_Q}~,  
\end{equation}
where the uncertainties, which come from the other parameters, are negligible. 
This ratio implies
\begin{equation}
 M_{\Omega_{bc}}- M_{\Xi_{bc}}= 41(7) ~{\rm MeV}~,
\label{omegabc}
\end{equation}
where one has used the QCDSR central value for the $\Xi_{bc}$ baryon: 
$M_{\Xi_{bc}} \simeq 6.86 \GeV$. 

Finally, all the results and predictions for the Doubly Heavy Baryons masses and splittings 
are summarized in Table (\ref{TabQQq}).\\

{\scriptsize
\begin{table}[hbt]
  \begin{center}
  \begin{minipage}[c]{13cm}
  \setlength{\tabcolsep}{0.45pc}
  \caption{\small DRSR predictions for the Doubly Heavy Baryon masses and splittings.
  These results are compared with the ones obtained from Potential Models (PM)\cite{Dosch,BRAC}.}
  \begin{tabular}{lllr}
  &\\
  \hline \hline
  DRSR & Input (\GeV) & Hyperfine & PM (\MeV)   \\
  && Splittings (\MeV)& \\
  \hline \hline
  ${{\Xi^*_{cc}}/ {\Xi_{cc }}}= 1.0167(19)$&${\Xi_{cc }}=3.52$\cite{pdg}& ${{\Xi^*_{cc }}- {\Xi_{cc}}}=59(7)$&70-93\\
  ${{\Xi^*_{bb }}/ {\Xi_{bb}}}=1.0019(3)$&${\Xi_{bb}}=9.94$\cite{BAGAN}&${{\Xi^*_{bb }}- {\Xi_{bb}}}= 19(3)$&30-38\\
  ${{\Omega_{cc}}/ {\Xi_{cc}}}=1.0260(70)$&${\Xi_{cc }}=3.52$\cite{pdg}& ${\Omega_{cc}}- {\Xi_{cc}}= 92(24) $&90-102\\
  ${\Omega_{bb}}/ {\Xi_{bb}}=1.0049(13)$&${\Xi_{bb}}=9.94$\cite{BAGAN}&${\Omega_{bb}}- {\Xi_{bb}}= 49(13)$&60-73\\
  ${{\Omega^*_{cc}}/ {\Xi^*_{cc}}}=1.0260(75)$&${\Xi^*_{cc }}=3.58$&$ {\Omega^*_{cc}}- {\Xi^*_{cc}}= 94(27)$&91-100\\
  ${{\Omega^*_{bb}}/ {\Xi^*_{bb}}}=1.0050(15)$&${\Xi^*_{bb}}=9.96$&${\Omega^*_{bb}}- {\Xi^*_{bb}}= 50(15)$&60-72\\
  ${{\Omega_{bc}}/ {\Xi_{bc}}}=1.0060(17)$&${\Xi_{bc}}=6.86$\cite{Dosch}&${\Omega_{bc}}- {\Xi_{bc}}= 41(7)$&70-89\\
  \hline \hline
  \label{TabQQq}
  \end{tabular}
  \end{minipage}
  \end{center}
\end{table}}

\cleardoublepage
\chapter{Conclusions}
In the present work, the QCD Sum Rules approach has been used to evaluate the masses of 
molecular states that could be associated with charmonium and bottomonium 
exotic states. One also estimated the heavy baryon masses in QCD, which have not yet been observed 
experimentally.

The studies on $\DsxDsx ~(0^{++})$ and $\DxDx ~(0^{++})$ molecular currents indicate
that the structure for the $Y(4140)$ state, observed by CDF collaboration in the decay 
$B^+ \rightarrow Y(4140) \:K^+ \rightarrow J/\psi \:\phi \:K^+$, could be quite well described 
by both currents since the obtained masses are given by 
$M_{\DsxDsx} = (4.19 \pm 0.13) \GeV$ and $M_{\DxDx} = (4.15 \pm 0.14) \GeV$, which are 
in a good agreement with the experimental mass for the $Y(4140)$ state. From this result, a possible 
interpretation would be that $Y(4140)$ state can be a mixture of these two molecular states. 
Although the authors in Ref.\cite{liu1} interpreted the $Y(3930)$ state as $\DxDx ~(0^{++})$ 
molecular scalar state, the results presented herein do not support this interpretation.

The QCDSR calculation for the $\DsxDso ~(1^{-+})$ and $\DxDo ~(1^{-+})$ molecular currents 
indicates that the narrow structure $X(4350)$, observed by Belle collaboration in the decay channel
$\gamma\gamma \rightarrow X(4350) \rightarrow J/\psi \:\phi$, cannot be described by any of 
these molecular currents since the obtained masses for the respective molecular states are given by:
$M_{\DsxDso} = (5.05 \pm 0.15) \GeV$ and $M_{\DxDo} = (4.92 \pm 0.12) \GeV$.

In chapter \ref{chap:QCDSR}, one explains the sum rule techniques used in the present 
work: QCDSR, FESR and DRSR. 

The QCDSR approach is the most usual for calculating the two-point correlation 
function and introduces the so-called Borel window: the region which establishes the pole 
dominance, OPE convergence and $\tau$-stability. With all these conditions satisfied, the results 
obtained from a QCDSR calculation are reliable. The FESR approach is constructed in such a way 
that it presents a concise relation between the low-lying hadronic mass and the continuum 
threshold $t_c$. 
Thus, it was proposed the study of the molecular currents $\DsxDso ~(1^{--})$, $\DxDo ~(1^{--})$, 
$\DsoDso ~(0^{++})$ and $\DoDo ~(0^{++})$, combining the results obtained from both QCDSR 
and FESR approaches. This technique leads to the following conclusion:
\begin{itemize}
\item obtaining the masses around the intersection point of the QCDSR and FESR, for all the 
molecular states, establish more robust criteria in defining the continuum threshold in 
sum rules. In current scientific literature, the criteria for fixing this parameter are often based 
on the ones used for the conventional charmonium systems, typically given by 
$\sqrt{t_c} \simeq M_{J/\psi} + 0.5 \GeV$. 
The present analysis for the $1^{--}$ states indicates values around $250-300 \MeV$ for the 
mass-splittings between the lowest ground state and the first radial excitation. Notice that these 
mass-splittings are roughly approximated by the value of the continuum threshold $\sqrt{t_c}$ 
at which the QCDSR and FESR match. 
Therefore, the ad-hoc choice for the continuum threshold $\sqrt{t_c}$ as 
$500 \MeV$ above the ground state mass could not be the most indicated for the exotic states.
\item the $0^{++}$ molecular scalar states:  $\DsoDso$, $\DoDo$, $\BsoBso$ and $\BoBo$ 
presented a mass $\sim 550 \MeV$ bigger than one obtained for the respective $1^{--}$ 
molecular vector states: $\DsxDso$, $\DxDo$, $\BsxBso$ and $\BxBo$. For the charmonium case, 
the mass observed for the $\chi_{c0} ~(0^{++})$ scalar state is bigger than the mass of the 
$J/\psi ~(1^{--})$.
\item according to this technique, none of the molecules $\DsxDso ~(1^{--})$ or 
$\DxDo ~(1^{--})$ is a good candidate to explain the charmonium exotic states $Y(4260)$, 
$Y(4360)$ and $Y(4660)$ since the obtained masses are much bigger than the 
experimental data.
\end{itemize}

In an attempt to explain at least the $Y(4260)$ exotic state, observed in the invariant mass 
spectrum of $J/\psi \:\pi\pi$, one has proposed the following molecular currents $J/\psi \:f_0(980)$ 
and $J/\psi \:\sigma(600)$ based on in a color singlet and octet configurations.
The obtained masses for both currents in a color singlet configuration are very similar, and 
the difference is only $\sim 30 \MeV$: 
$M_{J/\psi \:f_0} = (4.67 \pm 0.12) \GeV$ and $M_{J/\psi \:\sigma} = (4.65 \pm 0.13) \GeV$. 
For the molecular states in a color octet configuration, the obtained masses are $\sim 400 \MeV$ 
bigger than ones for color singlet currents. Therefore, one concludes that the mass of the 
molecular states in a color singlet configuration are very close to the mass of the 
$Y(4660)$ state. This result is not compatible with the proposition done in Ref.\cite{oset}, where 
the authors conclude that the $Y(4260)$ state could be described by the $J/\psi \:f_0(980)$ 
molecular state. Notice that the obtained mass in the present work, for the $J/\psi \:f_0(980)$ 
molecular state, is largely above the meson-meson threshold 
$E_{th} \left[J/\psi \:f_0(980) \right] \simeq 4.09 \GeV$ and, therefore, such molecule would not be bound. 
However, one has to consider that the current in Eq.(\ref{eq:SngJpsif0}) is written in terms of 
currents that couples not only with the ground state of $J/\psi$ and $f_0(980)$ mesons, 
but it also couples with all excited states with the $J^{PC} = 1^{--}$ quantum numbers. 
Then, an interesting interpretation becomes possible from QCDSR calculations: 
the current (\ref{eq:SngJpsif0}) could describe the ground state of a molecule bounded by 
the first excited state of $J/\psi$, the meson $\psi^\prime$, with the $f_0(980)$ meson. Therefore, 
probably the mass obtained in Eq.(\ref{eq:MassSngJpsif0}) corresponds to the 
$\psi^\prime \:f_0(980)$ molecular state.
The interpretation of the $Y(4660)$ as a $\psi^\prime \:f_0(980)$ molecular state was first 
proposed in Ref.\cite{ghm}. This molecule is also in agreement with the dominant decay 
channel of the $Y(4660)$ state: $Y(4660) \rightarrow \psi^\prime \:\pi^+ \pi^-$.

The results for the $\Upsilon \:f_0(980)$ and $\Upsilon \:\sigma(600)$ molecular states are 
compatible with the mass of $Y_b(10890)$ state, observed recently by Belle Collaboration. 
However, analyzing the following meson-meson thresholds: 
\begin{eqnarray*}
	E_{th}[\Upsilon(1S) \:f_0(980)] &\simeq& 10.44 \GeV ~~\mbox{ and }~~
	E_{th}[\Upsilon^\prime \:f_0(980)] \simeq 11.00 \GeV ~,
\end{eqnarray*}
and considering that the thresholds, which contain the $\sigma$ meson instead of $f_0$, are 
approximately $380 \MeV$ below these values. Then, the unique possible interpretation for the 
$Y_b(10890)$ as a molecular bound state, among the molecular currents studied in the 
present work, is that it could be only described by the $\Upsilon^\prime \:f_0(980)$ molecular state.

Only a few experimental data are available for the spectrum of $B_c ~(b \bar{c})$ mesons. 
Therefore, it was proposed the existence of exotic states in this mass region that could be 
observed in the near future experiments. These states would be described by molecular states 
composed of $D$ and $B$ mesons. Thus, the QCDSR were used to evaluate the masses of 
these molecular states: $DB ~(0^+)$, $D^\ast B^\ast ~(0^+)$, $D^\ast B ~(1^+)$ and 
$DB^\ast ~(1^+)$. 

One has found that for $DB ~(0^+)$ molecular state the mass is  
$M_{DB} = (6.75 \pm 0.14) \GeV$, which is $\sim400\MeV$ below the DB threshold 
$E_{th}[D \:B] \simeq 7.15 \GeV$. This result indicates that such molecular state would be 
tightly bound. 
For the $D B^\ast ~(1^+)$ molecular state, the obtained mass is 
$M_{DB^\ast} = (6.85 \pm 0.15) \GeV$, approximately $\sim330\MeV$ below the $DB^\ast$ 
threshold $E_{th}[D \:B^\ast] \simeq 7.18 \GeV$. This molecule would also correspond to a 
tightly bound state. 
These two results are not in agreement with the ones obtained in Ref.\cite{sunliu}, where the 
authors have used One Boson Exchange model (OBE) to study these molecules. Their findings 
suggest that both molecules would be loosely bound states with a binding energy in order of 
$10 \MeV$. 

For the other two molecules $D^\ast B^\ast ~(0^+)$ and $D^\ast B ~(1^+)$, the 
masses obtained are given by: $M_{D^\ast B^\ast} = (7.27 \pm 0.12) \GeV$ and 
$M_{D^\ast B} = (7.16 \pm 0.12) \GeV$, respectively. 
These central values lead approximately to the same predictions made in Ref.\cite{sunliu}, for the 
respective molecular states in a OBE model. As one can see from the respective meson-meson 
threshold $E_{th}[D^\ast B^\ast] \simeq 7.32 \GeV$ and $E_{th}[D^\ast \:B] \simeq 7.29 \GeV$, these 
molecules are compatible with loosely bound states only if one considers the uncertainties. 
Therefore, more efforts must be done in order to establish the existence of such molecules.

For heavy baryons in QCD, the DRSR approach has been used to evaluate the mass-splittings 
between two baryons, which differ only by SU(3) spontaneous breaking symmetry.
For Singly Heavy Baryons $(Qqq)$, one concludes that:
\begin{itemize}
\item one estimates an improved value to the ratio of the quark condensates $\qq[q]$ and 
$\qq[s]$, given by: $\kappa = 0.74 \pm 0.06$. 
This value for $\kappa$ was estimated from a DRSR calculation to the baryons which have already 
been experimentally observed. Then using the available experimental data, one could estimate a 
more accurate value for this quantity.
\item with DRSR approach one has made predictions for the masses of the $\Xi^\prime_{c,b}$, 
$\Omega_b$, $\Xi^\ast_{b}$ and $\Omega^\ast_b$ baryons, and the results are summarized in 
Table (\ref{TabBarionsQqq}).
\item the results found for the $\Omega_b$ baryon: $M_{\Omega_b} = (6075.6 \pm 37.2) \MeV$
favor the mass experimentally observed by CDF collaboration \cite{cdf3} and disagrees with 
the one by D0 collaboration \cite{d02}.
\item another fascinating point is the predictions for hyperfine mass-splittings of the Singly 
Heavy Baryons. The splittings $M_{\Omega^\ast_b} - M_{\Omega_b}$ and 
$M_{\Xi^\ast_b} - M_{\Xi^\prime_b}$ are in agreement, considering the uncertainties, with the 
results obtained by other theoretical models - see Table (\ref{TabHyperQqq}) - like potential models 
(PM) \cite{RICHARD} and $1/N_c$ expansion model \cite{JENKINS}. 
However, only with the sum rules is possible to estimate a more accurately result:
\begin{equation*}
  M_{\Xi^\ast_b} - M_{\Xi_b} ~~\simeq~~ M_{\Xi^\ast_c} - M_{\Xi_c} \nno ~~.
\end{equation*}
With the future discoveries on the baryonic spectra, like the observation of $\Xi^\prime_b$, 
$\Xi^\ast_b$ and $\Omega^\ast_b$ baryons, one expects to test all these values obtained for hyperfine 
mass-splittings, which were estimated in the present work.
\item according to these hyperfine mass-splittings, one also expects that the $\Omega^\ast_{c,b}$ 
baryons only can decay electromagnetically, since there is no phase space for the existence of 
hadronic decay channels. Whereas for the $\Xi^\ast_{c,b}$ baryons, there is the possibility for the 
existence of hadronic decay channel and the presence of pions in the final state could be even
observed. Thus, one can estimate the likely dominant decay channels:
\vspace{-0.3cm}
\begin{eqnarray*}
  \Omega^\ast_{Q} ~~\rightarrow~~ \Omega_{Q} + \gamma ~,&&~~~~~
  \Xi^\ast_{Q} ~~\rightarrow~~ \Xi_{Q} + \pi  ~.
\end{eqnarray*}
\item all of these predictions for the Singly Heavy Baryons will be an excellent test for the 
sum rule approach for estimating the masses of not yet observed heavy baryons.
\end{itemize}
\vspace{0.6cm}

Experimentally, only SELEX collaboration supports the observation of the first Doubly 
Heavy Baryons in nature. The $\Xi^+_{cc} \:(ccd)$ baryon was seen in the decay channel 
$\Xi^+_{cc} \rightarrow \Lambda^+_c \:K^- \pi^+$, with a mass: 
$M_{\Xi^+_{cc}} = (3518.9 \pm 0.9) \MeV$. One expects that the experiments, carried out by 
LHC, could provide a vast amount of data on heavy baryon spectroscopy. Therefore, there is an 
exciting scenario for estimating the masses of the Doubly Heavy Baryons using the DRSR approach. 
The mass-splitting between $\Xi^*_{cc}$ and $\Xi_{cc}$ baryons is around $60 \MeV$, which 
is compatible with the one from potential models \cite{RICHARD, Dosch}, but it is larger than 
the one of about $24 \MeV$ obtained in Ref.\cite{vijande}. 
The mass-splitting between $\Xi^*_{bb}$ and $\Xi_{bb}$ baryons also agrees with 
potential models \cite{RICHARD, Dosch}.
Another relevant observation is that the mass-splittings (\ref{chibcmass}) 
do not support a hadronic decay channel, for instance
$\Xi^*_{QQ} \rightarrow \Xi_{QQ} +\pi's$, since there is no enough energy for a pion production. 
Therefore, from a DRSR point of view, the dominant decay channel for $\Xi^\ast_{QQ}$ baryon 
should be into electromagnetic decays: 
$\Xi^*_{QQ} ~~\rightarrow~~ \Xi_{QQ} + \gamma$.

Future researches on ${\Xi^*_{cc}}$ and ${\Xi^*_{bb}}$ baryons could support (or not) the existence 
of this decay channel. All other results obtained with DRSR approach for Doubly Heavy Baryons 
are summarized in Table (\ref{TabQQq}).

\cleardoublepage

\appendix
\chapter{Useful Relations}\label{app:Useful}

In the present work, the natural units $\hbar = c = 1$ and Einstein convention have been 
considered. The latter implies summation over a set of repeated indices in a given equation. 
The space-time metric in four-dimension, in the Minkowski space, is given by 
\begin{eqnarray}
  d s^2 &=& d x_0^2 - dx_1^2 - dx_2^2 - dx_3^2
\end{eqnarray}
where $x_0$ is the time-like variable and $x_i ~(i=1,2,3)$ are the space-like coordinates.

\section{SU(N) Group}
The SU(N) group is an unitary group composed by $N \times N$ matrices and determinant 
equal to one. All of these matrices belong to the SU(N) group and can be represented by theirs
$(N^2 -1)$ generators. In particular, 
\begin{eqnarray*}
  \mbox{ SU(2)} & \rightarrow & \mbox{ the 3 generators are the Pauli's matrices} \\
  \mbox{ SU(3)} & \rightarrow & \mbox{ the 8 generators are the Gell-Mann's matrices} ~.
\end{eqnarray*}
The SU(N) generators satisfy the following algebra
\begin{eqnarray}
  \left[ t^A, t^B \right] &=& i f^{ABC} \:t^C  
\end{eqnarray}
where $f^{ABC}$ is the structure constant totally antisymmetric of the group. 
It is also possible to define the anti-commutation relation:
\begin{eqnarray}
  \left\{ t^A, t^B \right\} &=& d^{ABC} \:t^C + \frac{1}{3}\de^{AB}
\end{eqnarray}
where $d^{ABC}$ is the structure constant totally symmetric of the group. 
Some useful relations of the SU(N) group are given by:
\begin{eqnarray}
  \delta_{aa} &=& N \\
  f^{ABD} \:f^{ABD'} &=& N \:\de^{DD'} ~\equiv~ C_A \:\de^{DD'} \\
  t^A_{ab} \: t^A_{bd} &=& \left( \frac{N^2 - 1}{2N} \right) \delta_{ad} ~\equiv~ C_F \:\delta_{ad} \\
  t^A_{ab} \: t^A_{cd} &=& \frac{1}{2} \left( \de_{ad}\de_{bc} - \frac{1}{N} \de_{ab}\de_{cd} \right) \\
  t^A_{ab} \: t^A_{cd} &=& C_F \:\de_{ab}\de_{cd} - C_A \:t^A_{ac} \: t^A_{bd} \\
  \Tr \left[ t^A \right] &=& 0 \\
  \Tr \left[ t^A t^B \right] &=& \frac{\de^{AB}}{2} \\
  \Tr \left[ t^A t^A \right] &=& \frac{N^2 - 1}{2}
\end{eqnarray}
the capital roman indices denote the generators for the SU(N) group, while the non-capital 
roman indices denote the charge color.

\section{Algebra of Dirac Matrices}
The Dirac matrices, $\gamma^\mu$, satisfy the following algebra:
\begin{eqnarray}
  \left\{ \ga_\mu, \ga_\nu \right\} &=& 2g_{\mu\nu} ~,
\end{eqnarray}
and also the following relations:
\begin{eqnarray}
  \ga_\mu \ga^\mu &=& 4\\
  \ga^0 &=& \ga^{0\:T} ~=~ \ga^{0\:\dag} \\
  \ga^{\mu\:\dag} &=& \ga^0 \ga^\mu \ga^{0}\\
  \ga^\mu \ga_\alpha \ga_\mu &=& -2 \ga_\alpha \\
  \ga^\mu \ga_\alpha \ga_\beta \ga_\mu &=& 4g_{\alpha\beta} \\
  \ga^\mu \ga_\alpha \ga_\beta \ga_\rho \ga_\mu &=& -2 \ga_\rho \ga_\beta \ga_\alpha ~~.
\end{eqnarray}
Now considering the definitions
\begin{eqnarray}
  C &=& i\ga^2 \ga^0 \\
  \sigma_{\al\be} &=& i(\ga_\al \ga_\be - g_{\al\be})\\
  \ga_5 &=& i\ga^0 \ga^1 \ga^2 \ga^3
\end{eqnarray}
one can demonstrate that
\begin{eqnarray}
  C^{-1} ~=~ C^{T} &=& C^{\dag} ~=~ -C \\
  C \:\ga^T_\mu\: C^{-1} &=& -\ga_\mu \\
  C \:\sigma^T_{\al\be}\: C^{-1} &=& -\sigma_{\al\be} \\
  C \:\ga_5^T\: C^{-1} &=& \ga_5 
\end{eqnarray}

\begin{eqnarray}  
  C \:\sigma^T_{\al\be} \ga^T_\rho\: C^{-1} &=& \sigma_{\al\be} \ga_\rho \\
  C \:(\ga_5 \ga_\alpha)^T\: C^{-1} &=& \ga_5 \ga_\alpha\\
  \sigma^{\al\be} \sigma_{\al\be} &=& 12 \\
  \sigma_{\al\be} \ga^\mu \sigma^{\al\be} &=& 0 \\
  \sigma_{\al\be} (\ga_\mu  \ga_\nu) \sigma^{\al\be} &=& 16g_{\mu\nu} - 4\ga_\mu \ga_\nu \\
  \ga_5 &=& \ga^T_5 ~=~ \ga^\dag_5 \\
  \ga_5^2 &=& 1 \\
  \ga_5^\dagger &=& -\ga^0 \ga_5 \ga^0 \\
  \ga^\mu \ga_5 \ga_\mu &=& -4 \ga_5 \\
  \left\{ \ga^\mu, \ga_5 \right\} &=& 0 \\
  \left[ \ga_5, \sigma_{\al\be} \right] &=& 0 ~~.
\end{eqnarray}

The traces involving the Dirac matrices, which appear during the calculation of 
the correlation function are given by:
\begin{eqnarray}
	\Tr \left[ \bf{\hat{1}} \right] &=& 4 \\
	\Tr \left[ \ga_5 \right] &=& 0 \\
	\mbox{Trace of a odd mumber of} ~\ga'^s &=& 0 \\
	\Tr \left[ \ga_\mu \ga_\nu \right] &=& 4g_{\mu\nu} \\
	\Tr \left[ \ga_\mu \ga_\nu \ga_5 \right] &=& 0 \\
	\Tr \left[ \ga_\mu \ga_\nu \ga_\rho \ga_\sigma \right] &=& 
		4(g_{\mu\nu} g_{\rho\sigma} - g_{\mu\rho} g_{\nu\sigma} + g_{\mu\sigma} g_{\nu\rho}) \\
	\Tr \left[ \sigma_{\al\be} \right] &=& 0 ~.
	\label{traces}
\end{eqnarray}

\section{Integration Techniques}

\subsection{Dirac Delta Function}
The Dirac delta function in four-dimension, defined as $\de^{(4)}(x-y)$, is zero in any region of the 
space-time except in $x=y$ and satisfies:
\begin{eqnarray}
  \int \!d^4x ~\de^{(4)}(x - y) &=& 1 \\
  \int \!d^4x ~f(x) \de^{(4)}(x - y) &=& f(y) ~~.  
\end{eqnarray}

\subsection{Fourier Transform}
By convention, in a Fourier transform, the factor $(2 \pi)^4$ is always kept in the denominator 
of the momentum integrals. For example:
\begin{eqnarray}
  f(x) &=& \int \frac{d^4p}{(2\pi)^4} ~e^{-i p \cdot x} F(p) \\
  F(p) &=& \int d^4p ~e^{i p \cdot x} f(x) ~.
\end{eqnarray}

An important identity between Dirac delta function and Fourier transform 
is given by:
\begin{eqnarray}
  \de^{(4)}(x-y) &=& \int \frac{d^4p}{(2\pi^4)} ~e^{i p \cdot (x-y)} F(p)
\end{eqnarray}

\subsection{Wick Rotation}
The Wick rotations are particularly useful to find mathematical solutions in Minkowski space, for 
problems that could be easily solved in Euclidean space. Applying it onto the four-position 
$x$ and the four-momentum $p$ are defined respectively by:
\begin{eqnarray}
  \int \!d^4x &=& -i \int \!d^4 x_{_E} \\
  \int \!d^4p &=& +i \int \!d^4 p_{_E} 
\end{eqnarray}
where the subscript ${}_E$ indicates a four-vector defined in a Euclidean space.

\subsection{Schwinger and Feynman parameterization}
A very useful way to evaluate the four-momentum integrals which arise from the correlation function 
calculation is through the Schwinger or Feynman parameterizations:
\begin{enumerate}
\item Schwinger parameterization:
	\begin{equation}
	  \frac{1}{(M^2 - p^2)^n} = \frac{1}{(n-1)!} \int\limits^{+\infty}_{0} \! d\al
	  \:\al^{n-1} ~e^{-\al(M^2-p^2)} ~.
	  \label{sch}
	\end{equation}
\item Feynman parameterization:
	\begin{equation}
	  \frac{1}{a^\alpha \:b^\beta} = \frac{\Gamma(\alpha+\beta)}{\Gamma(\alpha) \Gamma(\beta)} 
	  \int\limits^{1}_{0} \!dx \:\frac{x^{\al-1} (1-x)^{\be-1}}{\left[ (a-b)x+b \right]^{\al+\be}} ~.
	  \label{fey}
	\end{equation}
\end{enumerate}

\subsection{Gaussian Integrals}
An example about using the Wick rotation is given by the Gaussian integrals that are well defined 
only in Euclidean space. Then, calculating the integral:
\begin{eqnarray}
  \int \!d^4x ~e^{a x^2 + iq \cdot x} && (a>0) ~~.
\end{eqnarray}
and using the Wick rotation, one obtains:
\begin{eqnarray}
  \int \!d^4x ~e^{a x^2 + iq \cdot x} &\stackrel{\mbox{\tiny Wick}}{=}&
  -i \int \!d^4x_{_E} ~e^{-a x_{_E}^2 + i Q \cdot x_{_E}} ~~ = ~~
  -\frac{i \pi^2}{a^2} ~e^{-\frac{Q^2}{4a}} ~~.
  \label{gaussx}
\end{eqnarray}
where $Q^2 = -q^2$. A similar calculation can be done to demonstrate that:
\begin{eqnarray}
  \int \!d^4p ~e^{a p^2 - ip \cdot x} &\stackrel{\mbox{\tiny Wick}}{=}&
  i \int \!d^4p_{_E} ~e^{-a p_{_E}^2 + i p_{_E} \cdot x_E} ~~ = ~~
  \frac{i \pi^2}{a^2} ~e^{-\frac{x_E^2}{4a}} ~~.
  \label{gaussp}
\end{eqnarray}

\subsection{Four-Momentum Integrals}
To calculate the correlation function is necessary to evaluate the following  
four-momentum integrals:
\begin{eqnarray}
  \int\! d^4p \:\frac{e^{-ip \cdot x}}{(p^2 - M^2)^k} \nno \\ \\
  \int\! d^4p \:\frac{p_\mu \:e^{-ip \cdot x}}{(p^2 - M^2)^k} ~. \nno
\end{eqnarray}
The results of these integrals are obtained as follows:
\begin{enumerate}
\item First integral: using the Schwinger parameterization
\begin{eqnarray*}
  \int\! d^4p \:\frac{e^{-ip \cdot x}}{(p^2 - M^2)^k} 
  &=& \frac{(-1)^k}{(k-1)!} \int\! d^4p ~e^{-ip \cdot x} 
  \int\limits^{+\infty}_{0} \! \frac{d\al}{\al^{1-k}} ~e^{-\al(M^2-p^2)} ~.
\end{eqnarray*}
		
Inverting the integrals and using the Gaussian integral (\ref{gaussp}), one obtains
\begin{eqnarray}
  \int\! d^4p \:\frac{e^{-ip \cdot x}}{(p^2 - M^2)^k} 
  &=& \frac{(-1)^k}{(k-1)!} \int\limits^{+\infty}_{0} \!\frac{d\al}{\al^{1-k}} ~e^{-\al M^2} 
  \int\!d^4p ~e^{\al p^2 - ip \cdot x} \nno\\
  &=& \frac{(-1)^k i \pi^2}{(k-1)!} \int\limits^{+\infty}_{0} \!\frac{d\al}{\al^{3-k}} 
  ~e^{-\al M^2 + \frac{x^2}{4\alpha}} 
  \label{i:mom1}
\end{eqnarray}

\item Second integral: analogously one has
\begin{eqnarray*}
  \int\! d^4p \:\frac{p_\mu \:e^{-ip \cdot x}}{(p^2 - M^2)^k} 
  &=& \frac{(-1)^k}{(k-1)!} \int\! d^4p ~p_\mu \:e^{-ip \cdot x} 
  \int\limits^{+\infty}_{0} \! \frac{d\be}{\be^{1-k}} ~e^{-\be(M^2-p^2)} ~.
\end{eqnarray*}
The four-momentum integral can be rewritten in terms of a derivative, then:
\begin{eqnarray*}
  \int\! d^4p \:\frac{p_\mu \:e^{-ip \cdot x}}{(p^2 - M^2)^k} 
  &=& \frac{(-1)^k \:i}{(k-1)!} \int\limits^{+\infty}_{0} \! \frac{d\be}{\be^{1-k}} 
  ~e^{-\be M^2} \frac{\partial}{\partial x^\mu} \left[ \int\! d^4\!p ~e^{\be p^2 - ip \cdot x} \right]\\
  &=& \frac{(-1)^{k+1} \pi^2}{(k-1)!} \int\limits^{+\infty}_{0} \! \frac{d\be}{\be^{3-k}} 
  ~e^{-\be M^2} \:\frac{\partial}{\partial x^\mu} ~e^{\frac{x^2}{4\be}} ~.
\end{eqnarray*}
Finally, one gets
\begin{eqnarray}
  \int\! d^4p \:\frac{p_\mu \:e^{-ip \cdot x}}{(p^2 - M^2)^k} 
  &=& \frac{(-1)^{k+1} \pi^2}{2(k-1)!} \int\limits^{+\infty}_{0} \! \frac{d\be}{\be^{4-k}} 
  ~x_\mu \:e^{-\be M^2 + \frac{x^2}{4\be}} ~.
  \label{i:mom2}
\end{eqnarray}
\end{enumerate}

\subsection{Gamma Function $\Gamma(N)$}
In QCDSR, it is necessary to evaluate integrals like 
\begin{eqnarray}
  \Gamma_N(Q^2) &=& \int\limits_0^{+\infty} \!d\la \:\la^N \ e^{-\la \:f}  ~~,
  \label{la0}
\end{eqnarray}
where $f \equiv f(Q^2)$ is an arbitrary positive function and $N$ is a positive integer.
These integrals can be put in terms of the Gamma function, making the change of variable: 
$u = \la \:f$. Then, one writes
\begin{eqnarray}
  \Gamma_N(Q^2) ~~=~~ 
  \frac{1}{f^{N+1}}\int\limits_0^{+\infty} \!du \:u^N \ e^{-u}
  &=& \frac{\Gamma(N+1)}{f^{N+1}} ~,~~~(N > 0) ~.
  \label{la1}
\end{eqnarray}
The Gamma function, also known as the Euler function, is defined as

\vspace{-0.3cm}
\begin{eqnarray}
  \Gamma(N+1) &=& \int\limits_0^{+\infty} \!du ~u^{N} \:e^{-u} ~~=~~ N! ~~,
  \label{fgamma}
\end{eqnarray}
where the relation with Factorial function is remarkable. The advantage of using Gamma function 
is the possibility to do the analytic extension of the divergent integrals (\ref{la0}), when considering 
negative values for $N$.
Usually, the variable $n$ is used instead of $N$ for indicating a negative integer number, so that: 
$|N| = -n$. Thus, the integral (\ref{la1}) can be rewritten as
\begin{eqnarray}
  \Gamma_n(Q^2) ~~=~~ f^{n-1} \:\int\limits_0^{+\infty} \!\frac{du}{u^n}  ~e^{-u}
  &=& f^{n-1} ~\Gamma(-n+1) ~,~~~(n > 0) ~.
  \label{la2}
\end{eqnarray}
The analytic extension of Gamma function can be evaluated using the following recursive relation:
\vspace{-0.5cm}
\begin{eqnarray}
  \Gamma(-n+1) &=& \frac{1}{(-n+1)} \Gamma(-n+2) ~.
\end{eqnarray}
As one can see, when $n=1$, it clearly diverges. Indeed, iterating this relation, it is possible to 
demonstrate that there are divergences for any $n> 0$:
\begin{eqnarray}
  \Gamma(-n+1) &=& \frac{1}{(-n+1)(-n+2)} \Gamma(-n+3) \nno\\
  &=& \frac{1}{(-n+1)(-n+2)(-n+3)} \Gamma(-n+4) \nno\\
  &=& \frac{(-1)^{n-1}}{(n-1)!} \Gamma(0) ~.
  \label{neg}
\end{eqnarray}
A very useful expression for the $\Gamma(0)$ function is given through the expansion of the 
$\Gamma(\epsilon)$ function, around $\epsilon = 0$, which results in
\vspace{-0.2cm}
\begin{eqnarray}
  \Gamma(0) ~~=~~ \lim_{\epsilon \rightarrow 0} ~\Gamma(\epsilon) &=& 
  \lim_{\epsilon \rightarrow 0} ~\bigg[
  \frac{1}{\epsilon} - \gamma_{_E} + \mathcal{O}(\epsilon) \bigg]~,
  \label{expGamma}
\end{eqnarray}

\noindent where $\gamma_{_E} \simeq 0.57723 $ is the Euler-Mascheroni number. This expansion 
allows one to rewrite the Eq.(\ref{neg}) as
\begin{eqnarray}
  \Gamma(- n + 1) ~~= ~~ \lim_{\epsilon \rightarrow 0} ~ \Gamma(\epsilon - n + 1) &=& 
  \frac{(-1)^{n-1}}{(n-1)!} ~~\lim_{\epsilon \rightarrow 0}
  \left[ \frac{1}{\epsilon} - \gamma_{_E} + \mathcal{O}(\epsilon) \right] ~.
\end{eqnarray}
Therefore, the integral (\ref{la2}) is given by 
\begin{eqnarray}
  \Gamma_n(Q^2)
  &=& \lim_{\epsilon \rightarrow 0} ~f^{n - 1 - \epsilon} ~\Gamma(\epsilon - n + 1) ~.
\end{eqnarray}
Considering the expansions around $\epsilon=0$, one gets:
\begin{eqnarray}  
  \Gamma_n(Q^2)
  &=& \lim_{\epsilon \rightarrow 0} \frac{(-1)^{n-1}}{(n-1)!} f^{n-1} 
  \bigg[ 1 - \epsilon \:\Log \:f  + \mathcal{O}(\epsilon^2) \bigg]
  \bigg[\:\frac{1}{\epsilon} - \gamma_{_E} + \mathcal{O}(\epsilon) \bigg] \nno\\
  &=& \frac{(-1)^{n-1}}{(n-1)!} f^{n-1} \lim_{\epsilon \rightarrow 0} \: \left[ 
  \frac{1}{\epsilon} -\gamma_{_E} - \Log \:f + \mathcal{O}(\epsilon) \right] ~.
\end{eqnarray}
Note that the integrals contain a divergence which can be removed, for instance, using 
the dimensional regularization ($N_\epsilon = \frac{1}{\epsilon} - \gamma_{_E}$). Therefore, 
the final result for the regularized integral (\ref{la2}) can be expressed by 
\begin{eqnarray}
  \Gamma_n(Q^2) &=& 
  \frac{(-1)^n}{(n-1)!} f^{n-1} \:\Log \:f ~~.
\end{eqnarray}
In summary, one has obtained the following results:
\begin{eqnarray}
\addtolength{\fboxsep}{10pt} 
\boxed{ \begin{gathered} 
  \begin{array}{rcl} 
    \displaystyle
    \int\limits_0^{+\infty} \!d\la ~\la^N \ e^{-\la f} ~~&=&
    \displaystyle ~~ \frac{N!}{f^{N+1}} ~,~~~\hfill(N \geq 0)\\
    \displaystyle
    \int\limits_0^{+\infty} \!\frac{d\la}{\la^n} \ e^{-\la f} ~~&=&~~ 
    \displaystyle \frac{(-1)^n}{(n-1)!} ~f^{n-1} \:\Log \:f ~,~~~\hfill(n>0)
  \end{array}
\end{gathered}} 
\label{i:gamma}
\end{eqnarray}

\subsection{Integral $I_{nml}(Q^2)$}
Consider the integral
\begin{equation}
  I_{nml}(Q^2) = \int\limits_{0}^{+\infty}\! \frac{d\al}{\al^n} 
  	\int\limits_{0}^{+\infty} \!\frac{d\be}{\be^m} 
  	\:\frac{e^{-M^2(\al+\be) \ - \ \left( \frac{\al\be}{\al+\be} \right) Q^2}}{(\al+\be)^l} ~.
\end{equation}
For evaluate it, one considers the change of variables
\begin{eqnarray}
  \al ~=~ \lambda \al', &~~~ \be ~=~ \lambda (1-\al'),  &~~~
  \left| \frac{\partial(\al,\be)}{\partial(\al', \lambda)} \right| ~=~ \lambda
\end{eqnarray}
which gives the following result:
\begin{eqnarray}
  I_{nml}(Q^2) &=& \int\limits^{1}_{0}\! \frac{d\al}{\al^n(1-\al)^m} 
  \int\limits^{+\infty}_{0} \!\frac{d\la}{\la^{n+m+l-1}} ~e^{-\la \left[ M^2 + \al(1-\al)Q^2 \right]} ~.
\end{eqnarray}
For simplicity, the apostrophe is omitted in the variable $\al'$. As seen before, the 
$\lambda$-integral has two results that depend whether the power in $\lambda$ is positive 
or negative. Therefore, the integral $I_{nml}(Q^2)$ admits the following results:
\begin{eqnarray}
\addtolength{\fboxsep}{10pt} 
\boxed{ \begin{gathered} 
  \begin{array}{rcl} 
    && \hspace{-2.5cm} \mbox{for } n+m+l-1 > 0:\\
    \displaystyle
    I_{nml}(Q^2) &=&
    \displaystyle
    (-1)^{n+m+l-1} \int\limits^{1}_{0}\! \frac{d\al}{\al^n(1-\al)^m} ~
    \frac{{\cal H}^{n+m+l-2}_\alpha}{(n-m-l-2)!} ~\Log \:{\cal H}_\alpha  ~~~~~\\ \\
    &&\hspace{-2.5cm} \mbox{for } n+m+l-1 \leq 0:\\
    \displaystyle
    I^\ast_{nml}(Q^2) &=&
    \displaystyle
    \int\limits^{1}_{0}\! \frac{d\al}{\al^n(1-\al)^m} \frac{(1-n-m-l)!}{{\cal H}^{2-n-m-l}_\alpha}
  \end{array}
\end{gathered}}
\label{i:Inml}
\end{eqnarray}
where ${\cal H}_\alpha = M^2 + \alpha(1-\alpha) Q^2$. 
\vfill

\subsection{Integral $I_{nmkl}(Q^2)$}
Another integral of great interest in QCDSR is given by
\begin{eqnarray}
  I_{nmkl}(Q^2) &=& \int\limits_0^{+\infty} \!\frac{d\al}{\al^n} \int\limits_0^{+\infty} \!\frac{d\be}{\be^m} 
  \int\limits_0^{+\infty} \!\frac{d\ga}{\ga^k}
  ~\frac{e^{-M^2(\al+\be) - \left(\frac{\al\be\ga}{\al\be + \be\ga + \ga\al} \right) Q^2}}
  {(\al\be + \be\ga + \ga\al)^l} ~.
\end{eqnarray}
To simplify this expression, one considers the change of variables
\begin{eqnarray}
  \al ~=~ \lambda \al', &~~~ \be ~=~\lambda \be',  ~~~~ \ga ~=~ \frac{\lambda \:\al' \be'}{(1-\al'-\be')}, &~~~
  \left| \frac{\partial(\al,\be,\ga)}{\partial(\al',\be',\lambda)} \right| ~=~ 
  \frac{\lambda^2 \al' \be'}{(1-\al' -\be')^2} ~~~~~
\end{eqnarray}
which results in the following expression
\begin{eqnarray}
  I_{nmkl}(Q^2) = \int\limits_0^1 \!\frac{d\al}{\al^{n+k+l-1}} 
  \int\limits_0^{1-\al} \!\frac{d\be}{\be^{m+k+l-1}} \:(1 \!-\! \al \!-\! \be)^{k+l-2}
  \int\limits_0^{+\infty} \!d\lambda \:
  \frac{e^{- \lambda \left[ M^2(\al+\be) + \al \be Q^2 \right] }}{\lambda^{n+m+k+2l-2}} ~~~~
\end{eqnarray} 
Again, one omits the apostrophe in the integration variables. Note that the 
$\lambda$-integral can be evaluated using the Gamma function. Thus, the integral 
$I_{nmkl}(Q^2)$ admits the following results:
\begin{eqnarray}
\addtolength{\fboxsep}{6pt} 
\boxed{ \begin{gathered} 
  \begin{array}{rcl} 
    && \hspace{-2.8cm} \mbox{for }  n+m+k+2l-2 > 0:\\
    \displaystyle
    I_{nmkl}(Q^2) &=&
    \displaystyle
     (-1)^{n+m+k} 
     \!\int\limits_0^1 \!\!\frac{d\al}{\al^{n+k+l-1}} 
     \!\int\limits_0^{1-\al} \!\!\frac{d\be}{\be^{m+k+l-1}} \\
     && \displaystyle \hspace{1cm}
     \times~ \frac{(1 \!-\! \al \!-\! \be)^{k+l-2} \:{\cal F}_{\al\be}^{\:n+m+k+2l-3}}{(n+m+k+2l-3)!}
     \:\Log \:{\cal F}_{\al\be} \\ \\
    &&\hspace{-2.8cm} \mbox{for }  n+m+k+2l-2 \leq 0:\\
    \displaystyle
    I^\ast_{nmkl}(Q^2) &=&
    \displaystyle
     \!\int\limits_0^1 \!\!\frac{d\al}{\al^{n+k+l-1}} 
     \!\int\limits_0^{1-\al} \!\!\frac{d\be}{\be^{m+k+l-1}} \:
     \frac{(1 \!-\! \al \!-\! \be)^{k+l-2} \:(2 \!-\! n \!-\! m \!-\! k \!-\! 2l)!}{{\cal F}_{\al\be}^{3-n-m-k-2l}}
  \end{array}
\end{gathered}} ~~
\label{i:Inmkl}
\end{eqnarray}
where ${\cal F}_{\alpha\beta} = M^2(\alpha+\beta) + \alpha \beta Q^2$.

\subsection{Special Gaussian Integral $G_n(Q^2)$}
Considering now the following special gaussian integral
\begin{eqnarray}
  G_n(Q^2) &\equiv& \int\! \frac{d^4x}{(x^2)^n} \:e^{a \:x^2 + iq \cdot x} ~~~~(a, n >0) ~~.
\end{eqnarray}
For evaluate it, one must use the Wick rotation and Schwinger parameterization, so that:
\begin{eqnarray}
  G_n(Q^2) &=& -i \int\! d^4\!x_{_E} \:\frac{e^{-a x_{_E}^2 +  i Q\cdot x_{_E}}}{(-x_{_E}^2)^n} \nno\\
	&=& \frac{(-1)^{n+1} \:i}{(n-1)!} \int\! d^4\!x_{_E} ~e^{iQ \cdot x_{_E}} 
	\int\limits^{+\infty}_{0} \! d\de \:\de^{n-1} ~e^{-(a+\de) x_{_E}^2} \nno\\
	&=& \frac{(-1)^{n+1} \:i}{(n-1)!} \int\limits^{+\infty}_{0} \! d\de \:\de^{n-1} 
	\int\! d^4\!x_{_E} ~e^{-(a+\de) x_{_E}^2 + iQ \cdot x_{_E}} ~.
\end{eqnarray}
Completing the square of quadratic term in the exponential, one obtains:
\begin{eqnarray}
  G_n(Q^2) &=& \frac{(-1)^{n+1} \:i}{(n-1)!} \int\limits^{+\infty}_{0} \! d\de \:\de^{n-1} 
  ~e^{-\frac{Q^2}{4(a+\de)}} \int\! d^4\!x_{_E} ~e^{-a \left[ x_{_E} + \frac{iQ}{2(a+\de)} \right]^2} \nno \\
  &=& \frac{(-1)^{n+1} \:i\pi^2}{(n-1)!} \int\limits^{+\infty}_{0} \! d\de 
  \:\frac{\de^{n-1}}{(a+\de)^2} ~e^{-\frac{Q^2}{4(a+\de)}} ~.
\end{eqnarray}
In general, during a QCDSR calculation, the usual form for this integral is obtained making 
the following change in the integration variable: $\de = \frac{1}{4\ga}$. The result is given by:
\begin{eqnarray}
  G_n(Q^2) &=& \frac{(-1)^{n+1} \:2^{4-2n} \:i\pi^2}{(n-1)!} \int\limits^{+\infty}_{0} \! \frac{d\ga}{\ga^{n-1}}
  \:\frac{1}{(1+4a \gamma)^2} 
  \:e^{ -\left(\frac{\gamma}{1+4a \gamma} \right) Q^2} ~.
\end{eqnarray}
Then, one could consider two particular cases for the choice of the parameter $a$:
\begin{eqnarray}
  \label{i:gaussA1}
  \left( a = \frac{1}{4\al} \right) \Rightarrow&& 
  G_n(Q^2) ~=~
  \frac{(-1)^{n+1} \:2^{4-2n} \:i\pi^2}{(n-1)!} \int\limits^{+\infty}_{0} \! \frac{d\ga}{\ga^{n-1}}
  \:\frac{\al^2}{(\al + \ga)^2} 
  \:e^{ -\left(\frac{\al\ga}{\al + \ga} \right) Q^2} ~~~~
\end{eqnarray}

$\left( a = \frac{1}{4\al} + \frac{1}{4\be} \right) \Rightarrow$  \vspace{-0.5cm}
\begin{eqnarray}
  \label{i:gaussA2}
  G_n(Q^2) &=&
  \frac{(-1)^{n+1} \:2^{4-2n} \:i\pi^2}{(n-1)!} \int\limits^{+\infty}_{0} \! \frac{d\ga}{\ga^{n-1}}
  \:\frac{\al^2 \be^2}{(\al \be + \be\ga + \ga\al)^2} 
  \:e^{ -\left(\frac{\al\be\ga}{\al \be + \be\ga + \ga\al} \right) Q^2} ~~.
\end{eqnarray}

\subsection{Special Gaussian Integrals $G_N(Q^2)$}
Another integral, often used in a QCDSR calculation, is given by
\begin{eqnarray}
  G_N(Q^2) &\equiv& \int\!d^4x ~(x^2)^N \:e^{a \:x^2 + i q \cdot x} ~~~~(a, N \geq 0) ~~.
\end{eqnarray}
Besides the Wick rotation and Schwinger parameterization, to solve this kind of integral 
one needs to use the Eq.(\ref{gaussx}):
$G_0(Q^2) = -\frac{i\pi^2}{a^2} \:e^{-\frac{Q^2}{4a}}$, 
and the recursive relation:
\begin{eqnarray}
  G_N(Q^2) &=& \frac{\partial}{\partial a} G_{N-1}(Q^2) ~.
\end{eqnarray}
With this, one can demonstrate that
\begin{eqnarray}
  G_N(Q^2)
  &=& \frac{(-1)^{N+1} \:i\pi^2}{a^{N+2}} \:e^{-\frac{Q^2}{4a}}
  \sum_{K=0}^{N} C_{NK} \left( \frac{Q^2}{4a} \right)^K
\end{eqnarray}
where the constants $C_{NK}$ are defined as follows:
\begin{eqnarray}
  C_{NK} &=& \frac{(-1)^K N!}{K!(N-K)!} \frac{(N+1)!}{(K+1)!} ~~.
\end{eqnarray}
Again, considering two particular cases for the variable $a$, one has:
\begin{eqnarray}
  \label{i:gaussB1}
  \left( a = \frac{1}{4\al} \right) &\Rightarrow& 
  (-1)^{N+1} \:4^{N+2} \:i\pi^2 \:\al^{N+2} \:e^{-\al \:Q^2}
  \sum_{K=0}^{N} C_{NK} \left( \al \:Q^2 \right)^K \\
  \label{i:gaussB2}
  \left( a= \frac{1}{4\al} + \frac{1}{4\be} \right) &\Rightarrow& 
  (-1)^{N+1} \:i\pi^2 \left(\frac{\al+\be}{4\al\be} \right)^{N+2}
  \:e^{-\left(\frac{\al+\be}{4\al\be}\right) Q^2}
  \sum_{K=0}^{N} C_{NK} \left( \frac{\al \be Q^2}{\al+\be} \right)^K ~~~~~~
\end{eqnarray}

\cleardoublepage

\chapter{Full Propagator of QCD} \label{app:Propagators}	
As seen in Chapter \ref{chap:QCDSR}, it is convenient to introduce the definition of the 
full propagator of QCD in order to include in the sum rule the non-perturbative effects from
 QCD vacuum. Its expression is given by:
\vspace{-0.3cm}
\begin{eqnarray}
  {\cal S}^{QCD}(x) &=& S^0_{ab} + S^{GG}_{\al\be}(x) + {\cal S}^{q\bar{q}}_{ab}(x) + 
  {\cal S}^{G}_{ab,\al\be}(x) + {\cal S}^{q\bar{q}G}_{ab, \al\be}(x) + 
  {\cal S}^{\GGi}_{ab}(x)
  \label{propQCD}
\end{eqnarray}
where $S^0_{ab}(x)$ and $S^{GG}_{\al\be}(x)$ are respectively the perturbative quark 
and gluons propagators. The non-perturbative propagators are given by:
\vspace{-0.3cm}
\begin{eqnarray}
  \label{vacqq}
  {\cal S}^{q\bar{q}}_{ab}(x) &=& \langle 0| \!: q_{a}(x) \bar{q}_{b}(0) :\! |0 \rangle \\
  \label{vacG}
 {\cal S}^{G}_{ab, \al\be}(x) &=& \frac{i \:t^{_N}_{cd}}{2} ~ \int \!d^4y 
  \:S^0_{ac}(x-y) \:\gamma^\alpha y^\beta \:S^0_{db}(y) \\
  \label{vacqGq}
  {\cal S}^{q\bar{q}G}_{ab, \al\be}(x) &=& 
  \langle 0| \!: q_a(x) g_s G^N_{\alpha\beta}(0) \bar{q}_b(0) :\! |0 \rangle \\
  \label{vacGG}
  {\cal S}^{\GGi}_{ab}(x) &=& \frac{t^{N}_{cd} \:t^{M}_{ef}}{8} 
  \:\langle 0| \!:\! g_s^2 G^N_{\alpha\rho}(0) \:G^M_{\beta\lambda}(0) \!:\! |0 \rangle \nno\\
  && \times~ \iint \!d^4y \:d^4z ~y^\rho \: z^\lambda ~S^0_{ae}(x-z) \:\gamma^\beta \:S^0_{fc}(z-y) 
  	\:\gamma^\alpha \:S^0_{db}(y)
\end{eqnarray}
and they are responsible for including the contributions of the quark condensates $\qq[q]$, 
the gluon condensates $\GG$ and the mixed condensates $\qGq[q]$ in the OPE. All of these 
contributions will be calculated throughout this appendix.
\vfill

\section{Quark and Mixed Condensates}
As the vacuum expectation values (VEV) are treated as local operators, it is reasonable to 
assume small values of $x$ in the quark fields $q_a(x)$, in Eqs.(\ref{vacqq}) and (\ref{vacqGq}). 
Then, expanding it in a Taylor series around $x = 0$, one obtains
\vspace{-0.3cm}
\begin{eqnarray}
  {\cal S}^{q\bar{q}}_{ab}(x) &\!\!=\!\!& \langle 0| \!:\! q_{a}(x) \bar{q}_{b}(0) \!:\! |0 \rangle \nno\\
  && \hspace{-1.5cm}  = \langle 0| \!:\! q_{a}(0) \bar{q}_{b}(0) \!:\! |0 \rangle + 
  x_\rho \langle 0| \!:\! [\partial^\rho q_{a}(0)] \: \bar{q}_{b}(0) \!:\! |0 \rangle
  + \frac{1}{2} x_\rho x_\lambda \langle 0| \!:\! [\partial^\rho\partial^\lambda q_{a}(0)] \: 
  \bar{q}_{b}(0) \!:\! |0 \rangle +  \ldots \nno\\ \\
  {\cal S}^{q\bar{q}G}_{ab, \al\be}(x) &\!\!=\!\!& 
  \langle 0| \!:\! q_a(x)\:g_s G^{N}_{\alpha\beta}(0) \:\bar{q}_b(0) \!:\! |0 \rangle \nno \\
  &\!\!=\!\!& \langle 0| \!:\! q_a(0)\:g_s G^{N}_{\alpha\beta}(0) \:\bar{q}_b(0) \!:\! |0 \rangle
  + x_\rho \langle 0| \!:\! [\partial^\rho q_a(0)] \:g_s G^{N}_{\alpha\beta}(0) \:\bar{q}_b(0) \!:\! |0 \rangle
  + \ldots ~~~ \nno\\
\end{eqnarray}
For the gluon condensate, consider the Fock-Schwinger gauge which guarantees the 
locality of the gluon field operator $G^{N}_{\alpha \beta} (0)$. As the correlation function 
is an invariant gauge object, the ordinary derivatives present in these equations must be 
replaced by the covariant derivatives as follows: $D_\rho = \partial_\rho - ig_s A_\rho$.
Notice that in Fock-Schwinger gauge, this replacement is naturally obtained from the 
following calculation
\vspace{-0.3cm}
\begin{eqnarray}
  x^\rho D_\rho &=& x^\rho \partial_\rho - ig_s \:x^\rho A_\rho
  ~~=~~ x^\rho \partial_\rho + \frac{i\:g_s G_{\rho\lambda}(0)}{2} \:x^\rho x^\lambda
  ~~=~~ x^\rho \partial_\rho
\vspace{-0.3cm}
\end{eqnarray}
since the product $G_{\rho\lambda}(0) \:x^\rho x^\lambda = 0$, due to symmetry properties.
Thus, the VEVs in Eqs.(\ref{vacqq}) and (\ref{vacqGq}) can be rewritten as
\begin{eqnarray}
  \label{vacqq1}
  {\cal S}^{q\bar{q}}_{ab}(x) &=& 
  \langle 0| : q_{a}(0) \bar{q}_{b}(0) : |0 \rangle ~+~ 
  x_\rho \langle 0| : [D^\rho q_{a}(0)] \: \bar{q}_{b}(0) : |0 \rangle \nno \\
  &&
  +~ \frac{1}{2} x_\rho x_\lambda \langle 0| : [D^\rho D^\lambda q_{a}(0)] \: 
  \bar{q}_{b}(0) : |0 \rangle ~+~  \ldots  \hspace{1cm} \\
  \label{vacqGq1}
  {\cal S}^{q\bar{q}G}_{ab, \al\be}(x) &=& 
  \langle 0| :q_a(0)\:g_s G^{N}_{\al\be}(0) \:\bar{q}_b(0): |0 \rangle \nno \\
  && +~ x^\rho \langle 0| : [D_\rho q_a(0)] \:g_s G^{N}_{\al\be}(0) \:\bar{q}_b(0): |0 \rangle
  ~+~ \ldots ~~~
\end{eqnarray}
As will be seen more fully later, solving each term of the Taylor expansion of Eqs. (\ref{vacqq1}) 
and (\ref{vacqGq1}), one obtains the quark condensate $\qq[q]$ and the mixed condensate 
$\qGq[q]$ expressions to the full propagator of QCD.

\subsection{Quark Condensate $\qq[q]$}
The first term of the expansion (\ref{vacqq1}) is given by
\begin{equation}
  \langle 0| \: :q^\eta_a(0) \: \bar{q}^\xi_b(0): \: |0 \rangle =  {\cal N}_1 \:\de_{ab} \de^{\eta\xi}
\end{equation}
where ${\cal N}_1$ is a normalization factor, $(a,b)$ color indices and $(\eta, \:\xi)$ the 
spinorial indices.
To determine ${\cal N}_1$, one multiplies both sides of the expression by 
$\de^{ab} \de_{\eta\xi}$ and rearranges the quark fields, so that:	
\begin{equation*}
  {\cal N}_1 = - \frac{1}{12}\lag 0 \left| :\bar{q}(0) q(0): \right| 0 \rag \equiv - \frac{1}{12} \:\qq[q] ~.
\end{equation*}
Then, the non-perturbative contribution of the quark condensate is given by
\begin{equation}\label{Sqq}
  {\cal S}^{\qq[q]}_{ab}(x) = -\frac{\de_{ab}}{12} \: \qq[q] ~~.
\end{equation}

\begin{figure}[t] \vspace{-0.5cm}
  \begin{center}
  \includegraphics[width=15.5cm]{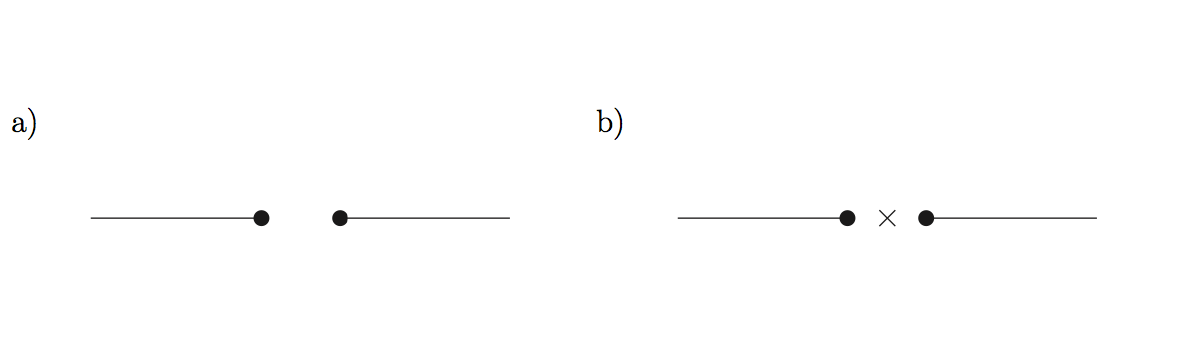}
  \vspace{-1.0cm}
  \caption{{\footnotesize a) quark condensate $\qq[q]$ and b) mass corrections to the quark propagator, 
  where the symbol $\times$ indicates the expansion in the first order in quark mass. 
  The low-momentum transferred through the condensates is represented by the broken lines in 
  the diagrams above.}}
  \label{DiagQQ}
  \end{center}
\end{figure}

The diagram that represents the term (\ref{Sqq}) is shown in Fig.(\ref{DiagQQ}a).
The numerical value of $\qq[q]$ can be estimated, for example, using the 
Partially Conserved Axial Current (PCAC) hypothesis \cite{pcac}:
\begin{equation}
 \qq[q] = - \frac{m_\pi^2 \:f_\pi^2}{2(m_u + m_d)}
\end{equation}
where $m_u$ and $m_d$ are respectively the $u$ and $d$ quark masses, $m_\pi$ the 
pion mass and $f_\pi$ the pion decay constant. Considering the values: 
$m_\pi = 138 \MeV$, $f_\pi = 132 \MeV$ and the usual relation for quark masses: 
$m_u + m_d \simeq 14 \MeV$, one gets
\begin{equation}
  \qq[q] = -(0.228 \GeV)^3  ~.
\end{equation}	
It is interesting to estimate the value for the quark condensate $\qq[s]$, due to SU(3) 
spontaneous symmetry breaking. From PCAC studies, one gets the relation: 
$m_s \qq[s] = - m^2_K \:f^2_K$. Then, using the available information about the $K$ meson, 
like mass and decay constant, one can evaluate the numerical value of $\qq[s]$. Therefore, 
one could introduce the parameter $\kappa$, which gives the quark condensate ratio as 
\begin{equation}
  \kappa = \frac{\qq[s]}{\qq[q]} ~,
\end{equation}
where $q = u, d$. It is still possible to extract more information about the quark condensates, 
considering the VEV of the second term in the expansion (\ref{vacqq1}), then:
\begin{equation}
  \langle 0| : [D^\rho q^\eta_a(0)] \: \bar{q}^\xi_b(0): |0 \rangle = 
  {\cal N}_2 \:\de_{ab} \left(\gamma^\rho \right)^{\eta\xi}
  \label{mqq}
\end{equation}
multiplying both sides by the term $\de_{ab} \left(\gamma_\rho \right)^{\xi\eta}$ and using
the Dirac equation for the quark field $q^\eta_a$: 
$\slashed{D} q^\eta_a = -im_q \:q^\eta_a$, one obtains the normalization factor ${\cal N}_2$:
\begin{eqnarray}
  {\cal N}_2 &=& - \frac{1}{48} \lag \bar{q}_a \slashed{D} q_a \rag ~~ = ~~ 
  \frac{i}{48} m_q\qq[q] ~.
\end{eqnarray}
According to Eq.(\ref{vacqq1}), multiplying Eq.(\ref{mqq}) by $x_\rho$, one obtains
the mass correction to the quark condensate:
\begin{equation}\label{Smqq}
{\cal S}^{m\qq[q]}_{ab}(x) = \frac{i \:\de_{ab} \slashed{x}}{48} \:m_q \qq[q]
\end{equation}
where this term will not contribute in the particular case $m_q \rightarrow 0$.
The diagram associated with this mass correction to the quark condensate is shown in 
Fig.(\ref{DiagQQ}b). Therefore, the quark condensate contribution for the correlation function 
is calculated using the expressions in Eqs.(\ref{Sqq}) and (\ref{Smqq}). 

\subsection{Mixed Condensate $\qGq[q]$}
\begin{figure}[t] 
  \begin{center}
  \includegraphics[width=15.5cm]{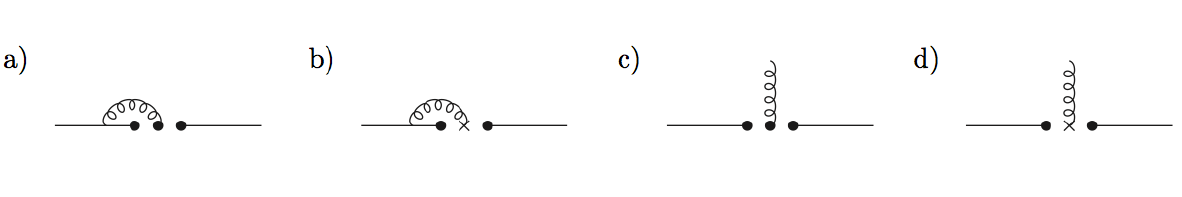}
  \vspace{-0.5cm}
  \caption{{\footnotesize Mixed condensate propagator $\qGq[q]$, considering the first 
  order corrections to the light quark masses $m_q$.}}
  \label{DiagQGQ}
  \end{center}
\end{figure}

Now considering the VEV of the third term of the expansion (\ref{vacqq1}):
\begin{equation}\label{expmix}
  \langle 0| \: :D^\rho D^\lambda q^\eta_a(x) |_{x=0} \:\bar{q}^\xi_b(0): \: |0 \rangle = 
  {\cal N}_3 \:\de_{ab} \de^{\eta\xi} g^{\rho\lambda} ~~,
\end{equation}
multiplying both sides of the equation by the term $\de_{ab} \de^{\eta\xi} g_{\rho\lambda}$, 
one obtains
\begin{equation}
  {\cal N}_3 = - \frac{1}{48} \lag \bar{q}_a D^2 q_a \rag ~.
\end{equation}
To evaluate the term $D^2 q_a$, one must consider the property of the commutator of covariant 
derivatives, $D_\rho$, which satisfies: 
\begin{equation}
  \left[ D_\rho, D_\lambda \right] = -i g_s \:G_{\rho\lambda}
\end{equation}
where $G_{\rho\lambda} \equiv t^N \:G^N_{\rho\lambda}$ is the gluon field tensor. After some 
algebraic manipulations, it is straightforward to prove that
\begin{eqnarray}
  \sigma^{\rho\lambda} \:\left[ D_\rho, D_\lambda \right]\: q_a &=& 2i \left( 
  \slashed{D} \slashed{D} - D^2 \right)q_a = -i g_s \:\sigma^{\rho\lambda} G_{\rho\lambda} \:q_a \nno\\
  \therefore ~~ 
  \left( \slashed{D} \slashed{D} - D^2 \right)q_a&=& -\frac{1}{2} \big( g_s \:\sigma \cdot G \:q_a \big) ~~.
  \label{comut}
\end{eqnarray}
For light quark propagators, only contributions from the terms which are linear 
in the quark mass are considered. Then, the first term of Eq.(\ref{comut}) can be neglected, since 
$\slashed{D} \slashed{D} q_a = -m^2_q \:q_a$. Thus, one obtains
\begin{equation}
  D^2 q = \frac{1}{2} \left( g_s \:\sigma^{\rho\lambda} G_{\rho\lambda} q \right)
  \label{D2q}
\end{equation}
which is an expression valid only for the light quarks. In this case, the normalization factor 
is given by:
\begin{equation}
  {\cal N}_3 = - \frac{1}{96} \lag \bar{q} g_s \sigma \cdot G q \rag 
  \equiv - \frac{1}{96} \qGq[q] ~.
\end{equation}
Remembering that, according to Eq.(\ref{vacqq1}), to calculate the mixed condensate 
contribution one should multiply Eq.(\ref{expmix}) by the factor
$\frac{1}{2} x_\rho x_\lambda$. Therefore, 
\begin{equation}\label{Sqmix}
  {\cal S}^{\qGq[q]}_{ab}(x) = -\frac{x^2 \:\de_{ab}}{192} \:\qGq[q] ~~.
\end{equation}	
This expression is represented in Fig.(\ref{DiagQGQ}a) and is valid only for the light quarks, 
in which the approximation $m_q \rightarrow 0$ can be evaluated. Usually, the mixed condensate 
is given in terms of quark condensate $\qq[q]$ through the relation:
\begin{equation}
  \qGq[q] = m^2_0 \qq[q]
\end{equation}
where $m^2_0 = 0.8 \GeV^2$ \cite{SNB}.

Considering the fourth term of the expansion (\ref{vacqq1}), one obtains another important 
contribution to the mixed condensate:
\vspace{-0.2cm}
\begin{equation}
   \langle 0| :D^\rho D^\lambda D^\mu q^\eta_a(x) |_{x=0} \: \bar{q}^\xi_b (0): |0 \rangle = 
   {\cal N}_4 \:\de_{ab} \:(g^{\rho\lambda} \ga^\mu + g^{\mu\lambda} \ga^\rho + 
   g^{\mu\rho} \ga^\lambda)^{\eta\xi}
   \label{exp4}
\end{equation}		
multiplying both sides by the term $g_{\rho\lambda} \:\de_{ab}(\ga_\mu)^{\xi\eta}$, 
using again the Dirac equation and Eq.(\ref{D2q}), one gets the expression
\vspace{-0.2cm}
\begin{eqnarray}
   {\cal N}_4 &=& -\frac{1}{3^2 \: 2^5} \langle \:\bar{q} \:D^2 \slashed{D} q\: \rangle
   ~~=~~ \frac{i \:m_q}{3^2 \: 2^6} \qGq[q] ~.
\end{eqnarray}
Then, multiplying Eq.(\ref{exp4}) by the factor $\frac{1}{6} x_\rho x_\lambda x_\mu$
one yields the mixed condensate contributions up to first order corrections to the light 
quark masses:
\begin{equation}\label{Sqmmix}
{\cal S}^{m \qGq[q]}_{ab}(x) =  \frac{i \:\de_{ab} \:x^2 \slashed{x}}{1152} \:m_q \qGq[q] ~.
\end{equation}

The next step is to consider the terms in the expansion (\ref{vacqGq1}), which were 
obtained from the VEV: 
${\cal S}^{q\bar{q} G}_{ab, \al\be} = \langle 0| \!: q_a(x) g_s G^N_{\al\be}(0) \bar{q}_b(0) :\! |0 \rangle$. 
In such a case, the most relevant contributions will be proportional to the mixed condensate. 
An extremely important observation to this VEV is concerning the presence of the free Lorentz indices, 
$\al$ and $\be$, which must be contracted - in the literature this kind of contribution is 
usually called as non-Factorizable Contributions. 
As discussed in Chapter \ref{chap:QCDSR}, see Eq.(\ref{vac1}), the contraction of these 
indices is done with terms of the propagator ${\cal S}^{G}_{ab, \be\al}(x)$ that will be 
further calculated. Having in mind the contraction of these indices, an useful compact notation 
to the non-factorizable contributions is given by: ${\cal S}^\ast(x)$.

The first term of the expansion (\ref{vacqGq1}) is given by
\vspace{-0.3cm}
\begin{equation}
  \langle 0| :q^\eta_a(0)\: g_s G^{_N}_{\al\be} \:\bar{q}^\xi_b(0): |0 \rangle = 
  {\cal N}_5 \:( \sigma_{\al\be} )^{\eta\xi} \:t^{_N}_{ab}
\end{equation}
where the factor $( \sigma_{\al\be} )^{\eta\xi}$ has been chosen in order to maintain the 
antisymmetry of both sides of the equation. Contracting with 
$( \sigma^{\al\be} )_{\xi\eta} \:t^{_N}_{ba}$ and rearranging the quark and gluon fields, 
one obtains the normalization factor:
\begin{eqnarray}
  {\cal N}_5 = - \frac{1}{192} \qGq[q] ~.
\end{eqnarray}
Then, the non-factorizable contribution of the mixed condensate, represented in 
Fig.(\ref{DiagQGQ}c), is given by:
\vspace{-0.8cm}
\begin{eqnarray}
  {\cal S}^{\ast \:\qGq[q]}_{ab, \al\be}(x) &=& -\frac{t^{_N}_{ab}
  \:\sigma_{\al\be}} {192} \qGq[q] ~.
  \label{Vqmix}
\end{eqnarray}
Finally, one calculates the second term of the expansion (\ref{vacqGq1}):
\vspace{-0.3cm}
\begin{equation}
  \langle 0| : [D_\rho q^\eta_a(0)] \: g_s G^{_N}_{\al\be} \:\bar{q}^\xi_b(0): |0 \rangle = 
  {\cal N}_6 \:\left( \sigma_{\al\be} \:\ga_\rho + \ga_\rho \:\sigma_{\al\be} \right)^{\eta\xi} 
  \:t^{_N}_{ab} ~.
\end{equation}
Multiplying both sides of this equation by the term
$(\sigma^{\al\be} \:\ga^\rho)_{\xi\eta} \:t^{_N}_{ba}$, one gets the normalization factor:
\begin{equation}
  {\cal N}_6 = \frac{i \:m_q}{768} \qGq[q] ~.
\end{equation}
According to Eq.(\ref{vacqGq1}), one must multiply this contribution by the factor 
$x^\rho$. Then, one obtains the mass corrections to the mixed condensate contribution, 
which are represented in Fig.(\ref{DiagQGQ}d), as follows
\vspace{-0.3cm}
\begin{equation}
  {\cal S}^{\ast \:m\qGq[q]}_{ab, \al\be} (x) = \frac{i \:t^{_N}_{ab}}{768} 
  ( \sigma_{\al\be} \slashed{x} + \slashed{x} \sigma_{\al\be} ) \qGq[q] ~~.
  \label{Vqmmix}
\end{equation}
Therefore, the Eqs. (\ref{Sqq}), (\ref{Smqq}), (\ref{Sqmix}), (\ref{Sqmmix}), (\ref{Vqmix}) 
and (\ref{Vqmmix}) provide the most relevant non-perturbative effects from the 
QCD vacuum to the propagators:
\begin{eqnarray}
  {\cal S}^{q\bar{q}}_{ab}(x) &=& {\cal S}^{\qq[q]}_{ab}(x) + {\cal S}^{m\qq[q]}_{ab}(x) +
  {\cal S}^{\qGq[q]}_{ab}(x) + {\cal S}^{m \qGq[q]}_{ab}(x) + \ldots \nno\\
  &=& - \frac{\de_{ab}}{12} ~\qq[q]  + \frac{i \:\de_{ab} \:\slashed{x}}{48} ~m_q \qq[q] 
  - \frac{\de_{ab} \:x^2}{192}\qGq[q] + \frac{i \:\de_{ab} \:x^2 \slashed{x}}{1152} ~m_q \qGq[q] + 
  \ldots ~~~~~~~~~ \\ 
%
  {\cal S}^{q\bar{q}G}_{ab, \al\be}(x) &=& {\cal S}^{\ast \:\qGq[q]}_{ab, \al\be}(x) + 
  {\cal S}^{\ast \:m\qGq[q]}_{ab, \al\be}(x) + \ldots \nno\\
  &=& - \frac{t^{N}_{ab} \:\sigma_{\al\be}}{192} ~\qGq[q] -\frac{i \:t^N_{ab}}{768}
  ( \sigma_{\alpha\beta} \:\slashed{x} + \slashed{x} \:\sigma_{\alpha\beta} ) ~m_q \qGq[q]  
  + \ldots ~~~~
\end{eqnarray}

\section{Gluon Emission}
The propagator associated with the gluon emission which contributes to the formation of 
$\GG$ and $\qGq[q]$ condensates is given by Eq.(\ref{vacG}). 
Its expression is similar to the vertex of interaction between quarks and gluons in QCD 
perturbative, however the gluon field is evaluated in Fock-Schwinger gauge. 
Thus, one gets:
\begin{eqnarray}
  {\cal S}^{G}_{ab, \al\be}(x) ~=~ \frac{i \:t^N_{cd}}{2} \int \!d^4y ~S^0_{ac}(x-y)
  \:\gamma^\alpha y^\beta \:S^0_{db}(y)
\end{eqnarray}
Writing the perturbative quark propagators $S^0_{ab}(x)$ in momentum space, one obtains:
\begin{eqnarray}
  {\cal S}^{G}_{ab, \al\be}(x) &\!\!\!=\!\!\!& \frac{i \:t^N_{cd}}{2} 
  \iiint \!d^4y \:\frac{d^4p_1}{(2\pi)^4} \:\frac{d^4p_2}{(2\pi)^4} ~
  e^{-i p_1 \cdot(x-y)} \:e^{-i p_2 \cdot(y)}
  \left[ \frac{i \:\de_{ac} (\slashed{p}_1 \!+\! m_q)}{p_1^2 - m_q^2} \right] \gamma^\alpha y^\beta
  \left[ \frac{i \:\de_{db} (\slashed{p}_2 \!+\! m_q)}{p_2^2 - m_q^2} \right] \nno
\end{eqnarray}
For simplicity, throughout the text the infinitesimal terms $i\epsilon$ were omitted from the 
denominators in the propagators. In the following, 
\begin{eqnarray}
  {\cal S}^{G}_{ab, \al\be}(x) &=& -\frac{i \:t^N_{ab}}{2} 
  \iint \!\frac{d^4p_1 \:d^4p_2}{(2\pi)^4} ~e^{-i p_1 \cdot x}
  \frac{\slashed{p}_1 \!+\! m_q}{p_1^2 - m_q^2} \cdot \gamma^\alpha \cdot
  \frac{\slashed{p}_2 \!+\! m_q}{p_2^2 - m_q^2} \int \!\frac{d^4y}{(2\pi)^4} \:
  y^\be \:e^{i y \cdot (p_1 - p_2)} \nno\\
  &&\hspace{-1.5cm} =~~ -\frac{i \:t^N_{ab}}{2} \iint \!\frac{d^4p_1 \:d^4p_2}{(2\pi)^4} 
  ~e^{-i p_1 \cdot x} ~
  \frac{\slashed{p}_1 \!+\! m_q}{p_1^2 - m_q^2} \cdot \gamma^\alpha \cdot
  \frac{\slashed{p}_2 \!+\! m_q}{p_2^2 - m_q^2} 
  \left(i\frac{\partial}{\partial p_{2}}\right)^\be \de^{(4)}(p_1 \!-\! p_2) \nno \\
  &&\hspace{-1.5cm} =~~ -\frac{t^N_{ab}}{2} \iint \!\frac{d^4p_1 \:d^4p_2}{(2\pi)^4} 
  ~e^{-i p_1 \cdot x} ~\de^{(4)}(p_1 \!-\! p_2) \:
  \frac{\slashed{p}_1 \!+\! m_q}{p_1^2 - m_q^2} \cdot \gamma^\alpha \cdot
  \left(\frac{\partial}{\partial p_{2}}\right)^\be
  \frac{\slashed{p}_2 \!+\! m_q}{p_2^2 - m_q^2} \nno\\
  &&\hspace{-1.5cm} =~~ -\frac{t^N_{ab}}{2} \iint \!\frac{d^4p_1 \:d^4p_2}{(2\pi)^4} 
  ~e^{-i p_1 \cdot x} ~\de^{(4)}(p_1 \!-\! p_2) \:
  \frac{\slashed{p}_1 \!+\! m_q}{p_1^2 - m_q^2} \cdot \gamma^\alpha \cdot
  \left[ \frac{\gamma^\be}{p_2^2 - m_q^2} \!-\! 
  \frac{2(\slashed{p}_2 + m_q) p_2^\beta}{(p_2^2 - m_q^2)^2} \right] \nno\\
  &&\hspace{-1.5cm} =~~ -\frac{t^N_{ab}}{2} \int \!\frac{d^4p}{(2\pi)^4} 
  ~\frac{e^{-i p \cdot x}}{(p^2-m_q^2)^2} ~
  (\slashed{p} + m_q) \cdot
  \left[ \gamma^\alpha \gamma^\be - 
  \frac{2 \gamma^\alpha p^\beta (\slashed{p} + m_q)}{p^2 - m_q^2} \right] ~.
\end{eqnarray}
Notice that the propagator ${\cal S}^G_{ab, \al\be}(x)$ must always be associated with 
an external totally antisymmetric gluon field tensor $G^N_{\al\be}$, so that their free Lorentz 
indices, $\alpha$ and $\beta$, could be contracted.
Then, a several number of simplifications can be done, due to the presence of the
$G^{N}_{\al\be}(0)$ and only antisymmetric terms will contribute to the above integral. 
Rearranging the terms of the integral by using the following relations:
\begin{eqnarray}
  \label{sym1}
  (\slashed{p} + m_q) \gamma^\al p^\be (\slashed{p} + m_q) &=&
  -\gamma^\al p^\be (p^2-m_q^2) + 2 p^\al p^\be (\slashed{p} + m_q) \\
  \label{sym2}
  \gamma^\al p^\be &=& \frac{1}{2}\left( \gamma^\al p^\be - \gamma^\be p^\al \right) +
  \frac{1}{2}\left( \gamma^\al p^\be + \gamma^\be p^\al \right)\\
  \label{sym3}
  \gamma^\al p^\be - \gamma^\be p^\al &=& 
  \frac{1}{2} \left( \gamma^\al \gamma^\be \slashed{p} - \slashed{p} \gamma^\al \gamma^\be \right)
\end{eqnarray}
one obtains the result
\begin{eqnarray}
  {\cal S}^{G}_{ab, \al\be}(x) &=& -\frac{t^{N}_{ab}}{2} \int \!\frac{d^4p}{(2\pi)^4} 
  ~\frac{e^{-i p \cdot x}}{(p^2-m_q^2)^2} ~
  \left[ (\slashed{p} + m_q) \gamma^\alpha \gamma^\be + 
  \gamma^\alpha p^\beta - \gamma^\beta p^\alpha \right] \nno \\
  &=& -\frac{t^{N}_{ab}}{4} \int \!\frac{d^4p}{(2\pi)^4} 
  ~\frac{e^{-i p \cdot x}}{(p^2-m_q^2)^2} ~
  \left[ \gamma^\alpha \gamma^\beta (\slashed{p} + m_q) + 
  (\slashed{p} + m_q) \gamma^\alpha \gamma^\be \right] ~~.
\end{eqnarray}
Making proper use of symmetry relations, it is convenient to rewrite the propagator in
terms of $\sigma^{\al\be} = i (\gamma^\al \gamma^\be - g^{\al\be})$. Thus, one gets 
\begin{equation}
  G^{N}_{\al\be}(0) \:\gamma^{\al} \gamma^{\be} ~=~ 
  G^{N}_{\al\be}(0) \: i\sigma^{\al\be} ~.
\end{equation}
Therefore, doing the change $\gamma^\al \gamma^\be \rightarrow i\sigma^{\al\be}$ 
into the propagator ${\cal S}^G_{ab, \al\be}(x)$, one obtains the final result, in the 
coordinate space, as 
\begin{eqnarray}
  {\cal S}^{G}_{ab, \al\be}(x) &=& - \frac{i \:t^{N}_{ab}}{4}
  \int\! \frac{d^4p}{(2\pi)^4} ~e^{-i p \cdot x} ~
  \left[ \frac{\sigma^{\al\be} (\slashed{p} + m_q) + (\slashed{p} + m_q) \sigma^{\al\be}}
  {(p^2-m_q^2)^2} \right]
  \label{VGluonX}
\end{eqnarray}
and, using the Fourier Transform, one can deduce in the momentum space 
\begin{eqnarray}
  {\cal S}^{G}_{ab, \al\be}(p) &=& - \frac{i \:t^{N}_{ab}}{4}
  \left[ \frac{\sigma^{\al\be} (\slashed{p} + m_q) + (\slashed{p} + m_q) \sigma^{\al\be}}
  {(p^2-m_q^2)^2} \right] ~~.
  \label{VGluonP}
\end{eqnarray}

Notice that it is possible to make a further simplification to these propagators, taking the limit 
in which the quark masses are too small. This is an appropriate approximation for the light quarks 
with isospin ($u$ and $d$) and the strange quark ($s$).

\subsection{Limit $m_q \rightarrow 0$}
When the propagators contain only light quarks ($u$, $d$ and $s$), it is convenient to do the
approximation $m_q \rightarrow 0$ to solve the momentum integral in Eq.(\ref{VGluonX}). 
Thus, 
\begin{eqnarray}
  {\cal \tilde{S}}^{G}_{ab, \al\be}(x) ~~\equiv~~ \lim_{m_q \rightarrow 0} \:{\cal S}^{G}_{ab, \al\be}(x)
  &=& - \frac{i \:t^{N}_{ab}}{4}
  \int\! \frac{d^4p}{(2\pi)^4} ~e^{-i p \cdot x} ~
  \left[ \frac{\sigma^{\al\be} \:\slashed{p} + \slashed{p}\: \sigma^{\al\be}}{p^4} \right] ~~.
\end{eqnarray}
This result can be simplified by using the integral 
\begin{eqnarray}
  \int \!d^4p \:\frac{\slashed{p}\:e^{-i p \cdot x}}{p^4} = \frac{2\pi^2 \slashed{x}}{x^2} ~.
\end{eqnarray}
Therefore, it is possible to demonstrate that 
\begin{eqnarray}\label{QLeve}
  {\cal \tilde{S}}^{G}_{ab, \al\be}(x) &=& - \frac{i \:t^{_N}_{ab}}{32 \pi^2 \:x^2}
  \left( \sigma^{\al\be} \:\slashed{x} + \slashed{x}\: \sigma^{\al\be} \right) ~,
\end{eqnarray}
which is the correct expression when the gluon is emitted by light quarks.

\section{Gluon Condensate}
As seen in Chapter \ref{chap:QCDSR}, in the case of the scalar current, the 
non-perturbative contributions due to the gluon condensate are obtained 
from Eq.(\ref{piGG}):
\begin{eqnarray}
  \Pi^{(2)}(x)
  &\!\!=\!\!& \Pi^{(2)}_{NLO}(x) + \Tr \left[ {\cal S}^{\GGi}_{ab}(x) S^0_{ba}(-x) \right] +
  \Tr \left[ S^0_{ab}(x) {\cal S}^{\GGi}_{ba}(-x) \right] \nno\\
  && +~ \frac{1}{2} \langle 0| \!:\! g_s^2 G^N_{\alpha\rho}(0) \:G^M_{\beta\lambda}(0) \!:\! |0 \rangle
  \:\Tr \left[ {\cal S}^{G}_{ab, \al\rho}(x) \:{\cal S}^{G}_{ba, \be\lambda}(-x) \right] ~.
  \label{app:GG}
\end{eqnarray}
Note that the last trace contains the propagators ${\cal S}^{G}_{ab, \al\rho}(x)$, which 
have already been calculated in the previous section. In a perturbation theory, this trace 
would be related to the diagram of the gluon exchange between two quarks. However, 
due to the presence of the VEV, 
the gluon exchange is indeed connected with the formation of the gluon condensate.
An important relation for this VEV of gluon fields is given by:
\begin{eqnarray}
  \langle 0| \!:\! g_s^2 G^N_{\alpha\rho}(0) \:G^M_{\beta\lambda}(0) \!:\! |0 \rangle &=&
  \frac{\de^{NM}}{96}(g_{\alpha\beta}g_{\rho\lambda} - g_{\alpha\lambda}g_{\rho\beta})
  \:\GG ~~.
  \label{relGG}
\end{eqnarray}
With this relation, it is possible to calculate the gluon condensate contributions which are 
formed by gluons emitted by two distinct quarks.

\begin{figure}[t] 
  \begin{center}
  \includegraphics[width=15.5cm]{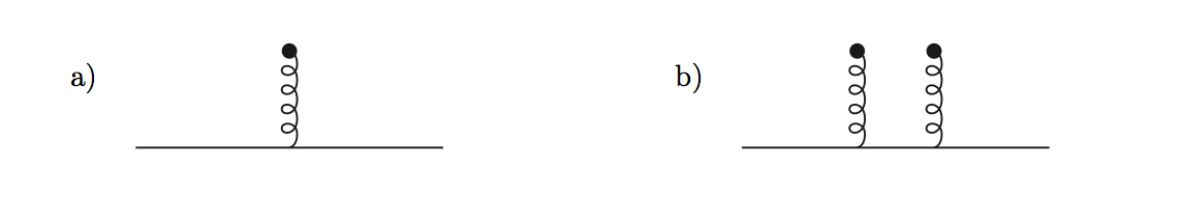}
  \vspace{-0.5cm}
  \caption{{\footnotesize Representation of the propagators {\bf a)} gluon emissions and  
  ${\bf b)}$ gluon condensate $\GG$.}}
  \label{DiagGG}
  \end{center}
\end{figure}

The other contributions, related to the first two traces of Eq.(\ref{app:GG}), contain
the propagator ${\cal S}_{ab}^{\GGi}$ defined as 
\begin{eqnarray}
  {\cal S}^{\GGi}_{ab}(x) &=& \frac{t^{N}_{cd} \:t^{M}_{ef}}{8} 
  \:\langle 0| \!:\! g_s^2 G^N_{\alpha\rho}(0) \:G^M_{\beta\lambda}(0) \!:\! |0 \rangle \nno\\
  && \times~ \iint \!d^4y \:d^4z ~y^\rho \: z^\lambda ~S^0_{ae}(x-z) \:\gamma^\beta \:S^0_{fc}(z-y) 
  	\:\gamma^\alpha \:S^0_{db}(y)  ~~.
\end{eqnarray}
Applying the Fourier transform to the free quark propagators $S^0(x)$, one obtains 
\begin{eqnarray}
  {\cal S}^{\GGi}_{ab}(x) &=& \frac{t^{N}_{cd} \:t^{M}_{ef}}{8} 
  	\langle 0| \!:\! g_s^2 G^N_{\al\rho}(0) G^M_{\be\lambda}(0) \!:\! |0 \rangle
  	\iiint \!\frac{d^4p_1 \:d^4p_2 \:d^4p_3}{(2\pi)^4} ~e^{-i p_1 \cdot x} ~
	S^0_{ae}(p_1) \:\gamma^\be \nno\\
  	&& \times~ S^0_{fc}(p_2) \:\gamma^\alpha \:S^0_{db}(p_3)
	\int \!\frac{d^4y}{(2\pi)^4} \:y^{\rho} ~e^{i (p_2 - p_3) \cdot y}  
	\int \!\frac{d^4z}{(2\pi)^4} \:z^{\lambda} ~e^{i (p_1-p_2) \cdot z} \nno \\
	&=& \frac{t^{N}_{cd} \:t^{M}_{ef}}{8} 
  	\langle 0| \!:\! g_s^2 G^N_{\al\rho}(0) G^M_{\be\lambda}(0) \!:\! |0 \rangle
  	\iiint \!\frac{d^4p_1 \:d^4p_2 \:d^4p_3}{(2\pi)^4} ~e^{-i p_1 \cdot x} ~
	S^0_{ae}(p_1) \:\gamma^\be \nno \\
  	&& \times~ S^0_{fc}(p_2) \:\gamma^\al \:S^0_{db}(p_3)
	\left( i \frac{\partial}{\partial p_3} \right)^\rho \bigg[ \delta^{(4)}(p_3 - p_2) \bigg] 
	\left( i \frac{\partial}{\partial p_2} \right)^\lambda \bigg[ \delta^{(4)}(p_2 - p_1) \bigg] \nno \\
	&=& -\frac{t^{N}_{cd} \:t^{M}_{ef}}{8} 
  	\langle 0| \!:\! g_s^2 G^N_{\al\rho}(0) G^M_{\be\lambda}(0) \!:\! |0 \rangle
  	\iiint \!\frac{d^4p_1 \:d^4p_2 \:d^4p_3}{(2\pi)^4} ~ e^{-i p_1 \cdot x} ~
	S^0_{ae}(p_1) \:\gamma^\be \nno\\
  	&& \times~ \delta^{(4)}(p_3 - p_2) \delta^{(4)}(p_2 - p_1) 
	\left( \frac{\partial}{\partial p_2} \right)^\lambda \left( \frac{\partial}{\partial p_3} \right)^\rho 
	\bigg[ S^0_{fc}(p_2) \:\gamma^\al \:S^0_{db}(p_3) \bigg] ~. ~~
\end{eqnarray}
To solve this equation, one uses the expression for the derivative of the propagator:
\begin{eqnarray}
  \left( \frac{\partial}{\partial p} \right)^\alpha S^0_{ab}(p) &=& 
  \frac{i \:\delta_{ab}}{p^2 - m_{Q}^2} \bigg[ \gamma^\alpha - 
  \frac{2(\slashed{p} + m_{Q})}{p^2 - m_{Q}^2} \:p^\alpha \bigg]
\end{eqnarray}
which allows to solve the integrals by using the Dirac Delta function. Then, 
\begin{eqnarray}
  	&=& -\frac{i \:t^{N}_{cb} \:t^{M}_{ac}}{8} 
  	\langle 0| \!:\! g_s^2 G^N_{\al\rho}(0) G^M_{\be\lambda}(0) \!:\! |0 \rangle
  	\iiint \!\frac{d^4p_1 \:d^4p_2 \:d^4p_3}
	{(2\pi)^4 (p_1^2 - m_Q^2)(p_2^2 - m_Q^2)(p_3^2 - m_Q^2)} ~ e^{-i p_1 \cdot x} \nno\\
  	&& \times~ (\slashed{p}_1 + m_Q) \:\gamma^\be
	\bigg[ \gamma^\lambda - \frac{2(\slashed{p}_2 + m_{Q})}{p_2^2 - m_{Q}^2} 
	\:p_2^\lambda \bigg] \:\gamma^\al \bigg[ \gamma^\rho - \frac{
	2(\slashed{p}_3 + m_{Q})}{p_3^2 - m_{Q}^2} \:p_3^\rho \bigg] 
	\delta^{(4)}(p_3 \!-\! p_2) \:\delta^{(4)}(p_2 \!-\! p_1) \nno\\
%
	%
	&=& -\frac{i \:t^{N}_{cb} \:t^{M}_{ac}}{8} 
  	\langle 0| \!:\! g_s^2 G^N_{\al\rho}(0) G^M_{\be\lambda}(0) \!:\! |0 \rangle
  	\int \!\frac{d^4p} {(2\pi)^4 (p^2 - m_Q^2)^3} ~ e^{-i p \cdot x} 
	(\slashed{p} + m_Q) \:\gamma^\be \nno\\
  	&& \times~ 
	\bigg[ \gamma^\lambda - \frac{2(\slashed{p} + m_{Q})}{p^2 - m_{Q}^2} 
	\:p^\lambda \bigg] \:\gamma^\al \bigg[ \gamma^\rho - \frac{
	2(\slashed{p} + m_{Q})}{p^2 - m_{Q}^2} \:p^\rho \bigg] ~.	
\end{eqnarray}
This result can be simplified by using symmetry properties, like the relations (\ref{sym1}), 
(\ref{sym2}) and (\ref{sym3}) for Dirac matrices, and (\ref{relGG}) for non-perturbative 
gluonic fields. After some algebraic manipulations, one finally obtains:
\begin{equation}
  {\cal S}^{\GGi}_{ab}(x) = \frac{i  \:\de_{ab}}{12} \:m_{Q} \GG
  \int \!\frac{d^4p}{(2\pi)^4} ~ \frac{e^{-i p \cdot x}}{(p^2 - m^2_{Q})^3} \nno\\
  \left[ 1 + \frac{m_{Q} (\slashed{p} + m_{Q})}{(p^2 - m^2_{Q})} \right]  ~.
\end{equation}

Therefore, in the momentum space, the expression for the gluon condensate propagator 
is given by
\begin{equation}\label{SQGG}
  {\cal S}^{\GGi}_{ab}(p) = \frac{i \:\de_{ab}}{12 (p^2 - m^2_{Q})^3}
  \left[ 1 + \frac{m_{Q} (\slashed{p} + m_{Q})}{(p^2 - m^2_{Q})} \right]  \:m_{Q} \GG  ~,
\end{equation}
which obviously does not contribute to the limit: $m_{Q} \rightarrow 0$.

The gluon condensate was first estimated in the analysis of the $\rho$ and $\phi$ leptonic 
decays and from the sum rules of the charmonium spectroscopy. Its numerical value is 
estimated in Refs. \cite{rry,SNB} as 
\begin{equation}
\lag g^2_s G^2 \rag = 0.88 \GeV^4 ~.
\end{equation}

Although it is possible to consider more terms for the full propagator of QCD (\ref{propQCD}), 
one must take into account the relevance of these terms to the OPE since the higher dimension 
condensates become increasingly difficult to be calculated. 
For the vast majority of the sum rules the quark, gluon and mixed condensates represent the 
most important non-perturbative contributions to the OPE. Then, it is very convenient truncate 
the OPE up to dimension-five condensates. However, in some cases the inclusion of higher 
condensate contributions, like the dimension-six condensates: four-quark condensate 
$\qq[q]^2$ and triple gluon condensate $\GGG$ could improve the OPE convergence and 
guarantee a more reliable result from a QCDSR calculation.

\section{Summary of Contributions}
Finally, the terms of the full propagator of QCD calculated in the previous section are 
summarized in Tables (\ref{tabSVZQ}) and (\ref{tabSVZq}). For the heavy quark, in the 
momentum space, the full propagator is given by:

\begin{equation}
  {\cal S}^{{Q}}_{ab}(p) ~=~ S^0_{ab}(p) + S^{GG}_{\al\be}(p) + {\cal S}^{G}_{ab, \al\be}(p) + 
  {\cal S}^{\GGi}_{ab}(p) + {\cal S}^{\GGGi}_{ab}(p) + \ldots
\end{equation}
and for the light quarks, in the coordinate space:

\begin{eqnarray}
  {\cal S}^{q}_{ab}(x) &=& S^0_{ab}(x) + S^{m}_{ab}(x) + {\cal S}^{\qq[q]}_{ab}(x) + 
  	{\cal S}^{m \qq[q]}_{ab}(x) + {\cal S}^{\qGq[q]}_{ab}(x) + {\cal S}^{m \qGq[q]}_{ab}(x) \nno \\
	&& +~ \tilde{\cal S}^{G}_{ab, \al\be}(x) + {\cal S}^{\ast \:\qGq[q]}_{ab, \al\be}(x) + 
		{\cal S}^{\ast \:m \qGq[q]}_{ab, \al\be}(x) + \ldots	
\end{eqnarray}
As well as Feynman rules provide the key pieces to the construction of the diagrams in QED, 
all of these expressions also allow the construction of the most relevant diagrams to the 
QCD sum rules.

\begin{table}[h]
\caption{\small Full propagator of QCD for the heavy quarks.}
\begin{center}
\begin{tabular}{l}
  \hline\hline
  \mbox{Representation}\\
  \hline\hline \\ \\
  \includegraphics[width=\textwidth]{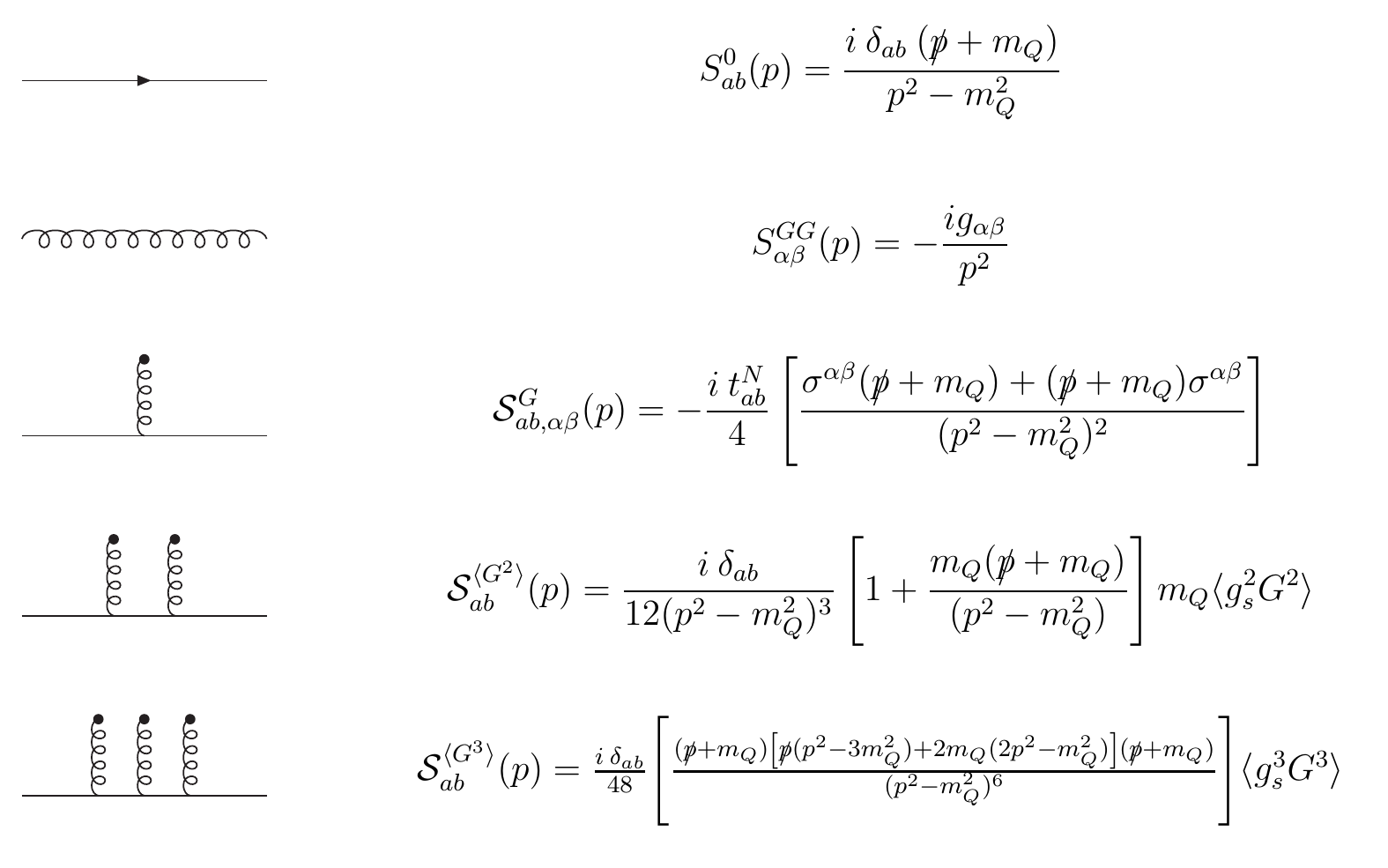}\\
  \\ \\
  \hline\hline
\end{tabular}
\end{center}
\label{tabSVZQ}
\end{table}

\begin{table}[p]
\caption{\small Full propagator of QCD for the light quarks.}
\begin{center}
\begin{tabular}{l}
  \hline\hline
  \mbox{Representation}\\
  \hline\hline \\ \\
  \includegraphics[width=\textwidth]{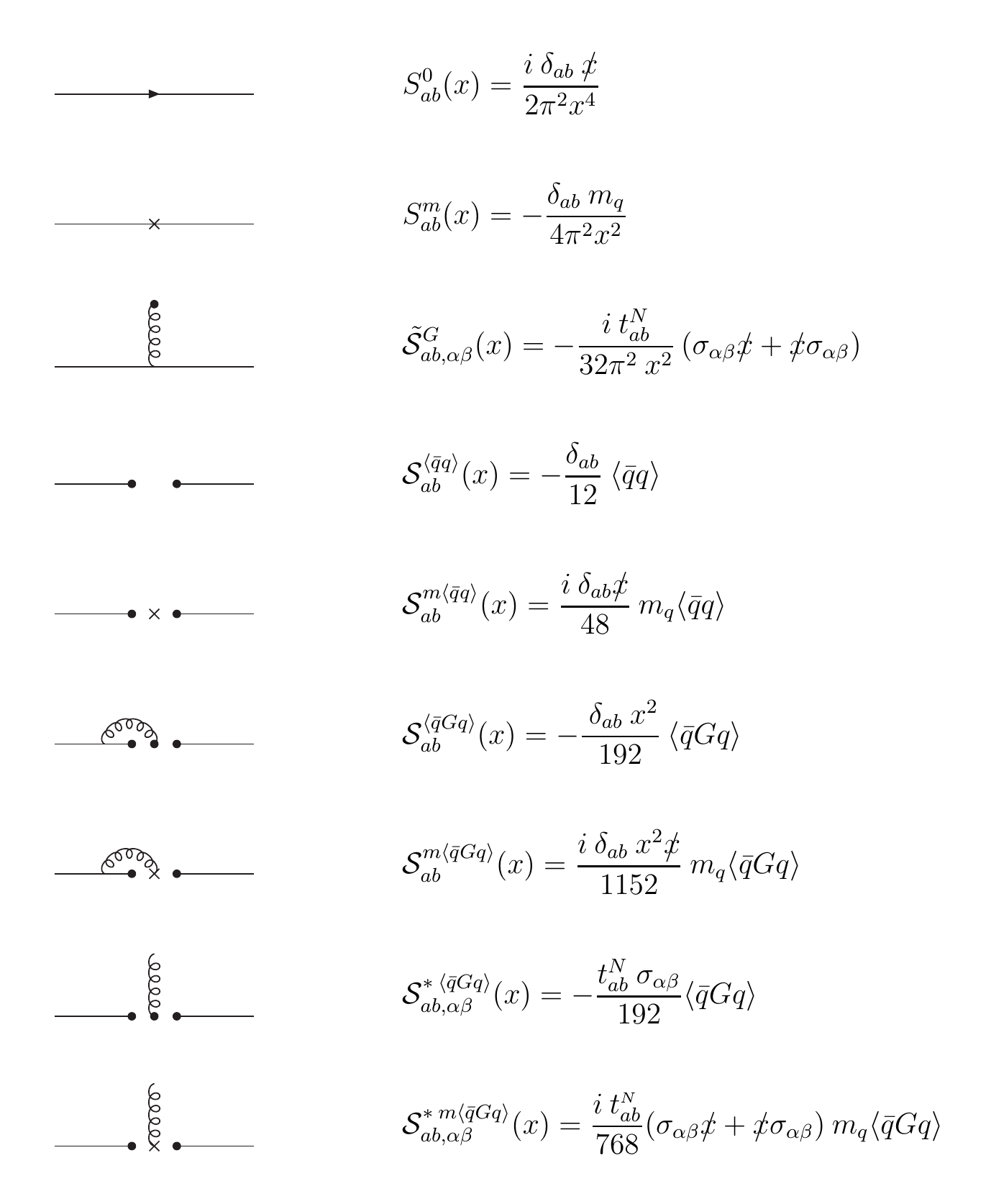}\\
  \\ \\
  \hline\hline
\end{tabular}
\end{center}
\label{tabSVZq}
\end{table}

\cleardoublepage


\addcontentsline{toc}{chapter}{List of Figures}
\listoffigures
\cleardoublepage
\addcontentsline{toc}{chapter}{List of Tables}
\listoftables
\cleardoublepage

\addcontentsline{toc}{chapter}{Bibliography}
\bibliographystyle{plain}

\cleardoublepage

%
\end{document}